\documentclass{aa}

\usepackage{pdflscape}
\usepackage{graphicx}
\usepackage{float}
\usepackage{txfonts}

\usepackage{rotating}
\usepackage{longtable,tabularx,xltabular}
\usepackage{array,multirow}
\usepackage{soul}
\usepackage{xspace}
\usepackage[usenames,dvipsnames]{color}
\usepackage{easyReview}
\usepackage{amssymb,amsmath}
\usepackage[breaklinks=true]{hyperref}

\usepackage{natbib}
\bibpunct{(}{)}{;}{a}{}{,}             

\newcolumntype{C}[1]{>{\centering\arraybackslash}p{#1}}
\let\oldAA\AA
\renewcommand*{\AA}{\,\oldAA\xspace}
\newcommand{\au}{\,AU\xspace}
\newcommand{\mm}{\,mm\xspace}
\newcommand{\mhz}{\,MHz\xspace}
\newcommand{\ghz}{\,GHz\xspace}
\newcommand{\kms}{\,\mbox{km\,s$^{-1}$}\xspace}
\newcommand{\cmcub}{\,\mbox{cm$^{-3}$}\xspace}
\newcommand{\kel}{\,K\xspace}
\newcommand{\mkel}{\,mK\xspace}
\newcommand{\msun}{\,\mbox{M$_\odot$}\xspace}
\newcommand{\msuna}{\,\mbox{M$_\odot$\,a$^{-1}$}\xspace}

\newcommand{\ic}{\mbox{IC 418}\xspace}
\newcommand{\ngc}{\mbox{NGC 7027}\xspace}

\newcommand{\coral}{\mbox{Co$^3$RaL}\xspace}
\renewcommand{\ga}{$\alpha$\xspace}
\newcommand{\gb}{$\beta$\xspace}
\renewcommand{\gg}{$\gamma$\xspace}
\newcommand{\gd}{$\delta$\xspace}
\renewcommand{\ge}{$\epsilon$\xspace}
\newcommand{\gz}{$\zeta$\xspace}
\newcommand{\gh}{$\eta$\xspace}
\newcommand{\gq}{$\theta$\xspace}
\newcommand{\gi}{$\iota$\xspace}
\newcommand{\gk}{$\kappa$\xspace}
\newcommand{\gl}{$\lambda$\xspace}
\newcommand{\gm}{$\mu$\xspace}
\newcommand{\gn}{$\nu$\xspace}
\newcommand{\Dn}{$\Delta n$\xspace}
\newcommand{\Dneq}{$\Delta n =$\xspace}
\newcommand{\Dnleq}{$\Delta n \leqslant$\xspace}
\newcommand{\Dngeq}{$\Delta n \geqslant$\xspace}

\newcommand{\hei}{\mbox{\ion{He}{i}}\xspace}
\newcommand{\trihei}{\mbox{\ion{\element[][3]{He}}{i}}\xspace}
\newcommand{\heii}{\mbox{\ion{He}{ii}}\xspace}
\newcommand{\triheii}{\mbox{\ion{\element[][3]{He}}{ii}}\xspace}
\newcommand{\ci}{\mbox{\ion{C}{i}}\xspace}
\newcommand{\cii}{\mbox{\ion{C}{ii}}\xspace}
\newcommand{\ciii}{\mbox{\ion{C}{iii}}\xspace}
\newcommand{\oi}{\mbox{\ion{O}{i}}\xspace}
\newcommand{\oii}{\mbox{\ion{O}{ii}}\xspace}
\newcommand{\oiii}{\mbox{\ion{O}{iii}}\xspace}

\begin{document}
	
\title{Deep survey of millimeter RRLs towards the planetary nebulae IC 418 and NGC 7027}
	
\author{T. Huertas-Roldán\inst{1}\fnmsep\inst{2} 
	\and J. Alcolea\inst{3}
	\and D. A. García-Hernández\inst{1}\fnmsep\inst{2}
	\and D. Tafoya\inst{4}
	\and J. P. Fonfría\inst{3}\fnmsep\inst{5}
	\and J. J. Díaz-Luis\inst{3}
	\and A. Manchado\inst{6}
	\and V. Bujarrabal\inst{3}
	\and R. Barzaga\inst{1}\fnmsep\inst{2}
	\and M. A. Gómez-Muñoz\inst{7}\fnmsep\inst{8}
}
	
\institute{Instituto de Astrofísica de Canarias (IAC),
	C/ Vía Láctea s/n, E-38205 La Laguna, Spain\\
	\email{thuertas@iac.es}
	\and 
	Departamento de Astrofísica, Universidad de La Laguna (ULL),
	E-38206 La Laguna, Spain
	\and
	Observatorio Astronómico Nacional (OAN, IGN/CNIG),
	C/ Alfonso XII 3, E-28014 Madrid, Spain
	\and
	Department of Space, Earth and Environment, Chalmers University of Technology,
	Onsala Space Observatory, 439 92 Onsala, Sweden
	\and
	Departamento de Astrofísica Molecular, Instituto de Física Fundamental (IFF, CSIC),
	C/ Serrano 123, E-28006 Madrid, Spain
	\and
	Consejo Superior de Investigaciones Científicas (CSIC), Spain
	\and
	Departament de Fïsica Quàntica i Astrofísica (FQA), Universitat de Barcelona (UB),
	C/ Martí i Franqués 1, E-08028 Barcelona, Spain
	\and
	Institut de Ciències del Cosmos (ICCUB), Universitat de Barcelona (UB),
	C/ Martí i Franqués 1, E-08028 Barcelona, Spain.
}
	
\date{Received n, 2025; accepted n, 2025}
	
	
\abstract
{The circumstellar environments of planetary nebulae (PNe) are wonderful chemically rich astrophysical laboratories in which the ionization of atoms and the formation of simple and complex molecules can be studied. The new generation of high-sensitivity receivers opens the possibility to carry out deep observations which are needed to unveil weak atomic and molecular spectra lying in the millimeter (mm) range.
}
{The main goal of this work is to study the emission lines detected in the spectra of the bright C-rich PNe \ic and \ngc and to identify all those emission features associated with radio recombination lines (RRLs) of light elements. We aim to analyze the RRLs detected on each source, and to model the sources and derive their physical parameters. This work allows us to provide the most complete and updated catalog of RRLs in space.
}
{We present the results of our very deep line survey of \ic and \ngc carried out at 2, 3 and 7\mm with the \mbox{IRAM 30m} and the \mbox{Yebes 40m} radio telescopes. We compare these observational data sets with synthetic models produced with the radiation transfer code \coral.
}
{Our observations towards the target PNe reveal the presence of several H and \hei RRLs at mm wavelengths in the spectra of \ic and \ngc and also of \heii RRLs in the spectrum of \ngc. Many of these lines had remained undetected until now due to their weakness and the lack of high-sensitivity observations at these frequencies. The data also confirm the absence of molecular emission towards \ic, above a detection level of $\sim 3$\mkel [$T_\mathrm{mb}$].
}
{These mm observations represent the most extended RRL line survey of two C-rich PNe carried out so far, with most of the lines never reported before. These extremely complete catalogs evidence the importance of high-sensitivity observations and are expected to be very helpful in the line identification process in mm observations, in particular for still unknown or poorly characterized molecular species existing in the vicinity of ionized environments.
}
	
\keywords{planetary nebulae: individual: IC 418, NGC 7027 -- radio lines: stars -- catalogs}
	
\maketitle
	
\section{Introduction} \label{sec:intro}
	
The planetary nebula (PN) phase is the last stage in the evolution of the circumstellar envelope (CSE) around a star with initial mass of $\sim 1 - 8$\msun. During the preceding asymptotic giant branch \citep[AGB; see e.g.,][for a review]{Herwig2005}, the star experiences a series of thermal pulses (or periodic He-shell flashes), which favor the nucleosynthesis of both light and heavy neutron-rich chemical elements \citep[e.g.,][]{Karakas2014}. Such newly synthesized chemical elements are mixed from the H-burning shell to the stellar surface via several third dredge-up events (TDUs). During each TDU, the C from the core is driven to the stellar surface, increasing its C/O ratio.

Throughout the AGB phase, stars also experience strong mass-loss (up to $\sim$ $10^{-4}$\msuna) produced by a stellar wind originated in the envelope. The mass-loss process originates a CSE around the central star that expands and efficiently enriches the interstellar medium (ISM) with atomic/molecular gas and dust grains \citep{Tielens2005a, Tielens2005b}. When this process abruptly stops, the star increases its effective temperature ($T_\mathrm{eff}$), entering into a short-lived ($\sim 10^{2} - 10^{4}\,$a) post-AGB stage \citep[see e.g.,][]{GarciaLario2006} towards the formation of a white dwarf. A PN emerges when the central star is hot enough ($T_\mathrm{eff} \geq 30,000$\kel) to photoionize the surrounding circumstellar material expelled during the preceding AGB phase.

The low-mass ($\sim 1.5 - 3.5$\msun) stars are converted to C-rich ($\mathrm{C/O} > 1$ at the stellar surface) by the end of the AGB phase, forming a C-rich CSE that provides a fantastic laboratory to detect C-based species. The detection of organic species in these environments provides invaluable insights to understand the complexity of the organic chemistry in space \citep[see e.g.,][and references therein]{Kwok2016, Pardo2022, Cabezas2023, Cernicharo2023, Pardo2023, Gupta2024}. Indeed, an important number of organic molecules have been detected towards the nearby and bright (prototypical) C-rich AGB star \mbox{IRC +10216} via their rotational lines in the radio domain \citep[see e.g.,][and references therein]{Agundez2014, Cernicharo2017, Cernicharo2019, McGuire2022, Tuo2024, Cernicharo2024a}.

In the subsequent PN phase, the hot central star emits very energetic UV photons, creating an environment that favors chemical reactions: simple and complex molecules coexist with radicals, ions, and atoms under extreme and rapidly changing conditions. However, even though the current list of molecules detected in PNe is significant \citep[see e.g.,][]{Bublitz2019, Schmidt2022}, the exact chemical pathways necessary to produce each species remain unclear. In particular, the largest organic molecules detected to date, such as C$_{60}$ and C$_{70}$ fullerenes \citep{Cami2010, GarciaHernadez2010}, are formed in (proto-)PNe through processes that are still under debate \citep[see e.g.,][and references therein]{GomezMunoz2024}.

In order to drive the field forward, more molecular species (i.e., the key molecular by-products or precursors) need to be detected in the radio domain. Such key possible molecular species may have remained undetected because of their weak rotational lines and the lack of accurate laboratory and theoretical spectroscopic information. The weakness of a rotational line may be due to several factors such as the low molecular abundance or low dipolar moments, but the size of the molecule is determinating. The larger the molecule the higher the number of rotational levels. This also implies a large rotational partition function even at low temperatures, and, consequently, low populated rotational levels. Therefore, the rotational lines associated with large C-based molecules are intrinsically weak. In this context, deep radio observations, which use  very large antennas and long exposure times, are essential to detect the weak rotational lines of the possible key molecular species. The significant noise reduction in deep radio observations may thus reveal new weak features, from both molecules and atoms, which have remained undetected to date.

In the ionized PN, many radio emission lines associated to atoms and molecules can be observed. Their radio spectra can also display emission due to new molecules in space that may remain mislaid among known molecular lines and atomic lines. This is particularly true for radio recombination lines (RRLs)\footnote{The RRLs are emission lines in the radio regime produced when an electron drops from a higher energy level of an atom or ion to a lower energy level.}, and therefore, the detection and identification of the numerous atomic RRLs is a crucial task to identify the spectra of new molecular species. The RRLs mainly originate in the ionized hydrogen rich region around the central star of the PN. However, it is to be noted here that \ci RRLs predominantly originate from the neutral hydrogen regions surrounding the PN. RRLs are not affected by dust opacity, so they turn out to be a precise tool to determine the electron temperature $T_\mathrm{e}$ and density $n_\mathrm{e}$, but also to analyze the relative abundance of He to H or to determine the kinematic distance to the \ion{H}{ii} region \citep[see][and references therein]{Mezger1968}. The latter has been a very useful technique during the first stages of the Radio Astronomy in order to study the large-scale distribution of \ion{H}{ii} regions in the Galaxy \citep[see e.g.,][]{Mezger1967, Dieter1967}. Another application/consequence of the study of RRLs is the development of precise numerical methods to derive and calculate the Einstein coefficients characterizing the absorption and emission processes on each observed line \citep[see e.g.,][and references therein for a complete overview of this complex numerical problem]{Kardashev1959a, Brocklehurst1970, Brocklehurst1971, Hummer1987, Hummer1992, Storey1988, Storey1995}.

The RRLs have been detected towards diverse astrophysical objects like \ion{H}{ii} regions and infrared (IR) dark clouds, among others, but PNe, emitting photons energetic enough to ionize the light elements H, He, C, and O, provide an ideal astrophysical environment to produce them. Previous studies have reported the detection of several molecular lines and RRLs in very bright PNe like those addressed in this work, \ic and \ngc \citep[see e.g.,][]{Goldberg1970, Bignell1974, Churchwell1976, Chaisson1976, Walmsley1981, Garay1989, Roelfsema1991, Zhang2008}. These works aimed to characterize the chemical environment of those objects with the radio astronomy facilities available; i.e., to detect molecular lines or to analyze the spatial distribution of the molecular and atomic emission across the nebula. These works used RRLs towards those sources to determine some physical parameters. However, the sensitivities achieved in those works ($\geq 8$\mkel [$T_\mathrm{a}^*$]) are low compared with the sensitivities obtained nowadays. Current receivers can provide higher sensitivities in the objects of interest (i.e., \ic and \ngc) and detect atomic and molecular lines that went unnoticed. Indeed, no RRLs weaker than Hn\gd and \hei n\gg \footnote{Greek letters correspond to the following electronic level gaps: $\alpha \rightarrow \Delta n=1$, $\beta \rightarrow \Delta n=2$, $\gamma \rightarrow \Delta n=3$, $\delta \rightarrow \Delta n=4$, $\epsilon \rightarrow \Delta n=5$, $\zeta \rightarrow \Delta n=6$, $\eta \rightarrow \Delta n=7$, $\theta \rightarrow \Delta n=8$, $\iota \rightarrow \Delta n=9$, $\kappa \rightarrow \Delta n=10$, $\lambda \rightarrow \Delta n=11$, $\mu \rightarrow \Delta n=12$, $\nu \rightarrow \Delta n=13$.} towards our sample of sources \citep[see e.g.,][]{Zhang2008, GuzmanRamirez2016} were observed because of the previous limited sensitivity.

In this paper, we present high-sensitivity ($\sim 1\,$mK [$T_\mathrm{a}^*$]) single-dish observations of H and He RRLs at 2, 3 and 7 mm towards the bright PNe \ic and \ngc. These two PNe are proto-typical sources displaying different properties (e.g., presence of C$_{60}$), geometries, and evolutionary stages. By comparing them, we may obtain key insights on possible chemical pathways as well as their atomic/molecular content and its influence, along with geometry, on the evolution of the PN. In Section \ref{sec:observations} we present the radio observations and the corresponding high-sensitivity data obtained. The procedure for the identification and modeling of the RRLs is presented in Section \ref{sec:identification_and_modeling}. The RRLs catalogs for each PN are presented in Sections \ref{sec:catalog_IC418} and \ref{sec:catalog_NGC7027}, together with the identification and modeling of the detected RRLs. Present and future possible applications and utilities of the RRLs catalogs are briefly discussed in Section \ref{sec:applications}. Finally, the conclusions of our work are summarized in Section \ref{sec:conclusions}.

\section{Observations} \label{sec:observations}

The spectra of both PNe have been obtained with the \mbox{IRAM 30m} and \mbox{Yebes 40m} radio telescopes. The 30m telescope is located at 2850$\,$m on the Pico Veleta in Sierra Nevada (Granada, Spain) and it is operated by the Institut de Radioastronomie Millimétrique\footnote{IRAM is an international research institute. Founded in 1979, it is participated in by the French {\it Centre National de la Recherche Scientifique} (CNRS), the German {\it Max-Planck-Gesellschaft} (MG), and the Spanish {\it Instituto Geográfico Nacional} (IGN) through the {\it Observatorio Astronómico Nacional} (OAN). IRAM operates the single-dish 30m telescope in Spain and the NOEMA interferometer in the French Alps.}. The observations were carried out during the 2021/2022 winter semester (proposal ID 158-21) using the Eight MIxer super-heterodyne Receiver  \citep[EMIR\footnote{\href{https://publicwiki.iram.es/EmirforAstronomers}{https://publicwiki.iram.es/EmirforAstronomers}},][]{Carter2012}. This receiver simultaneously provides an observing bandwidth of up to 32\ghz. We performed the observations in the dual-band (E090 and E130 receivers at 3\mm and 2\mm, respectively) single-sideband (LSB) dual-polarization mode (H+V). This setup allowed us to simultaneously cover two 8\ghz wide frequency ranges (81 to 89\ghz with the E090 and 131 to 139\ghz with the E130) in both polarizations.

The spectrometer used in this project was the fast Fourier transform (FFT) to sample the whole 32\ghz observed range (that is, 2 receivers $\times$ 2 polarizations $\times$ 8\ghz). The FFT was used in the 200$\,$kHz mode, which is equivalent to a resolution in velocity units of $\sim 0.43$\kms and $\sim 0.67$\kms at 2 and 3\mm, respectively. We improved the SNR of the 3\mm spectrum by smoothing to a final velocity resolution of $\sim 1.3$\kms. We calibrated the observations following the standard procedure at the 30m telescope. We first observed hot and cold loads (at room and liquid nitrogen temperature) to scale the spectrometer units into temperature units. Then, we observed the blank sky every 20$\,$min in the observing session to measure the transmission properties of the local atmosphere. The backend provides the final data in (atmosphere-corrected) antenna temperature ($T_\mathrm{a}^*$). The absolute calibration accuracy is \mbox{$\sim 10\%$} at 2 and 3\mm. After finishing the observations, we compared the continuum emission of every session. The measurements are stable through all the days, so the relative calibration between the two mm bands is good.

We observed the sources using the wobbler switching mode (WSW), which consists on nodding the sub-reflector between ON-source and OFF-source positions with a nodding frequency of $0.5\,$Hz. The OFF-position has to be larger than the beam size at each frequency band, so we used a separation of $\pm 60\,$arcsec in azimuth with respect to each target. The WSW mode guarantees an accurate removal of instrumental and atmospheric signal, providing very flat baselines. However, the strong continuum emission of our sources left a rippled baseline after the OFF-source position subtraction (see more information in Section \ref{sec:data_red}). Despite the good pointing of the telescope, with typical errors of about $2-4\,$arcsec, we observed strong continuum sources located very close to the targets in the sky to ensure that the pointing of the antenna was correct. This checking procedure was done every two hours or when a new target is selected. We used one of our sources, \ngc, as flux calibrator due to its high brightness. The total time spent for these observations were $42\,$ and $23\,$hours for \ic and \ngc, respectively.

We decided to change the frequency of the local oscillator (LO) during the observations by 100\mhz. This strategy allows to ensure that the detected signals come from the sky and are not due to local origin (parasites or interferences). Any signal from the source will appear in the spectrum at the same frequency position, independently of the frequency of the LO. However, signals produced elsewhere appear displaced depending on the frequency of the LO. This is an optimal procedure to locate radio frequency interferences (RFI) or signals with an undesired origin. We observed half of the total project time with each LO frequency.

The \mbox{Yebes 40m} antenna (RT40m)\footnote{\href{https://rt40m.oan.es/rt40m_en.php}{https://rt40m.oan.es/rt40m\_en.php}} is located at $980\,$m in the Centro Astronómico de Yebes (Guadalajara, Spain). This facility is operated by the Spanish {\it Instituto Geográfico Nacional} (IGN, Ministerio de Transportes y Movilidad Sostenible - MTMS). The observations took place during the 2021/2022 winter semester (proposal ID 22A011) using the Q-band receiver. This receiver allows to simultaneously observe a bandwidth of 18.5\ghz with a spectral coverage of $31.5 - 50$\ghz distributed in eight sub-bands. We carried out the observations in the dual polarization mode (H+V).

The backends used were the FFTs, that provide a resolution of 38$\,$kHz for a bandwidth of 2.5\ghz. To cover the whole Q-band, we used eight sub-bands, whose resolution is equivalent to \mbox{$\sim 0.35 - 0.26$\kms} from the first to the last sub-band. We smoothed the spectra of each sub-band to a final frequency resolution of \mbox{$\sim 300\,$kHz} \mbox{($\sim 2.8 - 1.8$\kms)} in order to improve the SNR. The standard calibration procedure at the RT40m is similar to the procedure in the 30m and the absolute calibration accuracy is $10\%-15\%$. We observed hot and cold loads to scale to temperature units. We checked the goodness of this absolute calibration by observing the continuum emission of one of the science targets, \ngc, and ensuring that it was roughly constant over the observing sessions.

The observations were performed using the position switching mode (PSW), which consists on moving the telescope between the ON-source and OFF-source positions. We selected an OFF-position with an offset of 400$\,$arcsec in horizontal direction away from the target. The PSW mode provides quite flat baselines after subtracting the OFF-position integrations. However, the final spectra show a rippled structure as in the 30m (see more information in Section \ref{sec:data_red}). Pointing and focus are good on the RT40m, but were checked every 1.5$\,$h with \mbox{Orion IRC2} to make corrections if necessary. The PNe \ic and \ngc were observed during 69 and 11$\,$hours, respectively. We used the same strategy to identify RFI as in the 30m telescope.

\subsection{Data reduction} \label{sec:data_red}

\begin{figure*}[ht]
	\centering
	\includegraphics[width=0.45\linewidth]{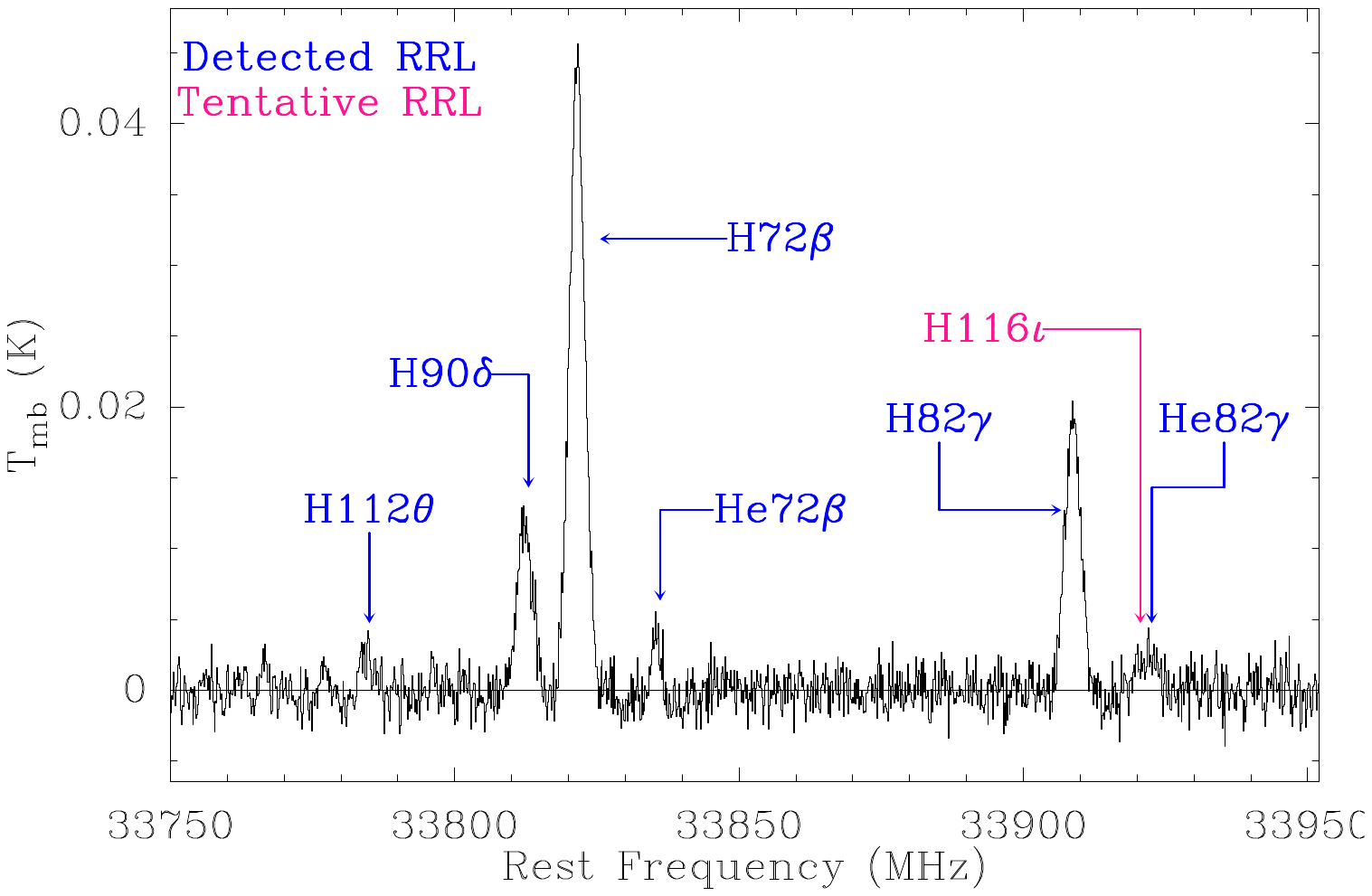}
	\hspace{-10mm}
	\includegraphics[width=0.45\linewidth]{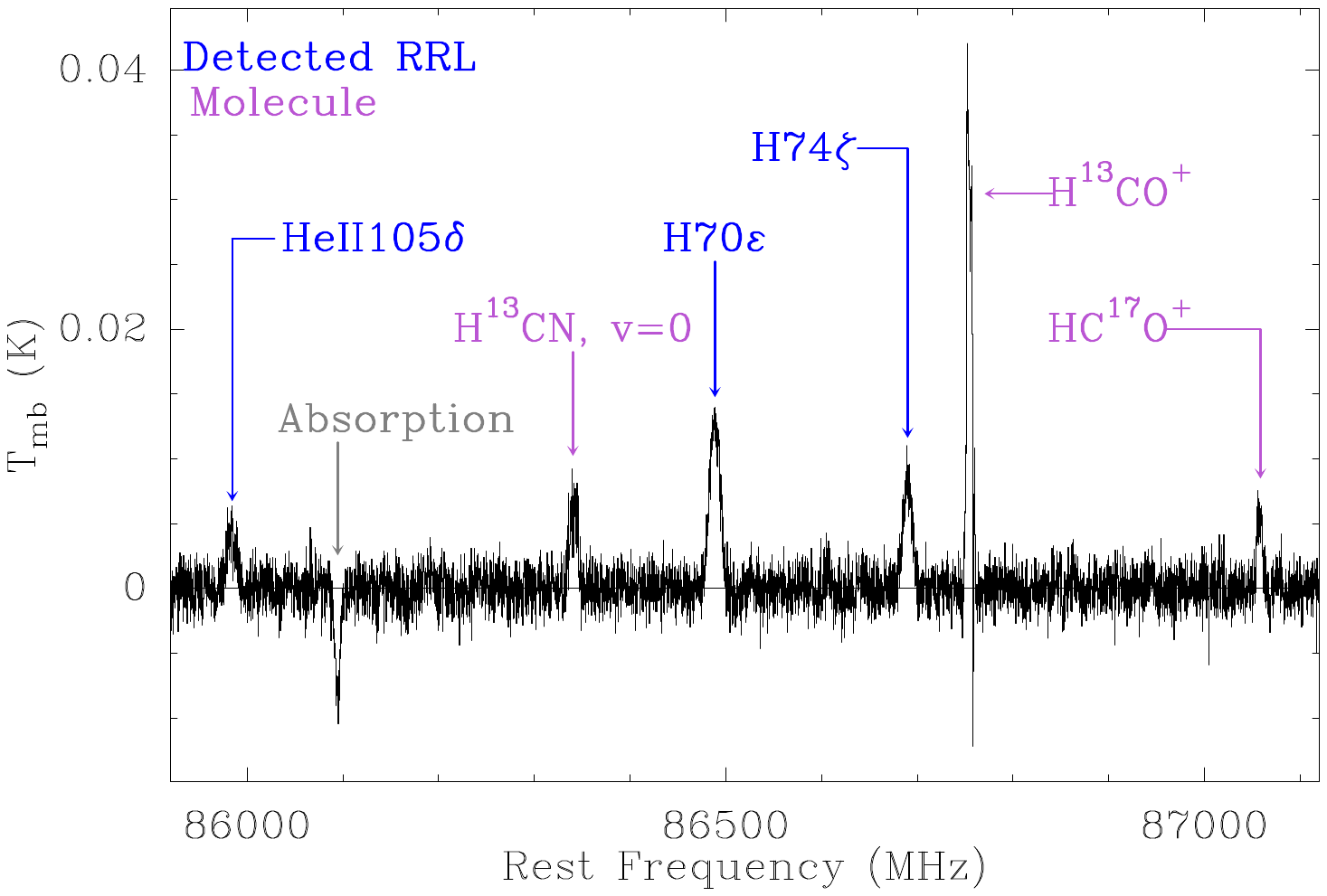}
	\vspace{-5mm}
	\caption{Examples of the RT40m spectrum of \ic (left) and the \mbox{IRAM 30m} spectrum of \ngc (right). The gray arrow on the right panel indicates an absorption feature produced by some molecule in the line of sight of \ngc (probably from a cold shell of its molecular envelope).}
	\label{fig:spectra_examples}
\end{figure*}

The data were reduced using the Continuum and Line Analysis Single-dish Software (CLASS) program of the GILDAS \footnote{GILDAS is a collection of state-of-the-art software oriented toward (sub-)millimeter radio-astronomical applications (either
single-dish or interferometric). The GILDAS developments are supported by IPAG ({\it Observatoire de Grenoble}) and IRAM, with additional contributions from LAB ({\it Observatoire de Bordeaux}) and LERMA ({\it Observatoire de Paris}). For more information, see \href{https://www.iram.fr/IRAMFR/GILDAS}{https://www.iram.fr/IRAMFR/GILDAS} (accessed on 24 September 2024).} astronomical software package (sept22a version). We first inspected the data to flag bad scans and channels and average the valid data. We checked the presence of RFI by comparing the average of each polarization for each observing session at each LO frequency, the average of both polarizations for each observing session at each LO frequency, the total average of each polarization at each LO frequency, and the average of both polarizations at each LO frequency.

The spectra of both sources measured with each telescope were rescaled to main-beam temperature ($T_\mathrm{mb}$) using
\begin{equation} \label{eq:Tmb_definition}
	T_\mathrm{mb} = T_\mathrm{a}^* \frac{\eta_\mathrm{f}}{\eta_\mathrm{mb}}\,,
\end{equation}
where $\eta_\mathrm{f}$ and $\eta_\mathrm{mb}$ are the forward and main beam efficiencies, respectively. We followed the procedure detailed for the 30m\footnote{\href{https://publicwiki.iram.es/CalibrationPapers?action=AttachFile&do=view&target=eb2013-v8.2.pdf}{C. Kramer, J. Peñalver, and A. Greve (2013): Improvement of the \mbox{IRAM 30m} telescope beam pattern.}} and the RT40m. We also provide the efficiencies used for each telescope at each frequency and polarization on Table \ref{tab:efficiencies}.

As we mentioned above, \ic and \ngc display a very strong continuum emission that produced a rippled final spectra, even though the WSW and the PSW usually provide very flat baselines. The width of these ripples is significantly larger than the width of the emission lines, so we could carefully remove it without modifying the shapes and intensities of real weak emission lines. The process consisted on slicing the spectra in small intervals ($\sim$\,200\,\mhz) and subtracting their baselines using low-degree polynomials (from zero to fourth order). Later, we stitched the flattened spectral segments to obtain a final flat and ripple-corrected spectrum for each source and frequency range. Examples of these final spectra are shown in Fig.\ref{fig:spectra_examples}.

\begin{table}
	\caption{Forward and beam efficiencies of the RT40m and \mbox{IRAM 30m} in 2022.}
	\label{tab:efficiencies}
	\centering
	\begin{tabular}{r | c | c c c | c}
		\hline \hline
		$\nu$ (GHz) & \multicolumn{2}{c}{Pol.} & $\eta_\mathrm{f}$ & \multicolumn{2}{c}{$\eta_\mathrm{mb}$} \\
		\hline
		32.4 & H & V & 0.97 & 0.66 & 0.65 \\
		34.6 & H & V & 0.97 & 0.62 & 0.62 \\
		36.9 & H & V & 0.97 & 0.62 & 0.60 \\
		39.2 & H & V & 0.97 & 0.59 & 0.58 \\
		41.5 & H & V & 0.97 & 0.58 & 0.56 \\
		43.8 & H & V & 0.97 & 0.56 & 0.54 \\
		46.1 & H & V & 0.97 & 0.54 & 0.51 \\
		48.4 & H & V & 0.97 & 0.51 & 0.49 \\
		86.0 & \multicolumn{2}{c}{H+V} & 0.95 & \multicolumn{2}{c}{0.81} \\
		115.0 & \multicolumn{2}{c}{H+V} & 0.94 & \multicolumn{2}{c}{0.78} \\
		145.0 & \multicolumn{2}{c}{H+V} & 0.93 & \multicolumn{2}{c}{0.73} \\
	\end{tabular}
	\tablefoot{The RT40m has different main beam efficiencies for each polarization on each sub-band. Main beam efficiencies at \mbox{IRAM 30m} are equal for both polarizations, so only one value is indicated.}
\end{table}

\section{Identification, fitting, and modeling of RRLs} \label{sec:identification_and_modeling}

Preliminary identification of the RRLs detected in our sample PNe was made by calculating the theoretical frequencies and comparing them with the frequencies of the emission lines in the spectra. We followed the approximation described in \cite{BaezRubio2015}. The frequency of each RRL transition can be calculated assuming the Rydberg approximation of the atom. For hydrogen-like atoms, each electronic transition occurs at a frequency
\begin{equation}\label{eq:Rydberg_freq_Hlike-atom}
	\nu_{m,n} = \frac{Z^2}{m_e}c\,R_{\infty}\,\mu(Z)\left(\frac{1}{n^2} - \frac{1}{m^2}\right)\,,
\end{equation}
where $Z$ is the atomic number, $c$ the speed of light, $R_{\infty}$ the Rydberg constant for an atom with infinite mass of the nucleus, $n$ and $m$ the quantum numbers associated to the upper and lower energy level, respectively, of the transition, and $\mu(Z)$ is the reduced mass of the system
\begin{equation}\label{eq:red_mass_Hlike-atom}
	\mu(Z) = \frac{(M(Z) - m_e\,Z)m_e}{M(Z) - (Z - 1)m_e}\,,
\end{equation}
were $M(Z)$ is the atomic mass and $m_e$ the electron mass. For hydrogen-like atoms, the energy of a given level $n$ in the Rydberg approximation corresponds to a quantum energy degeneracy of $g_n = 2\,n^2$ levels. However, in the case of highly excited electrons (i.e., $n\gtrsim 50$), this external electron experiences the nucleus as a point charge shielded by the remaining electrons, which are much closer to the nucleus. Therefore, we can use Eq. (\ref{eq:Rydberg_freq_Hlike-atom}) to predict the frequencies of the H, \hei, \trihei, \ci, \oi RRLs at mm frequencies.

For ionized atoms, the frequency of the transition may be calculated as:
\begin{equation}\label{eq:Rydberg_freq_ions}
	\nu_{(Z_2,i_2)_{m,n}} = \nu_{(Z_1,i_1)_{m,n}} \left(\frac{i_2}{i_1}\right)^2 \left(\frac{\mu(Z_2,i_2)}{\mu(Z_1,i_1)} - 1\right)\,,
\end{equation}
where $i_j$ is the number of electrons of the j-th ion, $Z_j$ is its atomic number, and $\mu(Z_j,i_j)$ is the reduced mass for $j = 1,2$
\begin{equation}
		\mu(Z_j, i_j) = \frac{M(Z_j) - m_e\,(i_j+1)}{M(Z_j) - i_j\,m_e}\,.
\end{equation}

After calculating all the possible RRL transition frequencies for the neutral atoms H, \trihei, \hei, \ci, \oi, and the ions \triheii, \heii, \cii, \ciii, \oii, and \oiii in the observed frequency ranges, we compared them with the frequencies of the emission lines detected (those above the $3\,\sigma$ limit) in our spectra. The most intense radio emission lines were identified without doubts and we could easily fit them in most of the cases with Gaussian profiles using the fitting procedure in CLASS.

Partially blended lines were fitted together using independent Gaussian profiles to obtain the individual fitting parameters of each line. Those strongly blended lines were fitted following a two-step iterative process. First, we fixed the distance between the two lines leaving free all the parameters of the first one. We varied the initial parameters until we obtained a final set of parameters comparable to those of the lines of the same atom and level difference. The second step consisted on fixing the parameters of the first line, using those we already obtained, and leaving the second line parameters free. We again varied the initial parameters until the final Gaussian parameters of the second line were comparable to those of the lines of the same atom and level difference.

No signs of  \trihei, and \triheii, \cii, \ciii, \oii, and \oiii are present in the spectra at any frequency band. In the ISM, RRLs from different species are well resolved because they are intrinsically narrow and, e.g., the C and O lines are clearly detected at frequencies slightly higher than the He lines for a given transition. However, in PNe the RRLs are wider, making impossible to clearly separate the several species. In consequence, the C and O RRL emission in PNe could contribute to the He emission as an asymmetry to the Gaussian profile of the He line. Nevertheless, we did not observed such asymmetric profile in any of the He lines detected.

The emission line fitting with Gaussian profiles showed the fact that some lines are not single but multi-component. This was a possibility to check for any fit with a resulting FWHM $\gtrsim 30$\kms. Moreover, the weakest hydrogen lines (\Dngeq 9) and those of helium (\Dngeq 3) are extremely weak and their detection can not be confirmed without ambiguity by just checking the frequency coincidence of the emission with those in our RRL catalog. Therefore, a more careful and precise analysis was needed.

The Code for Computing Continuum and Radio-recombination Lines \citep[\coral,][]{SanchezContreras2024a} is a C-based program designed to model the free-free continuum and RRL emissions from a 3D asymmetric ionized nebula. It operates on user-defined geometrical shapes specified analytically and relies on density, temperature, and velocity fields provided by analytical functions. The outputs are ASCII tables containing data needed to generate continuum images, spectral energy distribution (SED) plots, spectral cubes, and spectral line profiles.

\coral computes the radiative transfer solution for a non-homogeneous source by subdividing it into small cells where temperature and density are considered constant. The program iteratively solves the radiation transfer problem for each cell along rays, using the solution from the previous cell as input for the next one, resulting in the intensity for each line-of-sight, $I_\nu(x,y)$. This procedure can be applied to multiple velocity channels, producing spectral cubes that are convolved with a synthetic Gaussian beam to model the response of a given radio telescope and create the spectral line profiles of the RRLs.

\coral calculates the emission of RRLs under both LTE and non-LTE conditions. For non-LTE conditions, the calculation relies on departure coefficients $b_n$, which relate the true population of a level, $N_n$, to the LTE population $N_n^{\mathrm{*}}$, using the relationship $N_n = b_n N_n^{\mathrm{*}}$. The departure coefficients $b_n$ are defined for temperatures and densities ranging from 500 to 30\,000 K and $10^{2}$ to $10^{14}\,\mathrm{cm}^{-3}$, for quantum numbers from $\mathsf{n_{u}} = 8$ to $\mathsf{n_{u}} = 100$, and for arbitrary principal quantum number change ($\Delta \mathsf {n_{u}}$). Outside these ranges, \coral provides solutions in LTE. The code can compute the emission of RRLs for various atomic species, including hydrogen, helium, ionized helium, and other hydrogen-like or helium-like elements such as oxygen and carbon, along with their respective ionized versions. The abundances used to model the observational data are listed in Table \ref{tab:coral_param}.

\subsection{Modeling of IC 418 and NGC 7027} \label{sec:model_IC418_NGC7027}

\begin{table*}[ht!]
	\caption{Summary of initial parameters and outputs from \coral.}
	\label{tab:coral_param}
	\centering
	\begin{tabular}{ l l c c }
		\hline\hline
		\multicolumn{2}{c}{\bf Parameters} & \textbf{\ic} & \textbf{\ngc}  \\
		\hline
		\multicolumn{4}{c}{\textit{Input parameters}} \\
		Distance (pc) & $d$ & 1\,200 $^\mathrm{(a)}$ & 860 $^\mathrm{(b)}$ \\
		Inclination angle & $i$& $-20^\circ$ $^\mathrm{(c)}$ & $-55^\circ$ $^\mathrm{(d)}$ \\
		{\it Ellipsoid semi-axis} & A & 7\,500 & 4\,000 \\
		                               & B & 9\,750 & 6\,000 \\
		                               & C & 7\,500 & 4\,000 \\
		{\it Radial structure} (AU) & $R_{\rm in}$ & 10 & 10 \\
		                                              & $R_0$ & 500 & 300 \\
		                                              & $R_1$ & 5\,800 & 2\,800 \\
		                                              & $R_2$ & 6\,500 & 3\,400 \\
		                                              & $R_{\rm out}$ & 9\,750 & 6\,000 \\
		{\it Electron density} (\cmcub) & $N_{\rm e}(r) \in [R_{\rm in}, R_0]\,^{(e)}$ & 27\,240 & 8\,000 \\
		                                                      & $N_{\rm e}(r) \in [R_0, R_1]\,^{(e)}$ & $27\,240 \left( \dfrac{r}{R_0} \right)^{-0.9}$ & 8\,000 \\
		                                                      & $N_{\rm e}(r) \in [R_1, R_2]\,^{(e)}$ & 23\,000 & 150\,000 \\
		{\it Electron temperature} (K) & $T_{\rm e}(r) \in [R_{\rm in}, R_0]\,^{(e)}$ & 11\,000 & 2\,5000 \\
		                                                     & $T_{\rm e}(r) \in [R_0, R_1]\,^{(e)}$ & 11\,000 & $25\,000 -3\,000\left( \dfrac{r-R_0}{R_1-R_0} \right)$ \\
		                                                     & $T_{\rm e}(r) \in [R_1, R_2]\,^{(e)}$ & 11\,000 & 22\,000 \\
		{\it Expansion velocity} (\kms) & $v_{\rm exp,0}$ & 1.0 & 5.0 \\
		                                                      & $v_{\infty}$ & 13.0 & 25.0 \\
		& $v_{\rm exp}(r) \in [R_{\rm in}, R_2]^{(e)}$ & \multicolumn{2}{c}{$v_{\rm exp,0} - (v_{\infty}-v_{\rm exp,0})\left(\dfrac{r - R_{\rm in}}{R_2 - R_{\rm in}}\right)\,^{(f)}$} \\ 
		{\it Turbulent velocity} $^\mathrm{(g)}$ (\kms) & $v_{\rm turb}$ & 8.0 & 10.0 \\
		{\it Abundances} & H & 0.90 & 0.90 \\
		                               & \hei & 0.09$^\mathrm{(h)}$ & 0.064$^\mathrm{(i)}$ \\
		                               & \heii & 0 & 0.042$^\mathrm{(i)}$ \\
		\hline
		\multicolumn{4}{c}{\textit{Derived parameters}} \\
		\multicolumn{2}{l}{Ionized mass (\msun)} & $5.12 \cdot 10^{-2}$ & $2.92 \cdot 10^{-2}$ \\
		\multicolumn{2}{l}{Scalar momentum (g\,cm\,s$^{-1}$)} & $9.97 \cdot 10^{37}$ & $1.19 \cdot 10^{38}$ \\
		\multicolumn{2}{l}{Kinetic energy (erg)} & $5.07 \cdot 10^{43}$ & $1.21 \cdot 10^{44}$ \\
		\multicolumn{2}{l}{Mech. luminosity ($L_\odot$)} & 0.143 & 1.203 \\
		Kinematic timescale (a) & Aver & 2\,930 & 828 \\
		                                           & Min & 317 & 90 \\
		                                           & Max & 3\,565 & 1\,141 \\
		\multicolumn{2}{l}{Mass-loss rate (\msuna)} & $1.75 \cdot 10^{-5}$ & $3.53 \cdot 10^{-5}$ \\
		\hline
	\end{tabular}
	\tablefoot{$^{(a)}$ Intermediate value from \cite{GuzmanRamirez2009} and \cite{Ali2015}, also similar to \cite{GonzalezSantamaria2021}. $^{(b)}$ \cite{Ali2015}. $^{(c)}$ Combination of observations at different wavelengths from \cite{Garay1989, GuzmanRamirez2009, RamosLarios2012}. $^{(d)}$ \cite{Rodriguez2009}. $^{(e)}$ These parameters are defined between two ellipsoidal shells labeled using their characteristic radius. $^{(f)}$ Same expression for both sources. $^\mathrm{(g)}$ This parameter is defined as the $\sigma$ of a Voigt profile \citep[see Eqs. B.9, B.22 of][]{SanchezContreras2024a} $^{(h)}$ \cite{Pottasch2004}, Table 19. $^{(i)}$ \cite{BernardSalas2001} Table 6.}
\end{table*}

\begin{figure*}
	\centering
	\includegraphics[width=0.25\textwidth]{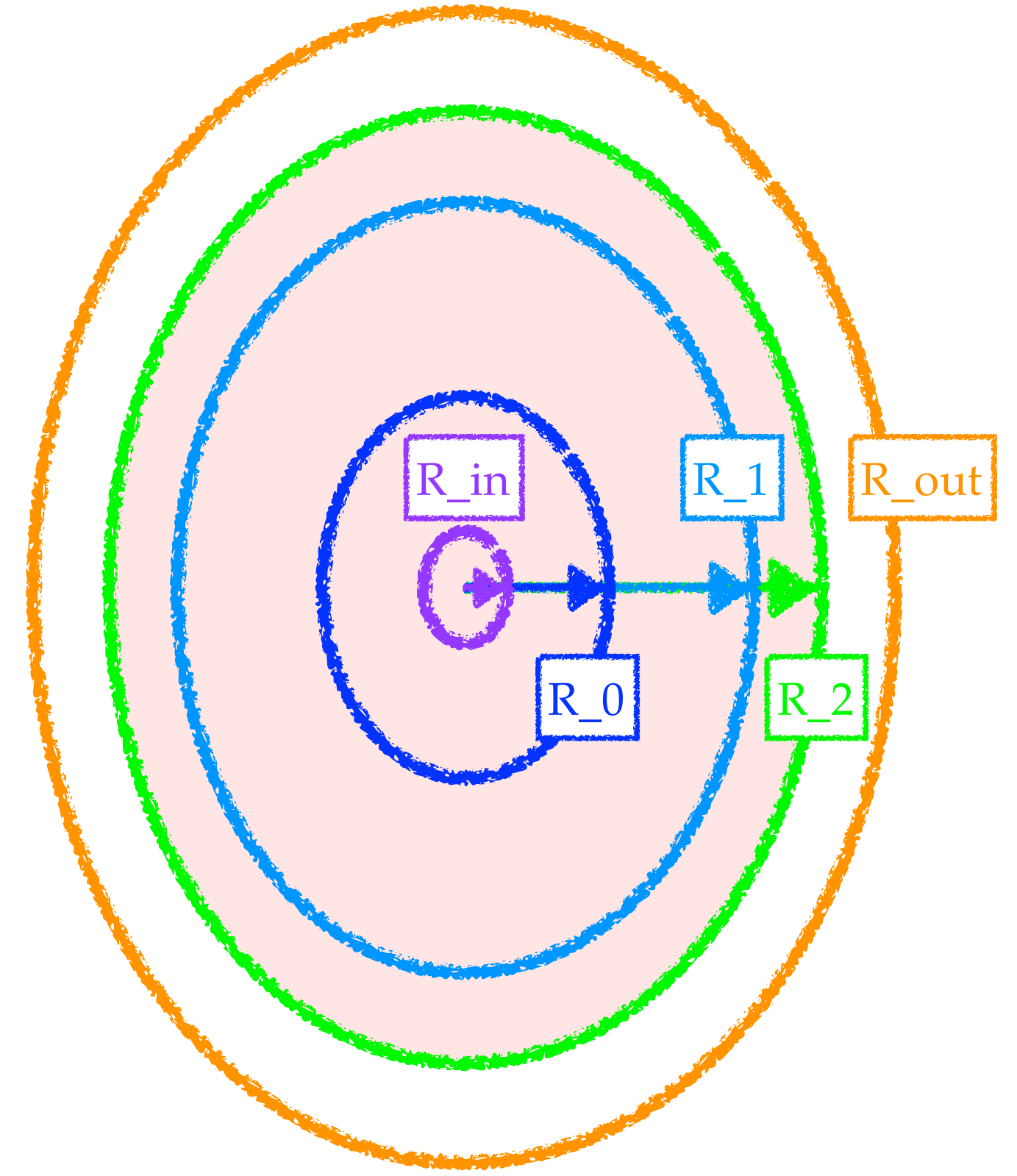}
	\includegraphics[width=0.7\linewidth]{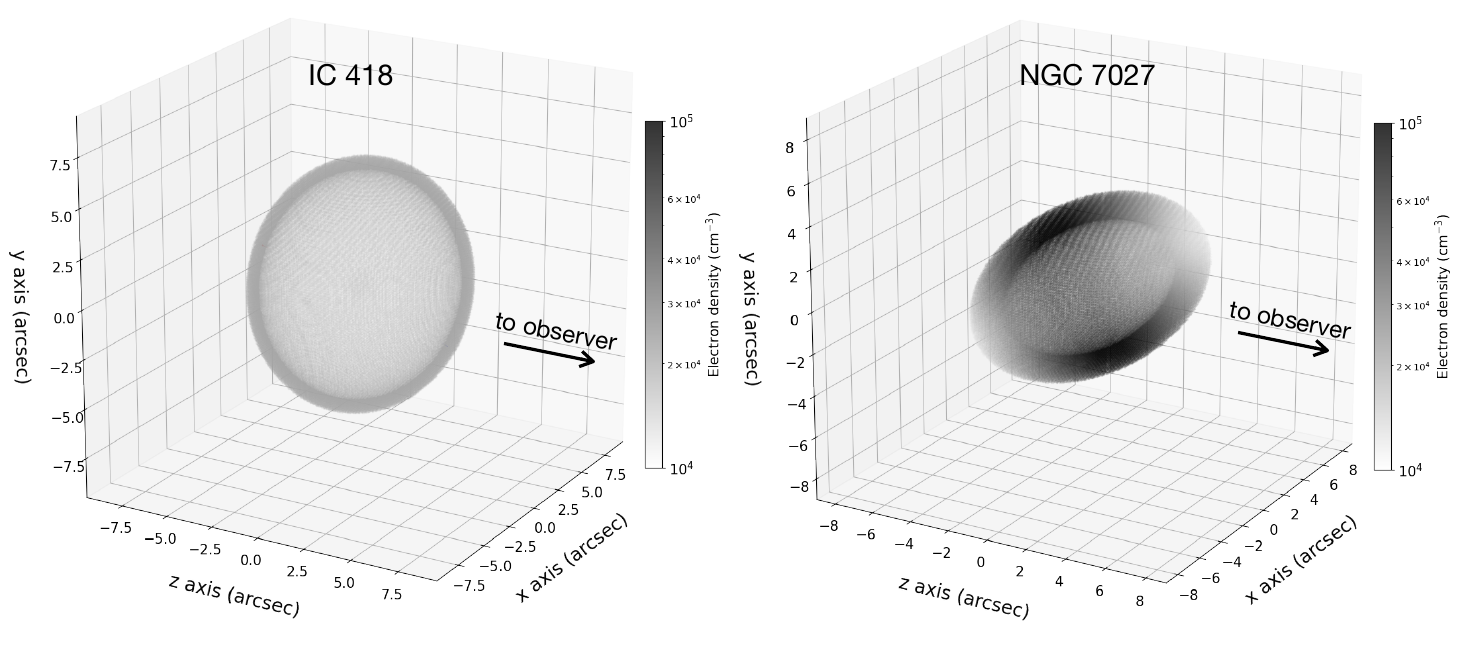}
	\caption{The left panel shows schematically the definition of the different ellipsoidal shells to model a PN with \coral. The light red area (up to $R_2$) shows the filled shells of the PN. The central and right panels are 3D models of \ic and \ngc (see text on Section \ref{sec:model_IC418_NGC7027} for further details).}
	\label{fig:IC418_NGC7027_3d}
\end{figure*}

The geometry of the emitting region assumed in the model is that of an ellipsoid of revolution, with an axial ratio defined by the values of B and A in Table \ref{tab:coral_param}. The electron density ($n_\mathrm{e}$), electron temperature ($T_\mathrm{e}$), and expansion velocity ($v_\mathrm{exp}$) vary as a function of the radial distance from the star. The variation along the semi-minor axis is modeled using a set of characteristic radii (see Table \ref{tab:coral_param}), which result in concentric ellipsoidal surfaces and shells where the physical parameters are either constant (\ic) or vary sinusoidally with latitude (\ngc). The explicit functional dependence of each parameter on these characteristic radii is provided in Table \ref{tab:coral_param}. The observed line widths could not be reproduced by expansion and thermal broadening alone, requiring the inclusion of a turbulent velocity component. The adopted turbulent velocities are listed in Table \ref{tab:coral_param}.

The physical parameters used by \coral that yielded the best fit to the observations are presented in Table \ref{tab:coral_param}. These parameters are the best ones that simultaneously fit both the SED and all H and He radio spectral lines. We select the initial parameters from previous works \citep[see][]{Garay1989, BernardSalas2001, Pottasch2004, GuzmanRamirez2009, Rodriguez2009, RamosLarios2012, Ali2015} and then modified them in subsequent \coral runs in order to reproduce, at the same time, the SED and all the observed radio spectral lines of all species. It is to be noted here that previous estimations of the $T_\mathrm{e}$ for \ngc could not reproduce the SED and all radio spectral lines simultaneously. Thus, we modified $T_\mathrm{e}$ until a good simultaneous match to all observations was obtained (see below).

\ngc is modeled as a filled ellipsoidal shell (region between $R_1$ and $R_2$; see left panel of Fig. \ref{fig:IC418_NGC7027_3d}) enclosing a filled region, with semi-axes of 6\,000\au $\times$ 4\,000\au and a shell thickness of 600\au (see right panel of Fig. \ref{fig:IC418_NGC7027_3d}). The inclination of the major axis of the ellipsoid relative to the plane of the sky is $55^\circ$ and has a position angle of $-30^\circ$. The density of the shell is $1.5\cdot10^{5}$\cmcub at the equator and decreases toward the poles following a sinusoidal dependence, reaching $1.2\cdot10^{4}$\cmcub at the poles. The density of the material within the region enclosed by the shell is $8 \cdot 10^{3}$\cmcub at the equator, decreasing toward the poles to a value of $8 \cdot 10^{2}$\cmcub. The expansion velocity of the plasma increases linearly from 5\kms to 25\kms with distance from the center, remaining constant on each ellipsoidal surface. Similarly, the electron temperature decreases linearly from 25\,000\kel to 22\,000\kel, also remaining constant across individual ellipsoidal surfaces (see Table \ref{tab:coral_param}).

\ic is also modeled as an ellipsoidal shell (region between $R_1$ and $R_2$; see left panel of Fig. \ref{fig:IC418_NGC7027_3d}) enclosing a filled region of $9\,750\,\times\,7\,500\,\mathrm{AU}^2$ (see central panel of Fig. \ref{fig:IC418_NGC7027_3d}). The major axis is inclined at $20^\circ$ relative to the plane of the sky and a position angle of $-22^\circ$. The shell has a thickness of 700\au. The density of the shell is $2.3\cdot10^{4}$\cmcub. For the region enclosed by the shell, the density decreases with distance as a power low from $2.7\cdot10^{4}$\cmcub to $2\cdot10^{3}$\cmcub, remaining constant across individual ellipsoidal surfaces. The expansion velocity increases linearly from 1\kms to 13\kms with distance from the center. The electron temperature is constant across the nebula with a value of 11\,000\kel.

\ic and \ngc have a radio continuum angular size of 12\,arcsec and 14\,arcsec, respectively. As the HPBW of an antenna depends on the frequency, a given source can be extended or unresolved when observed in different bands. In the first case, a single pointing to the nebula center will not receive all the emission from the source. In the opposite case, i.e., if the source is smaller than the telescope beam, the measured brightness temperatures are affected by the beam-dilution factor. The \coral code takes into account both effects, lost-flux and  beam dilution where they apply, so the output of the code can be directly compared with observations regardless the size of the sources and of the telescope beam.

\begin{figure*}
	\centering
	\includegraphics[width=0.7\linewidth]{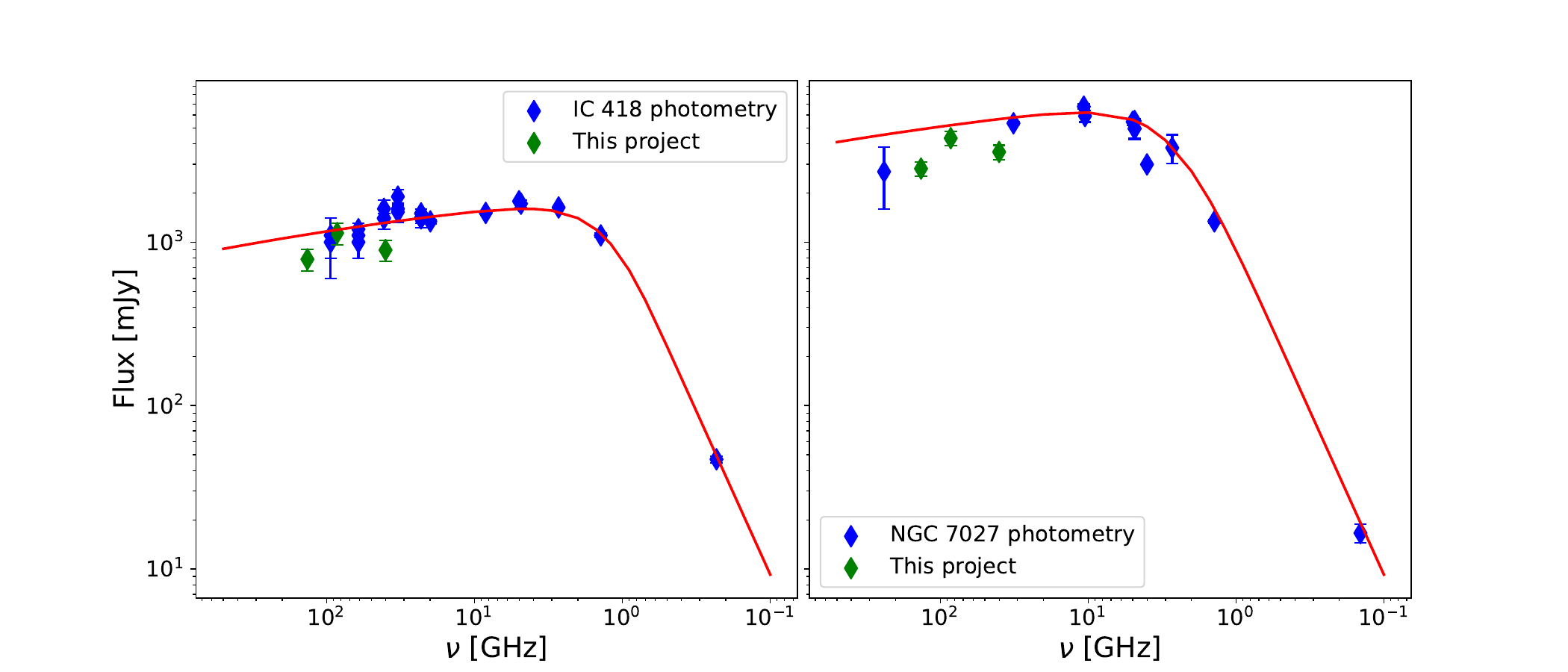}
	\caption{Spectral Energy Distribution of \ic (left panel) and \ngc (right panel). Continuous red lines represent the \coral predictions; diamonds display the observational measurements and corresponding errors. The \ic data has been taken from \cite{Vollmer2010, Condon1998a, Griffith1994, Murphy2010, Wright2009, Massardi2009, Bennett2003, Gold2011, Chen2009}, while the \ngc data has been taken from \cite{Frew2016, Condon1998a, Vollmer2010, Gordon2021, Becker1991, Gregory1991, Reich2014, Vollmer2008, Kellermann1973, SanchezContreras1998}.}
	\label{fig:IC418_NGC7027_SED}
\end{figure*}

Even though the modeling of 3D objects is a complex task, \coral provides really good results for both objects. The model reproduces the free-free observations at several frequencies, which means that the outputs of the two PNe are physically viable. For \ic (see Fig.\ref{fig:IC418_NGC7027_SED} left panel), the modeling nicely matches the free-free observations and the largest discrepancies of the model results with respect to the observations are found at $\nu > 30$\ghz. We measured the continuum emission from our observations averaging the spectra for each range before subtracting the baseline and measuring the rms noise. The \coral values are in reasonable agreement with our own predictions. In the case of \ngc (see Fig.\ref{fig:IC418_NGC7027_SED} right panel), the \coral results also reproduce quite well the observations. Here again, the larger discrepancies of \coral with respect to the observations appear at $\nu > 30$\ghz in our own measurements and at 1.3\mm, but all these measurements follow the general decreasing tendency predicted by the \coral model.

Regarding the RRLs emission, the agreement between the observations and the \coral predictions is good. The modeled lines match perfectly their observational expected central frequencies and their profiles are symmetric. So, by superimposing the spectrum of a line and its \coral prediction, it helps to confirm asymmetric profiles of RRLs. In short, the \coral models helped us to: (i) confirm the identification of RRLs by comparing the peak fluxes and line shapes; (ii) confirm those RRLs with several components; and (iii) discard some tentative RRL identifications because they do not display the predicted line shape and/or peak fluxes.

The first runs with \coral included the modeling of the whole set of species we searched in the spectra (i.e., H, \hei, \trihei, \ci, \oi, and their corresponding ions \heii, \triheii, \cii, \ciii, \oii, and \oiii), even though we did not find any observational evidence of the presence of \trihei, \ci, \oi, and their ionized species. We modeled them in both sources using the abundances in the literature. The maximum peak flux intensities (i.e., the \Dneq 1 transitions) predicted for \trihei, \ci, \oi, and their ionized species were well below the rms noise ($\sim 10^{-3}\sigma$), in good agreement with their absence in the observations.

\section{Results for IC 418} \label{sec:catalog_IC418}

The C-rich PN \ic has been studied all along the electromagnetic spectrum. The central star is surrounded by an elliptical main nebula, which is enclosed by several low surface brightness structures (shells, rings) and two detached haloes \citep[see e.g.,][and references therein]{GuzmanRamirez2009, RamosLarios2012}. The \ic IR spectrum displays both aliphatic and aromatic molecular bands \citep[see][]{Otsuka2014}, unidentified infrared features (UIRs) \citep[see e.g.,][]{Forrest1981, Hony2002} and C$_{60}$ emission bands \citep{Otsuka2014}. However, all previous attempts to detect any molecular emission at radio frequencies have failed so far \citep[see e.g.,][where they tried to detect the $J = 1-0$ and $J = 2-1$ lines of $^{12}$CO after reaching an rms noise of 5\mkel in $T_\mathrm{mb}$]{Dayal1996}.

Previous RRL works on this source have reported the detection of H and \hei radio spectral lines with $\Delta n = 1-4$: H38\ga, H39\ga, H76\ga, H85\ga, H91\ga, H92\ga, H110\ga, H95\gb, H114\gb, H115\gb, H16\gb, H132\gg, H145\gd, and \hei76\ga, \hei91\ga, \hei92\ga, \hei114\gb, \hei115\gb, \hei116\gb, and, \hei132\gg \citep{Terzian1972, Walmsley1981, Garay1989, Dayal1996, GuzmanRamirez2016}. With the exception of the H38\ga and H39\ga lines, all previous observations have been made at frequencies $< 15$\ghz. Therefore, the mm range has thus remained largely unexplored. In Table \ref{tab:summary_rrls} (and Table  \ref{tab:rrls_parameters} in the Appendix) as well as in Figs. \ref{fig:IC418_H_observational}, \ref{fig:IC418_HeI_observational}, we report for the first time detection of 158 previously undetected H and \hei RRLs at 2, 3, and 7\mm in the PN \ic as well as the tentative detection of 25 RRLs of the same atoms (the H38\ga and H39\ga lines were previously detected, but we do not cover that range). See examples of individual line profiles in Figs. \ref{fig:IC418_Hga} - \ref{fig:IC418_HeIgg}. In Table \ref{tab:rrls_parameters} we also report the upper limits for those H and \hei lines that are not detected.

The results (in terms of peak fluxes and FWHM values) from the radiative transfer model \coral are compatible with the observations within the observational standard deviation. The average FWHM is $28 \pm 5$\kms for H lines and $16 \pm 7$\kms for \hei lines. The global results of both the observations and the \coral models are summarized in Fig.\ref{fig:IC418_Co3RaL-Obs}.

\begin{table}
	\caption{Summary of the RRL detections (D) and tentative detections (T) by element, source and mm band.}
	\label{tab:summary_rrls}
	\centering
	\begin{tabular}{r r | c c | c c c}
		\hline \hline
		& & \multicolumn{2}{c|}{\ic} & \multicolumn{3}{c}{\ngc} \\
		& & H & \hei & H & \hei & \heii \\
		\hline
		D & 2\mm	& 13  & 3   & 10 & 3 & 5 \\
			& 3\mm	& 21	& 5   & 20 & 3 & 10 \\
			& 7\mm	& 93	& 23 & 53 & 8 & 22 \\
		T & 2\mm	& 3 	& 0   & 3   & 0 & 1 \\
			& 3\mm	& 4      & 1    & 0   & 2 & 3\\
			& 7\mm	& 13	& 4    & 8 & 9 & 2 \\
	\end{tabular}
\end{table}

\subsection{Neutral Hydrogen (H) RRLs} \label{sec:RRLs_IC418_H}

\begin{figure*}[ht]
	\centering
	\includegraphics[width=\linewidth]{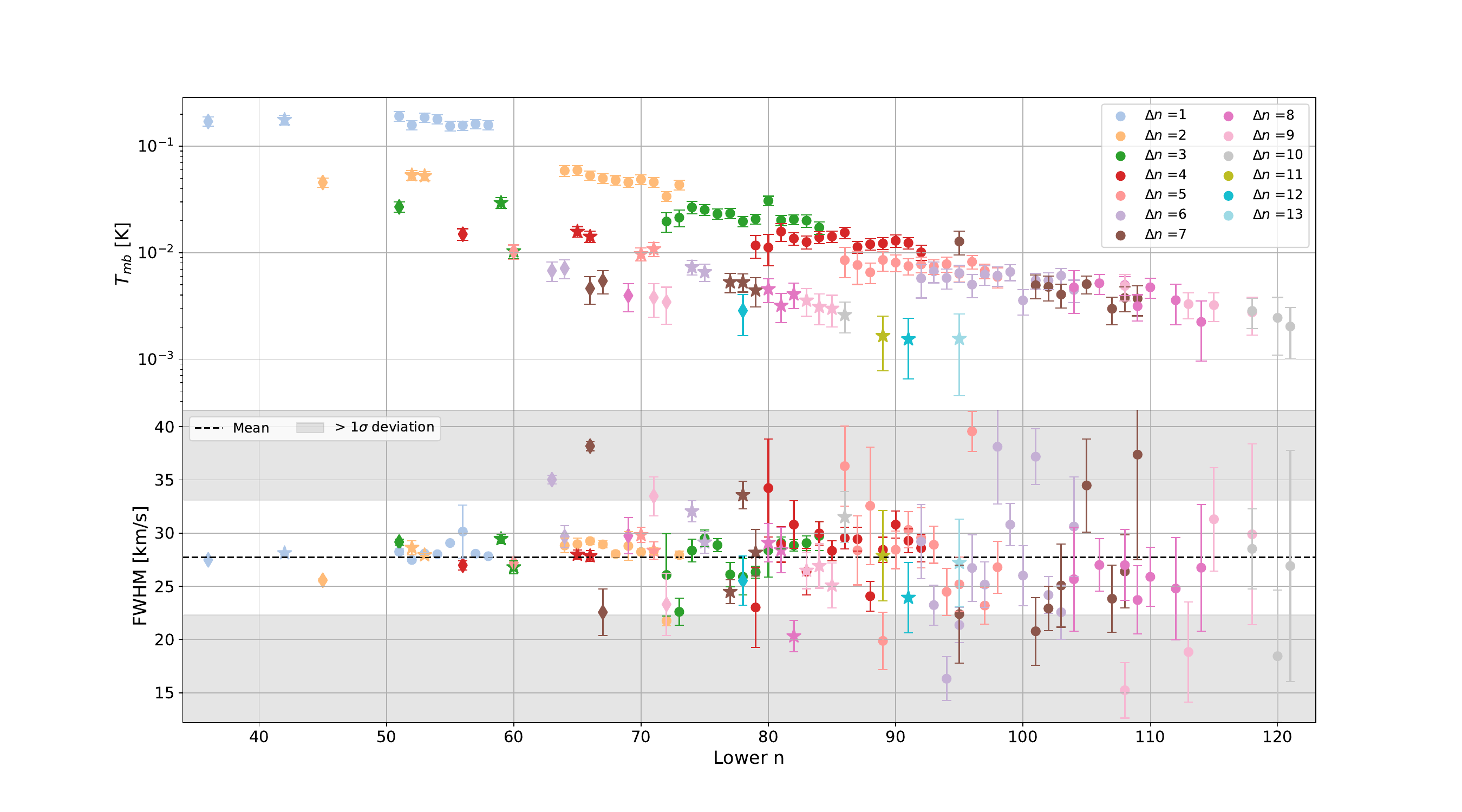}
	\caption{Peak intensity (in T$_\mathrm{mb}$ [\kel] units, top panel) and FWHM (in \kms, bottom panel) of the observed H lines in \ic. Diamonds represent the 2\mm data, stars the 3\mm data, and dots the 7\mm data.}
	\label{fig:IC418_H_observational}
\end{figure*}

The number of detected lines is quite large, so we first compared the lower level involved in the transition, $n$, with the peak flux and FWHM of each line (see Fig.\ref{fig:IC418_H_observational}).

We detect lines for \Dnleq 13 at 2 and 3\mm. All lines with the same \Dn have similar intensities and the intensities decrease as \Dn increases. At 7\mm, we can only reach \Dneq 10, but we find a strong dispersion of the peak intensities for sets of lines with \Dngeq 6. We tentatively detect lines with \Dneq $10-13$ at 2\mm, \Dneq $10-11$ at 3\mm, and \Dneq $7-10$ at 7\mm (see Table \ref{tab:rrls_parameters} in the Appendix for more detailed information).

The FWHMs also show a great dispersion as $n$ grows up (see Fig.\ref{fig:IC418_H_observational}). We have considered the 1$\sigma$ standard deviation of the mean value to differentiate between too narrow and broad lines. At 2\mm we find two outliers that display a higher FWHM. These two lines are merged in the spectrum with each other, which makes difficult to find a good Gaussian fit for both. The detections at 3\mm have FWHM values quite close to the mean, except for two cases. The broader line is a line contaminated by an unidentified emission (UF), while the narrower one is a line with a very sharp profile and no contamination. At 7\mm, a large dispersion of the FWHM values starts around $n = 90$ for \Dngeq 7, which are, by far, the weakest lines in this spectral range.

By comparing the observational peak flux values with the modeling, we observe a really good agreement (see Fig.\ref{fig:IC418_Co3RaL-Obs} upper panel) within the error bars (i.e., a maximum flux calibration uncertainty of 10\% combined the Gaussian fit error); even for the weakest lines at 2, 3, and 7\mm.

\begin{figure*}[h]
	\centering
	\includegraphics[width=\linewidth]{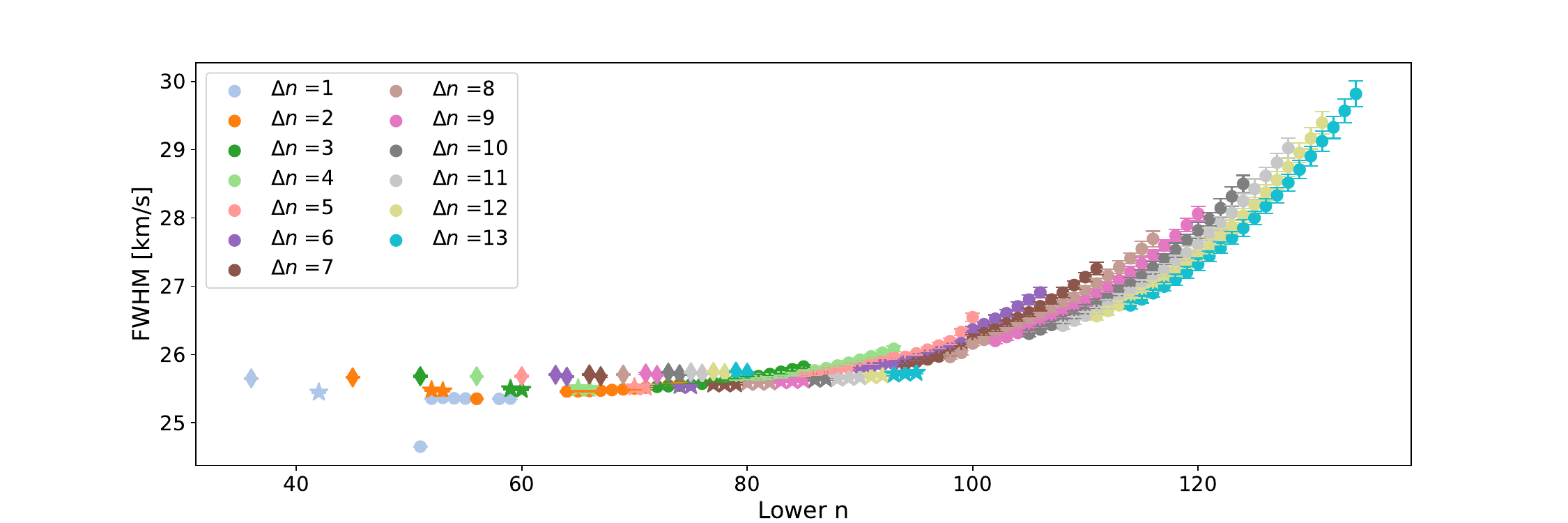}
	\caption{FWHM values (in \kms) from \coral modeling of the RRLs in \ic as a function of the lower quantum number n. Note the step increase in line width for n larger than 100. Such broadening is not seen/detected in the observations (see Section\ref{sec:applications} for more detailed explanations).}
	\label{fig:IC418_FWHM-nlow}
\end{figure*}

The FWHM values predicted by the \coral model are very homogeneous. Considering the standard deviation of the observational data, \coral shows a good agreement with the detections at 2 and 3\mm (see Fig.\ref{fig:IC418_Co3RaL-Obs} upper panel). At 7\mm, our observations show more dispersion for high \Dn values, while the FWHM model values tend to increase with \Dn (see also Fig. \ref{fig:IC418_FWHM-nlow}). However, the model is still compatible with the observations.

\subsection{Neutral Helium (\ion{He}{i}) RRLs} \label{sec:RRLs_IC418_He}

\begin{figure*}[ht]
	\centering
	\includegraphics[width=\linewidth]{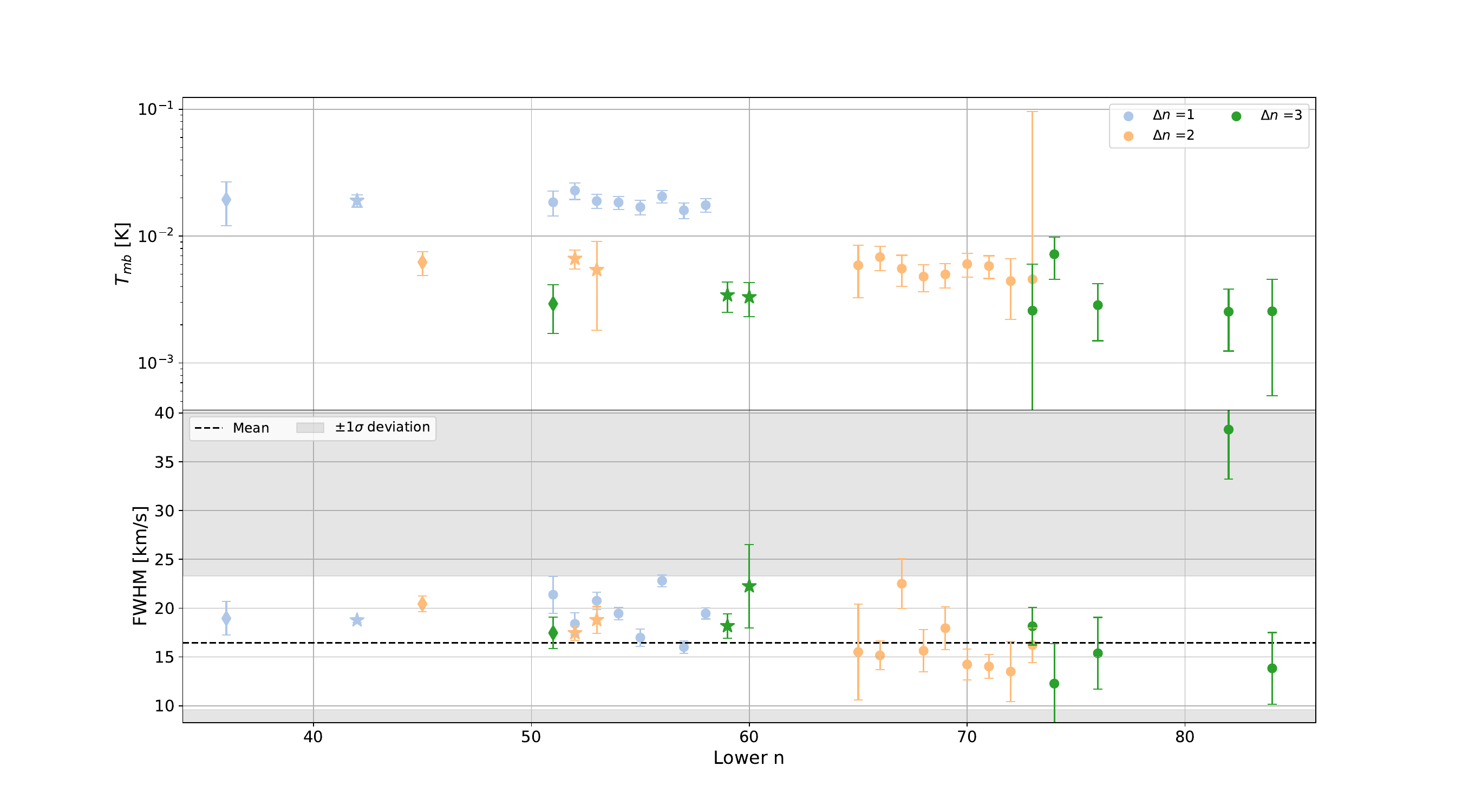}
	\caption{Peak intensity (in T$_\mathrm{mb}$ [\kel] units, top panel) and FWHM (in \kms, bottom panel) of the observed \hei lines in \ic. Diamonds represent the 2\mm data, stars the 3\mm data, and dots the 7\mm data.}
	\label{fig:IC418_HeI_observational}
\end{figure*}

The Fig.\ref{fig:IC418_HeI_observational} shows the observed \hei lines in the \ic radio spectra. The number of lines is significantly lower than for H (see also Table \ref{tab:summary_rrls} and Table \ref{tab:rrls_parameters} in the Appendix) because the intensity of the lines depends on their atomic number and mass, but also because the \hei abundance is 10 times lower than H abundance.

At 2\mm we clearly detect three \hei lines. All of them show FWHMs close to the mean value (see bottom panel of Fig.\ref{fig:IC418_HeI_observational}). We did not expect to observe more lines in this range, considering the list of frequencies we previously calculated for \hei using Eq. (\ref{eq:Rydberg_freq_Hlike-atom}) and the rms level of our 2\mm \ic observations.

At 3\mm there are three detections and one tentative detection. The detected lines display FWHMs very close to the mean with the exception of the broad \hei60\gg line. However, this line is blended with H87\gk and, as we mentioned above, and the parameters of the fit are affected by high uncertainties.

At 7\mm the dispersion of the peak intensities is larger for lines with \Dneq 3, which correspond to the weakest \hei lines. We only detect half of the \hei n\gg lines in this frequency range. The FWHMs do not strongly deviate from the mean value.

\begin{figure*}[ht]
	\centering
	\includegraphics[width=\linewidth]{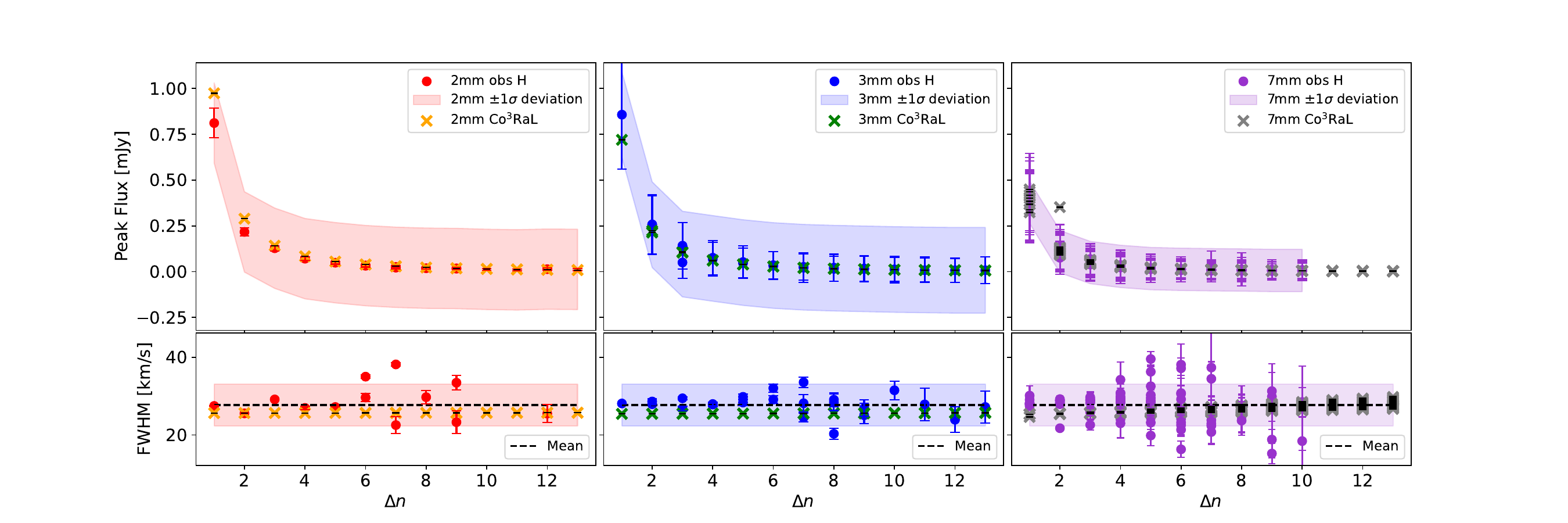}
	\includegraphics[width=\linewidth]{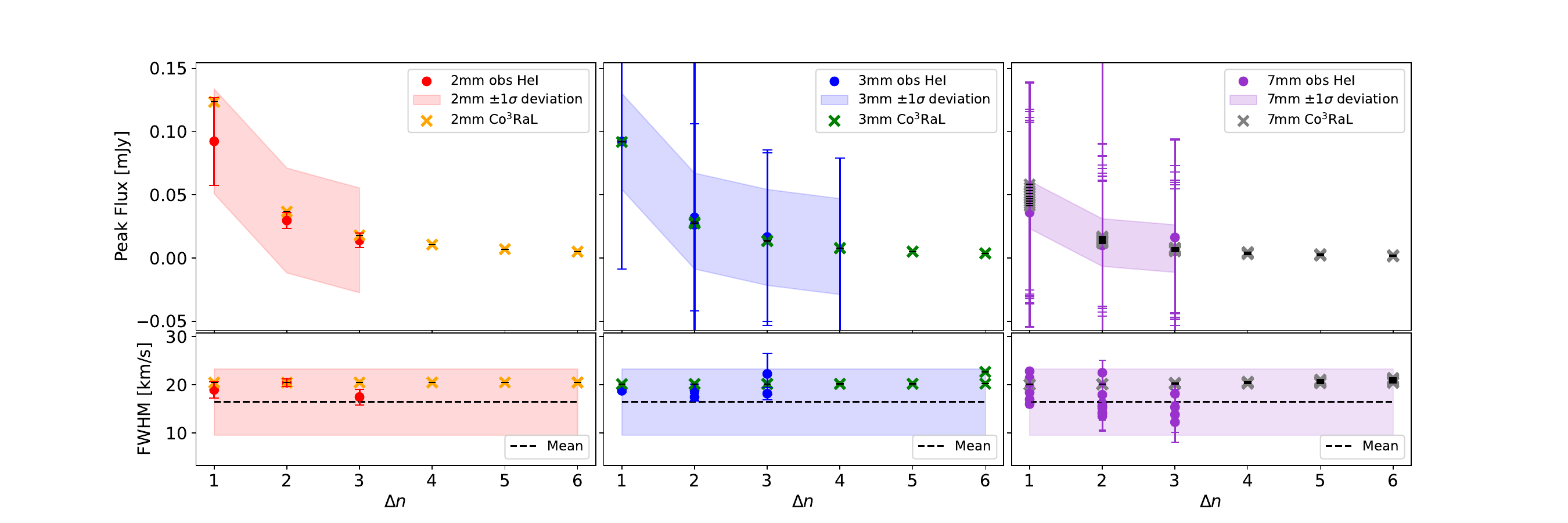}
	\caption{Compatibility of the \coral model data (peak fluxes and FWHM in mJy and \kms, respectively) with the observational ones for H (upper panel) and \hei (lower panel) in PN \ic. The dots represent the peak flux and FWHM of each observed emission line obtained with the Gaussian fit using the available procedure in CLASS with the respective errors, while the crosses represent the peak flux and FWHM values simulated by \coral. All data are separated by frequency bands. The colored regions on the peak flux plots represent the $\pm 1\,\sigma$ deviation of the data at each frequency band. The colored regions on the FWHM plots represent the $\pm 1\,\sigma$ deviation from the observational FWHM mean (see Section \ref{sec:observations} for more details).}
	\label{fig:IC418_Co3RaL-Obs}
\end{figure*}

The agreement of the observational intensity values with the \coral predictions is also very good (see Fig.\ref{fig:IC418_Co3RaL-Obs} lower panel). The predicted peak fluxes are statistically compatible with our line detections. The \coral FWHMs are roughly similar to the observed ones. The larger discrepancies appear again at 7\mm (see Section \ref{sec:applications} for a more detailed explanation).
	
\section{Results for NGC 7027} \label{sec:catalog_NGC7027}

The bipolar/multipolar C-rich PN \ngc is an intriguing and extensively studied source. Many molecules have been detected in this complex PN \citep[see e.g.][]{Zhang2008}, which is thought to host a central binary system \citep{MoragaBaez2023}. The existence of such a binary central star has contributed to its complex morphology. Three main axes or outflows have been confirmed with optical observations, each corresponding to a different ejection jet from the central system. The X-ray emission aligns with the youngest outflow \citep{Montez2018, MoragaBaez2023} and also enhances the formation of molecules such as \element[+]{HCO} \citep{Bublitz2023}. Even though the central star has a high effective temperature ($T_\mathrm{eff} = 200,000$\kel), the rich molecular content seems to be protected by a dusty equatorial belt confirmed by, e.g., \cite{Nakashima2010, MoragaBaez2023}.

The PN \ngc has also been observed in the IR range. Previous studies, such as \cite{Beintema1996}, confirm the presence of both aliphatic and aromatic species together with additional UIR features \citep[see e.g.][]{Kwok2011}. However, the four strongest IR emission bands of C$_{60}$ fullerenes are not detected in its ISO/SWS IR spectrum, being classified as a non-C$_{60}$ PN \citep{Otsuka2014}.

Over the last sixty years, \ngc has been extensively observed at sub-mm frequencies, resulting in the detection of several H RRLs; e.g., H113\ga, H85\ga, H66\ga, or H107\gb \citep[see][]{Rubin1971, Terzian1972, Chaisson1976, Churchwell1976, Terzian1978, Walmsley1981, Ershov1989, Roelfsema1991}. Additionally, a couple of \hei and \heii lines have been reported: \hei76\ga, \heii105\ga, and \heii121\ga \citep{Chaisson1976,Terzian1980,Mezger1980, Walmsley1981}. The first mm detections were made by \cite{Vallee1990}, who observed the source with the \mbox{IRAM 30m} at 99 and 232\ghz, reporting the detection of H30\ga, \hei30\ga, H40\ga, and \hei40\ga. The most recent and extended survey of \ngc was conducted by \cite{Zhang2008}, who reported nine Hn\ga, nine Hn\gb, eight Hn\gg, and  eight \hei n\ga lines. Their work was a significant step towards larger and high-sensitivity surveys, highlighting the importance of these kind of deep radio line surveys for both molecular and RRL detections.

\begin{figure*}[ht]
	\centering
	\includegraphics[width=\linewidth]{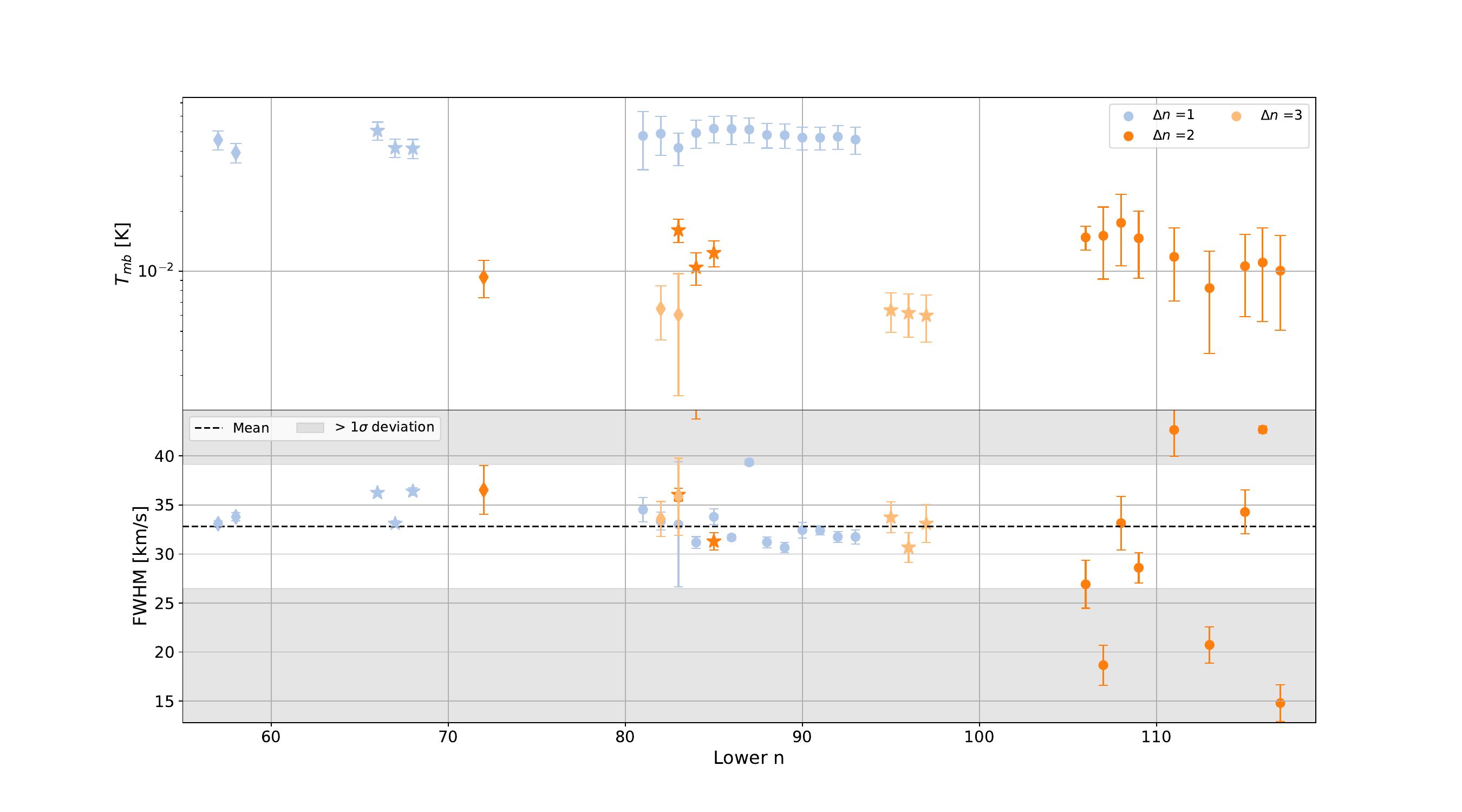}
	\caption{Peak intensity (in $T_\mathrm{mb}$ [\kel] units, top panel) and FWHM (in \kms, bottom panel) of the observed \heii lines in \ngc. Diamonds represent the 2\mm data, stars the 3\mm data, and dots the 7\mm data.}
	\label{fig:NGC_7027_HeII_observational}
\end{figure*}

In Table \ref{tab:summary_rrls} and Figs. \ref{fig:NGC_7027_H_observational}, \ref{fig:NGC_7027_HeI_observational}, and \ref{fig:NGC_7027_HeII_observational}, we report the detection and tentative detections of, respectively, 134 and 28 RRLs of H, \hei, and \heii RRLs (see some examples of individual lines in Figs. \ref{fig:NGC7027_Hga}	- \ref{fig:NGC7027_HeIIgg} and see Table \ref{tab:rrls_parameters} for more details). All H lines with \Dngeq 4, \hei lines for \Dngeq 2, as well as all \heii, are new detections towards PN \ngc.

The radio spectra of this PN are populated by both RRLs and molecular emission lines. Indeed, in some cases, the molecular lines are blended with the RRLs. The peak temperature and the FWHM of the detected lines as well as the comparison between the observational measurements and the \coral model predictions are displayed in Figs.\ref{fig:NGC_7027_H_observational}, \ref{fig:NGC_7027_HeI_observational}, \ref{fig:NGC_7027_HeII_observational}, and \ref{fig:NGC7027_Co3RaL-Obs}. The average FWHMs for the H, \hei, and \heii lines are $38 \pm 10$\kms, $29 \pm 7$\kms, and $33 \pm 6$\kms, respectively. Similarly to PN \ic (see Fig.\ref{fig:IC418_FWHM-nlow}), the FWHM values predicted by the \coral \ngc model slightly increase, for a fixed \Dn, with decreasing frequency.

\subsection{ Neutral Hydrogen (H) RRLs} \label{sec:RRLs_NGC7027_H}

\begin{figure*}[ht]
	\centering
	\includegraphics[width=\linewidth]{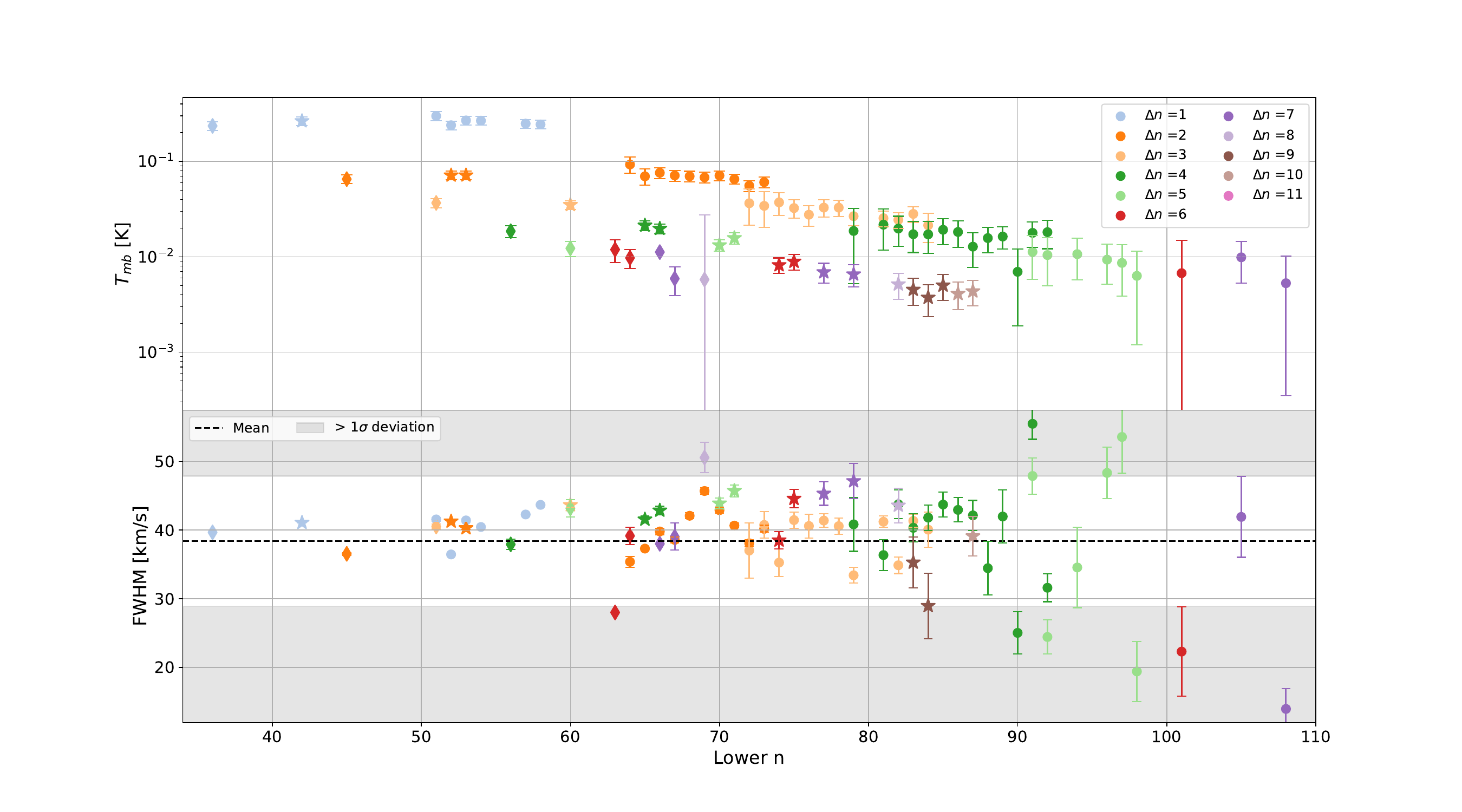}
	\caption{Peak intensity (in $T_\mathrm{mb}$ [\kel] units, top panel) and FWHM (in \kms, bottom panel) of the observed H lines in \ngc. Diamonds represent the 2\mm data, stars the 3\mm data, and dots the 7\mm data.}
	\label{fig:NGC_7027_H_observational}
\end{figure*}

The number of H RRLs detected in \ngc is smaller than for \ic (see Table \ref{tab:summary_rrls} and Table \ref{tab:rrls_parameters}). All this information is plotted in Fig.\ref{fig:NGC_7027_H_observational}.

The intensities of all the detected lines at 2\mm decrease as the quantum number $n$ grows up. The FWHMs of almost all the lines are close to the mean value. The only exception in this mm band displaying large peak flux error bars and a significantly large FWHM is H69\gq, a line that may be contaminated with extra molecular emission that affects its Gaussian fit.

At 3\mm, we detect all the expected lines with the exception of one line, which is below the detection limit. The high SNR of the H lines in this band results in high-quality line profile fits. The H RRLs with the same \Dn value show similar peak fluxes within the error bars. The FWHMs are statistically compatible between the different observational bands; except for the H86\gk line. The latter line clearly deviates from the average, displaying a FWHM \mbox{$\sim 65$\kms}. Such broadening is probably due to strong blending with an unidentified molecular line that modifies the RRL shape.

At 7\mm, we find again more variability, specially for \Dngeq 5. The higher rms level at sub-bands 5 to 8 (these correspond to $\sim 42 - 50$\ghz and it is produced by resonant signals in the receiver) hides several high-\Dn H lines that we clearly detect at the lower frequency sub-bands. In general, clear detections deviate less from the mean than the tentatives ones, but there are exceptions along the full frequency range.

The \coral model predicts peak intensities similar to those observed (see Fig.\ref{fig:NGC7027_Co3RaL-Obs} upper panel). The synthetic FWHMs match up quite well with the observational results at 2 and 3\mm. Once more, the model predictions at 7\mm show an increasing FWHM value with \Dn that it is not evident in the observations. This effect is the same that we observed above in the \coral model results for \ic (see Fig.\ref{fig:IC418_FWHM-nlow}).

\subsection{Neutral Helium (\ion{He}{i}) RRLs} \label{sec:RRLs_NGC7027_He}

\begin{figure*}[ht]
	\centering
	\includegraphics[width=\linewidth]{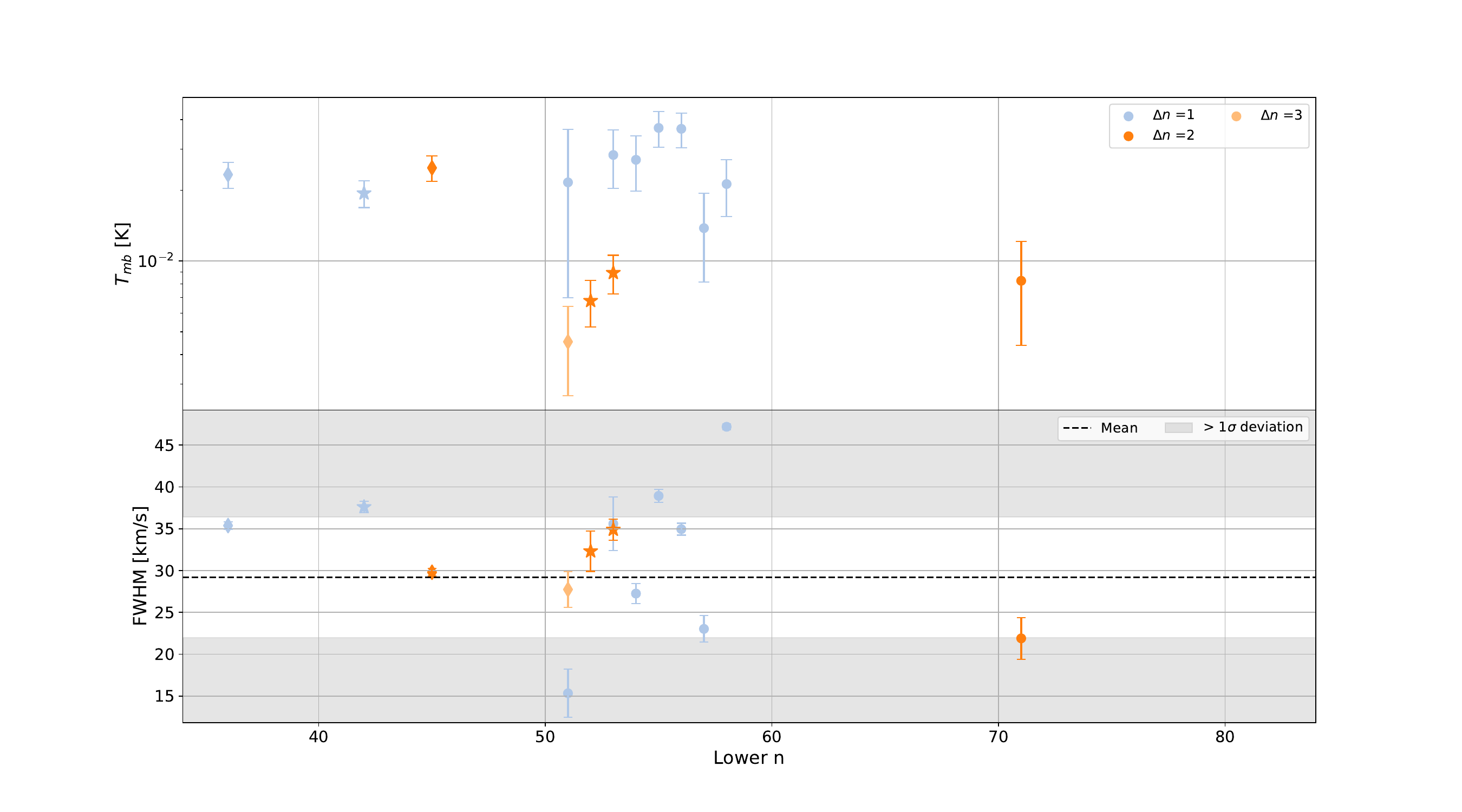}
	\caption{Peak intensity (in $T_\mathrm{mb}$ [\kel] units, top panel) and FWHM (in \kms, bottom panel) of the observed \hei lines in \ngc. Diamonds represent the 2\mm data, stars the 3\mm data, and dots the 7\mm data.}
	\label{fig:NGC_7027_HeI_observational}
\end{figure*}

The 2\mm band is the only one in which we detect \hei lines for \Dnleq 3. At 3 and 7\mm, we only observe lines with \Dnleq 2 (see Fig.\ref{fig:NGC_7027_HeI_observational}). In the 2\mm range, the stronger lines are blended with other H or \heii RRLs and their line profiles were difficult to fit. However, their FWHM values do not deviate much from the mean and remain inside the 1$\sigma$ region in Fig. \ref{fig:NGC_7027_HeI_observational}. At 3\mm, the three detected lines are blended with \heii lines. Only one of them has a FWHM value clearly outside the region delimited by the standard deviation, suggestive of the blending just mentioned before. At 7\mm, we only have detected clear \hei n\ga lines, but only one \hei n\gb line due to an insufficient sensitivity. This fact naturally implies a larger dispersion from the mean standard deviation value for most of the 7\mm \hei detections.

Finally, the \coral predictions are compatible with the peak intensities of all detected \hei lines, even the weaker ones (see Fig.\ref{fig:NGC7027_Co3RaL-Obs} middle panel). The predicted FWHMs are also statistically compatible with the observations.

\subsection{Ionized Helium (\ion{He}{ii}) RRLs} \label{sec:RRLs_NGC7027_HeII}

The strong UV and X-ray emission in \ngc allows the presence of \heii lines, which are a bit stronger than the \hei ones. At 2 and 3\mm, we only find clear detections of \heii lines with \Dnleq 3. They display FWHM values very close to the mean, except for \heii84\gb. This individual line has a FWHM close to 50\kms, but it is strongly blended with \hei42\ga, which is slightly stronger and affects the line fit.

At 7\mm, we detect all \heii n\ga lines in the frequency range and most of the \heii n\gb lines. They display similar peak intensities with larger error bars toward the upper end of the band range. The FWHM values of the \heii n\ga lines do not strongly deviate from the mean. The only exception is \heii87\ga, partially blended with H78\gg. The \heii n\gb lines are weaker, so their FWHMs display larger deviations from the mean value. At the upper end of the 7\mm range ($\sim\,$42.5\ghz), we cannot find clear detections because all lines lie below the $3\sigma$ detection limit. The dispersion around the mean is larger than at 2 and 3\mm, probably caused by the intrinsic weakness of the lines.

\begin{figure*}[ht]
	\centering
	\includegraphics[width=\linewidth]{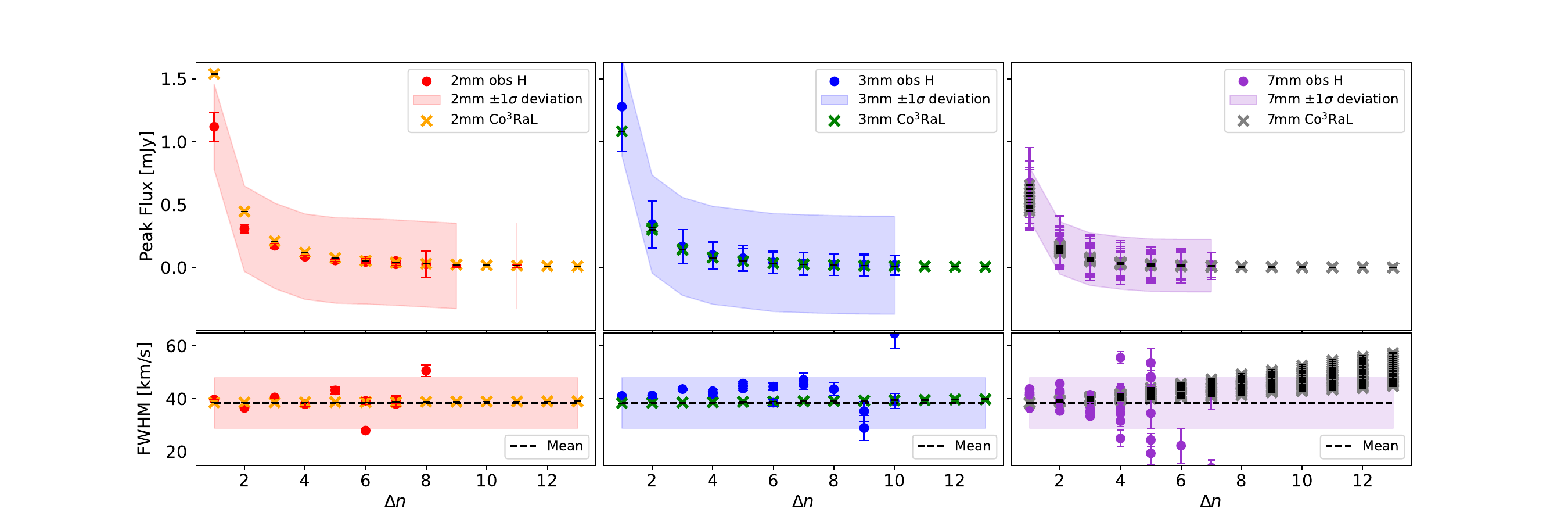}
	\includegraphics[width=\linewidth]{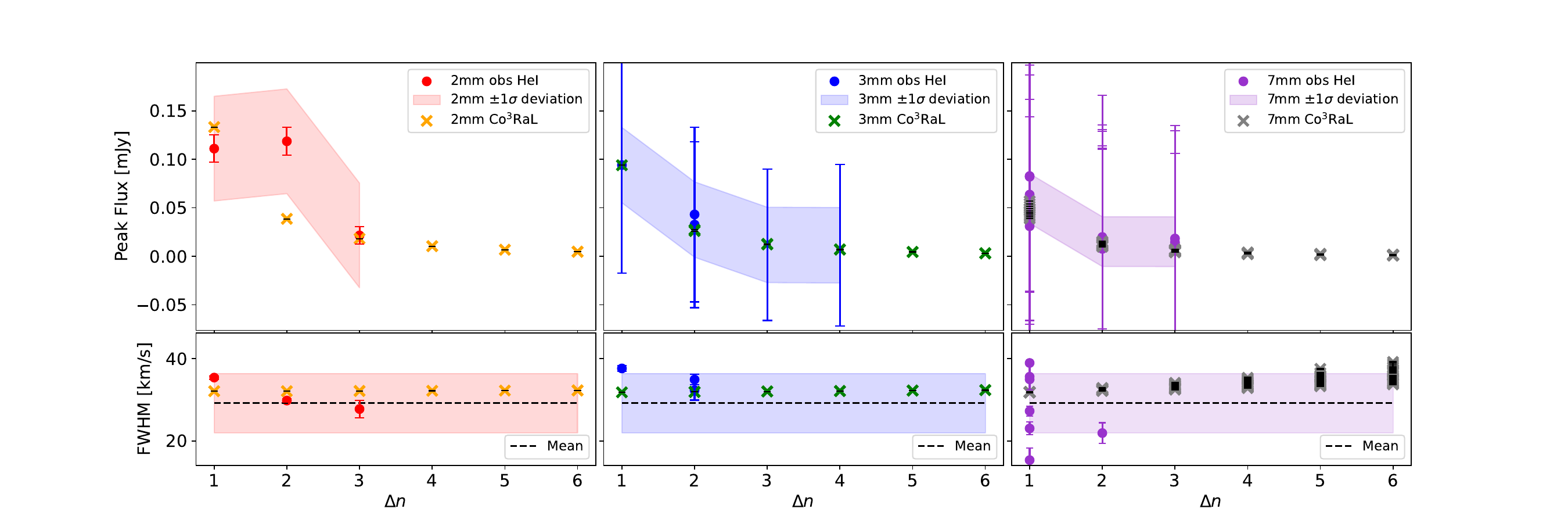}
	\includegraphics[width=\linewidth]{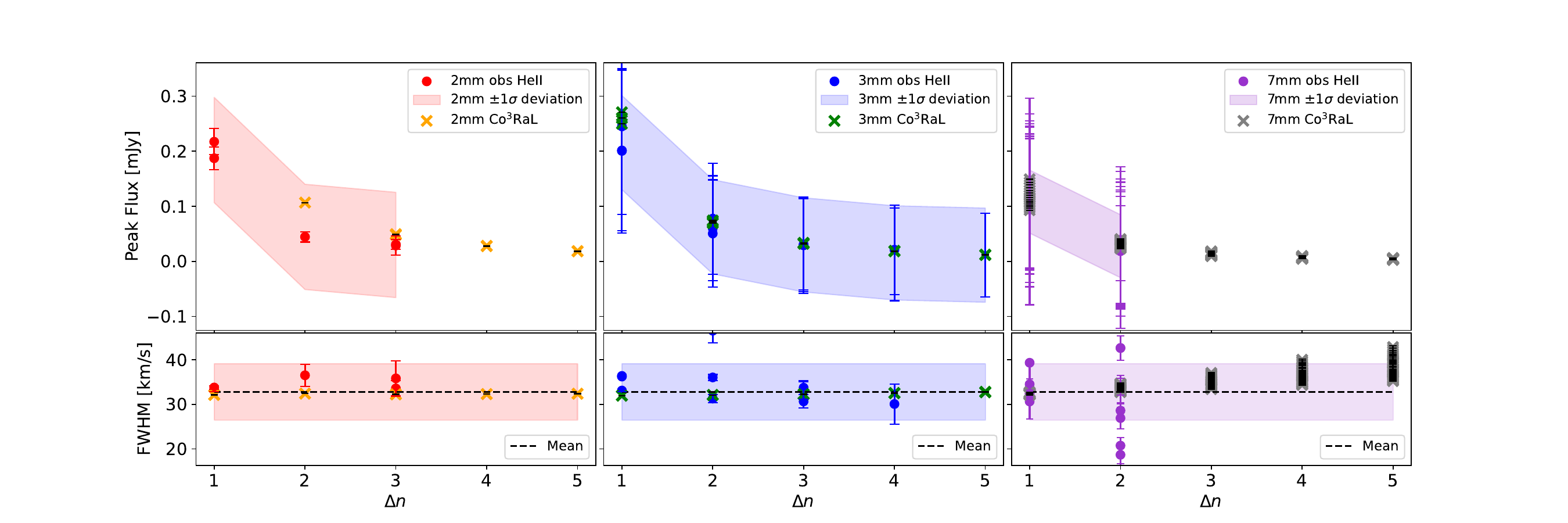}
\caption{Compatibility of \coral model data (peak fluxes and FWHM in mJy and \kms, respectively) with the observational ones for H (upper panel), \hei (middle panel), and \heii (lower panel) in PN \ngc. Color and label codes are the same as in Fig.\ref{fig:IC418_Co3RaL-Obs}.}
\label{fig:NGC7027_Co3RaL-Obs}
\end{figure*}

For \heii, the \coral modeling reproduces quite well the observational peak fluxes at 3 and 7\mm, but at 2\mm the predicted peak fluxes are higher than the observed ones (see Fig. \ref{fig:NGC7027_Co3RaL-Obs} lower panel). The FWHMs are in good agreement for all frequency ranges, right in the standard deviation limit. At 7\mm, the progressive broadening of the \coral synthetic lines is observed, as in the previous cases.

\section{Discussion and Applications to Radio Spectroscopy} \label{sec:applications}

The different number of detections on each source for each element (see Table \ref{tab:summary_rrls} and Table \ref{tab:rrls_parameters} in the Appendix) is mainly due to the higher sensitivity, but it may also be a direct consequence of the effective temperature of their respective central star and/or their intrinsic brightness. The central star (CS) of \ic has a lower effective temperature ($\sim 30\,000$\kel), so it provides less energetic photons to its surrounding nebula. Even though we have detected a higher number of H RRLs than in \ngc, there is no sign of \heii RRLs in \ic. This finding is completely logic: the higher effective temperature of the \ngc CS generates a stronger UV radiation, providing enough energetic photons to ionize \hei to \heii and \ion{He}{iii}. The nebular age for each PN is also important. \ngc is significantly younger than \ic, with its hotter CS producing more energetic UV radiation and the previously expelled circumstellar material being closer to the CS.

We do not observe RRLs of species heavier than He. The ionization potential of \ci recombination lines (11.26\,eV) is lower than that of H (13.6\,eV), so in the case of ionization-bounded PNe, only the photons with energies in the range 11.26\,-\,13.6\,eV can escape and photoionize C atoms outside the \ion{H}{ii} region. \ci RRLs would be expected to be wide spread throughout the PN because C is ejected from the CS simultaneously than H and He. However, the C abundance \citep[$3 \cdot 10^{-4} - 7 \cdot 10^{-4}$;][]{Pottasch2004} is relatively low compared to H and He \citep[which have/display abundances of 0.9 and 0.09, respectively;][]{Pottasch2004}. With this abundance, detecting \ci RRLs requires data with significantly lower noise levels (the modeling predicts intensities of $\sim 10^{-3}\,$mK)or a meaningful opacity enhancement related, for instance, to high density shells. Indeed, the \ci lines have been detected coming from outside the \ion{H}{ii} regions of \mbox{NGC 7009} \citep{Akras2024} showing a bipolar structure, which is likely associated to the interaction of low expanding  neutral shells and more recent higher velocity outflows. Our beam is probably smaller than the neutral shells outside the \ion{H}{ii} region of \ngc and the high velocity winds known to exist in this PN expand in a different direction to the line-of-sight. The emission related to possible interactions like those in \mbox{NGC 7009} are likely absent in our \ngc data but mapping this PN in a wider field-of-view could unveil \ci RRLs emission.

Interestingly, the observed He lines are narrower than the H ones for both sources. In the case of \ic, the range of FWHMs slightly overlaps ($28 \pm 5$\kms and $16 \pm 7$\kms for H and \hei, respectively), but the FWHM distributions for H and He can be taken as statistically different. This clear difference may indicate that the emission from the atomic H and He lines comes from specific locations across the nebula. The first ionization energy for He is $24.6\,\mathrm{eV}$ \citep{Kandula2011}; i.e., almost twice as high as the H ionization energy, while the second ionization energy for He is $54.4\,\mathrm{eV}$ \citep{Yerokhin2015}. This means that only the most energetic photons can ionize the He atoms and these energetic photons should be mostly absorbed closer to the CS. This is because the opacity of the gas increases for shorter wavelengths. As we move away from the CS, there are less photons energetic enough to ionize other species than H atoms. The majority of the emission of ionized He producing RRLs is located close to the CS because the energetic photons are absorbed there. Therefore, H RRLs mostly trace the outer shells of the ionized nebula, while He RRLs trace the inner parts (with a smaller solid angle than the outer layers). Moreover, having different layers traced by different species relates the distinct FWHMs with a radial velocity gradient.

The broadening of RRLs is the convolution of two main components: the Gaussian contribution and the gas expansion velocity effects \citep[see e.g.,][]{Kielkopf1973, BaezRubio2013}. The Gaussian contribution is produced by thermal and turbulent motions. For the \ic H RRLs, this gives FWHM $\sim 10$\kms; i.e., H RRLs narrower than observed and predicted, with FWHM $\sim 28 \pm 5$\kms. Thus, the total broadening of RRLs is mainly dominated by the expansion velocity. Therefore, wider lines come from outer shells than narrower ones. In the specific case of \ic, the gas closer to the CS (traced by \hei) expands slower than the gas from the outer layers (traced by H).

The situation for \ngc is apparently more complicated. The three FWHM mean values are $38 \pm 10$\kms, $29 \pm 7$\kms, and $33 \pm 6$\kms for H, \hei, and \heii lines, respectively, so the ranges overlap. In contrast with \ic, in which given a FWHM value of any line we can determine without doubts if it is a H or a \hei RRL, in \ngc the lines are statistically similar. Consequently, most of the expected gas acceleration occurs within the region where He is ionized and beyond the gas can be assumed to expand at velocities close to the terminal value. One of the main differences between both PNe is that the central star of \ngc has an effective temperature of \mbox{$\sim 200\,000$\kel} (while \mbox{$T_\mathrm{eff} \sim 30\,000$\kel} for \ic CS), so there are more energetic photons. As they move away from the central star, they interact with the gas around the star ionizing it and losing energy. However, the photons emitted through the recombination processes are still energetic enough at larger distances than for \ic. Therefore, He can be ionized and recombined at further distances and there is not a clearly differentiated region where one species dominates over the other.

The good agreement of the \coral results with the detected lines shows the good quality of the models when modeling and predicting an unprecedented amount of RRLs. The \coral predictions point out that the tentative lines in the spectra are indeed below the detection level, which means that these lines are very likely real and could be finally confirmed if we performed yet deeper integrations. The model reproduces quite well the FWHM values of the H and He lines for both sources. We remark that only $T_\mathrm{e}$ values of $22\,000 - 25\,000$\kel can reproduce simultaneously both the SED and all RRLs of H and He in \ngc. The $T_\mathrm{e}$ values derived from other lines like [\ion{O}{iii}] do not reproduce the SED and/or all RRLs. See, for example, the work by \cite{Zhang2005} and their temperature estimates based on collisionally excited lines and optical recombination lines in this source. Their results using collisionally excited lines show values of 10\,400\kel ([\ion{O}{i}] $\lambda 5577/(\lambda 6300 + \lambda 6363$)) or 8\,100\kel ([\ion{C}{i}] ($\lambda 9824 + \lambda 9850)/\lambda 8727$), which also differ from the results of the Balmer jump temperature obtained using other lines in the optical range radiatively excited (12\,600\kel from the [\ion{O}{iii}] ($\lambda 4959 + \lambda 5007)/\lambda 4363$ or 18\,700\kel from the [\ion{O}{ii}] ($\lambda 7320 + \lambda 7330)/\lambda 3726$). This kind of discrepancies were noticed long ago and that collisionally excited temperatures are lower than those derived from continuum measurements \citep[see e.g.,][]{Miller1972}. Recently, authors as \cite{SanchezContreras2024a} have also warned about the impossibility of deriving $T_\mathrm{e}$ values from the collisionally excited lines reproducing the SED of the pre-PN \mbox{M2-9} at mm frequencies. The discrepancy between $T_\mathrm{e}$ derived from the RRL and from collisionally excited lines as [\ion{O}{iii}] is still an open question. Nevertheless, \cite{Krabbe2005} showed a strong dependence of $T_\mathrm{e}$ with $n_\mathrm{e}$, which suggests that $n_\mathrm{e}$ and associated spatial structures could be playing a relevant role. Therefore, further research is needed to solve these discrepancies and provide a robust explanation.

Furthermore, no maser amplification is expected according to our models. After obtaining the set of model parameters that best fit both the RRLs and the SED, we looked for the combination of $T_\mathrm{e}$ and $n_\mathrm{e}$ producing maser emission in \ic. Independently of $T_\mathrm{e}$, the maser amplification effect is predicted for $N_\mathrm{e} > 3 \cdot 10^4$\cmcub, which is higher than our best estimate. We have to point out that maser amplification also requires optical paths along the line of sight producing enough coherence in velocity to allow an exponential amplification. In our models, most of the emission is produced in a relatively thin shell and there is not any combination of $T_\mathrm{e}$ and $n_\mathrm{e}$ producing maser amplification. Therefore, considering both the required conditions for maser amplification and model results, we can confidently rule out any maser emission in \ic. Similar results and conclusions are obtained for \ngc.

The predicted FWHMs display an interesting behavior as frequency decreases or as $n_\mathrm{lower}$ increase (see Fig.\ref{fig:IC418_FWHM-nlow}). The RRLs seem to be wider, but we do not observe this tendency in the observational results. This broadening is a feature of the model. The intensity of the lines depends directly on their Einstein coefficients, which are calculated using the Gaunt factors. Considering LTE or non-LTE provides slightly different results for the Gaunt factors. The \coral code uses the non-LTE Gaunt factors only for $n_\mathrm{lower} \leq 100$. Lines with $n_\mathrm{lower} \geq 100$ (observed in the 7\mm range) use the LTE Gaunt factors to calculate their Einstein coefficients. Even though the difference is quite small, the opacity of the lines with the LTE Gaunt factor is slightly larger than that of the lines using non-LTE Gaunt factors. This is, the opacity of lines with $n_\mathrm{lower} \geq 100$ (LTE) is slightly larger than the opacity of lines with $n_\mathrm{lower} \leq 100$ (non-LTE). The final effect of this changing of the thermodynamic regime is a broadening of the lines with $n_\mathrm{lower} \geq 100$, which are only visible at 7\mm. Finding a computationally efficient way to obtain the non-LTE Gaunt factor for $n_\mathrm{lower} \geq 100$ would correct this effect.

The outputs (i.e., physical parameters) of \coral give useful information about the physical characteristics of both PNe. This information is given in Table \ref{tab:coral_param}. The outputs of \ic are compatible with the bibliography \citep[see e.g.,][]{Morisset2009, RamosLarios2012}. We obtain a mean kinematic time scale compatible with the timescale of \cite{Dopita2017}. The mass loss rate is similar, but higher than that derived by \cite{Dopita2017} using post-AGB evolutionary models. This fact also justifies the different time scales. Our model only reproduces the observed peak intensities of all H, \hei RRLs with a $T_\mathrm{e}$ higher than previous works \citep[see e.g.,][]{Pottasch2004, Morisset2009, RamosLarios2012, Dopita2017}. In the case of \ngc, the physical parameters derived from \coral are also compatible with previous results. Time scales vary from $600 - 700\,$a \citep{Masson1989, Latter2000}, making our kinematic time scale results compatible with previous models \citep[see also][]{MoragaBaez2023}. There has been a wide range of $T_\mathrm{e}$, which strongly depend on the spectral line used to calculate them \citep[see e.g.,][]{Chaisson1976, Atherton1979, Walmsley1981, BernardSalas2001}. It is interesting to point out that we obtained very similar ionized masses for both PNe as \cite{SantanderGarcia2022}, which used quite different data analysis methods.

The RRLs are a very useful tool to study ionized gas regions, and more precisely, PNe. The measurement of RRLs of different elements is also a fundamental tool to analyze the atomic emission and its role in the evolution of the system. However, the knowledge of their existence could be very useful for molecular radio spectroscopy in ionized regions, where RRLs are present and may be mixed up with molecules.

The unambiguous detection of a molecular species in space is a difficult task. The list of identified individual species (including isotopologues) grows up to $\sim 300$ \citep{McGuire2022}. However, there are hundreds of UFs that have not been assigned to any molecular transition yet. The main difficulty is to estimate the spectra of molecules that are commonly hard to be synthesized in the laboratory or which cannot be determined with enough accuracy by means of complex ab initio calculations. The geometry of the molecule, a large number of atoms, or many overlapped electronic states increase the complexity of this task. Therefore, the largest molecules identified in space through their rotational transitions up to date have no more than 25 atoms \citep[see][]{Cernicharo2024b}.

The number of possible molecular carriers that may be associated with UFs is thus very large. The limitation of the number of truly UFs is crucial and this is usually done in two ways: (i) by comparing the UF frequencies with (public or private) molecular databases and (ii) identifying atomic emission lines. The RRL catalogs presented here may play a key role on this problem of the identification of the UF carriers. The identification of RRLs  in our sample of sources allowed us to correctly associate the atomic emission lines with their atomic carriers (i.e., H and He) and list the remaining emission lines (i.e., UFs) to be associated with their still unknown molecular carriers. By ignoring the RRLs in \ic, a peculiar PN without any known molecule detected at radio wavelengths, would have lead us to deal with dozens of emission lines (UFs) that do not correspond to molecular transition frequencies. In the case of \ngc, a PN with confirmed molecular emission at mm wavelengths, we could correctly identify some RRLs emission contamination to well known molecular emission lines as well as some molecular emission contamination to the RRLs. In short, the detection of the most intense RRLs of H and He made possible to correctly model both sources and predict the peak intensities of much weaker lines, as well as to distinguish their atomic or molecular origin.

\section{Conclusions} \label{sec:conclusions}

The unprecedented high-sensitivity mm data obtained towards PNe \ic and \ngc at three frequency bands has allowed us to report for the first time the detection of 323 RRLs, many of them never detected before in a PN and even in any astrophysical source. Additionally, other 50 weak features along the three frequency bands and below the 3$\sigma$ detection limit can be tentatively identified as RRLs. Every emission line was carefully compared with a complete list of RRL frequencies of H, \hei, and \heii. Those emission lines coincident with the frequency of a RRL have been carefully fitted with a Gaussian function and modeled with the \coral radiative transfer model. The spectra derived from \coral are compatible with the observational results within the error bars across the frequency ranges observed with the \mbox{IRAM 30m} and RT40m radio telescopes. We have also obtained the main nebular physical parameters from \coral to describe the physical characteristics of the sources.

These catalogs of detected H, \hei, and \heii RRLs towards two C-rich PNe constitute a significant breakthrough for a better understanding and characterization of the chemical environment of a PN. The reported lines confirm the presence of many RRLs that have not been observed or identified in previous studies. A revision on ionized sources should be done to be sure if some UFs are RRLs or some molecular emission lines are contaminated by RRLs.

The detection of this large number of RRLs leaves some open questions that have to be confirmed with interferometric maps and lower frequency observations. Is the spatial distribution of the species similar to that derived from the modeling? Are RRLs of heavier elements detectable towards denser regions of the PNe? Do RRLs become broader at lower frequencies, so there is a limit on differentiating them with ripples in the spectra? Do RRLs at lower frequencies provide different/more information of the sources? It would be interesting to have a better knowledge on the behavior of RRLs at lower frequencies and check if a limit on the detections exists.

The study and characterization of RRLs seems to be regaining importance in the field. New developments in receivers and radio telescopes provide the proper tools to perform this kind of sensitive and high-precision analysis of astronomical sources, which are essential to correctly confirm atomic and molecular emission lines in space. A better understanding of RRLs will improve our knowledge of the physics and chemistry of the emitting regions. The RRL catalogs presented here provide a tool to identify radio emission lines and to know which ones are truly produced by molecules; something that will help for the identification of new molecular species, improving our present understanding of the molecular formation pathways in space. In particular, the results presented in this paper have provided us with lists of weak molecular UFs towards both PNe (i.e., still unidentified emission lines that are not RRLs). Our next efforts will focus on the identification of the molecular carriers producing such weak and unidentified radio emission lines in PNe.

\begin{acknowledgements}
	The authors acknowledge Miguel Santander-García from the OAN for carrying out the observations at the Yebes RT40m. THR, DAGH, AMT, and RB acknowledge the support from the State Research Agency (AEI) of the Ministry of Science, Innovation and Universities (MICIU) of the Government of Spain, and the European Regional Development Fund (ERDF), under grants PID2020-115758GB-I00/AEI/10.13039/ 501100011033 and PID2023-147325NB-I00/AEI/10.13039/501100011033. THR acknowledges support from grant PID2020-115758GB-I00/PRE2021-100042 financed by MCIN/AEI/10.13039/501100011033 and the European Social Fund Plus (ESF+). JA, JPF, JJDL, and VB are partially supported by I+D+i projects PID2019-105203GB-C21 and PID2023-146056NB-C21,  funded by the Spanish MCIN/AEI/10.13039/501100011033 and EU/ERDF. JPF also acknowledges support from grants PID2023-147545NB-I00 and PID2023-146056NB-C22. MAGM acknowledge to be funded by the European Union (ERC, CET-3PO, 101042610). Views and opinions expressed are however those of the author(s) only and do not necessarily reflect those of the European Union or the European Research Council Executive Agency. Neither the European Union nor the granting authority can be held responsible for them. This work is based on observations carried out under project number 158-21 with the IRAM 30m telescope. IRAM is supported by INSU/CNRS (France), MPG (Germany) and IGN (Spain). Based on observations carried out with the Yebes 40 m telescope (22A011). The 40 m radio telescope at Yebes Observatory is operated by the Spanish Geographic Institute (IGN; Ministerio de Transportes y Movilidad Sostenible). This publication is based upon work from COST Action CA21126 - Carbon molecular nanostructures in space (NanoSpace), supported by COST (European Cooperation in Science and Technology).
\end{acknowledgements}

\bibliographystyle{aa} 
\bibliography{RRLsPaper}

\begin{appendix}
	
\section{Examples of RRL profiles}

In this section we present some examples of H, \hei, and \heii profiles, along with their Gaussian fittings and the modeled profile.

\subsection{IC 418} \label{sec:profiles_IC418}

\begin{figure*}[!h]
	\centering
	\includegraphics[width=0.24\textwidth]{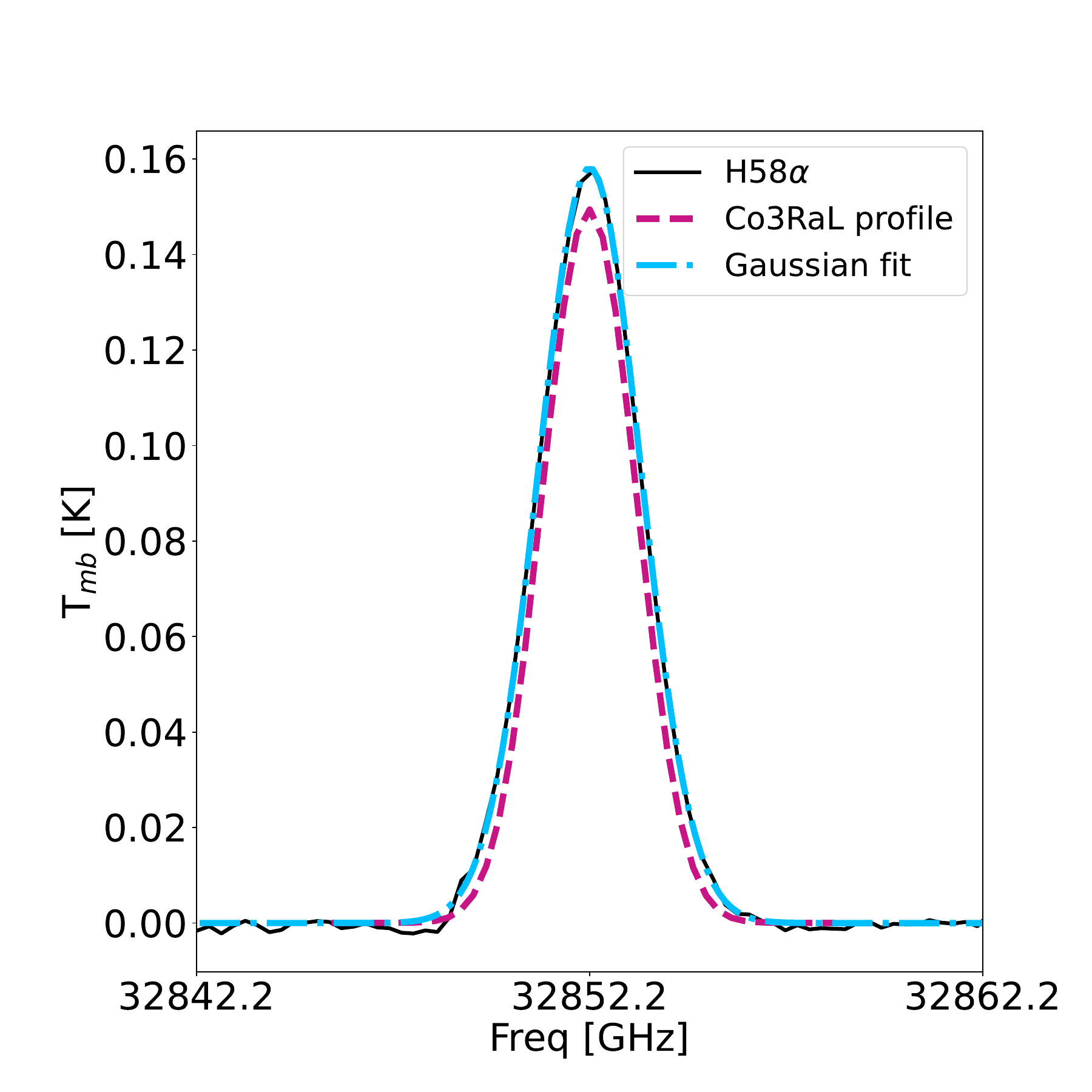}
	\includegraphics[width=0.24\textwidth]{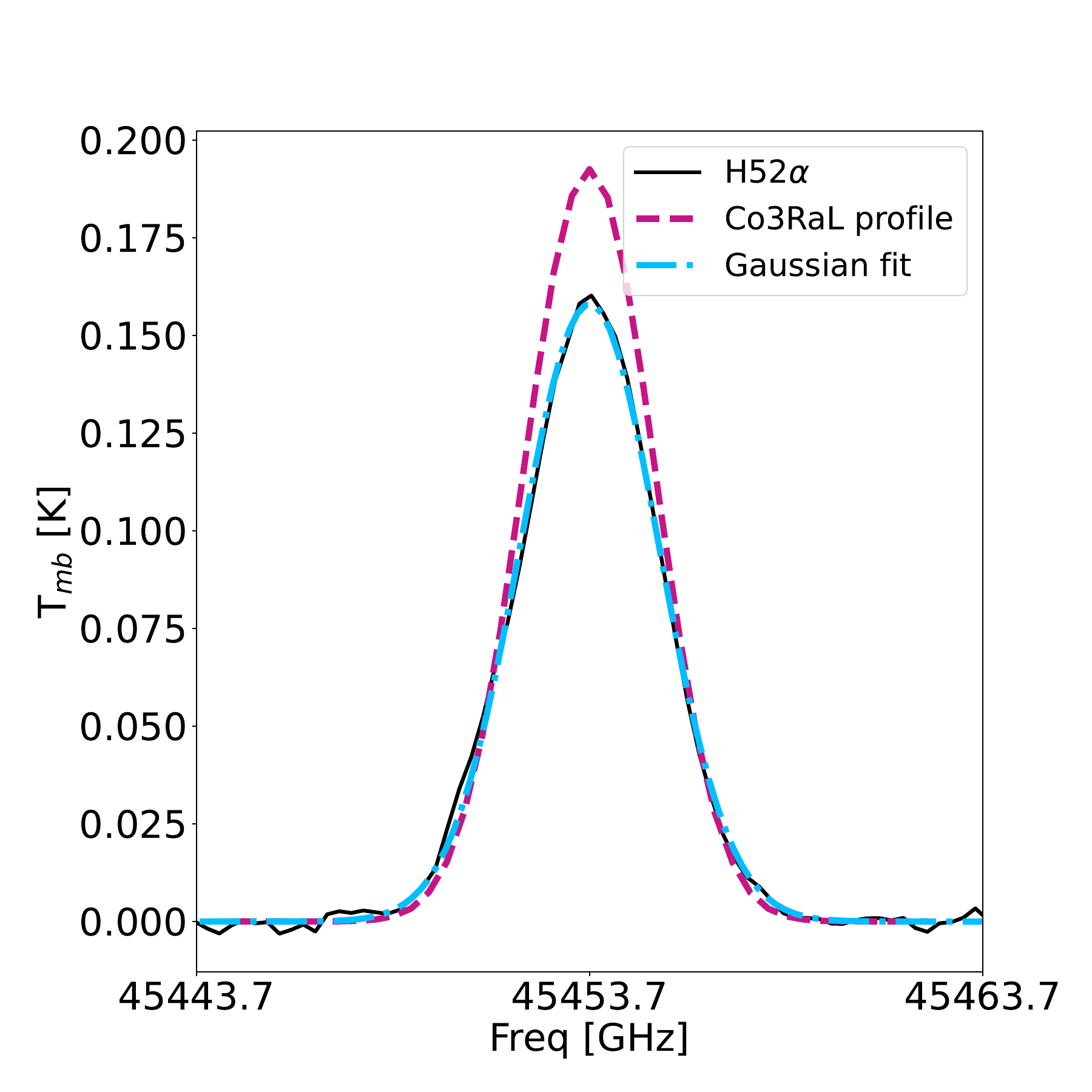}
	\includegraphics[width=0.24\textwidth]{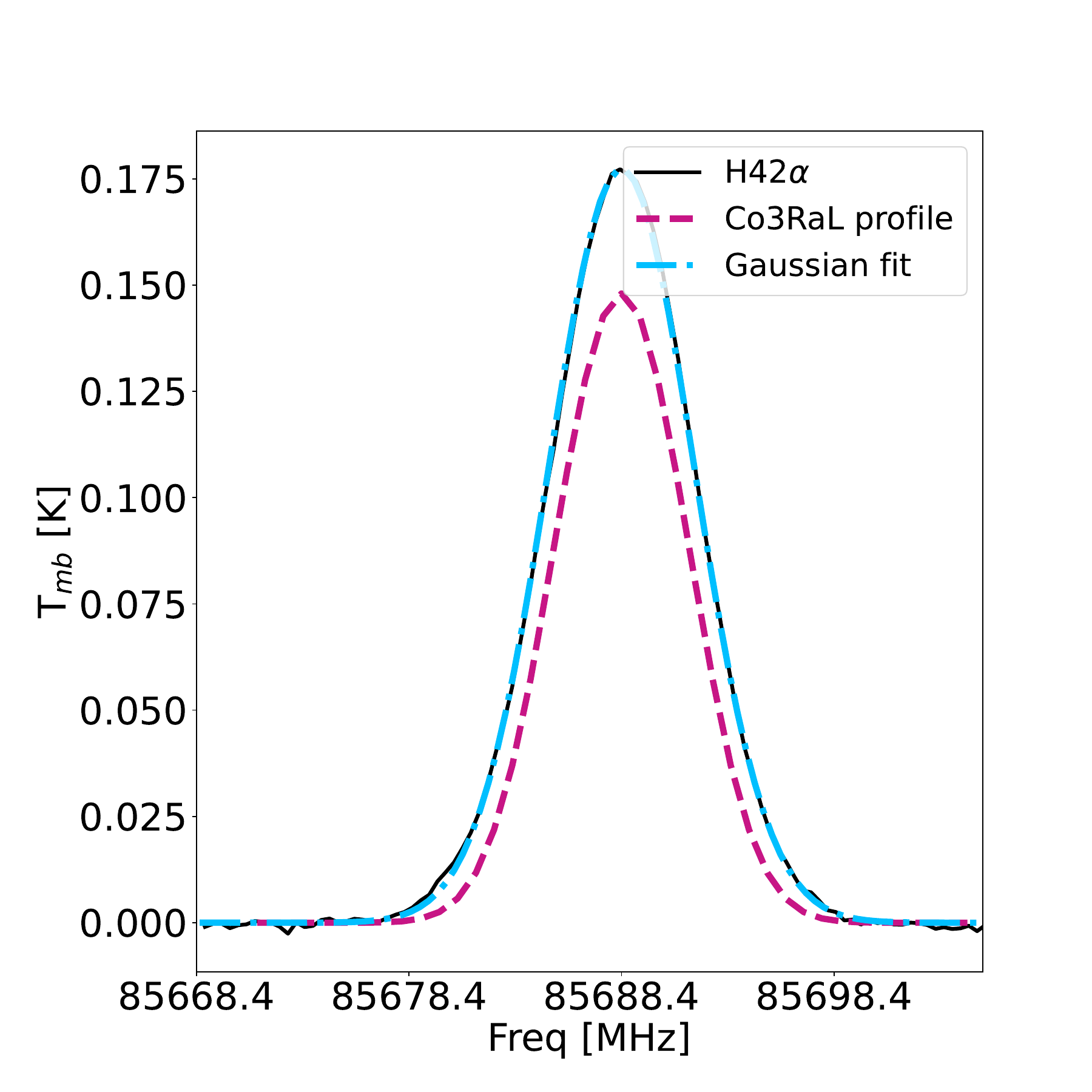}
	\includegraphics[width=0.24\textwidth]{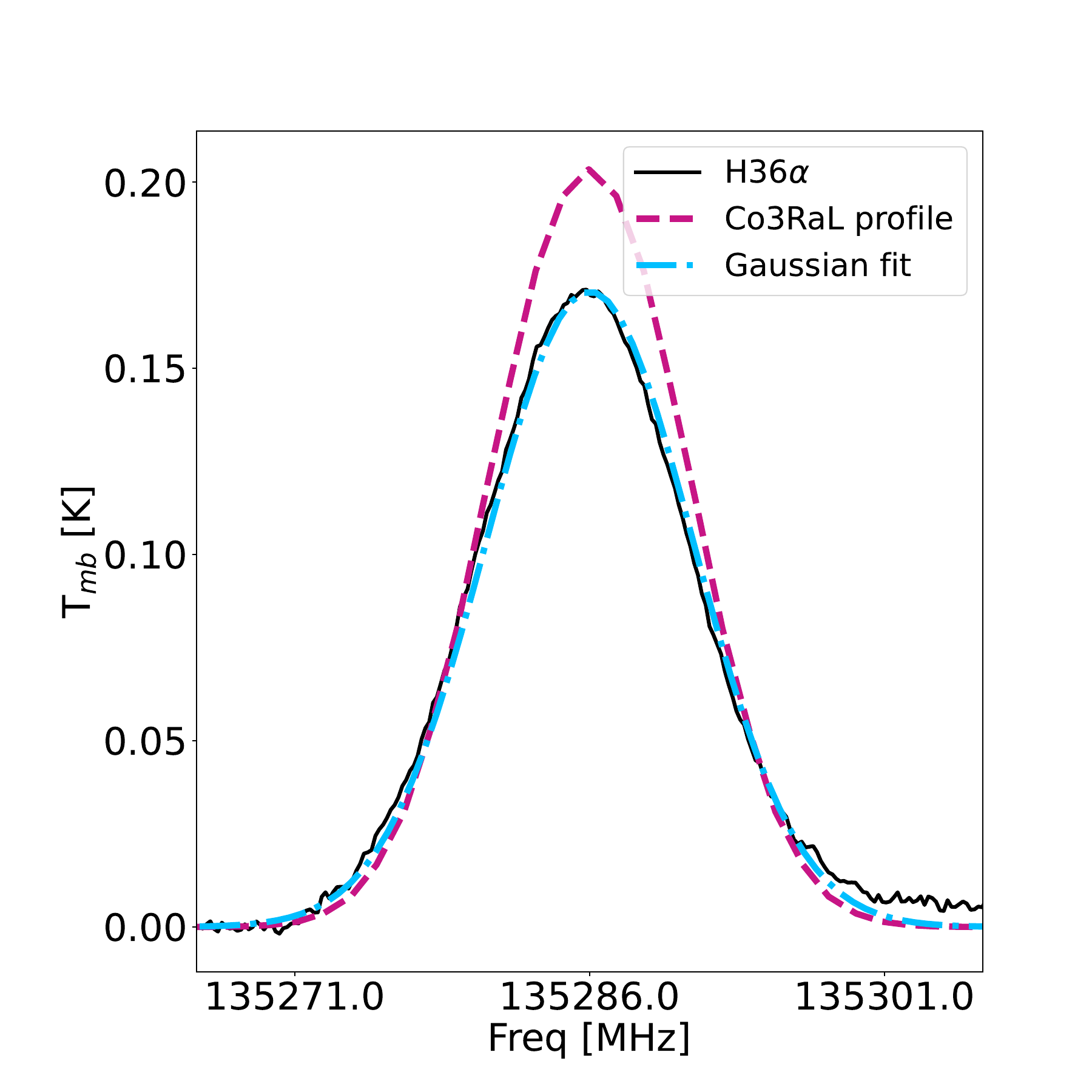}
	\caption{H\ga lines in \ic.} \label{fig:IC418_Hga}
\end{figure*}

\begin{figure*}[!h]
	\centering
	\includegraphics[width=0.24\textwidth]{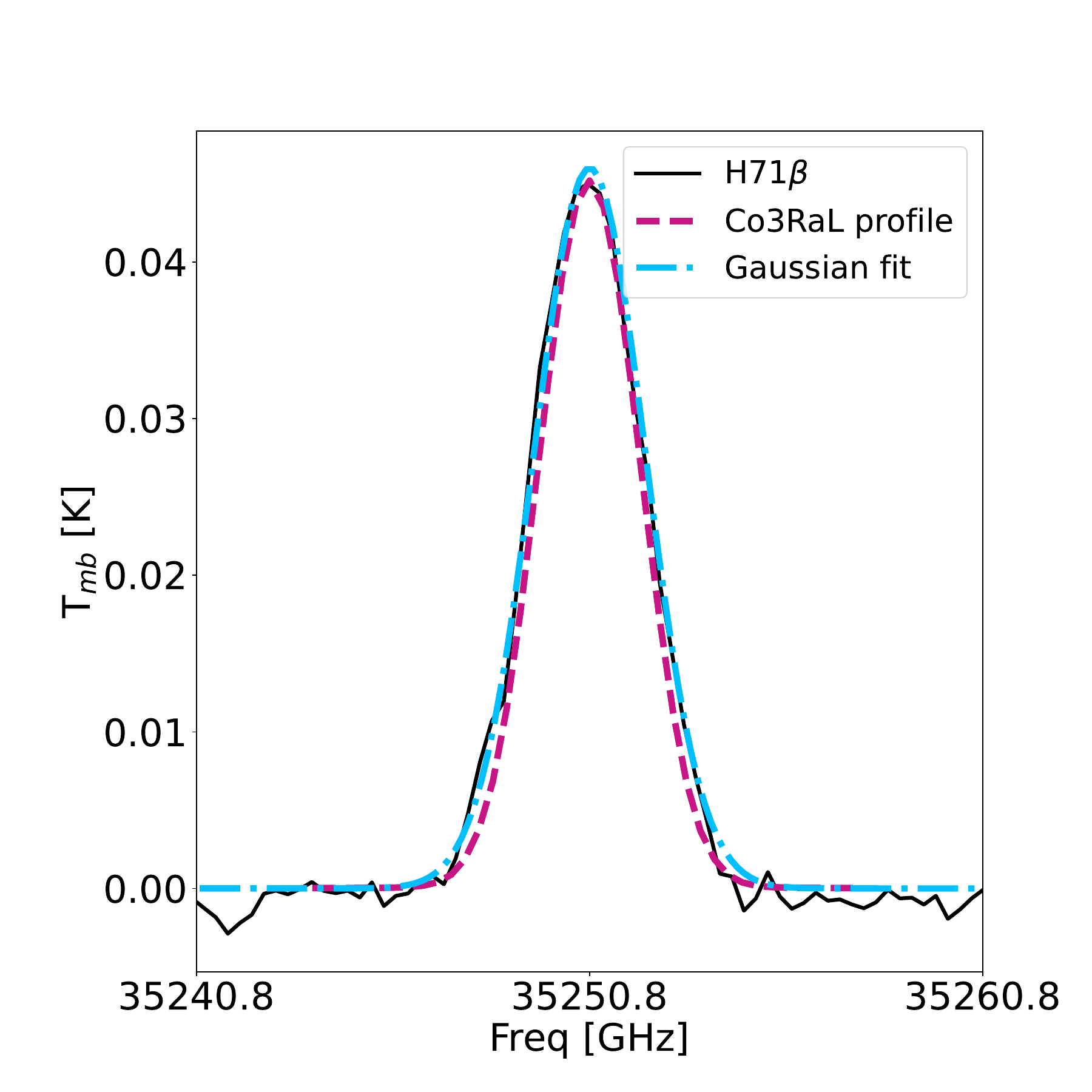}
	\includegraphics[width=0.24\textwidth]{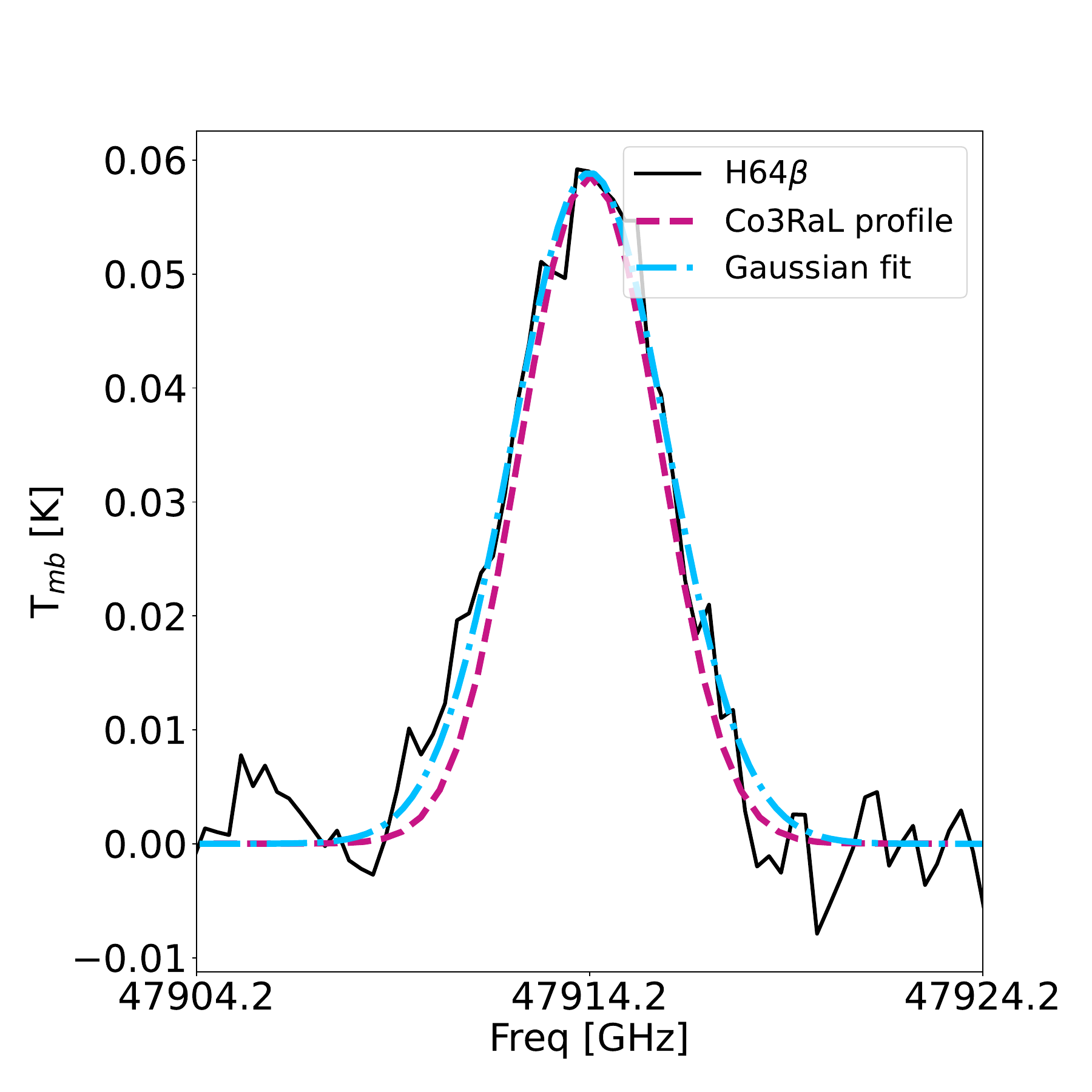}
	\includegraphics[width=0.24\textwidth]{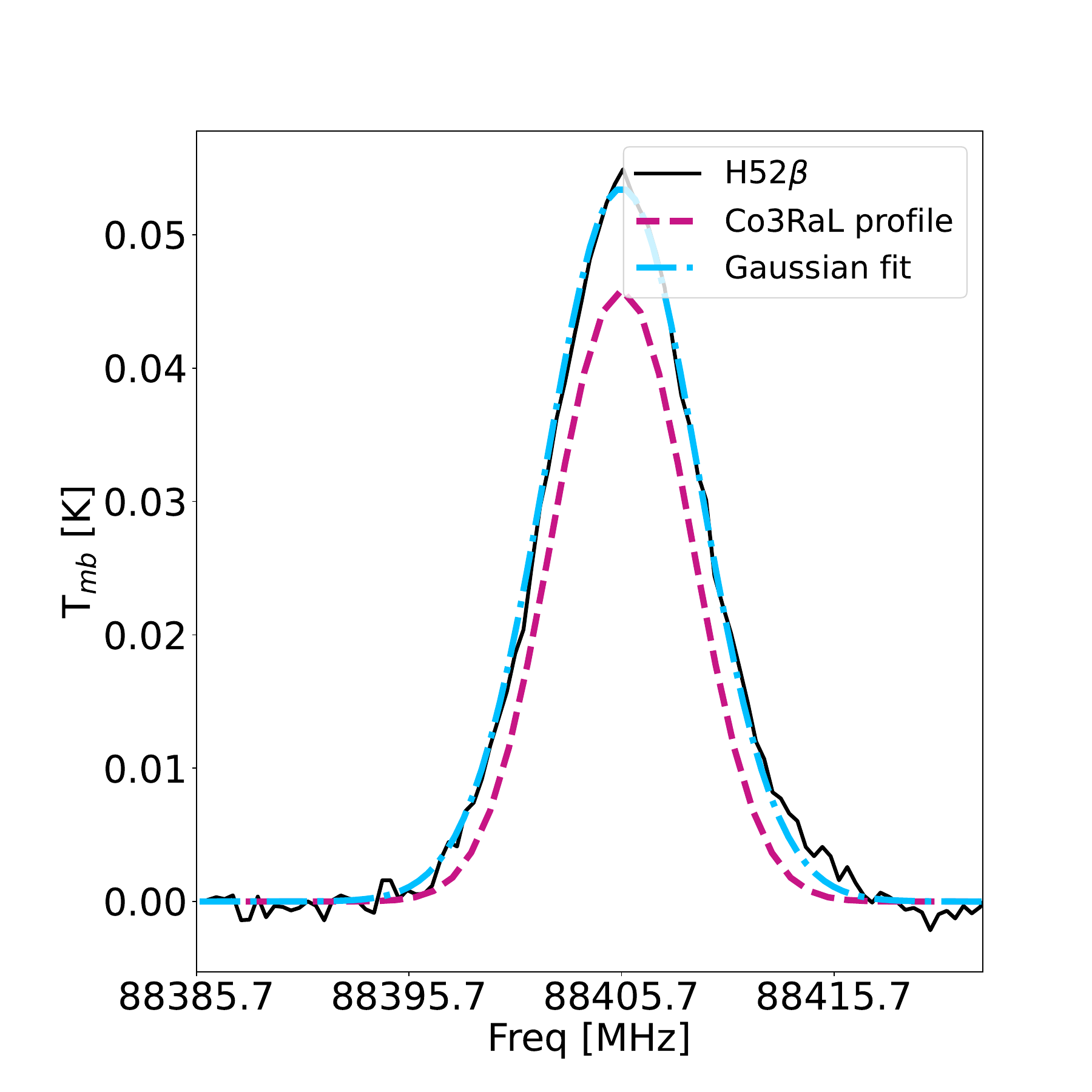}
	\includegraphics[width=0.24\textwidth]{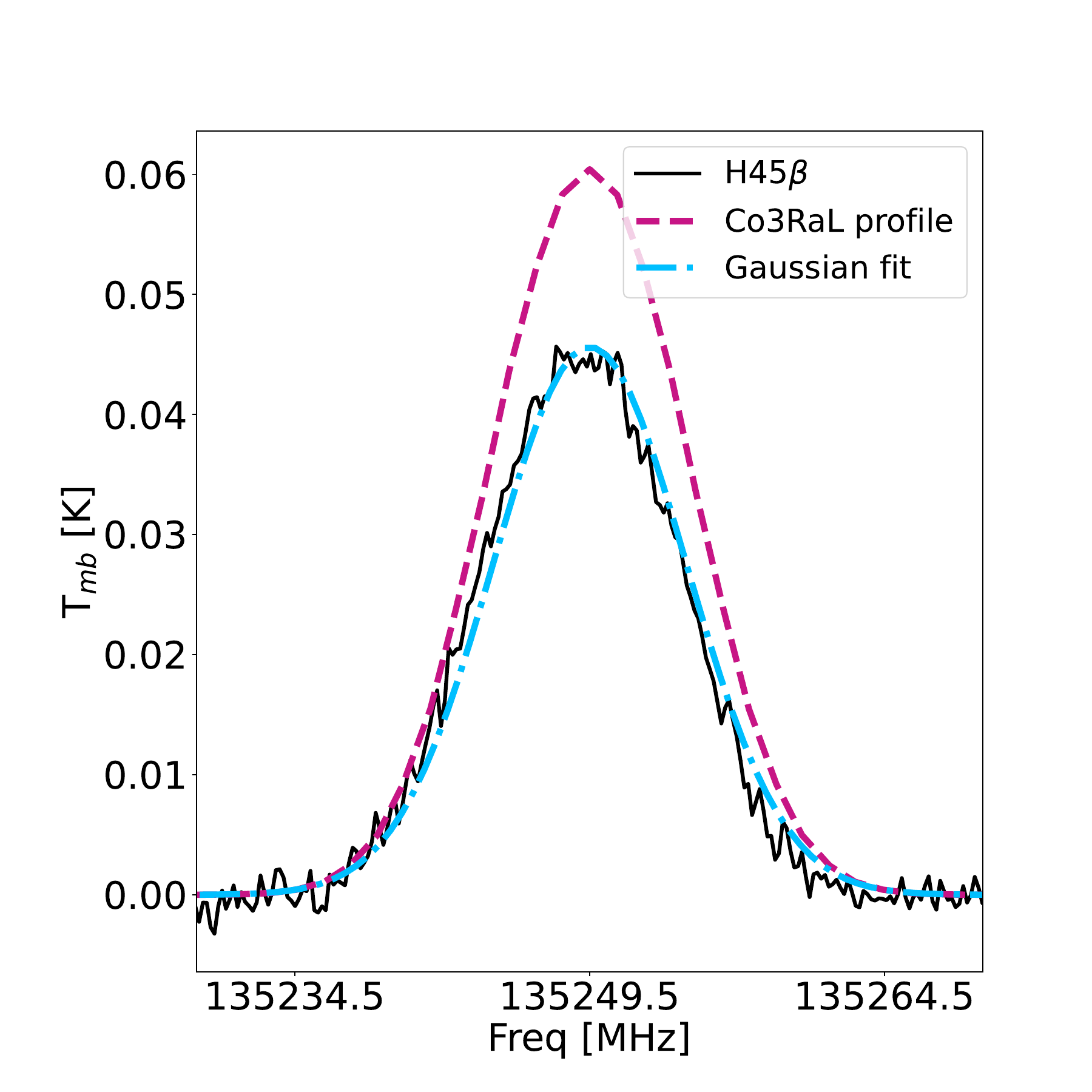}
	\caption{H\gb lines in \ic.} \label{fig:IC418_Hgb}
\end{figure*}

\begin{figure*}[!h]
	\centering
	\includegraphics[width=0.24\textwidth]{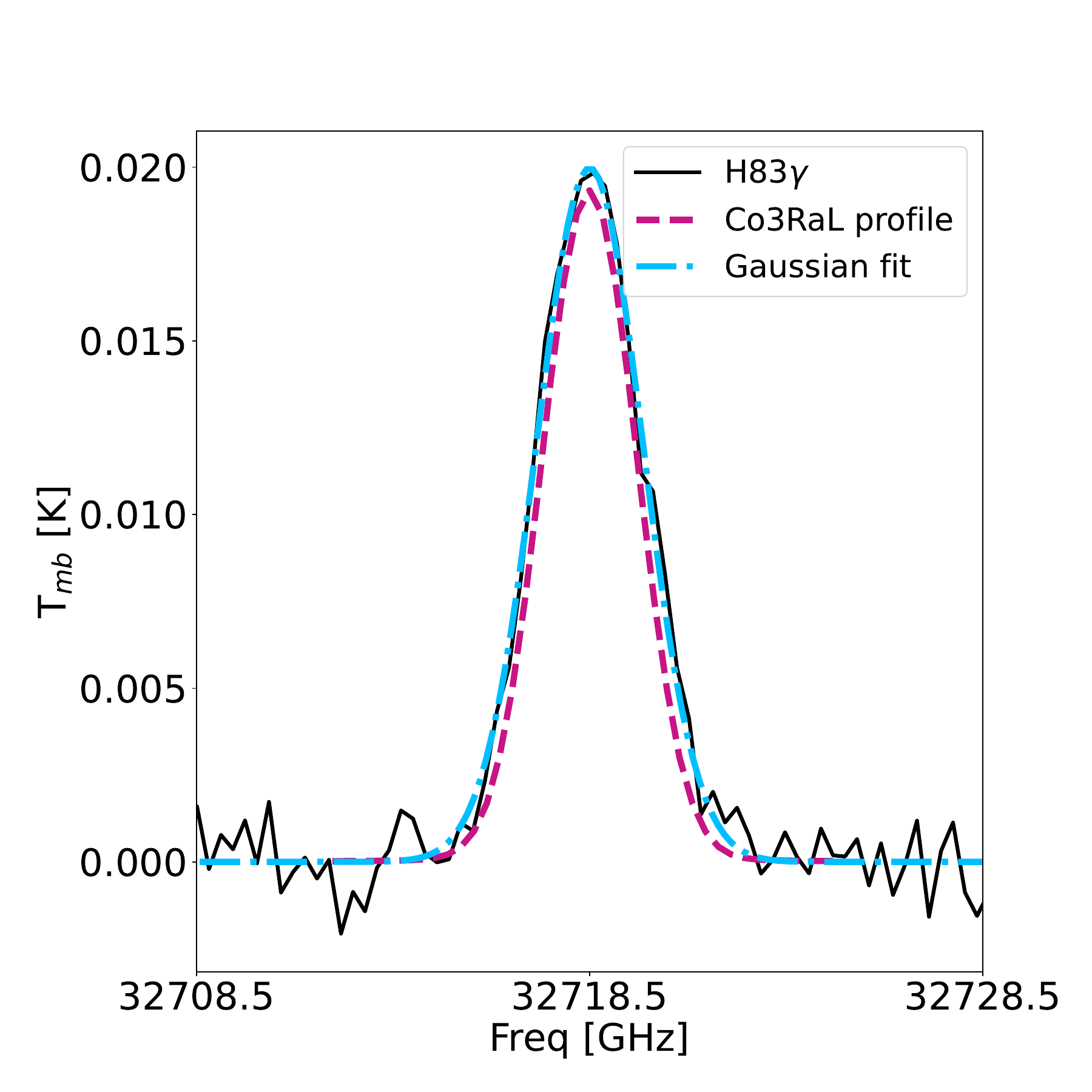}
	\includegraphics[width=0.24\textwidth]{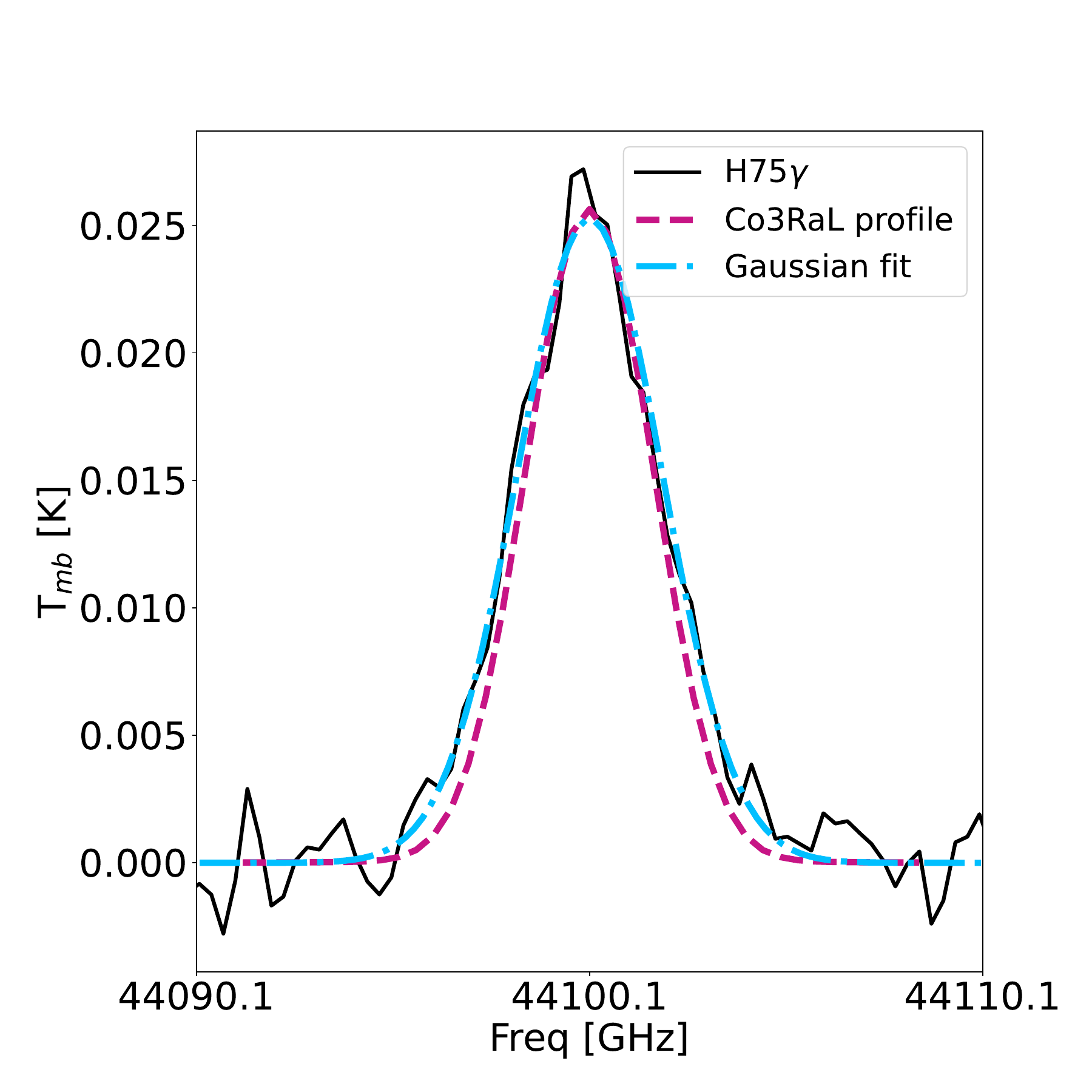}
	\includegraphics[width=0.24\textwidth]{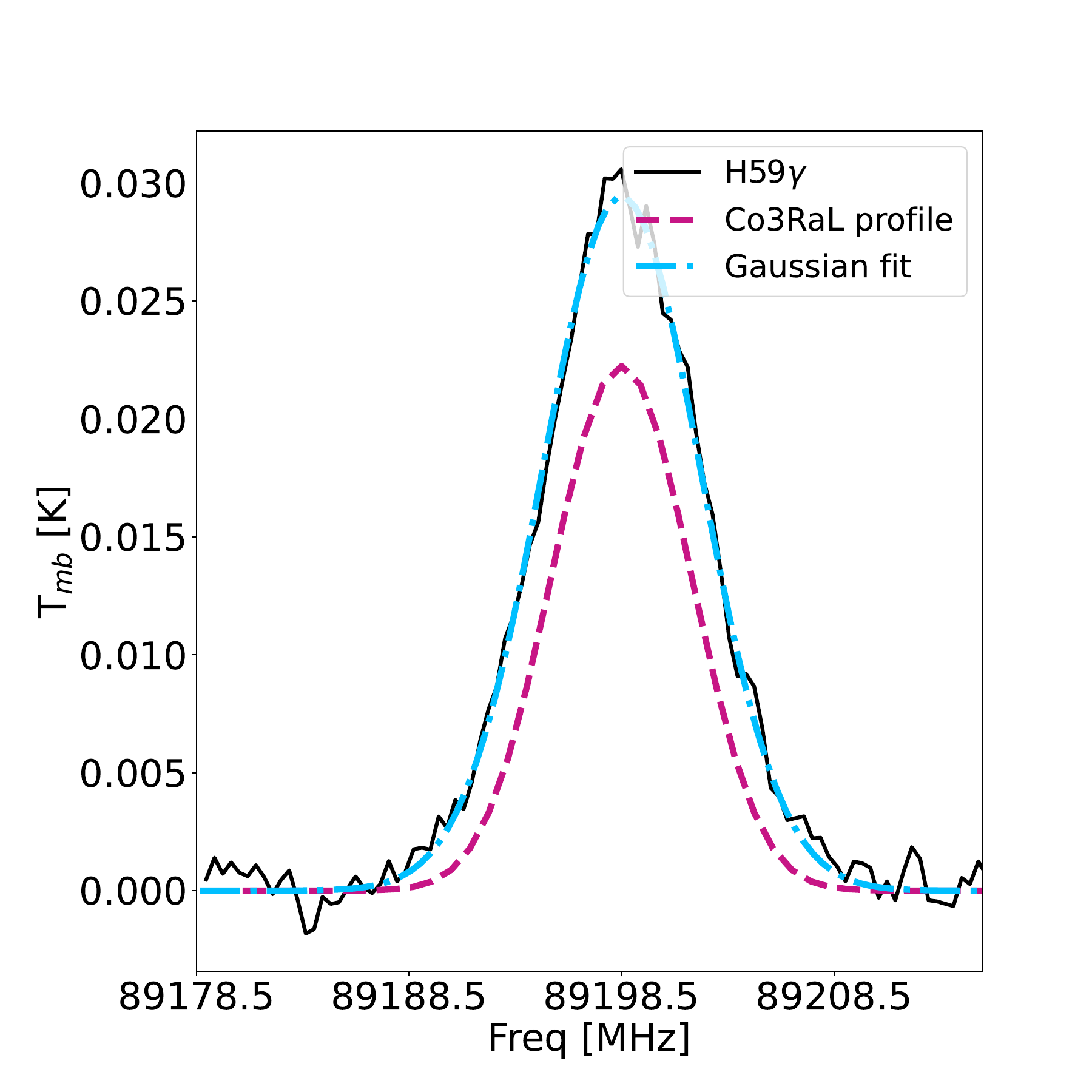}
	\includegraphics[width=0.24\textwidth]{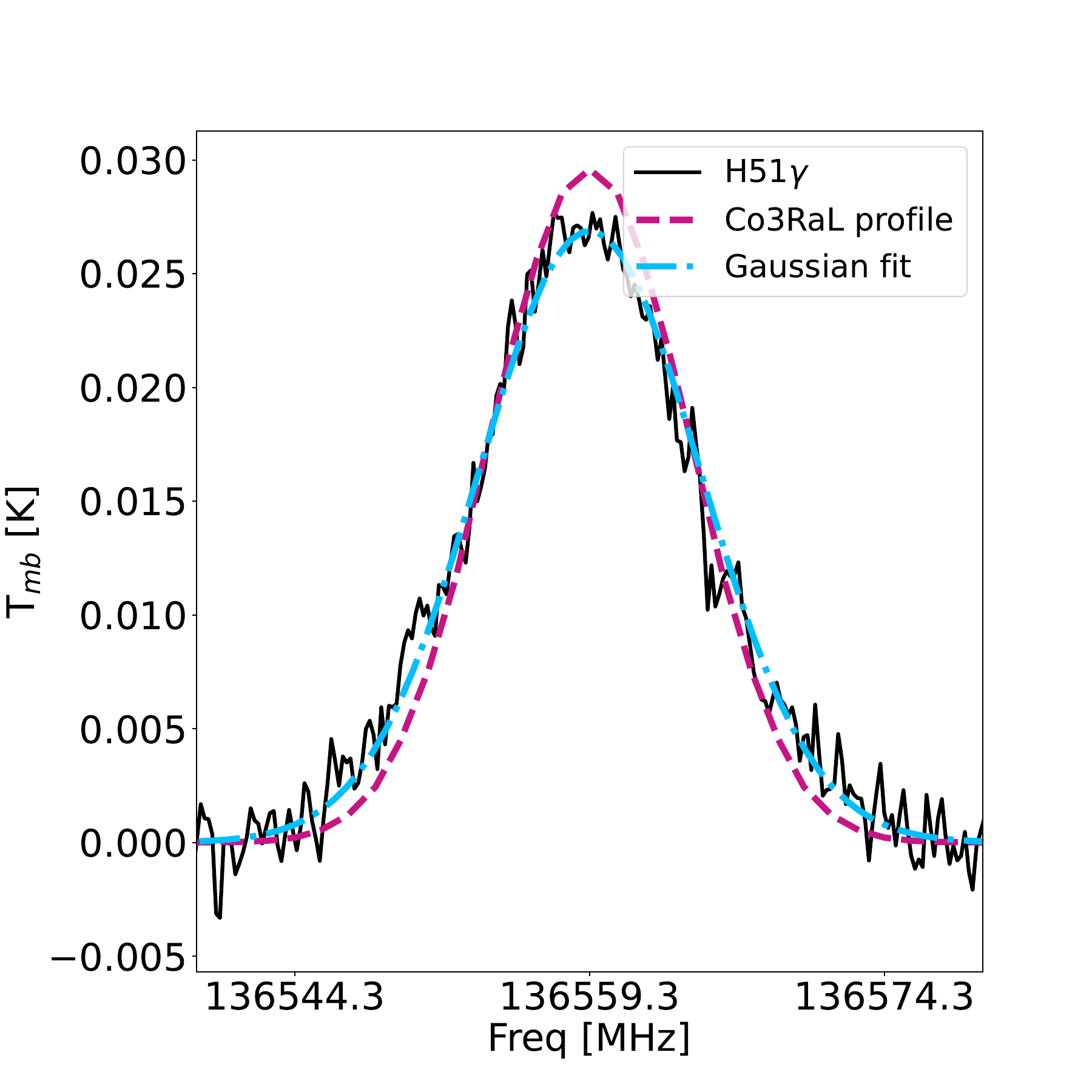}
	\caption{H\gg lines in \ic.} \label{fig:IC418_Hgg}
\end{figure*}

\begin{figure*}[!h]
	\centering
	\includegraphics[width=0.24\textwidth]{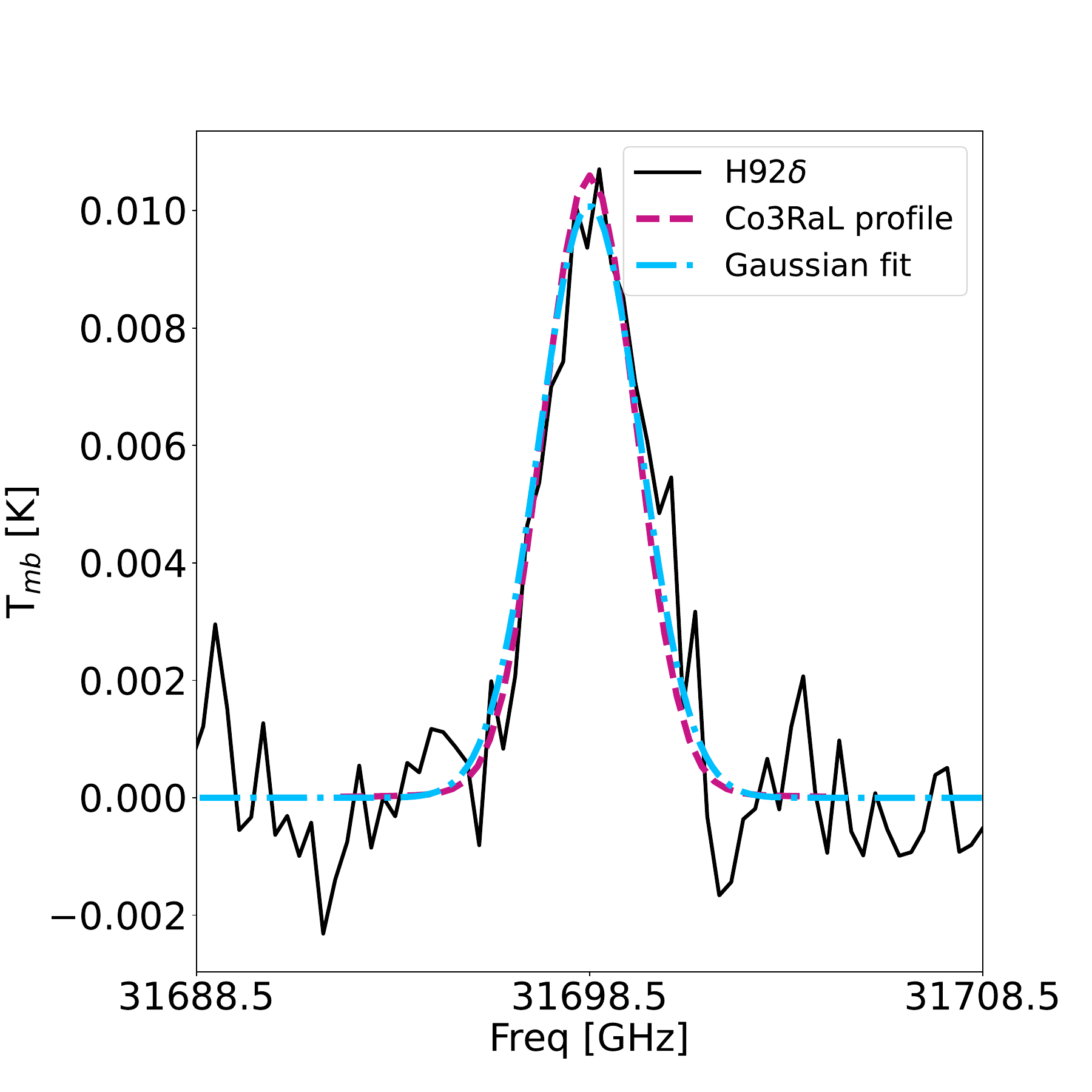}
	\includegraphics[width=0.24\textwidth]{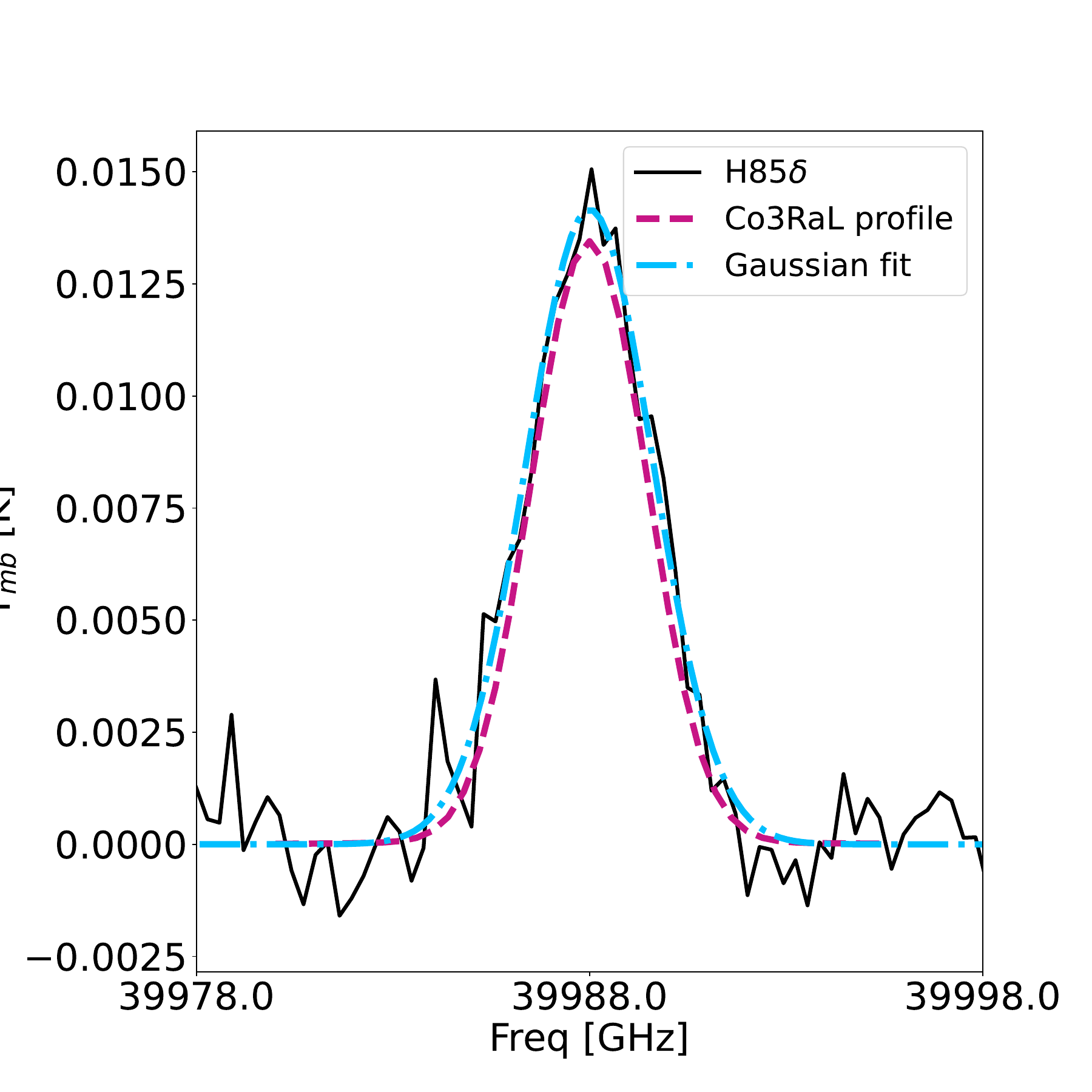}
	\includegraphics[width=0.24\textwidth]{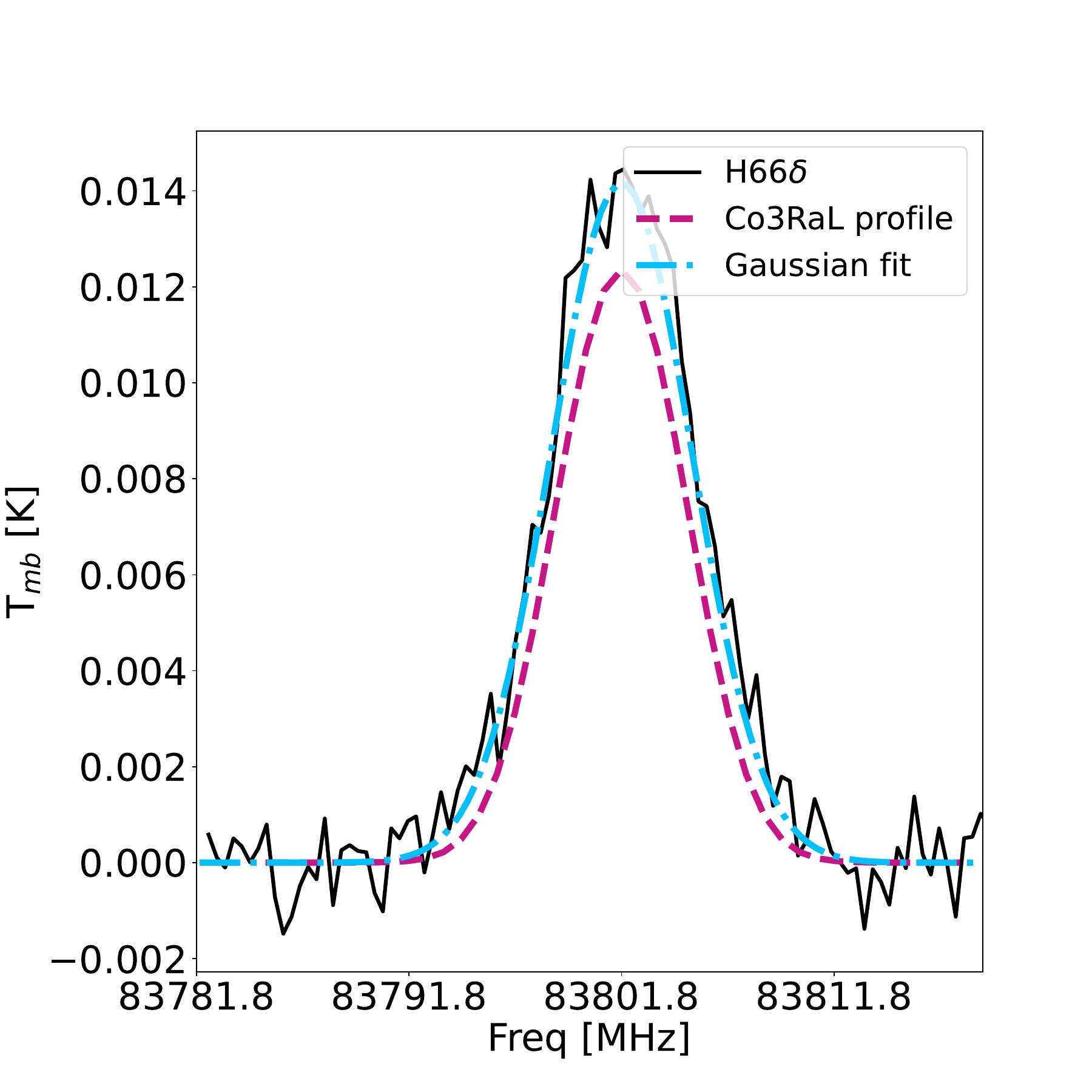}
	\includegraphics[width=0.24\textwidth]{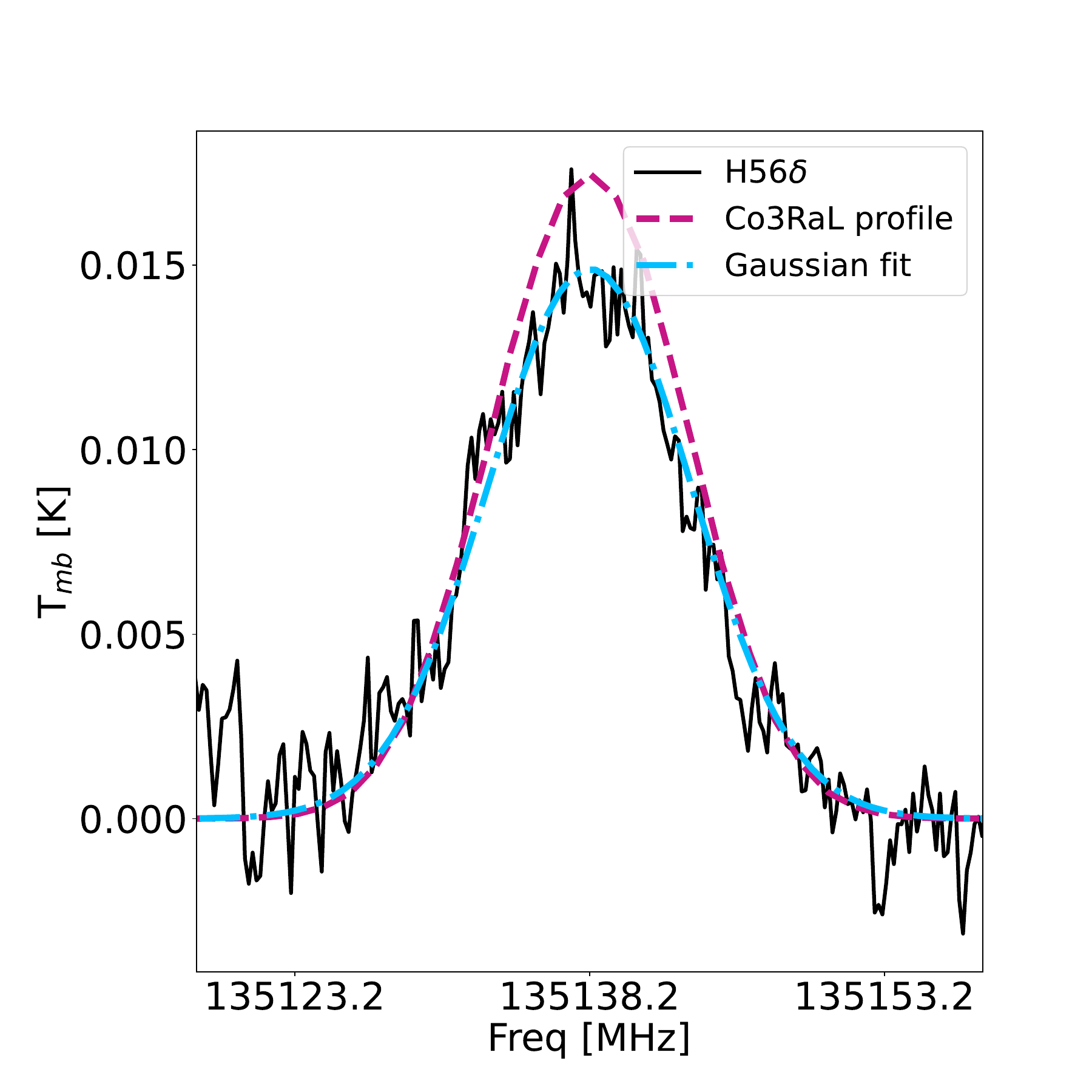}
	\caption{H\gd lines in \ic.} \label{fig:IC418_Hgd}
\end{figure*}

\begin{figure*}[!h]
	\centering
	\includegraphics[width=0.24\textwidth]{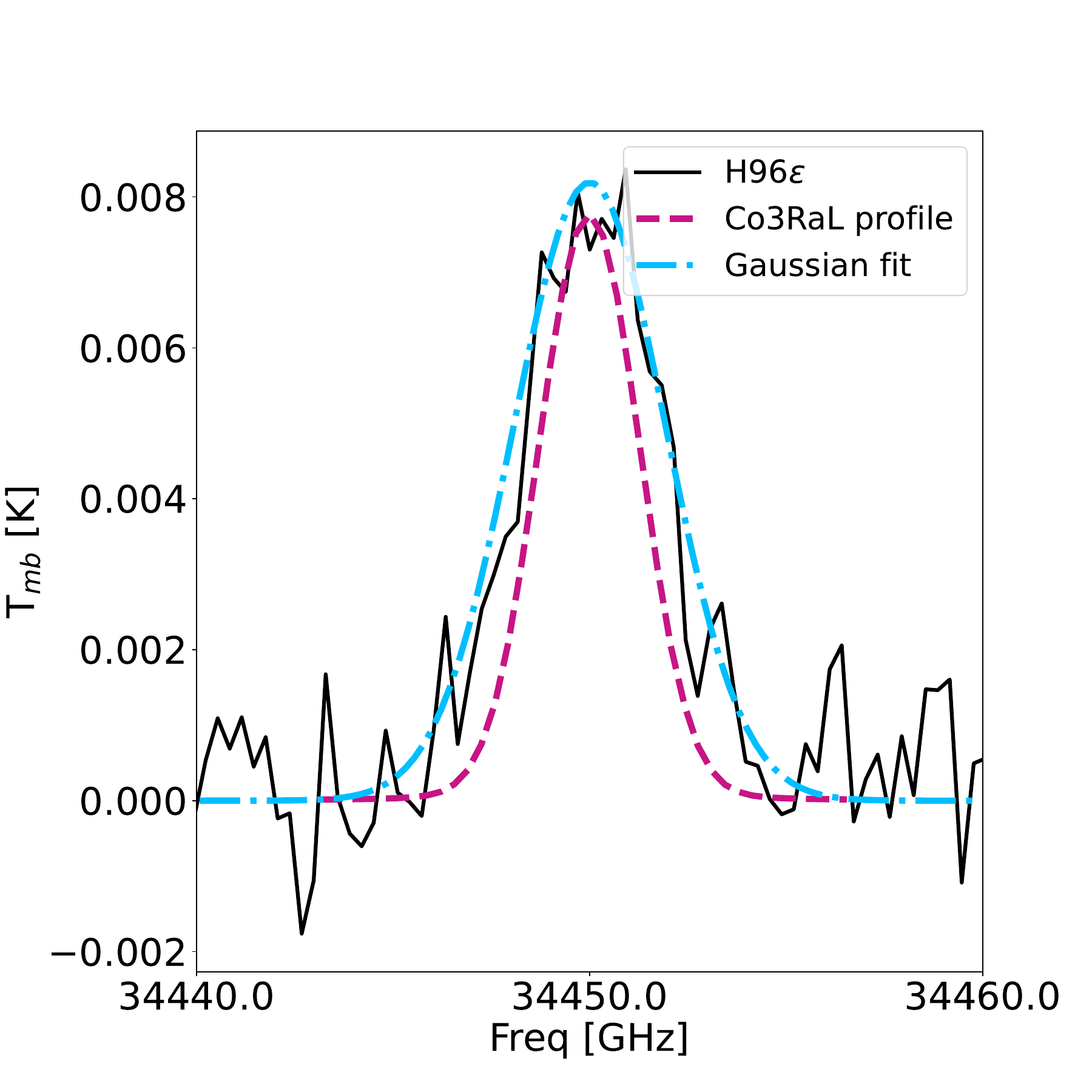}
	\includegraphics[width=0.24\textwidth]{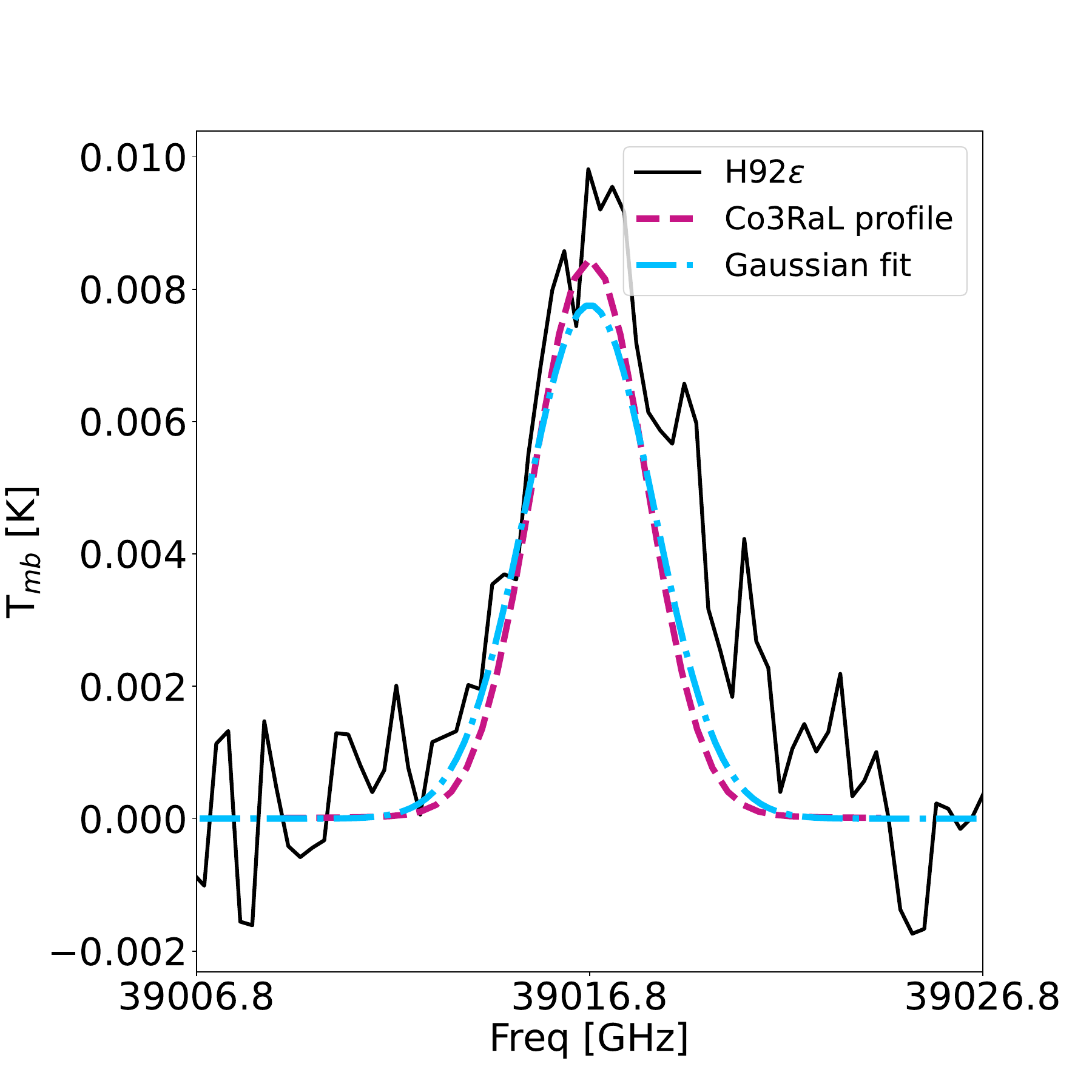}
	\includegraphics[width=0.24\textwidth]{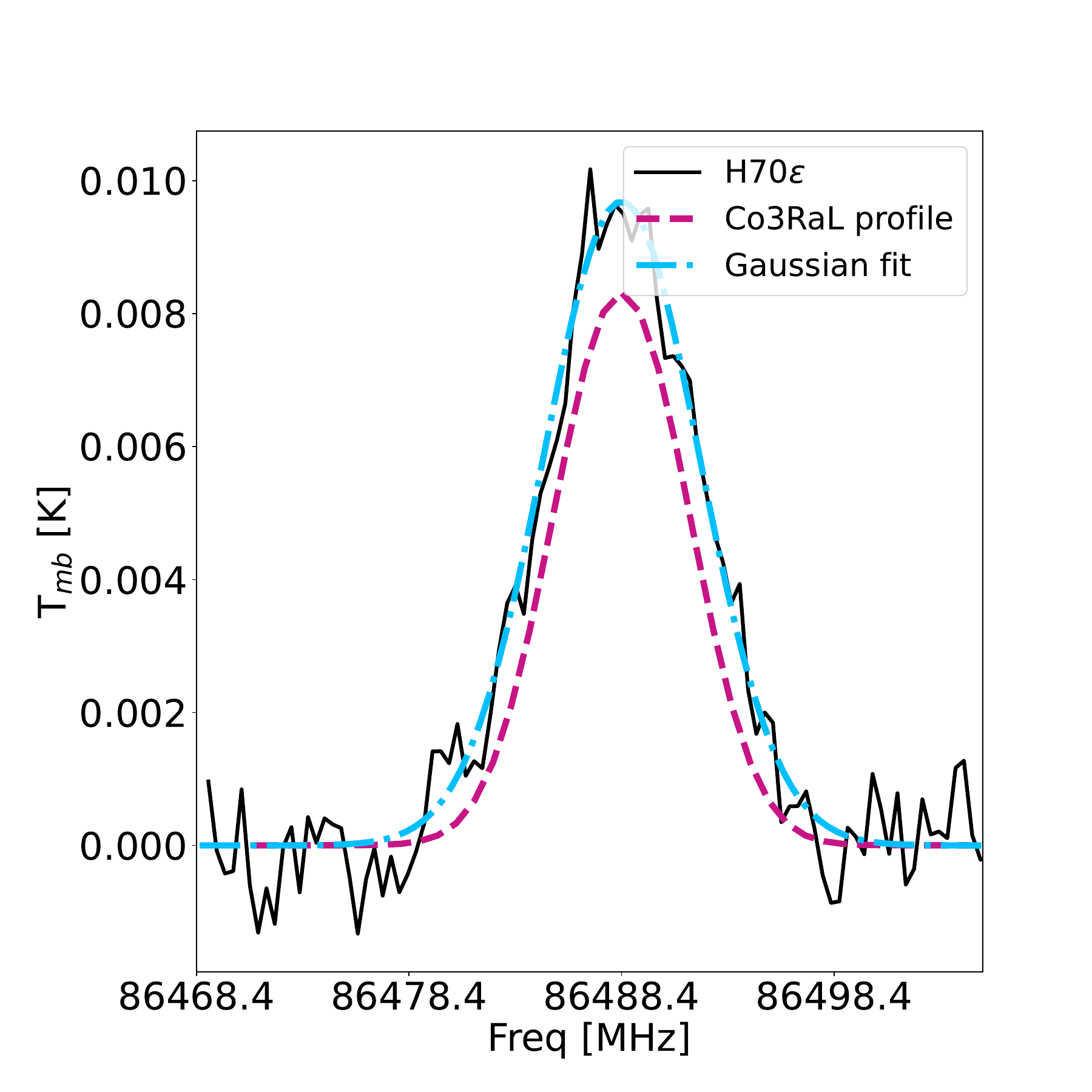}
	\includegraphics[width=0.24\textwidth]{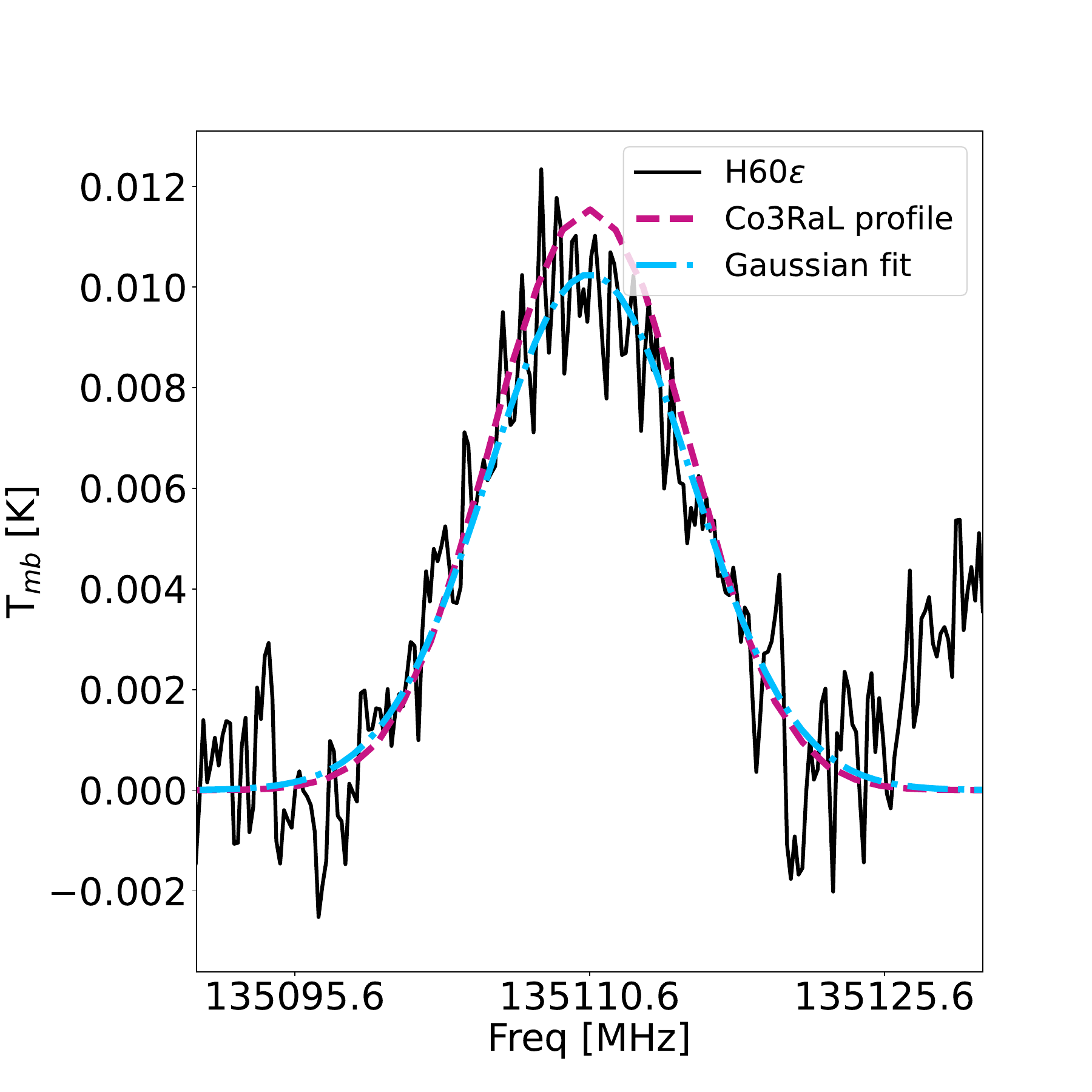}
	\caption{H\ge lines in \ic. H92\ge shows an asymmetric profile due to contamination with an UF (see Table \ref{tab:rrls_parameters}).} \label{fig:IC418_Hge}
\end{figure*}

\begin{figure*}[!h]
	\centering
	\includegraphics[width=0.24\textwidth]{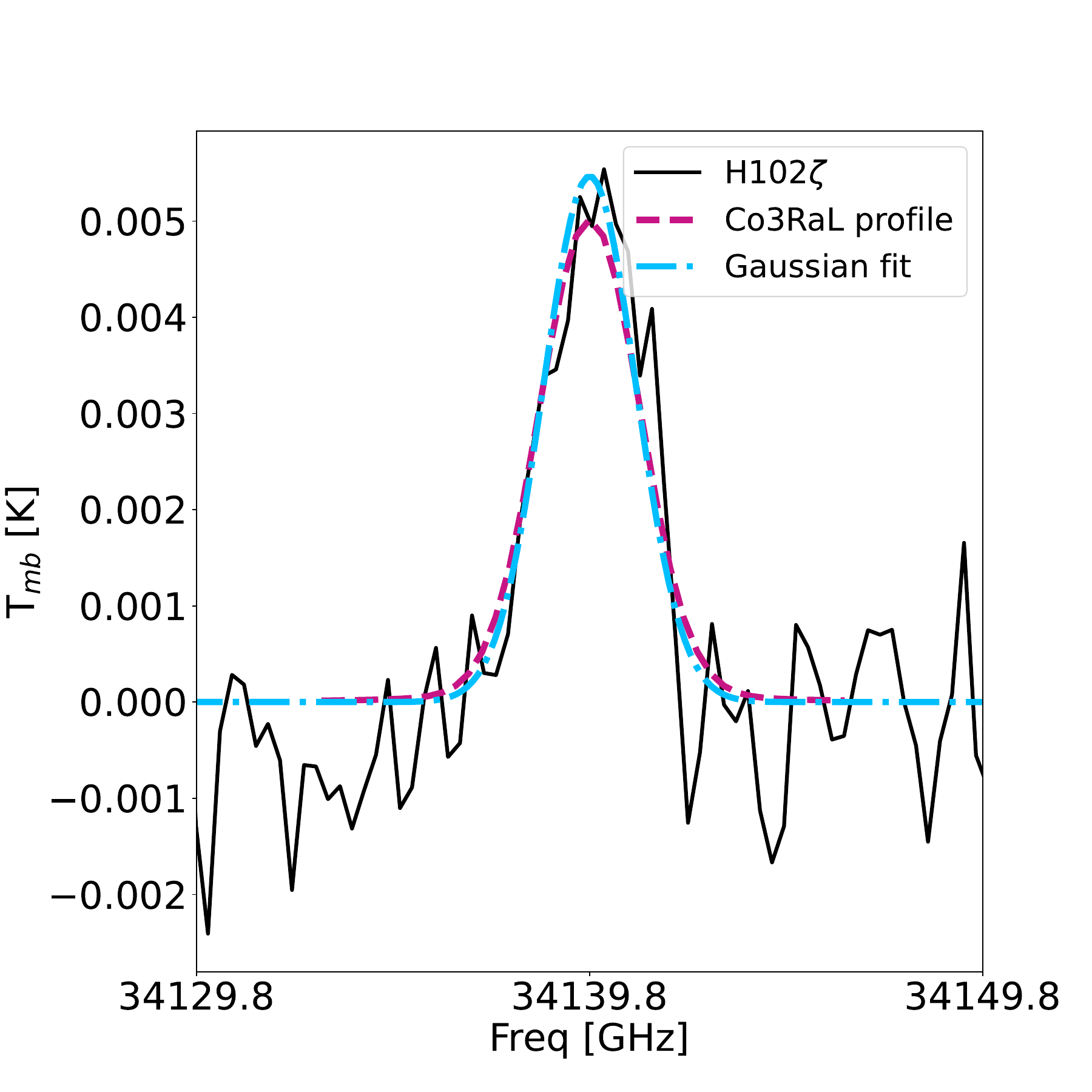}
	\includegraphics[width=0.24\textwidth]{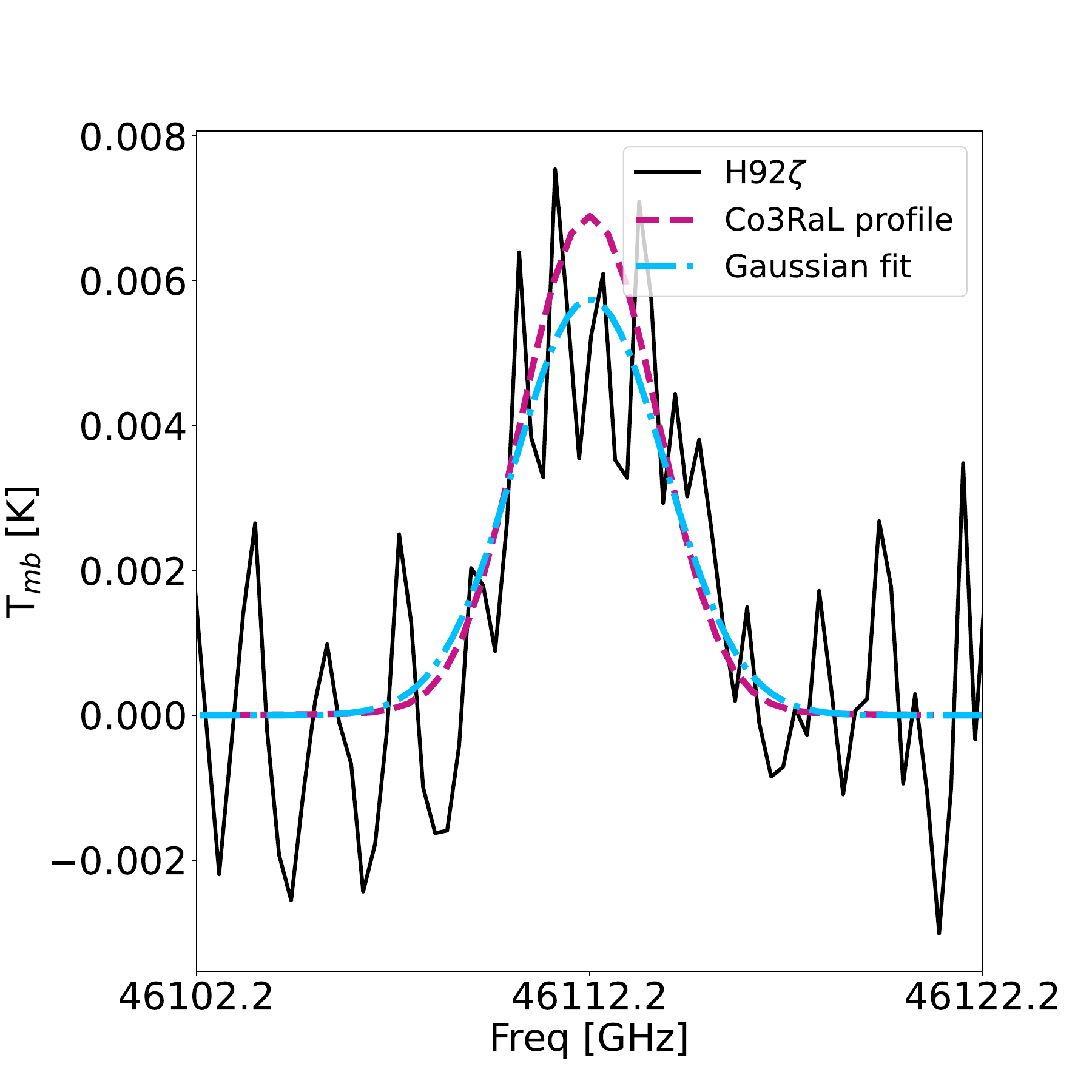}
	\includegraphics[width=0.24\textwidth]{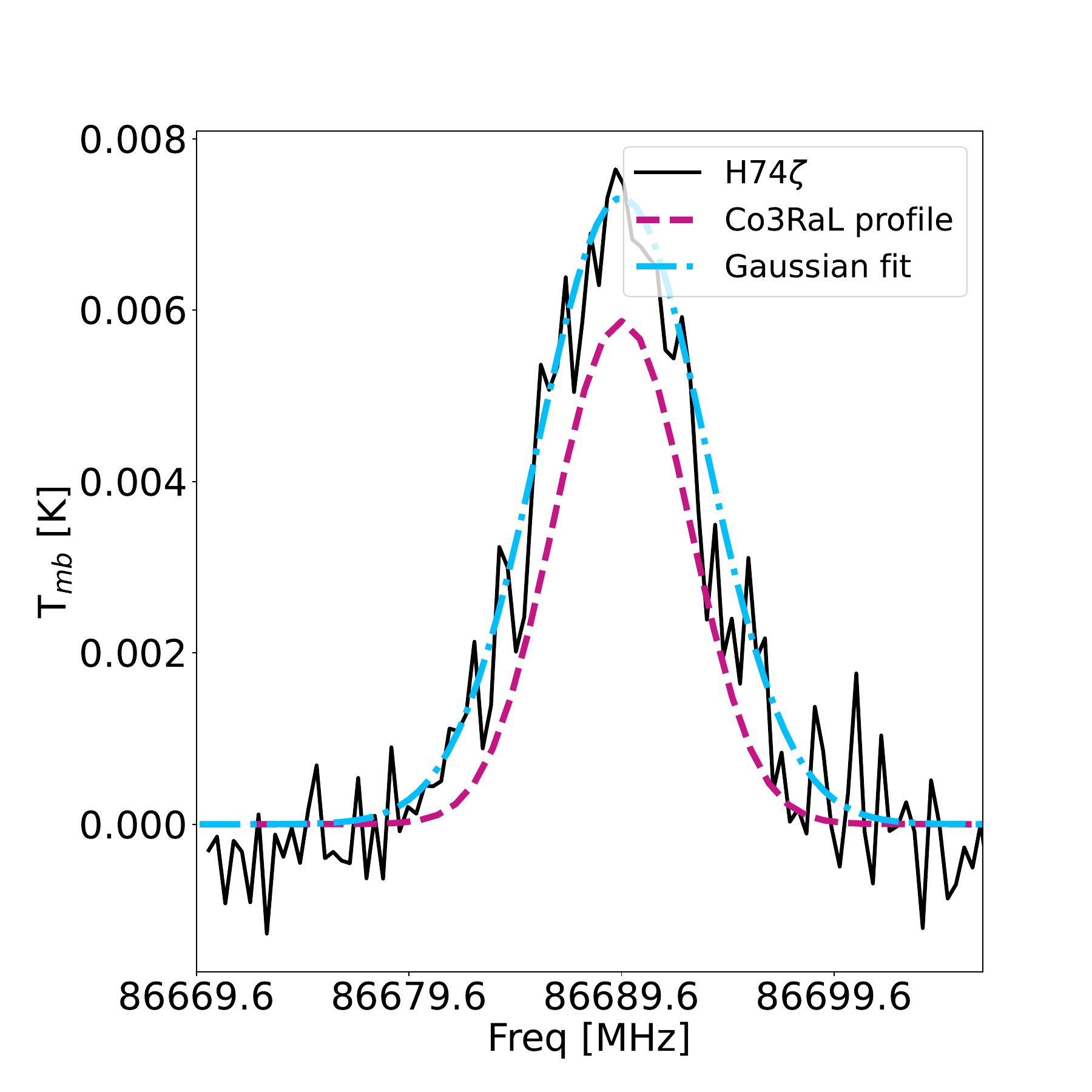}
	\includegraphics[width=0.24\textwidth]{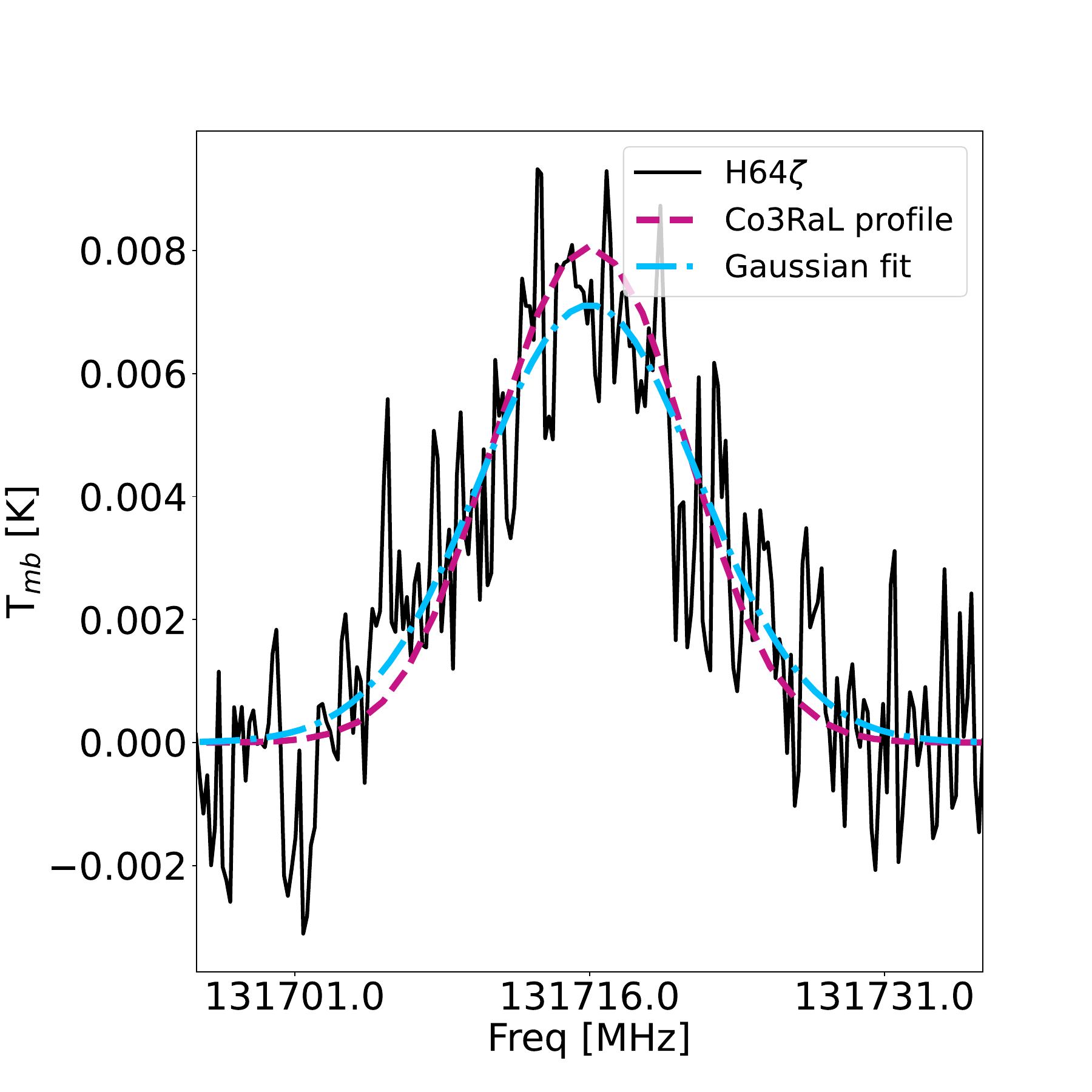}
	\caption{H\gz lines in \ic.} \label{fig:IC418_Hgz}
\end{figure*}

\begin{figure*}[!h]
	\centering
	\includegraphics[width=0.24\textwidth]{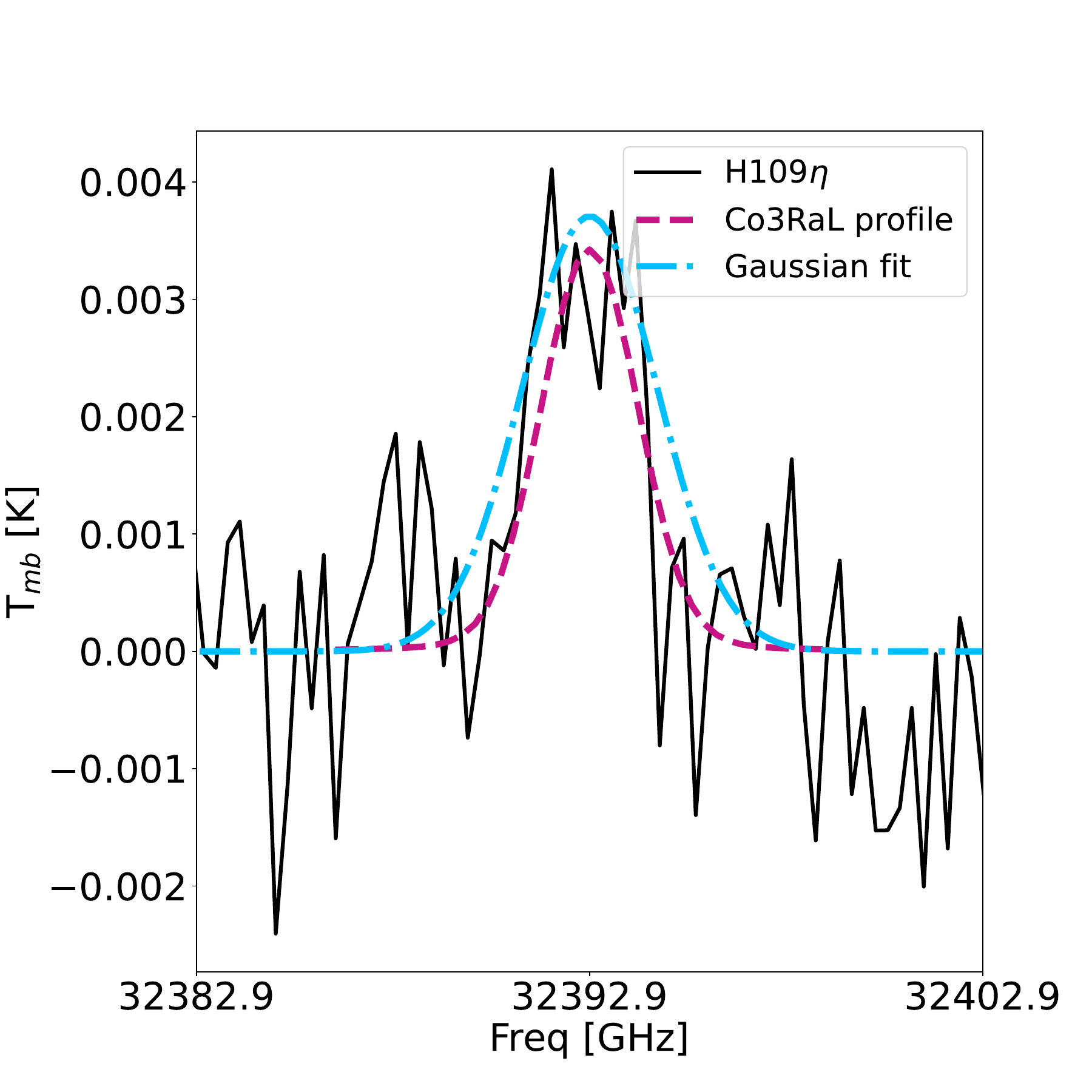}
	\includegraphics[width=0.24\textwidth]{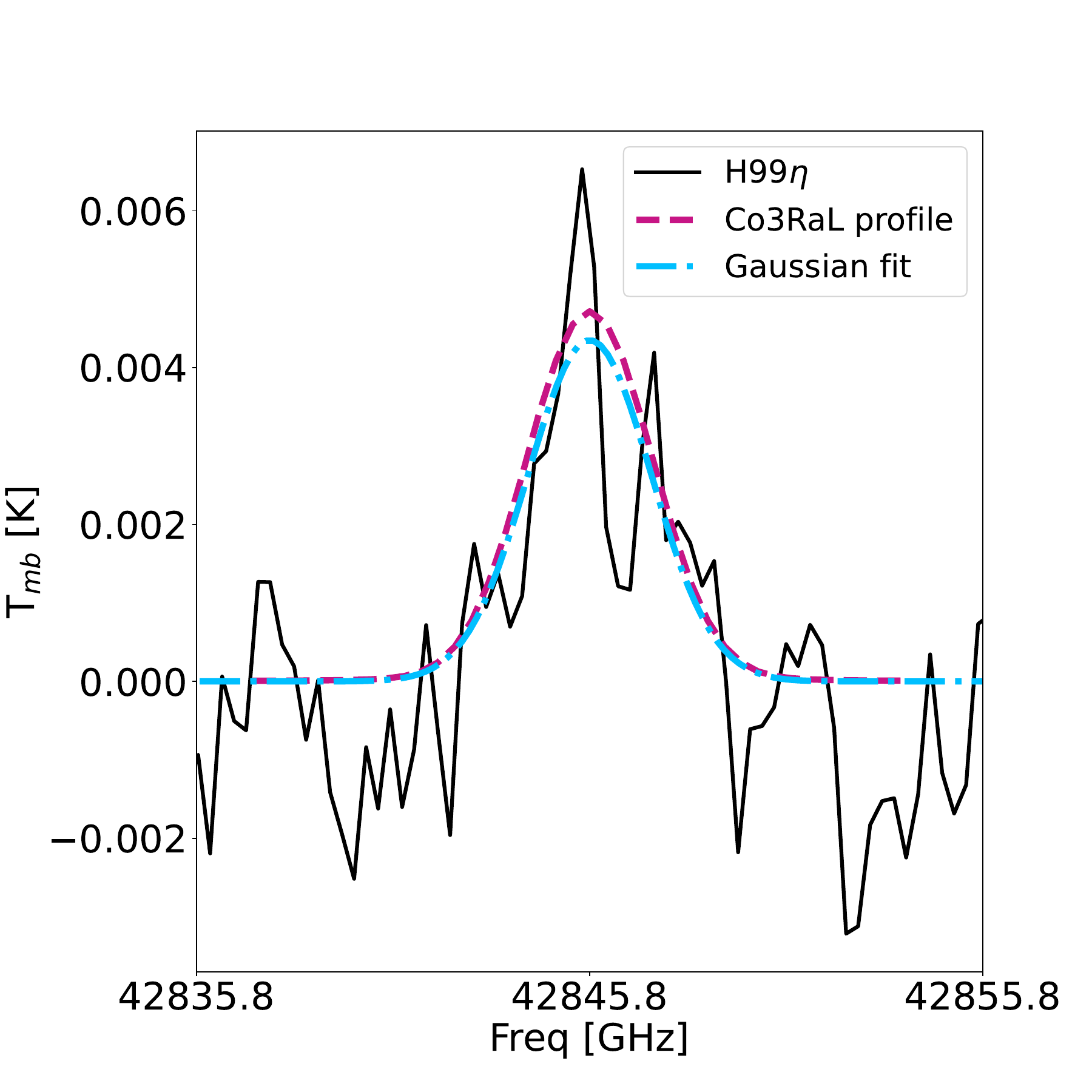}
	\includegraphics[width=0.24\textwidth]{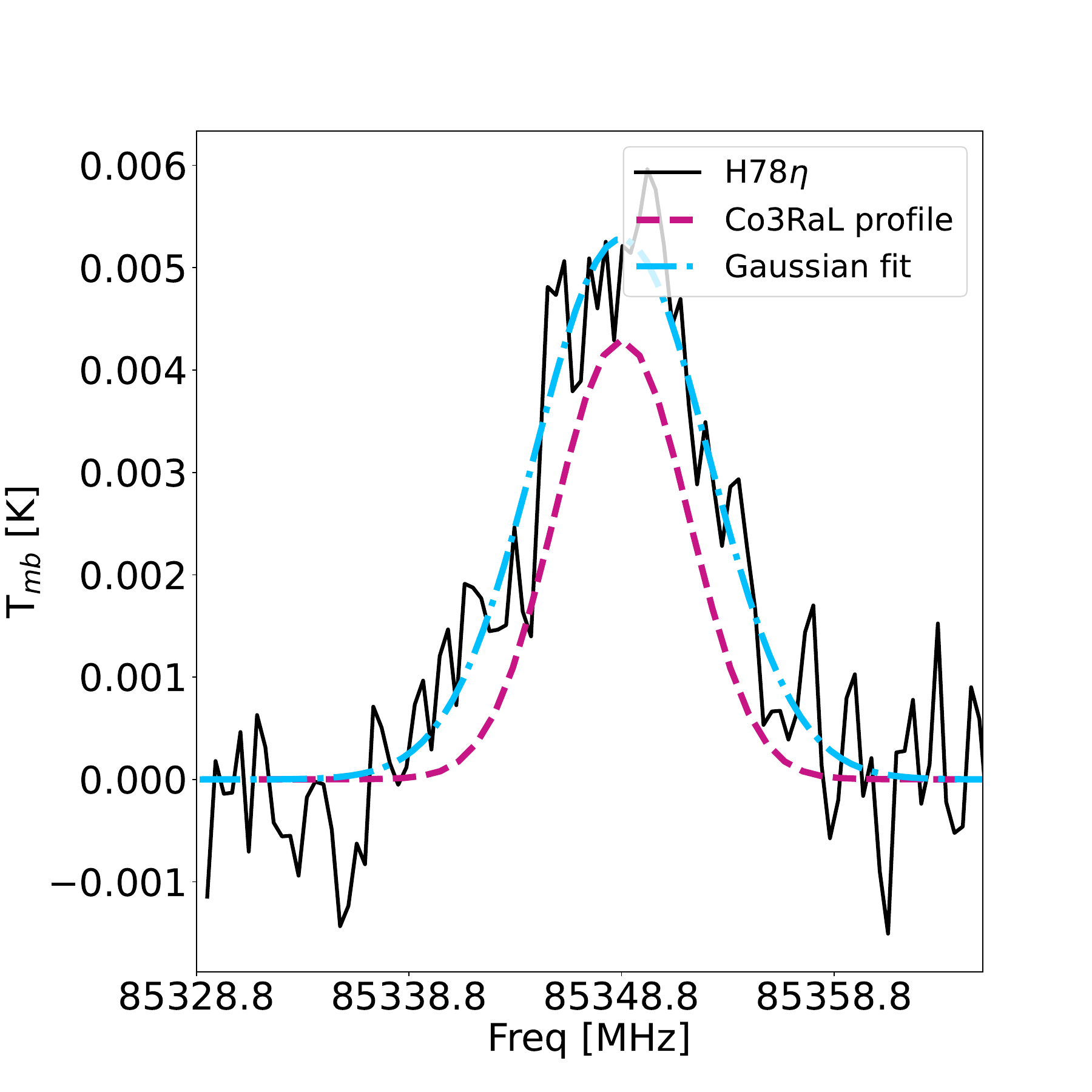}
	\includegraphics[width=0.24\textwidth]{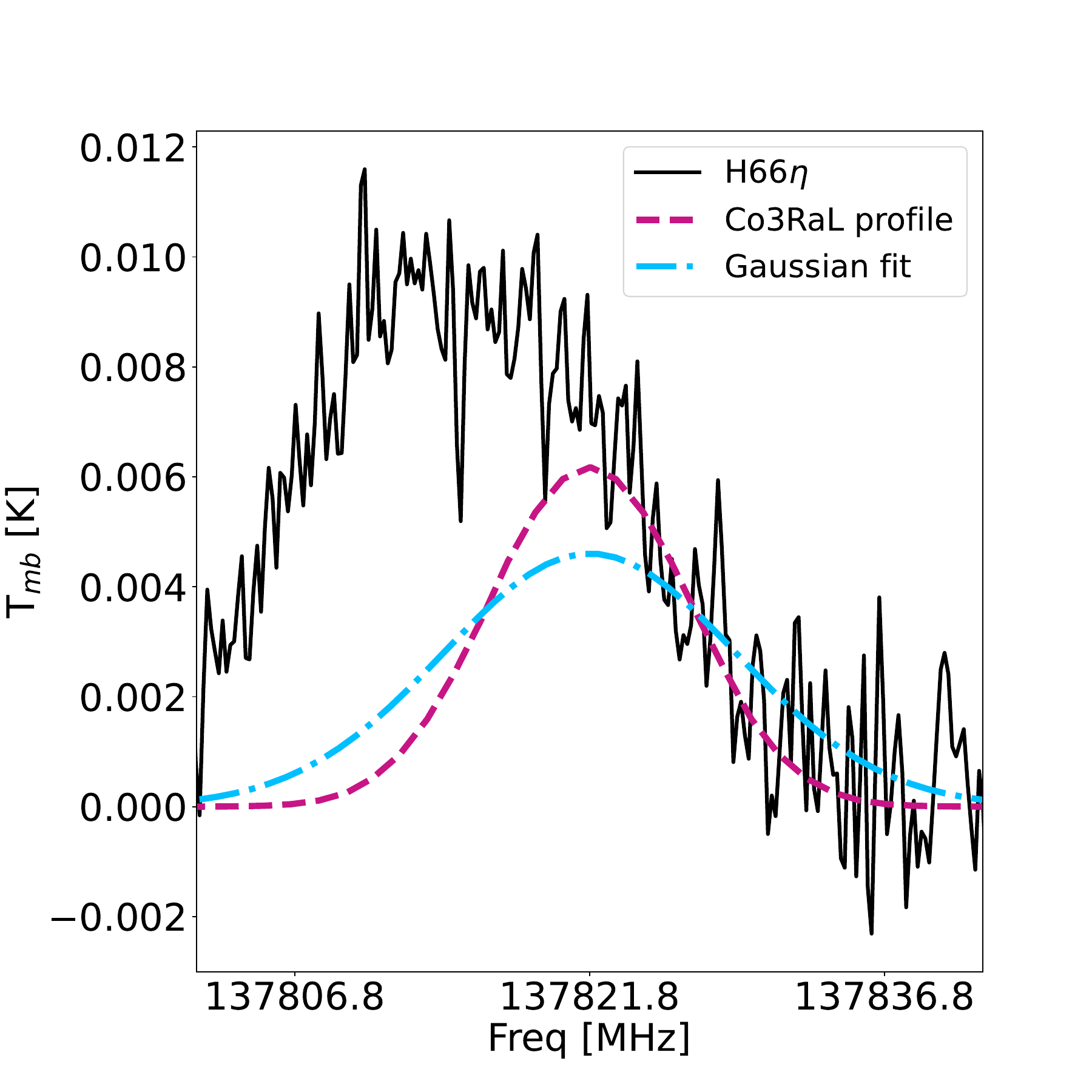}
	\caption{H\gh lines in \ic. H66\gh blended with H63\gz (see Table \ref{tab:rrls_parameters}).} \label{fig:IC418_Hgh}
\end{figure*}

\begin{figure*}[!h]
	\centering
	\includegraphics[width=0.24\textwidth]{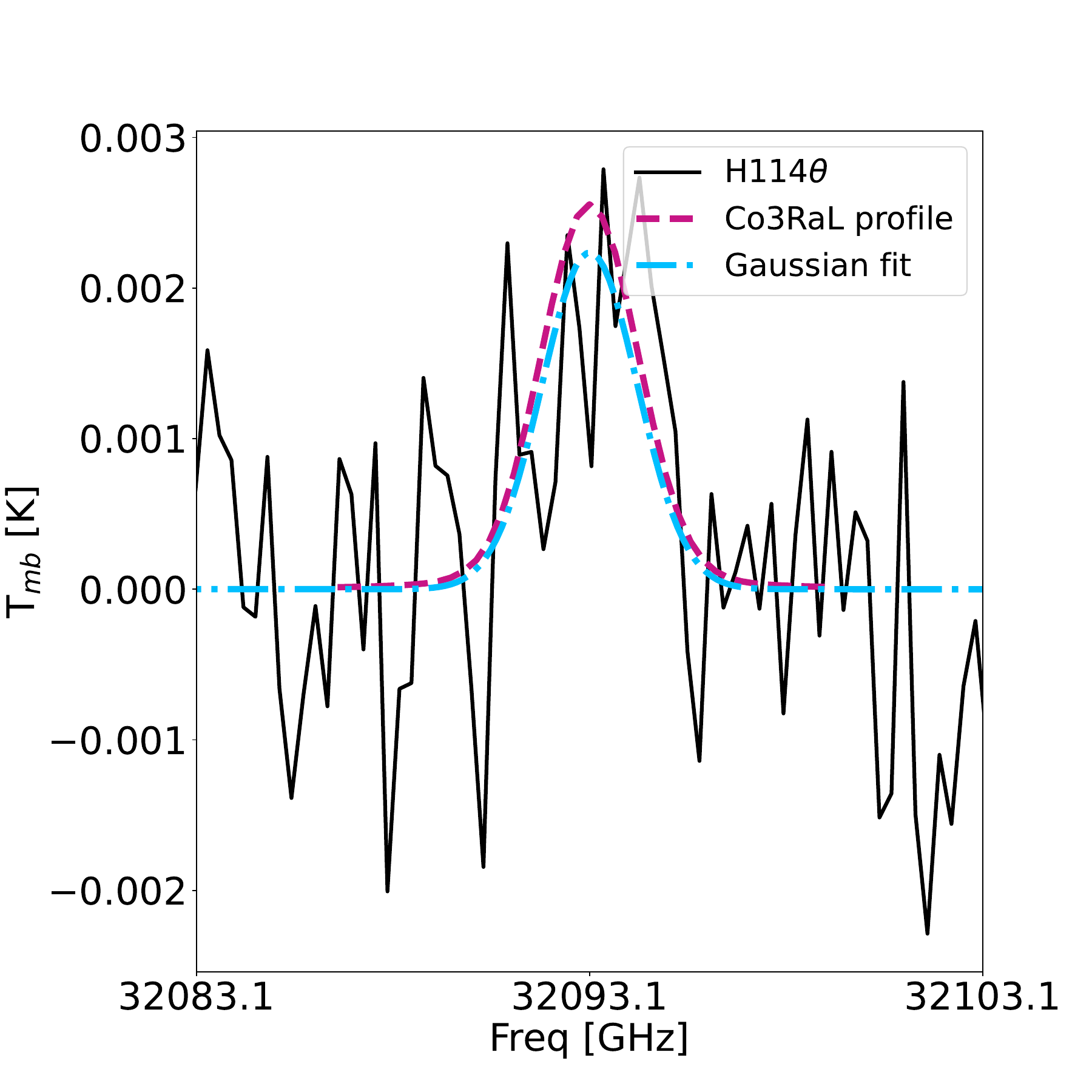}
	\includegraphics[width=0.24\textwidth]{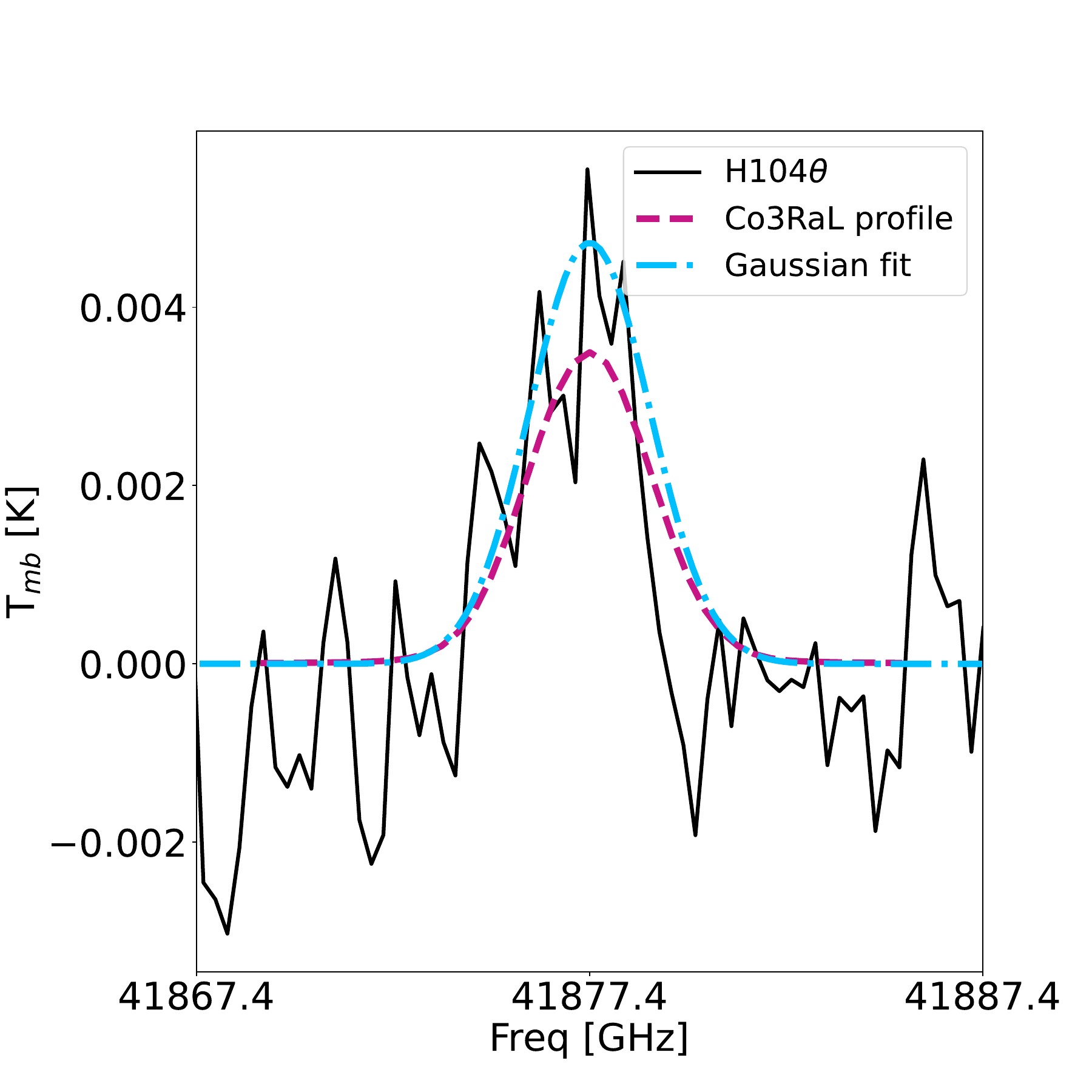}
	\includegraphics[width=0.24\textwidth]{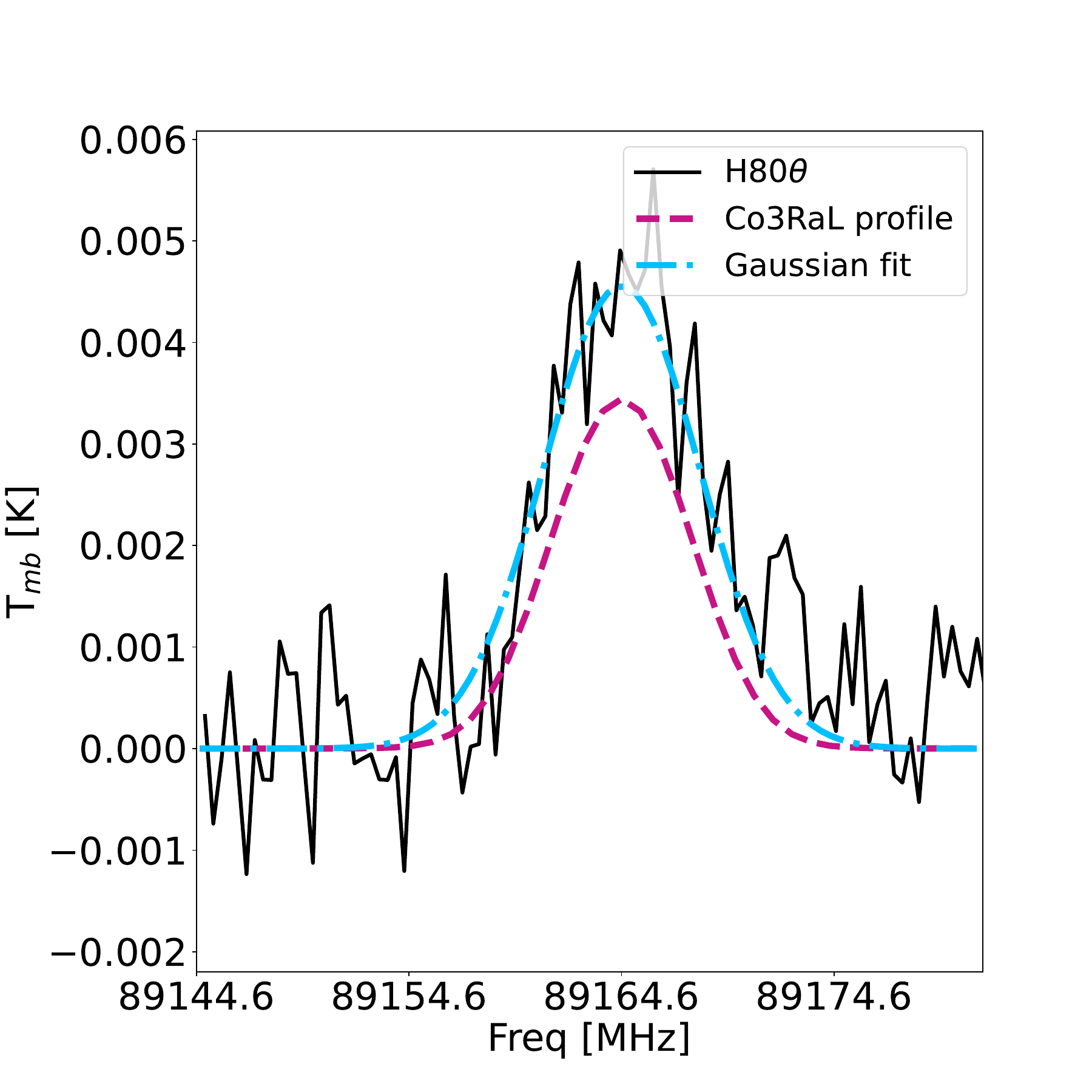}
	\includegraphics[width=0.24\textwidth]{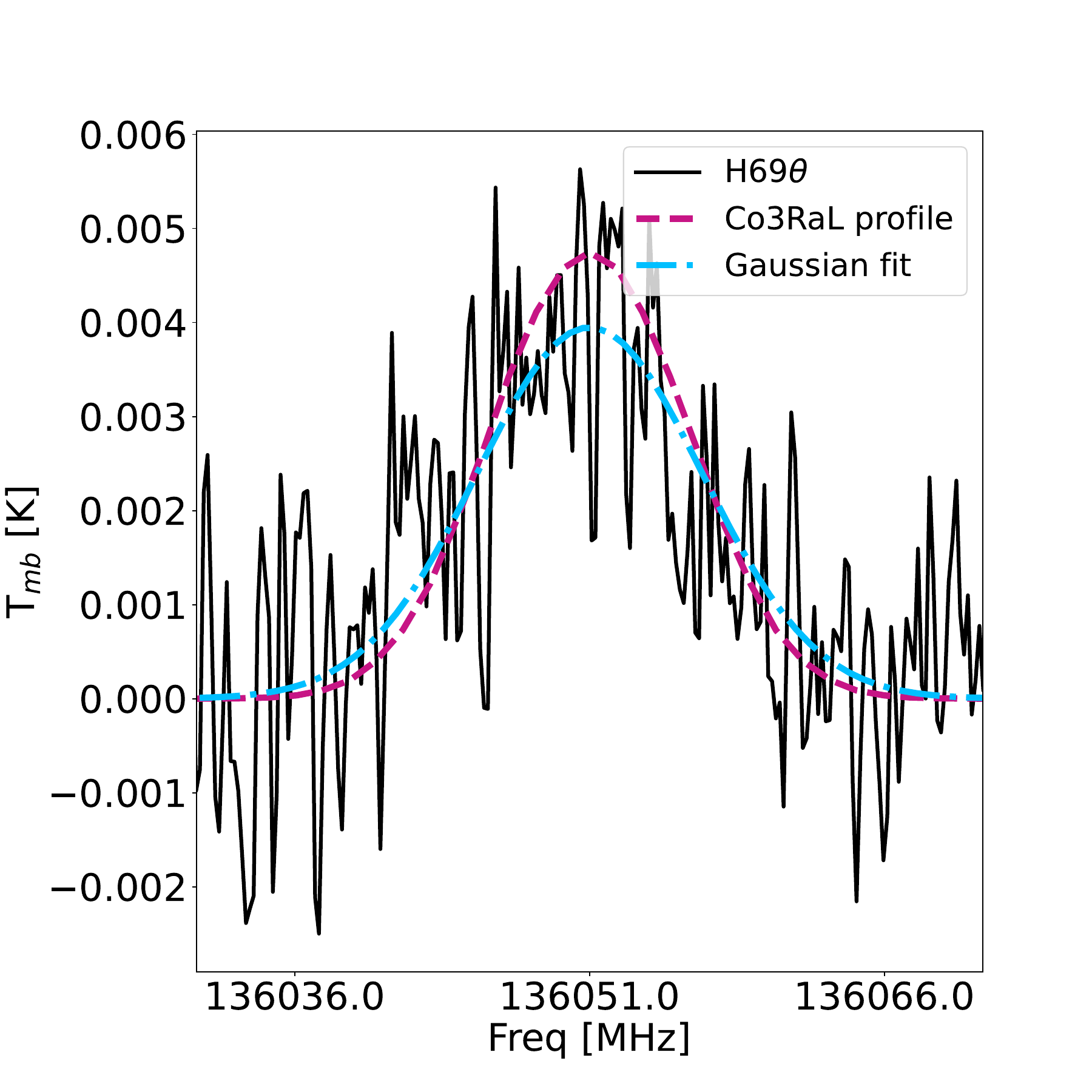}
	\caption{H\gq lines in \ic.} \label{fig:IC418_Hgq}
\end{figure*}

\begin{figure*}[!h]
	\centering
	\includegraphics[width=0.24\textwidth]{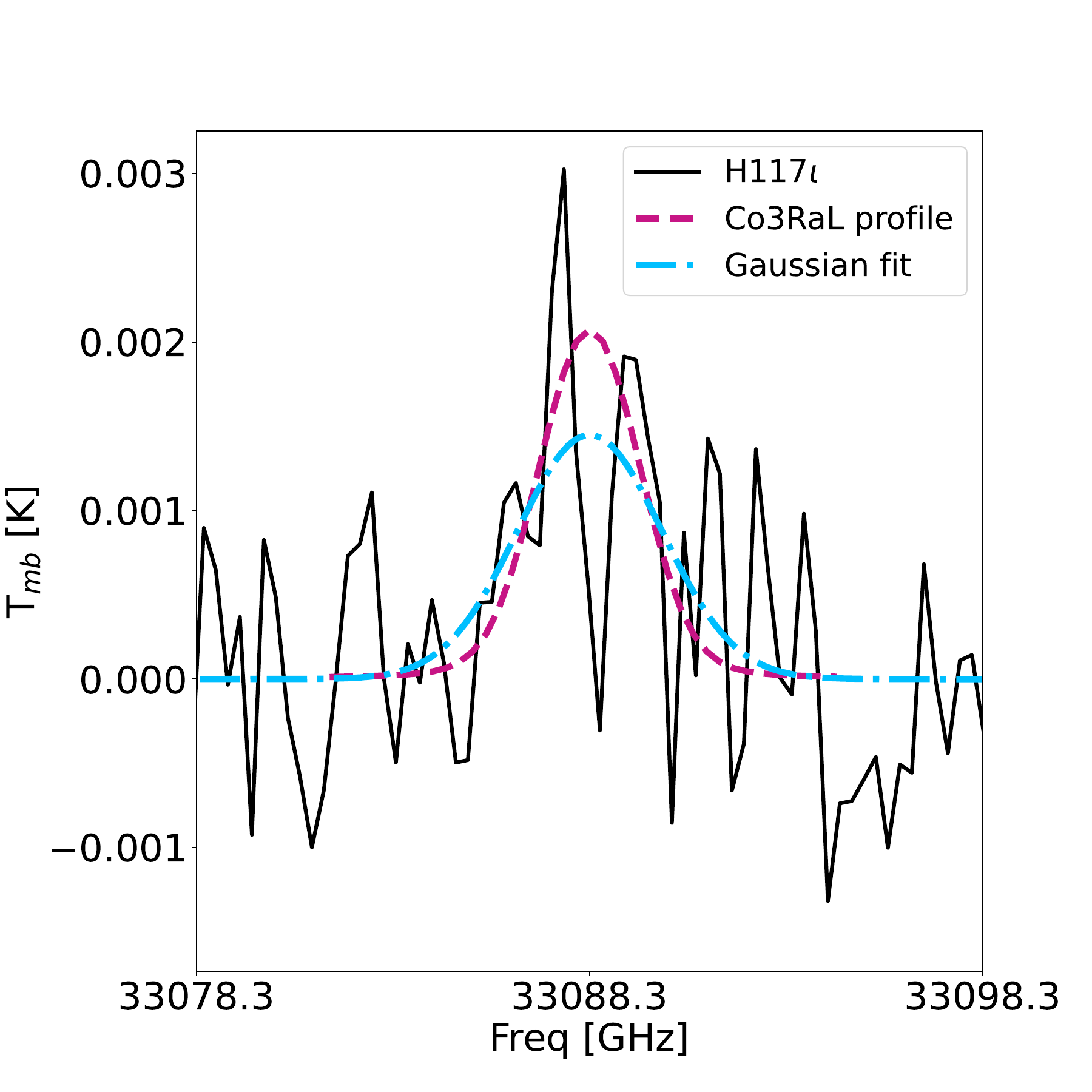}
	\includegraphics[width=0.24\textwidth]{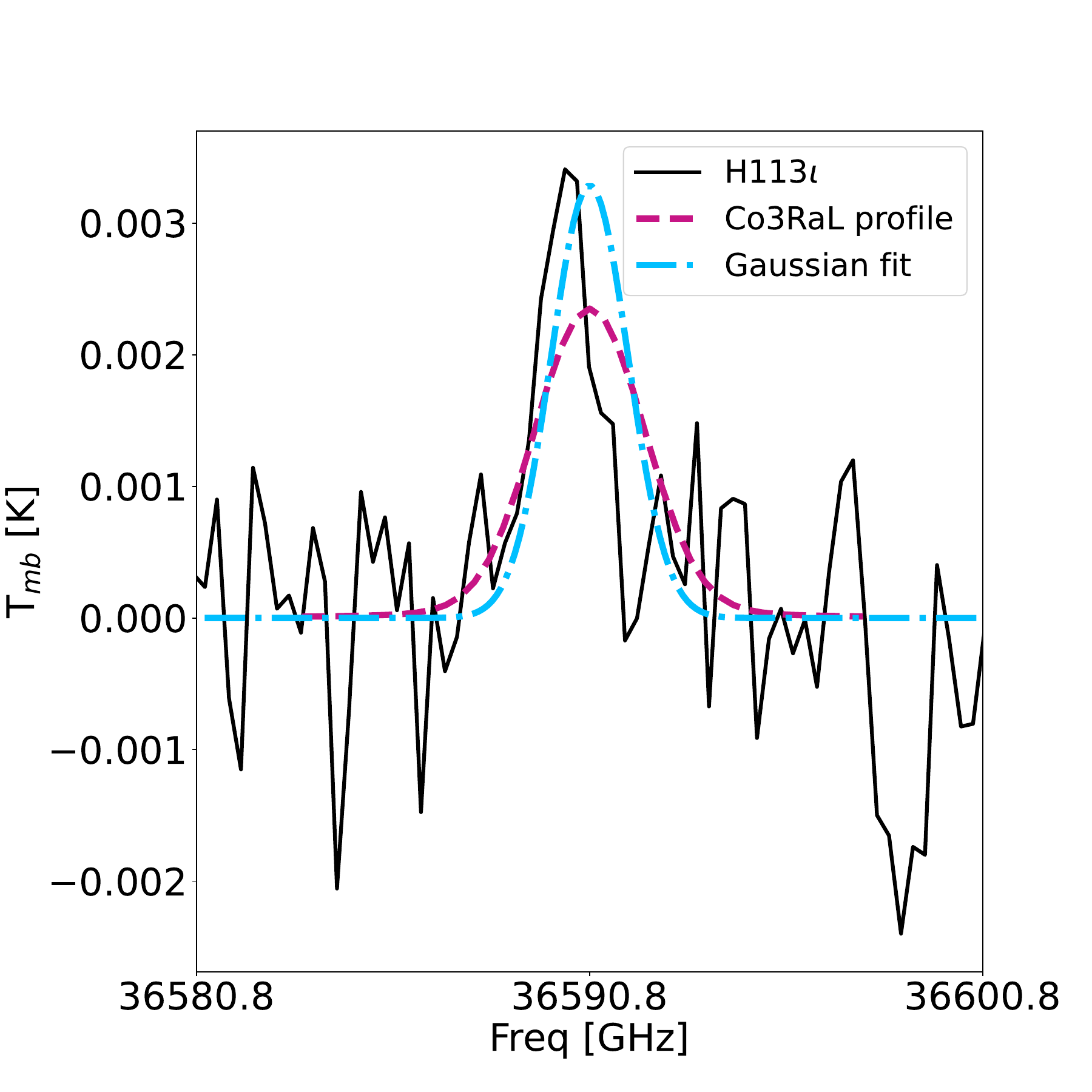}
	\includegraphics[width=0.24\textwidth]{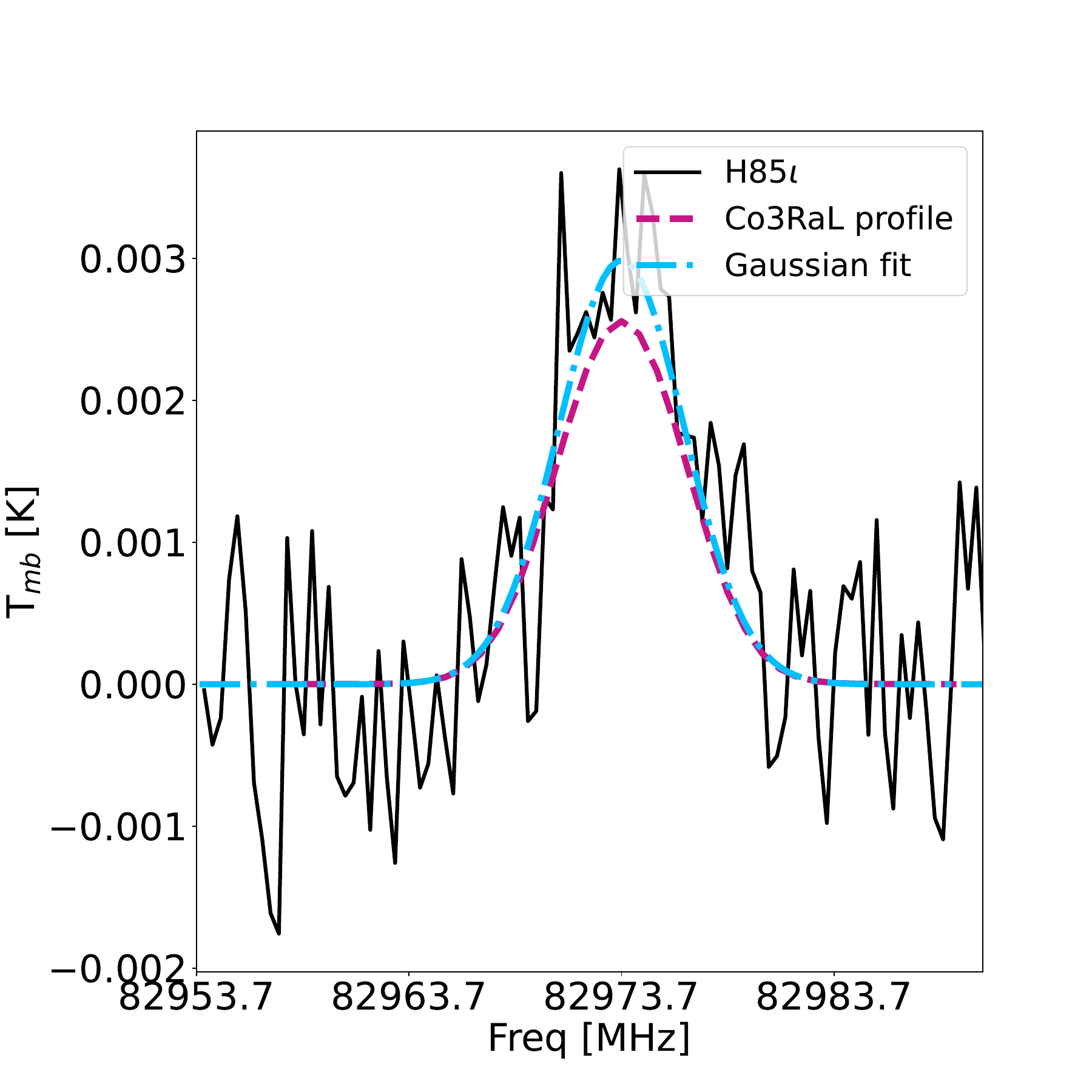}
	\includegraphics[width=0.24\textwidth]{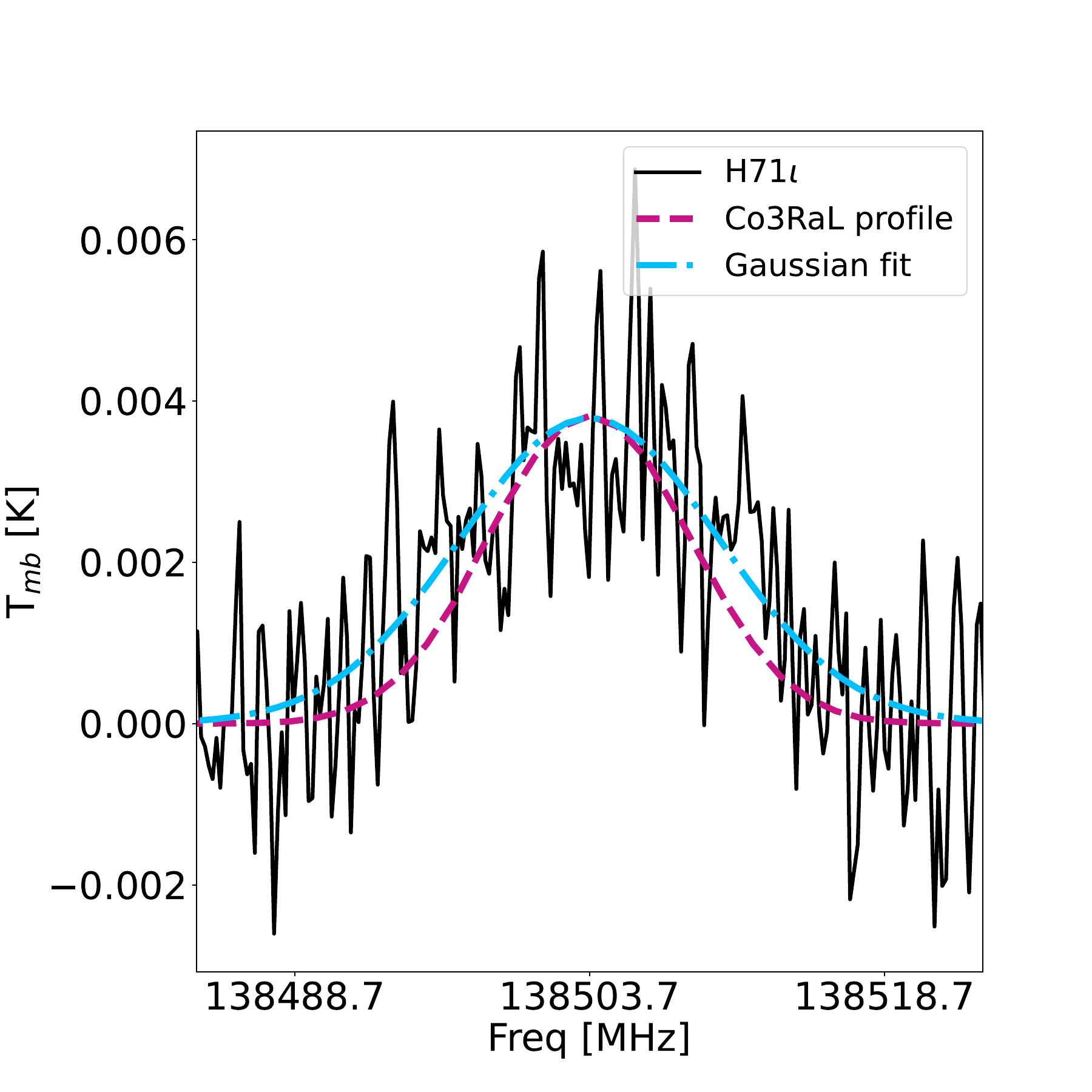}
	\caption{H\gi lines in \ic.} \label{fig:IC418_Hgi}
\end{figure*}

\begin{figure*}[!h]
	\centering
	\includegraphics[width=0.24\textwidth]{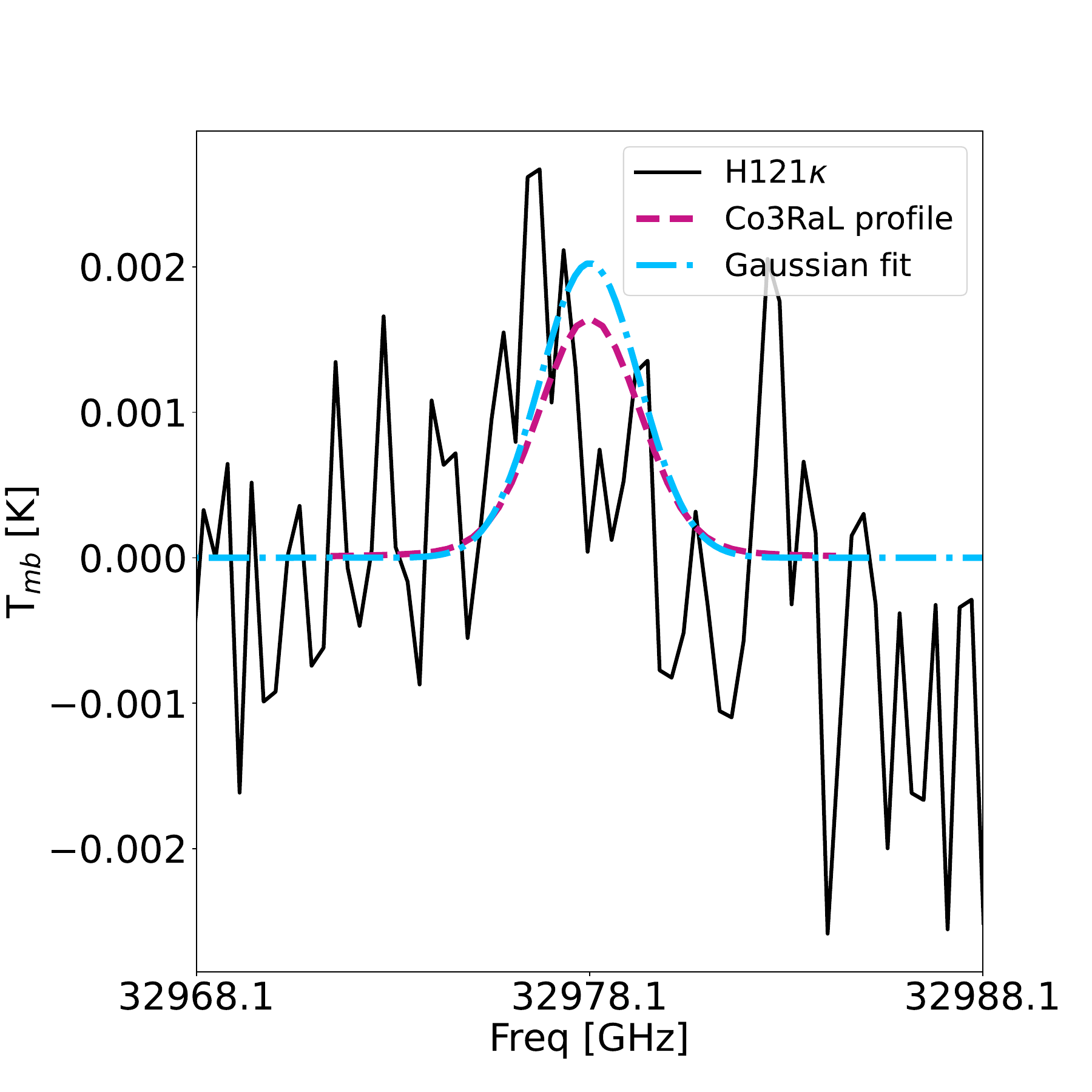}
	\includegraphics[width=0.24\textwidth]{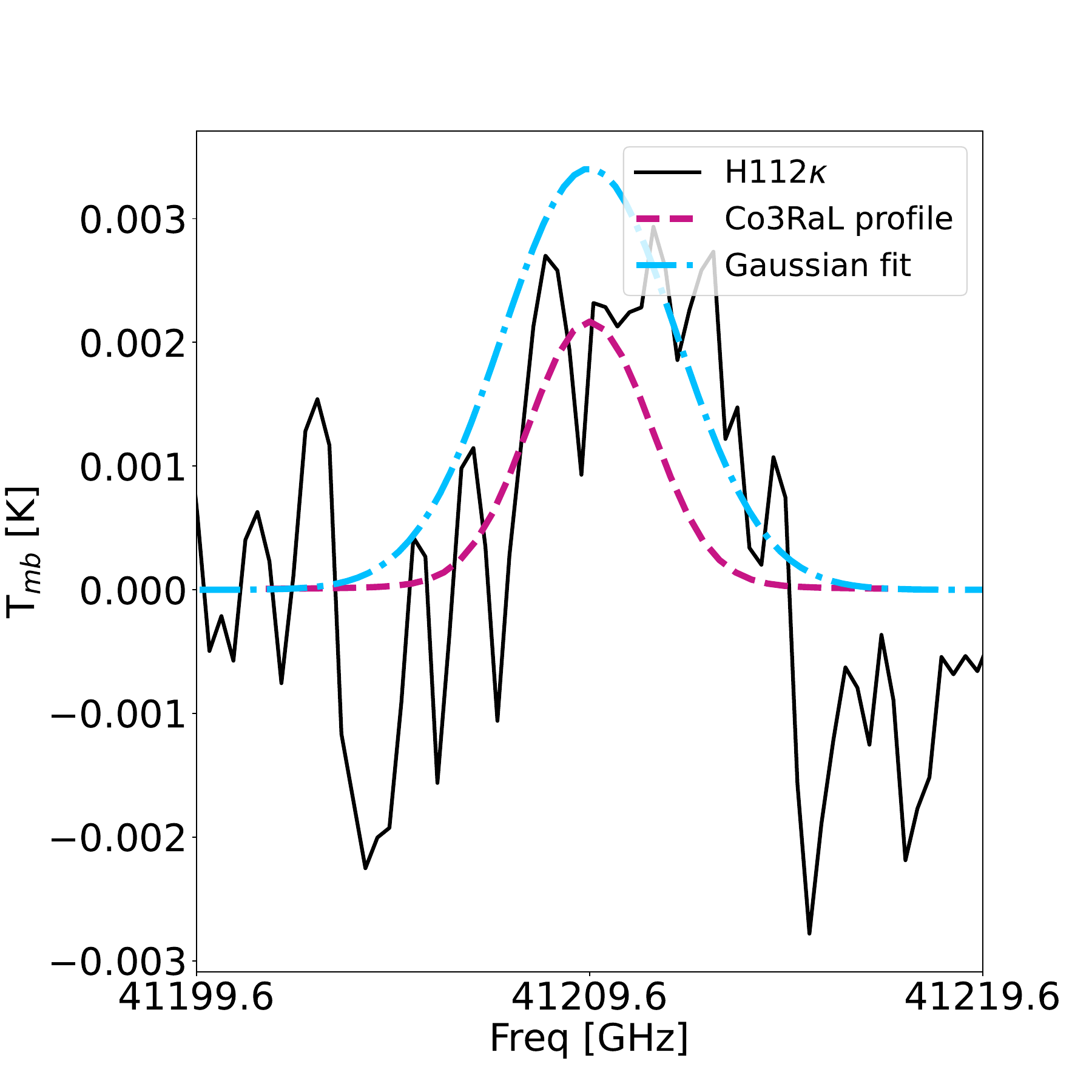}
	\includegraphics[width=0.24\textwidth]{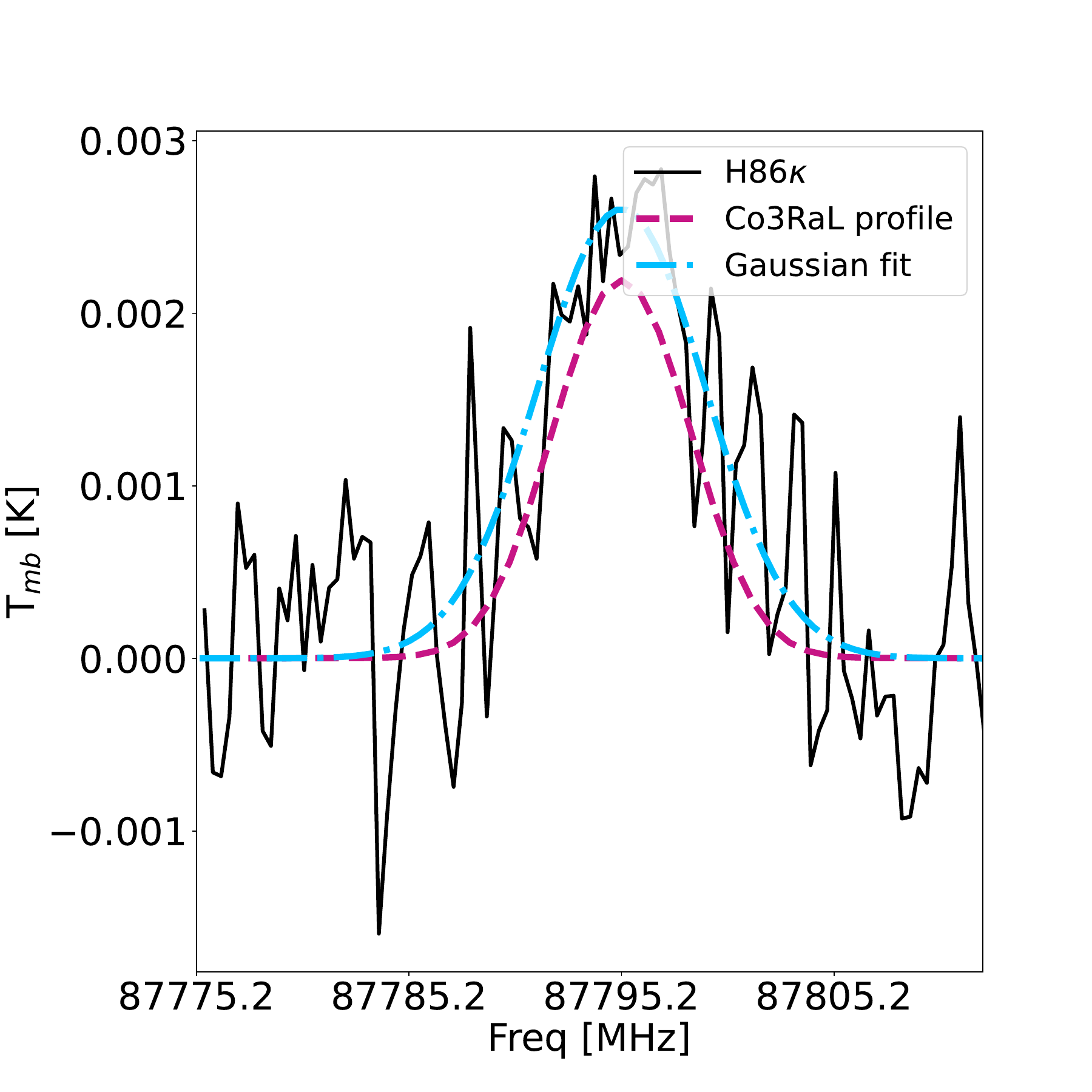}
	\includegraphics[width=0.24\textwidth]{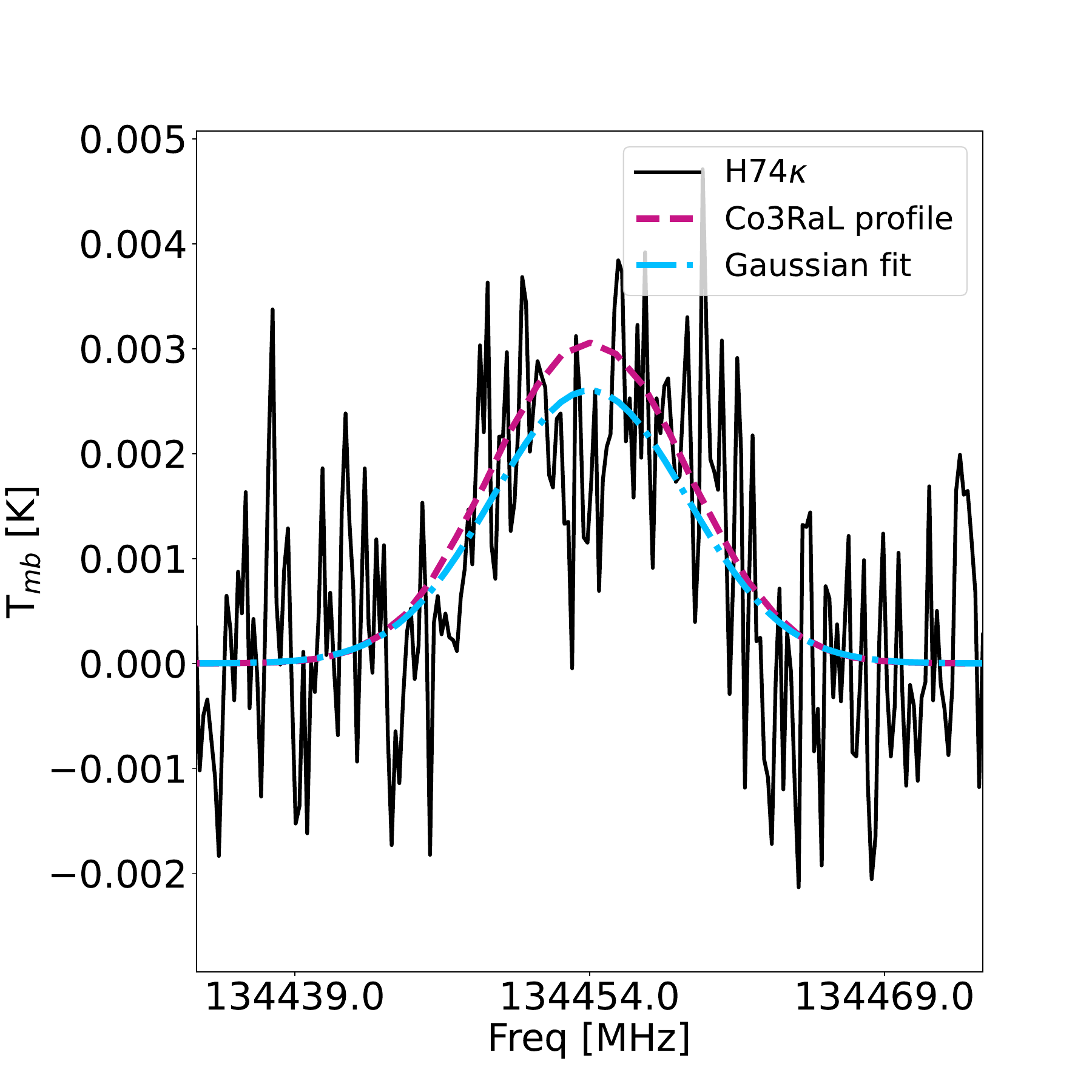}
	\caption{H\gk lines in \ic.} \label{fig:IC418_Hgk}
\end{figure*}

\begin{figure*}[!h]
	\centering
	\includegraphics[width=0.24\textwidth]{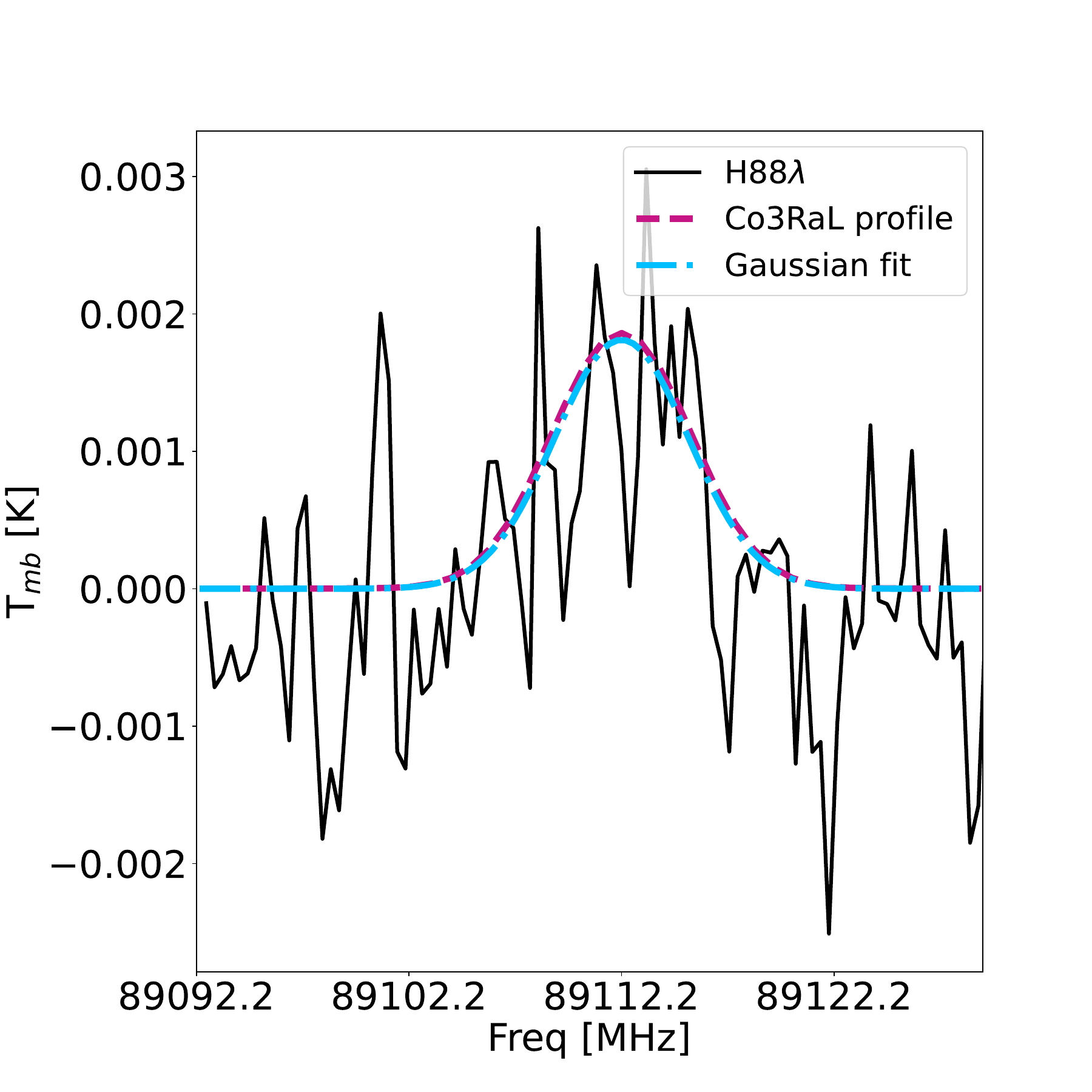}
	\includegraphics[width=0.24\textwidth]{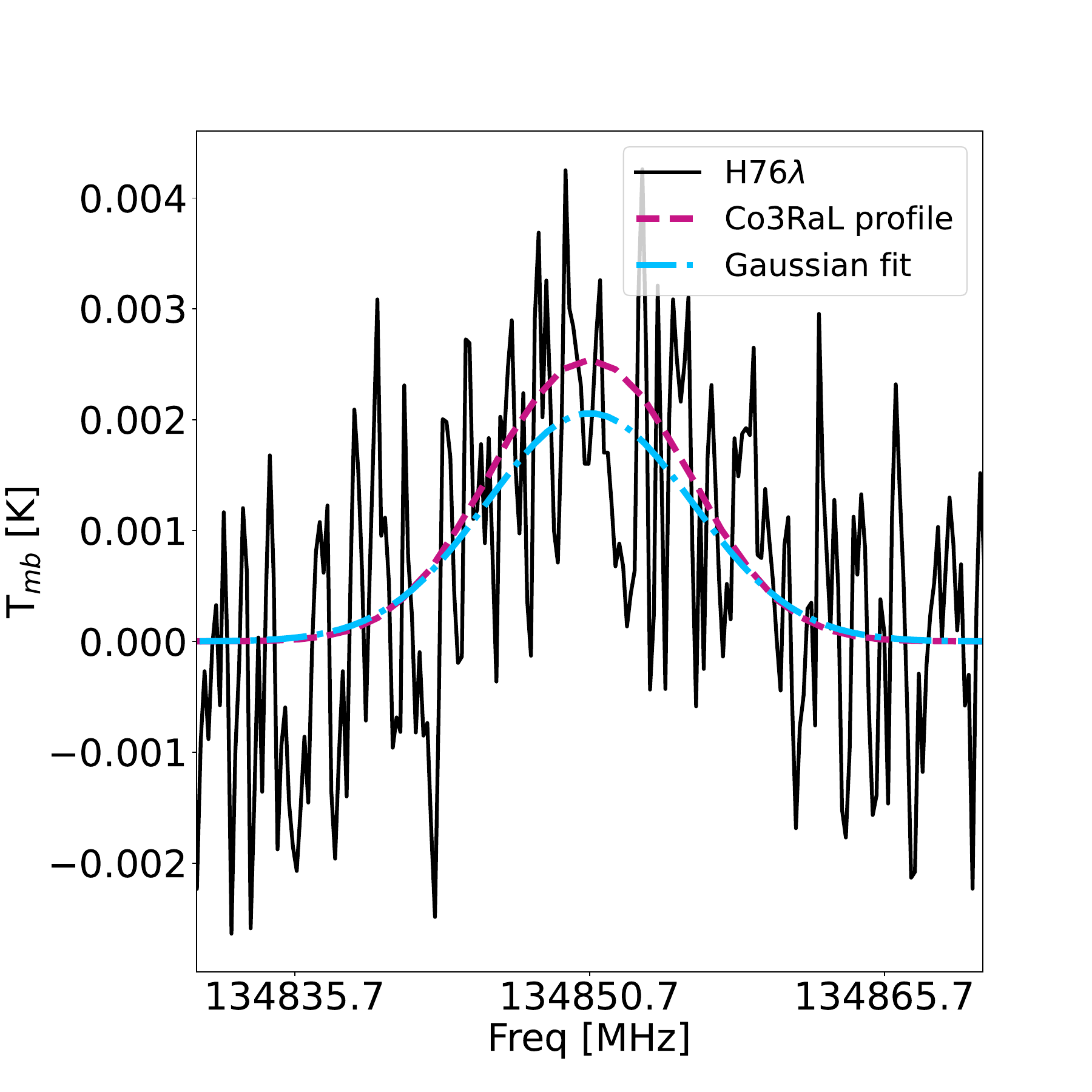}
	\includegraphics[width=0.24\textwidth]{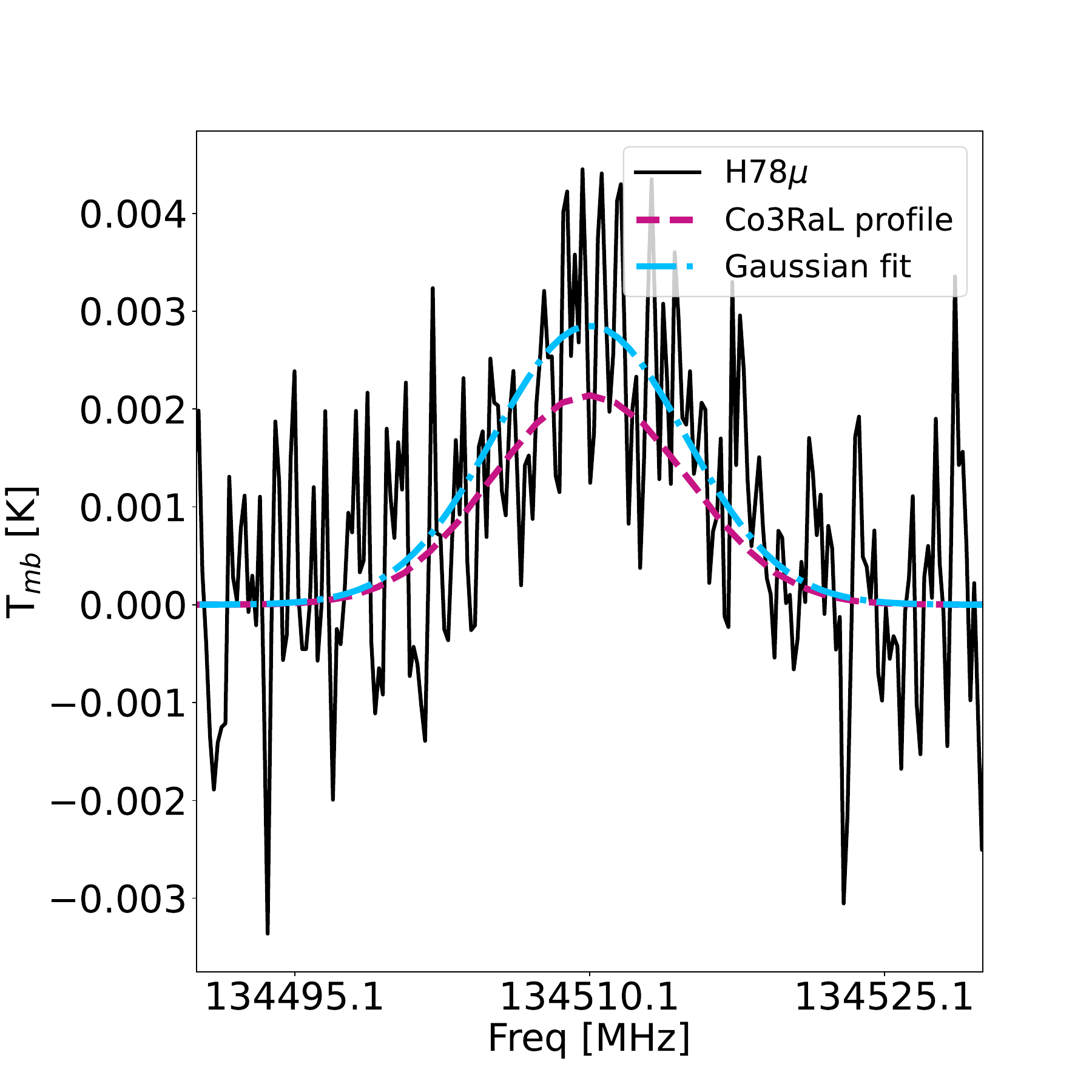}
	\includegraphics[width=0.24\textwidth]{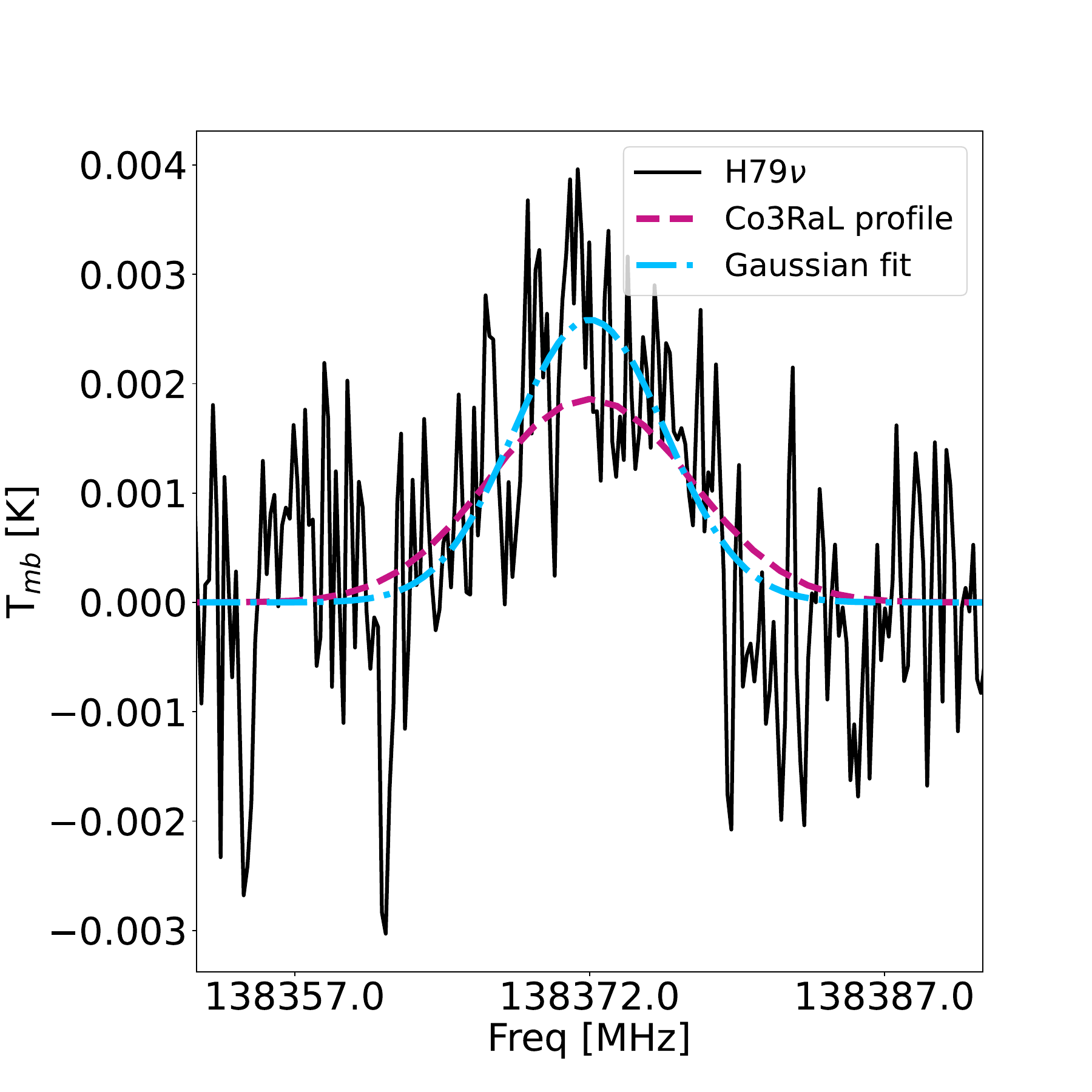}
	\caption{H\gl (both tentatives), H\gm, and H\gn lines in \ic.} \label{fig:IC418_Hglgmgn}
\end{figure*}

\begin{figure*}[!h]
	\centering
	\includegraphics[width=0.24\textwidth]{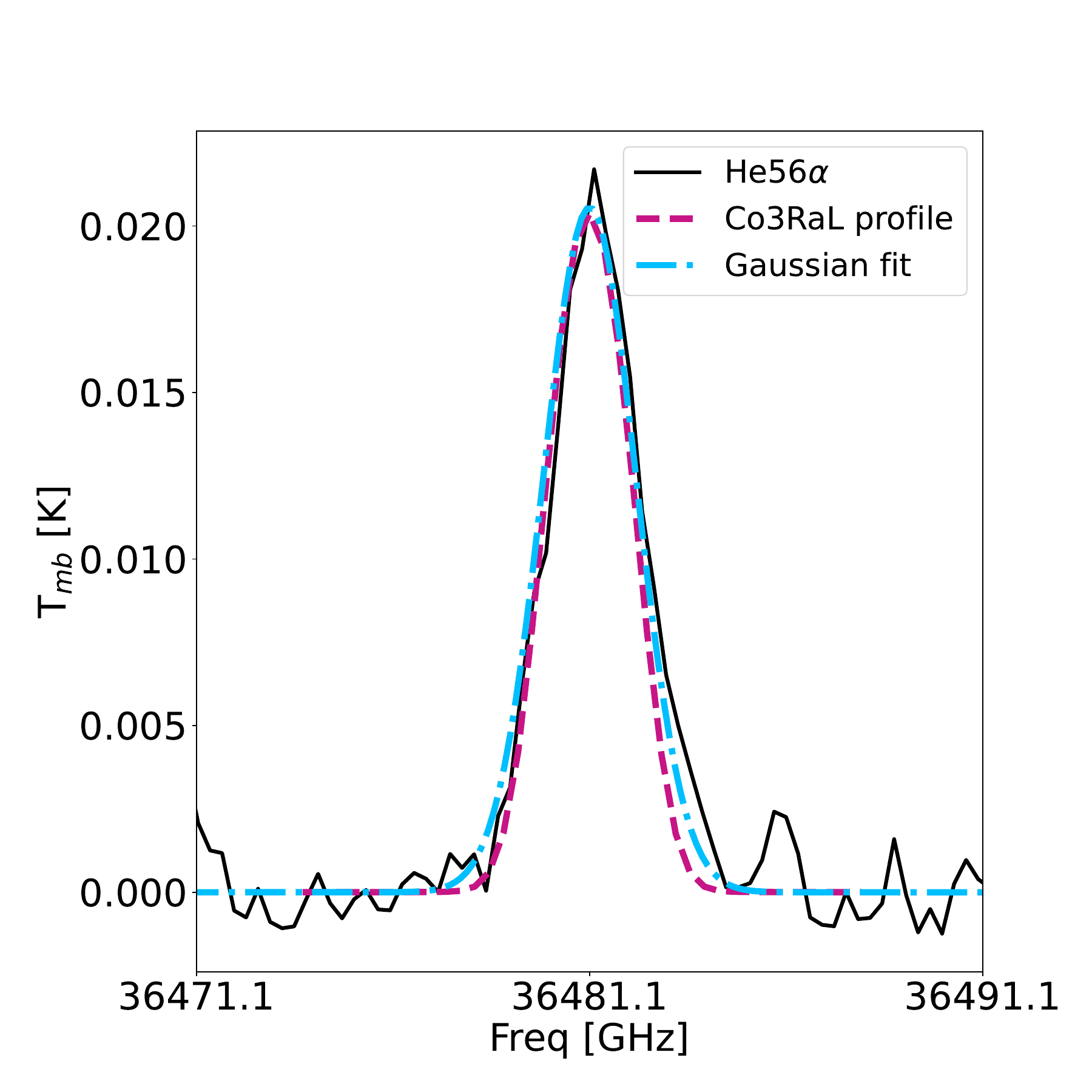}
	\includegraphics[width=0.24\textwidth]{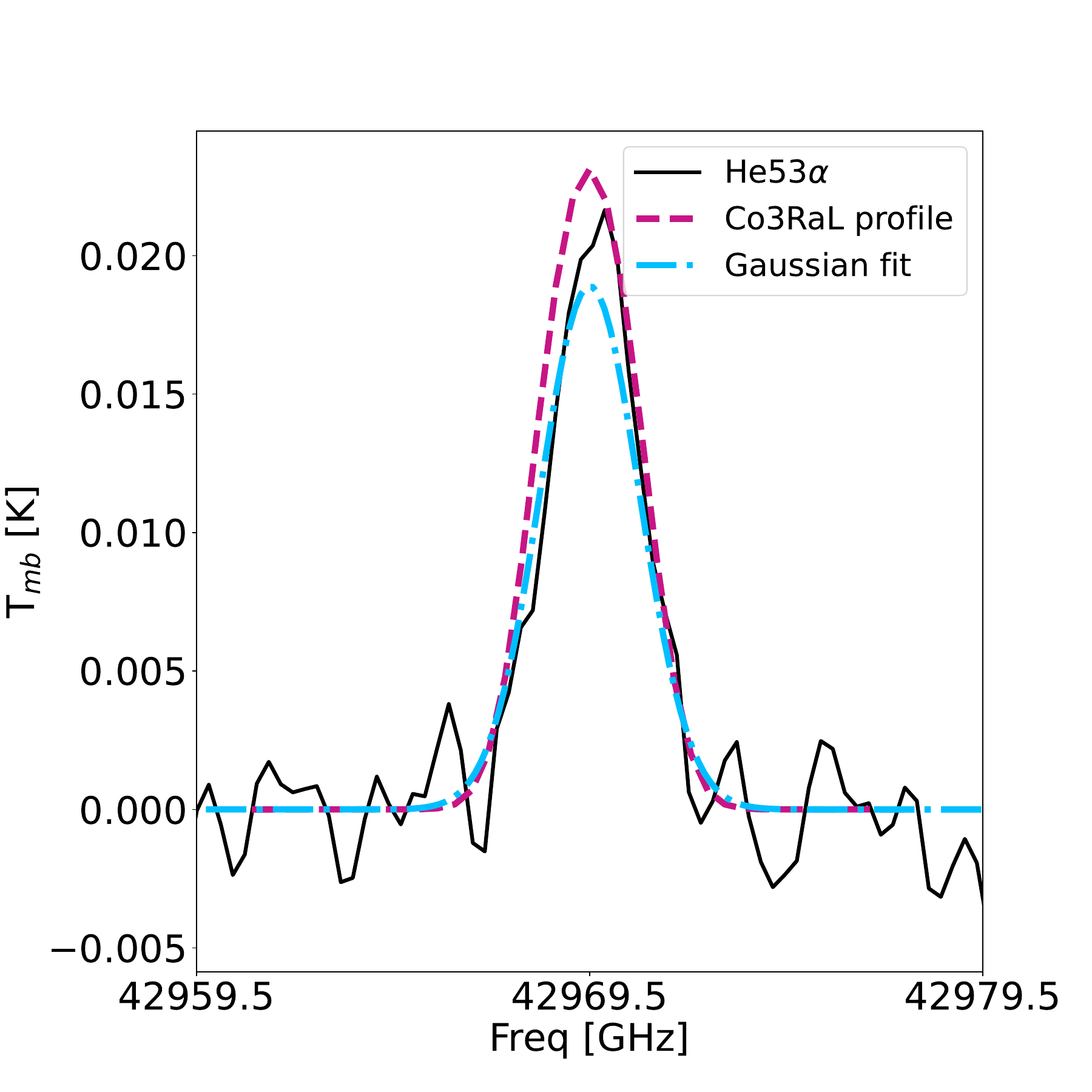}
	\includegraphics[width=0.24\textwidth]{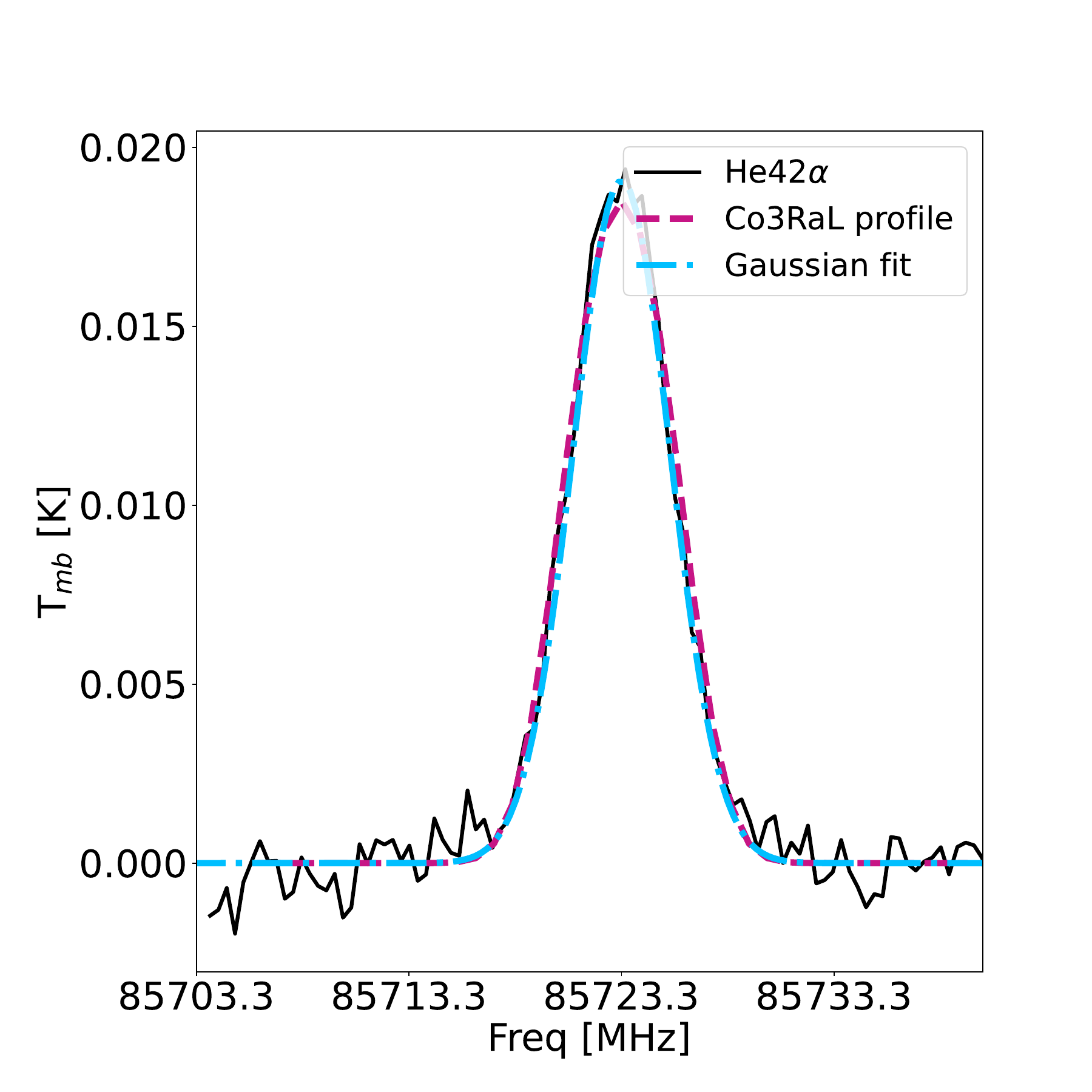}
	\includegraphics[width=0.24\textwidth]{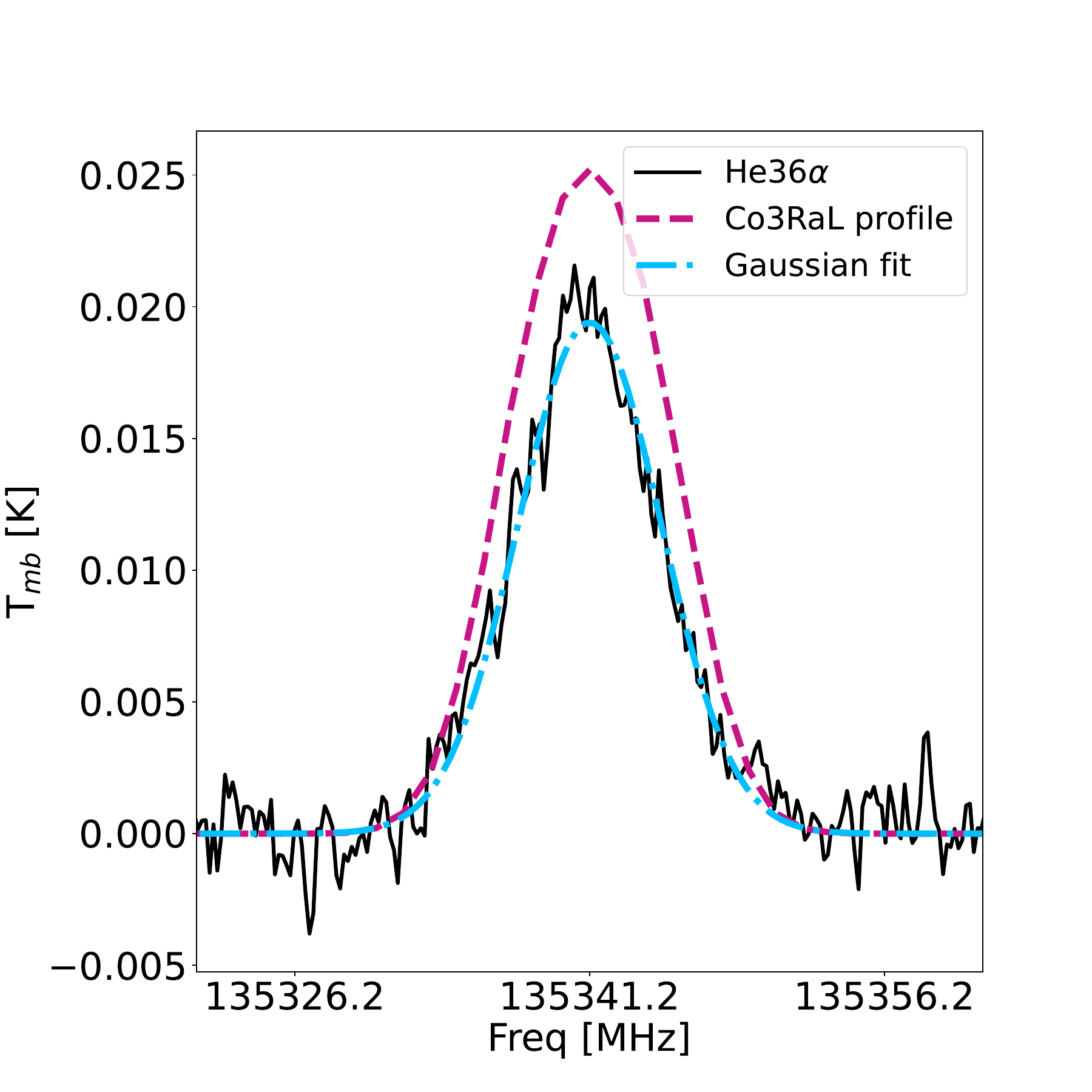}
	\caption{\hei\ga lines in \ic.} \label{fig:IC418_HeIga}
\end{figure*}

\begin{figure*}[!h]
	\centering
	\includegraphics[width=0.24\textwidth]{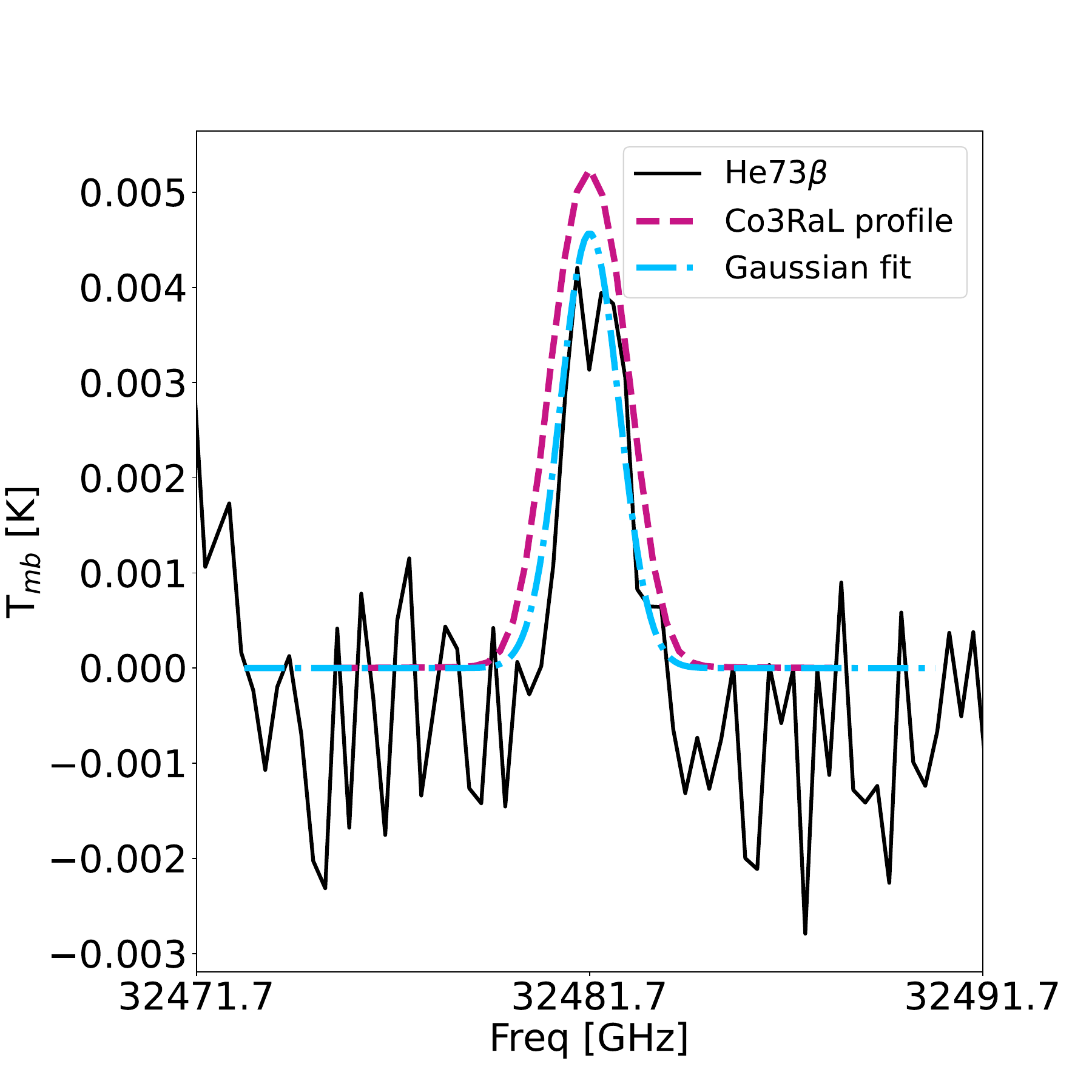}
	\includegraphics[width=0.24\textwidth]{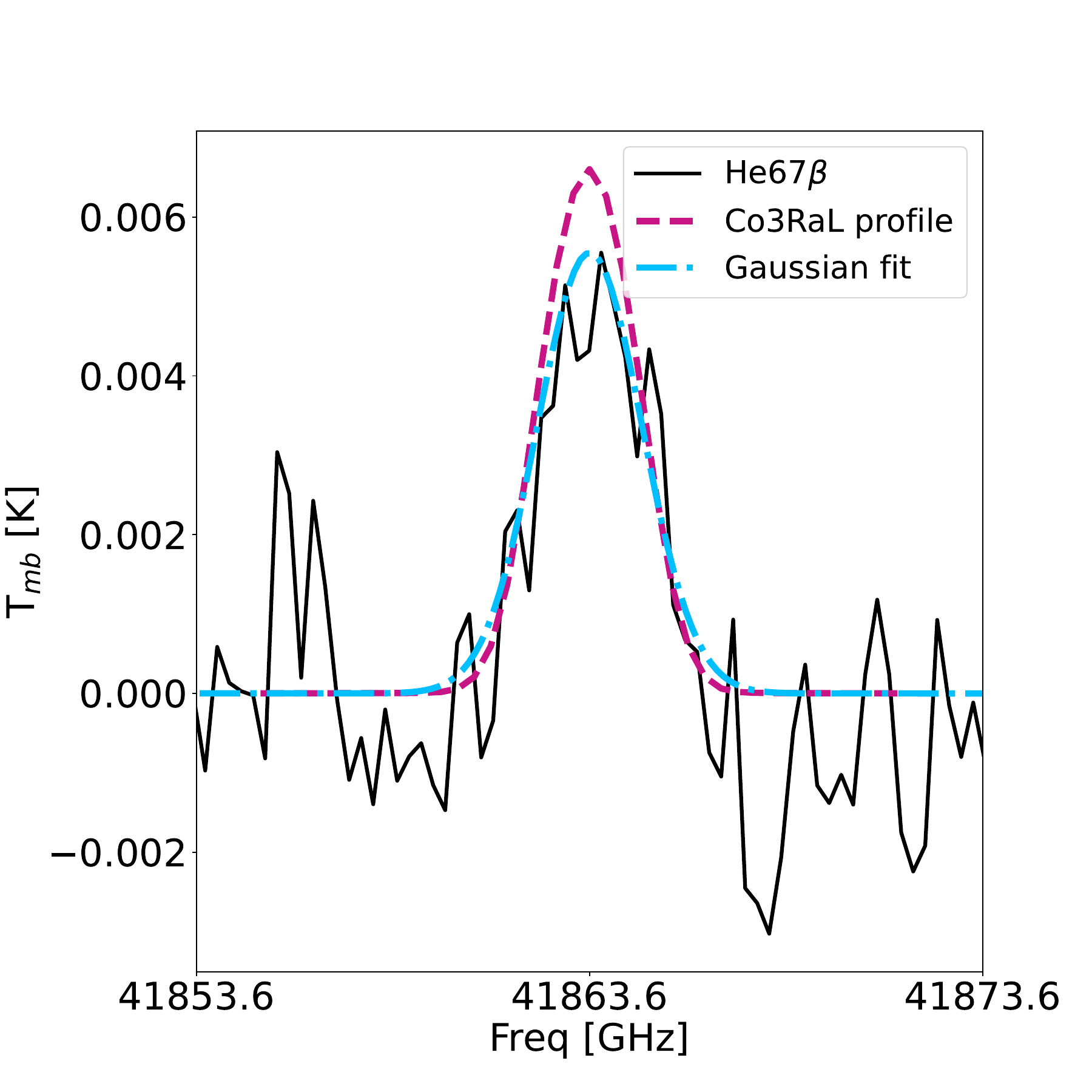}
	\includegraphics[width=0.24\textwidth]{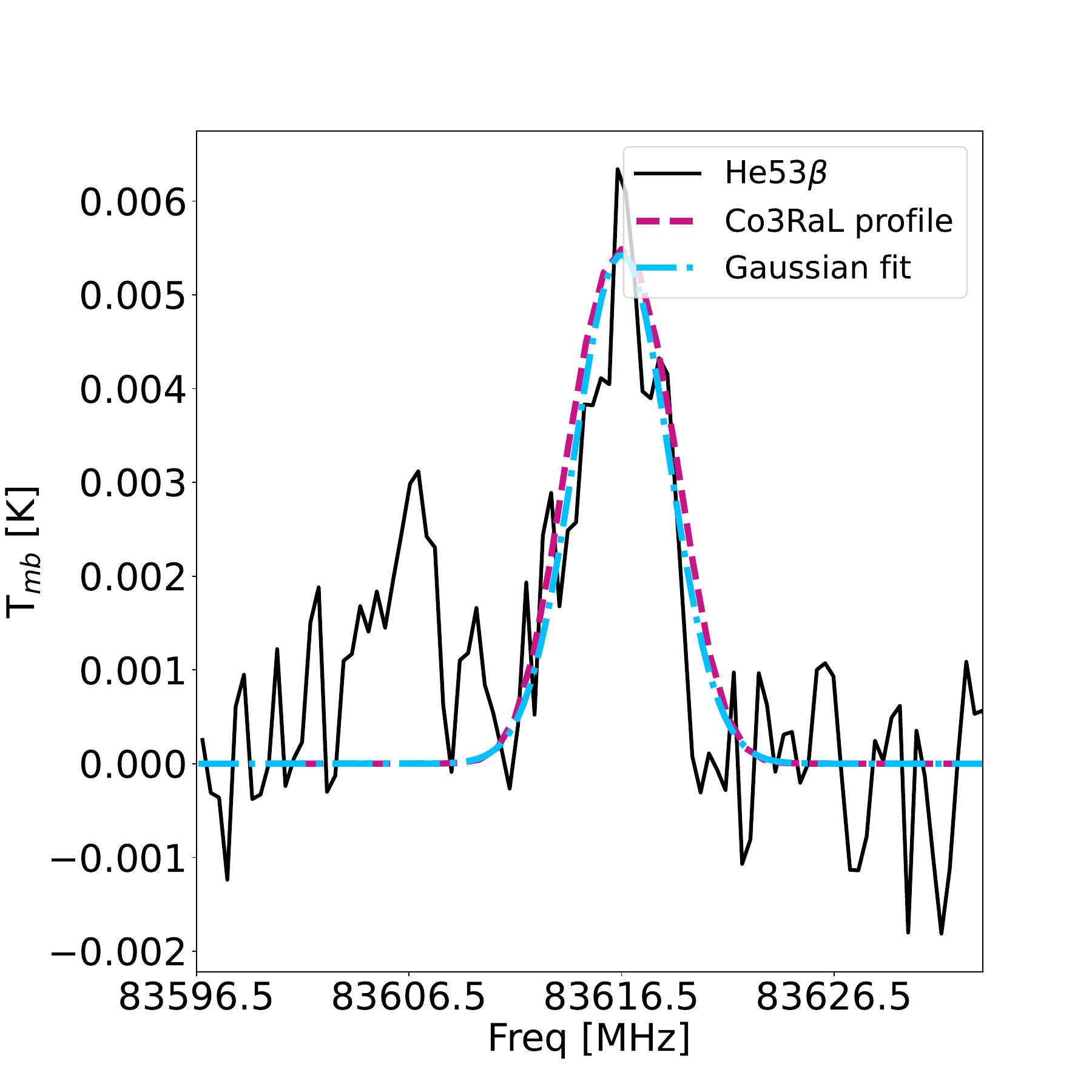}
	\includegraphics[width=0.24\textwidth]{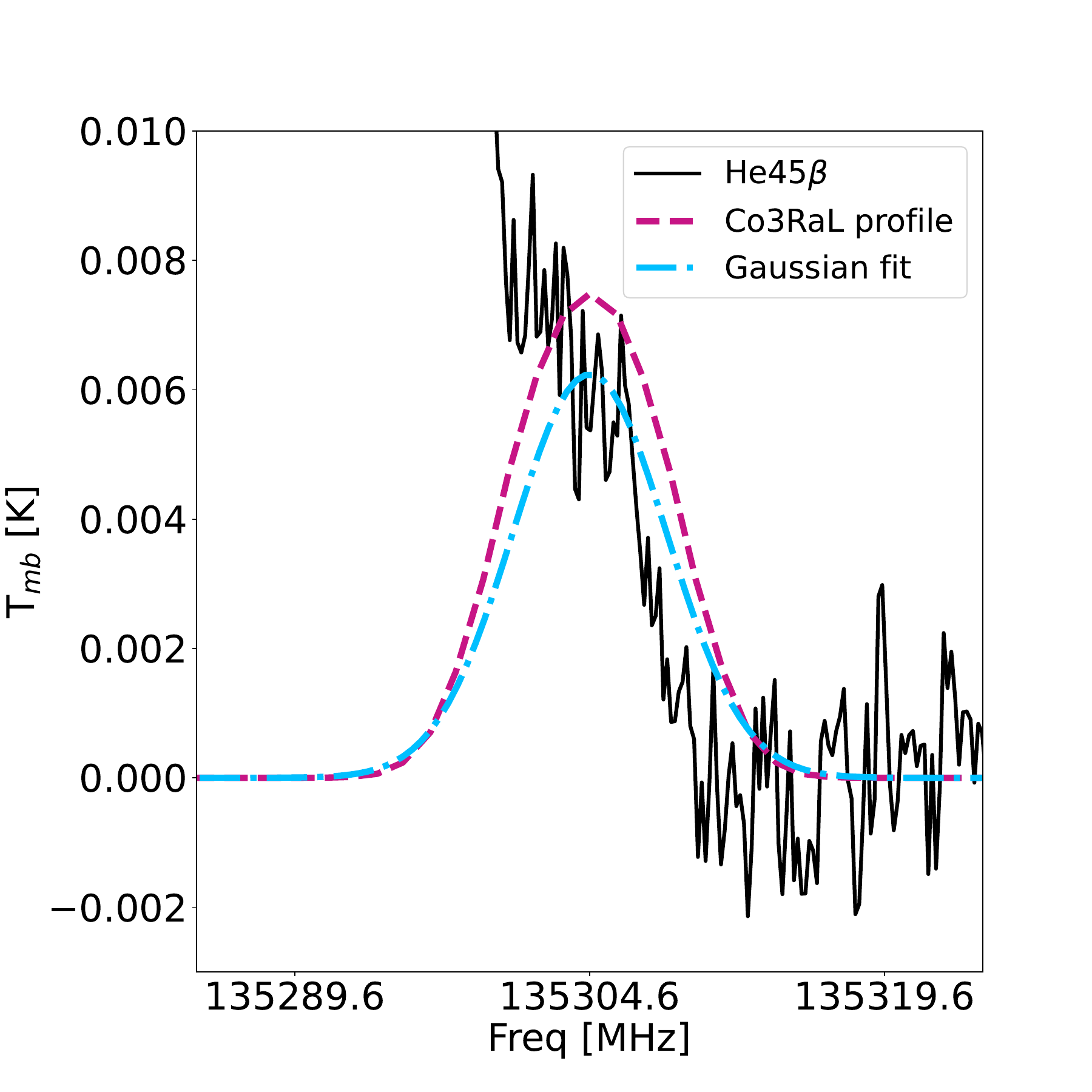}
	\caption{\hei\gb lines in \ic.} \label{fig:IC418_HeIgb}
\end{figure*}

\begin{figure*}[!h]
	\centering
	\includegraphics[width=0.24\textwidth]{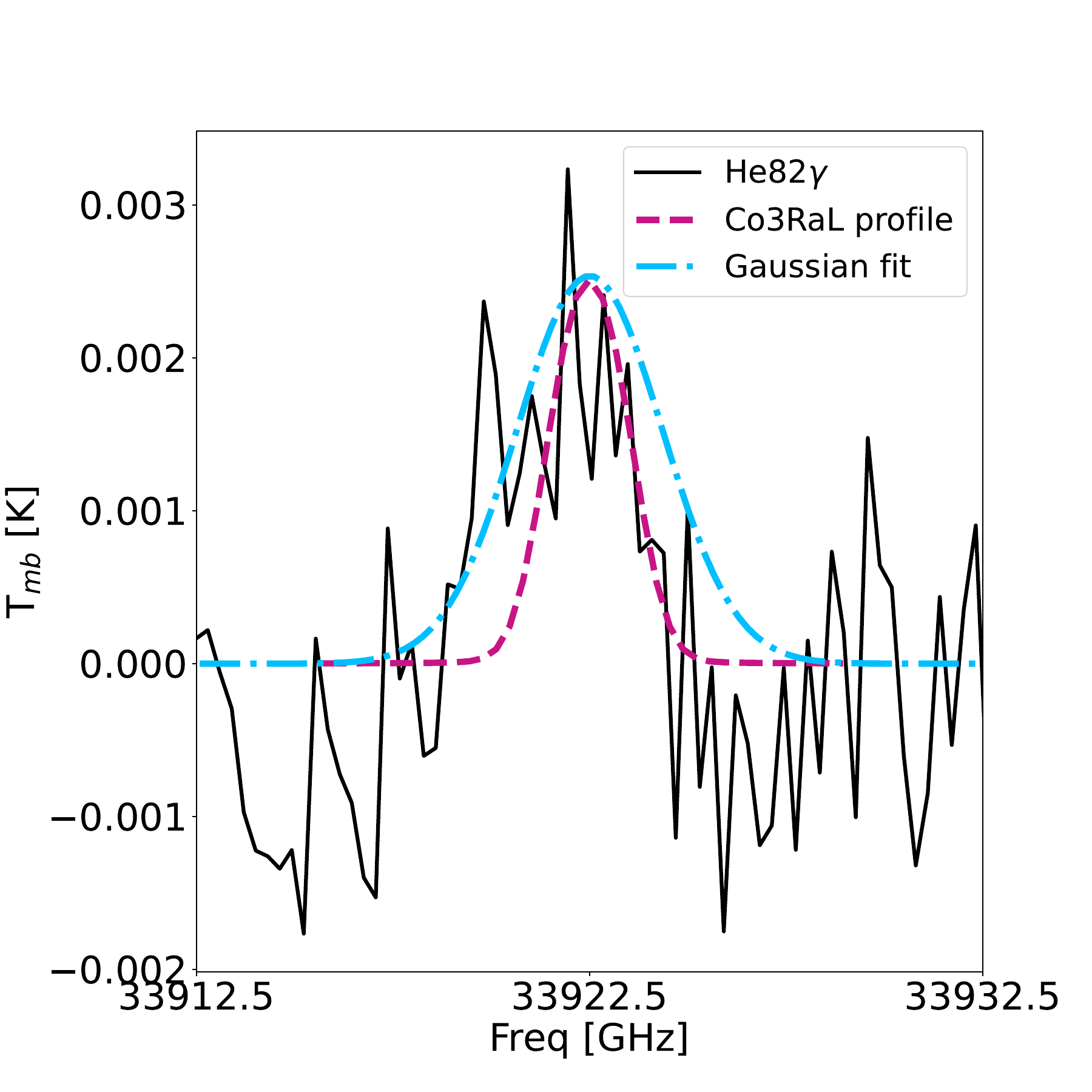}
	\includegraphics[width=0.24\textwidth]{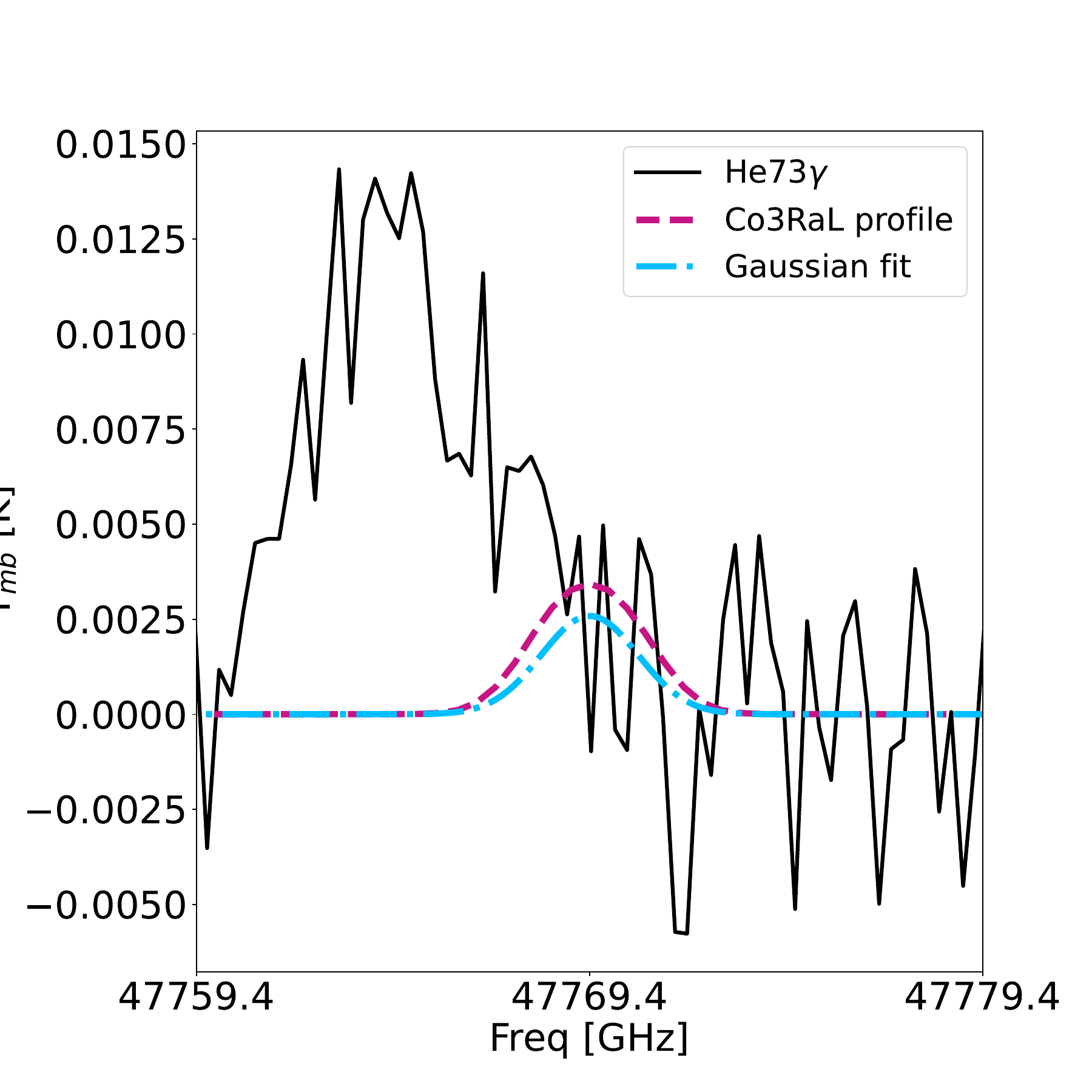}
	\includegraphics[width=0.24\textwidth]{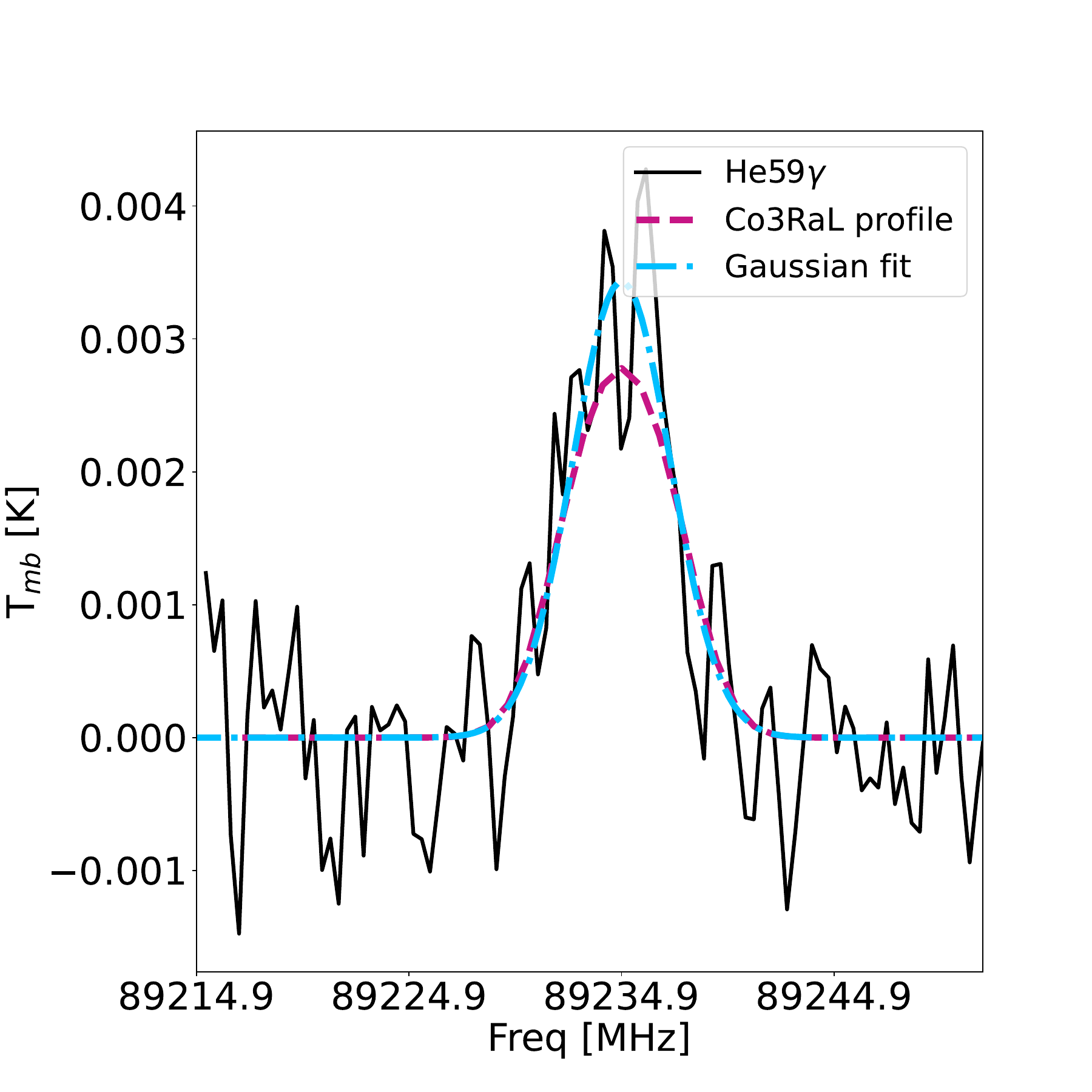}
	\includegraphics[width=0.24\textwidth]{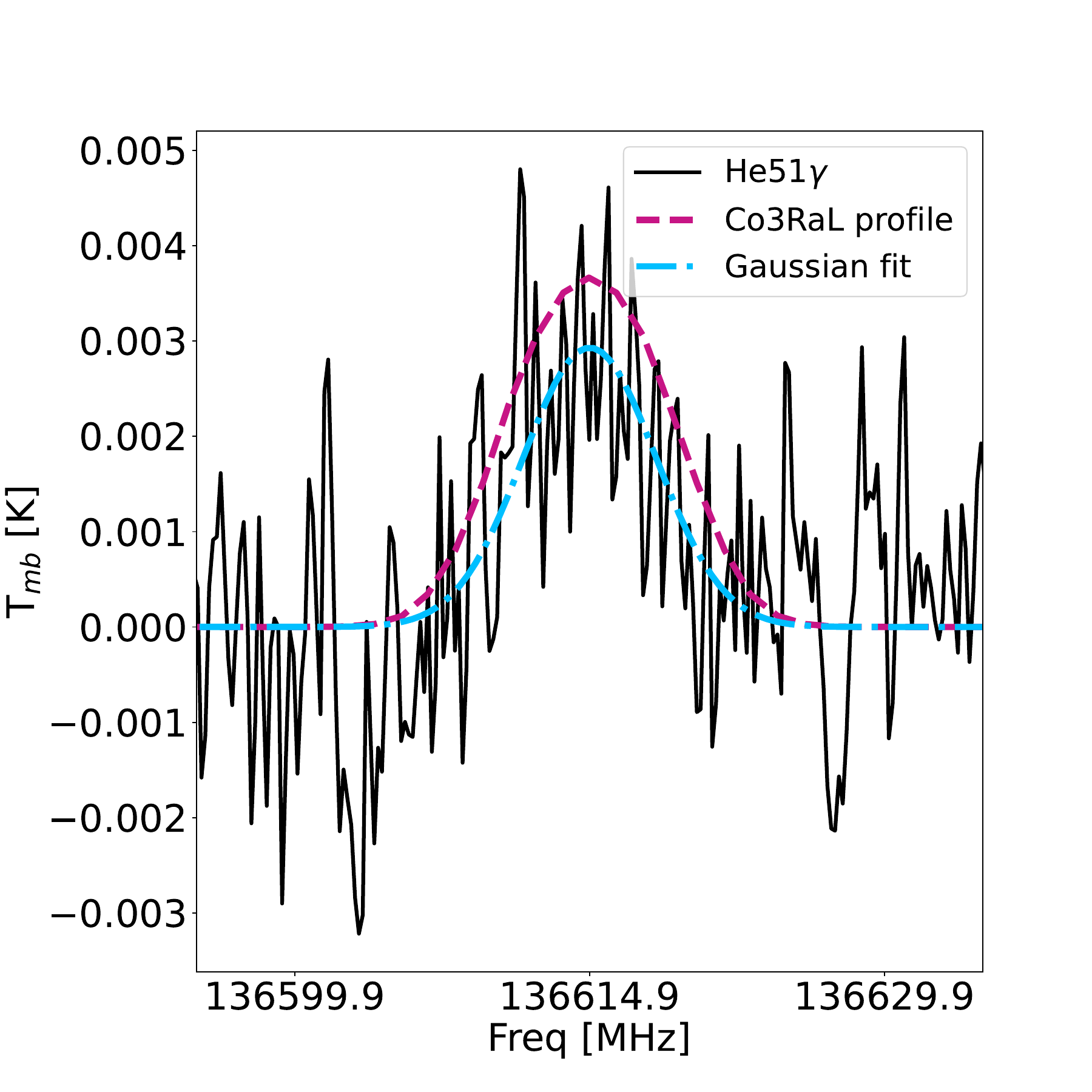}
	\caption{\hei\gg lines in \ic. \hei82\gg blended with H116\gi, \hei73\gg blended with H90\gd (see Table \ref{tab:rrls_parameters}).} \label{fig:IC418_HeIgg}
\end{figure*}

We present here examples of the H and \hei RRLs found on \ic. Four examples are displayed for each \Dn and species when data is available: two lines at 7\mm, one line at 3\mm, and one lines at 2\mm.See Figs. \ref{fig:IC418_Hga}, \ref{fig:IC418_Hgb}, \ref{fig:IC418_Hgg}, \ref{fig:IC418_Hgd}, \ref{fig:IC418_Hge}, \ref{fig:IC418_Hgz}, \ref{fig:IC418_Hgh}, \ref{fig:IC418_Hgq}, \ref{fig:IC418_Hgi}, \ref{fig:IC418_Hgk}, \ref{fig:IC418_Hglgmgn}, \ref{fig:IC418_HeIga}, \ref{fig:IC418_HeIgb}, \ref{fig:IC418_HeIgg}.

\subsection{NGC 7027} \label{sec:profiles_NGC7027}

\begin{figure*}[!h]
	\centering
	\includegraphics[width=0.24\textwidth]{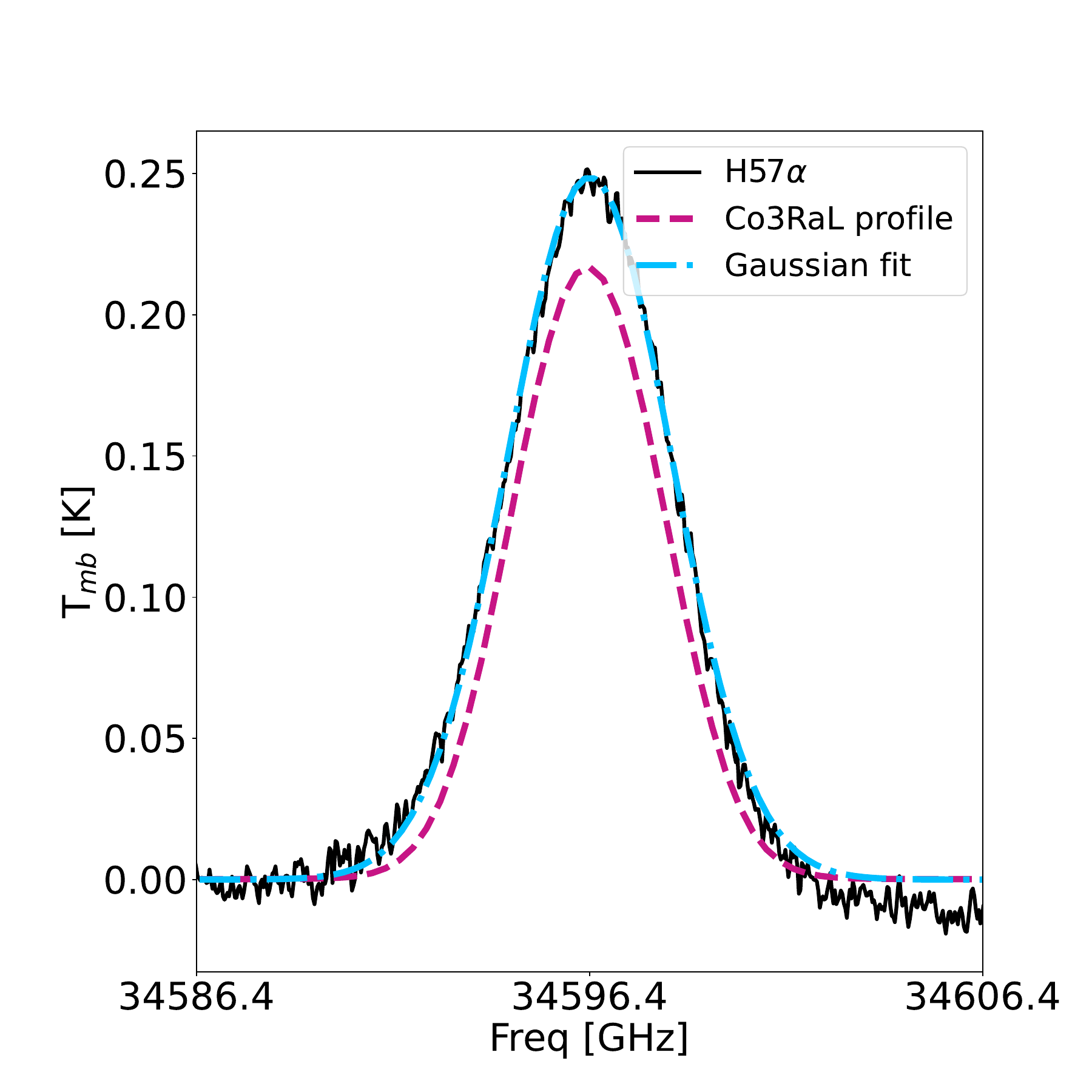}
	\includegraphics[width=0.24\textwidth]{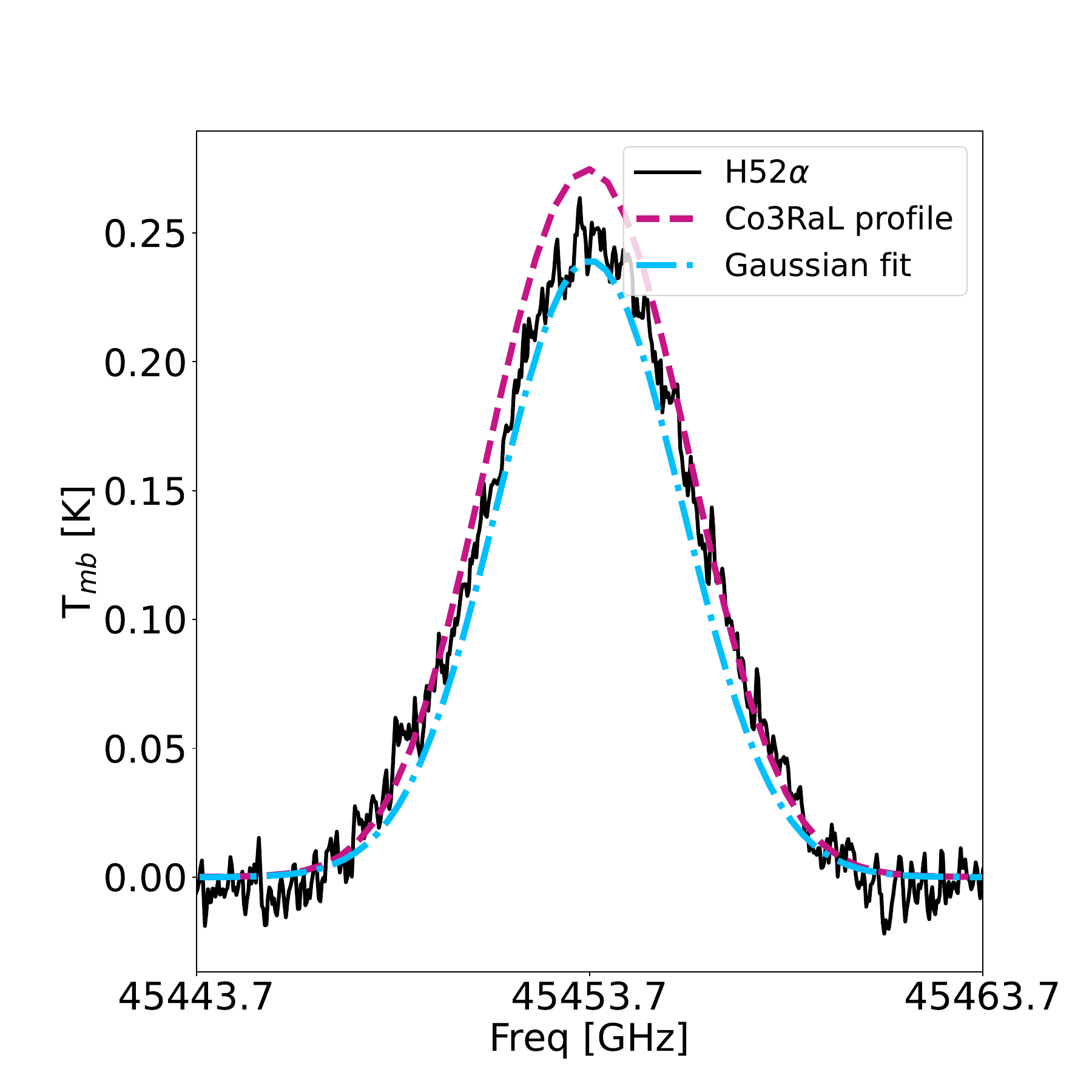}
	\includegraphics[width=0.24\textwidth]{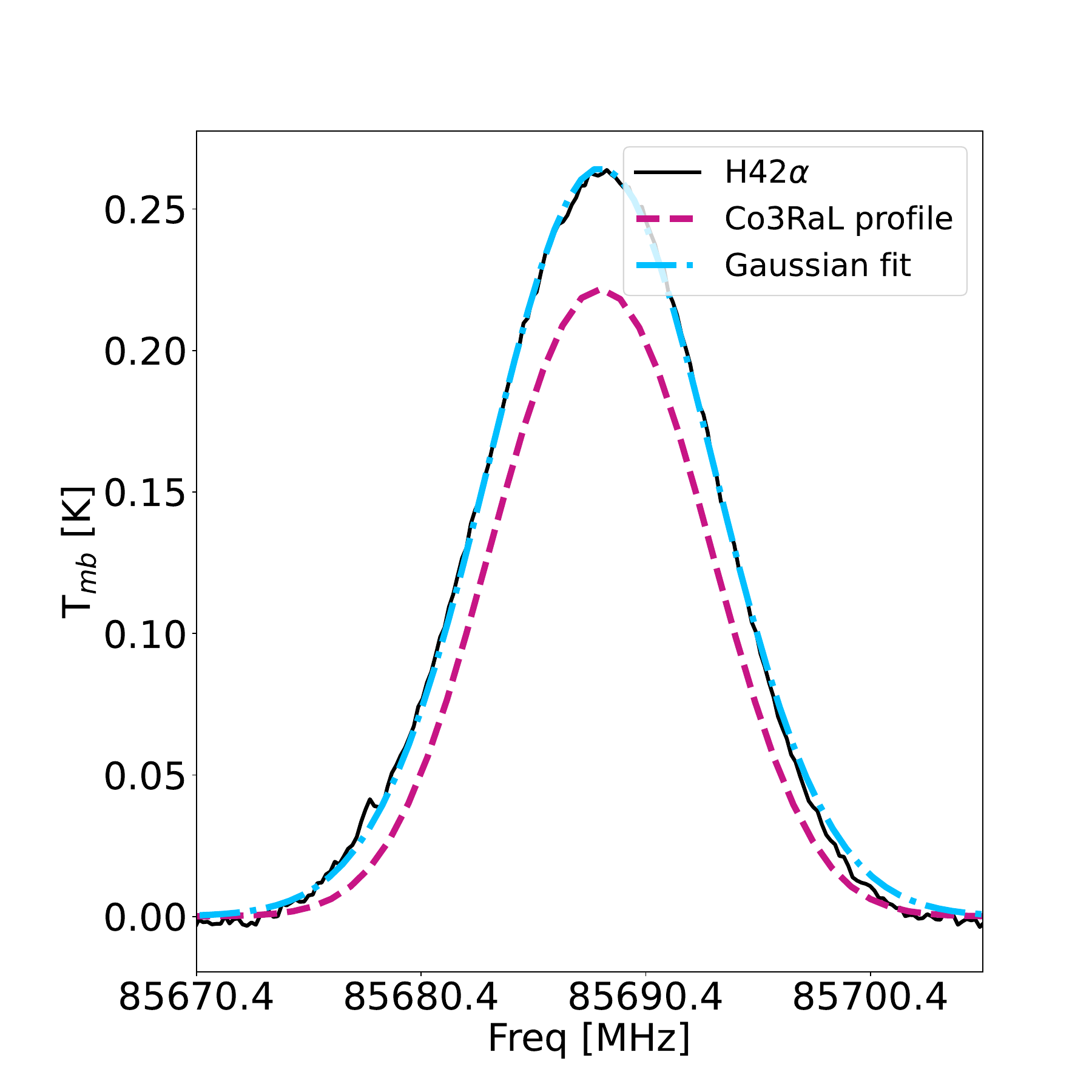}
	\includegraphics[width=0.24\textwidth]{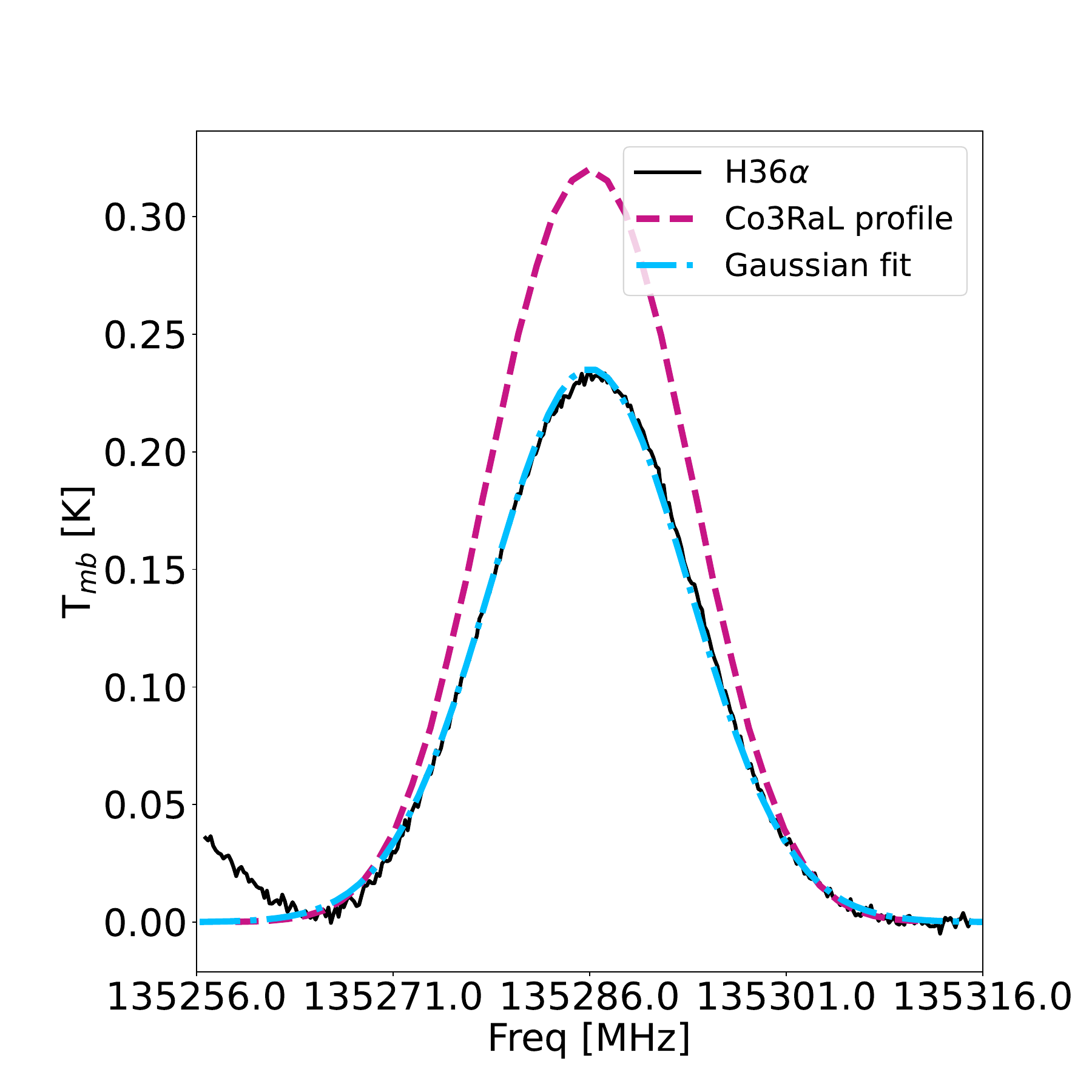}
	\caption{H\ga lines in \ngc. H36\ga blended with \hei45\gb (see Table \ref{tab:rrls_parameters}).} \label{fig:NGC7027_Hga}
\end{figure*}

\begin{figure*}[!h]
	\centering
	\includegraphics[width=0.24\textwidth]{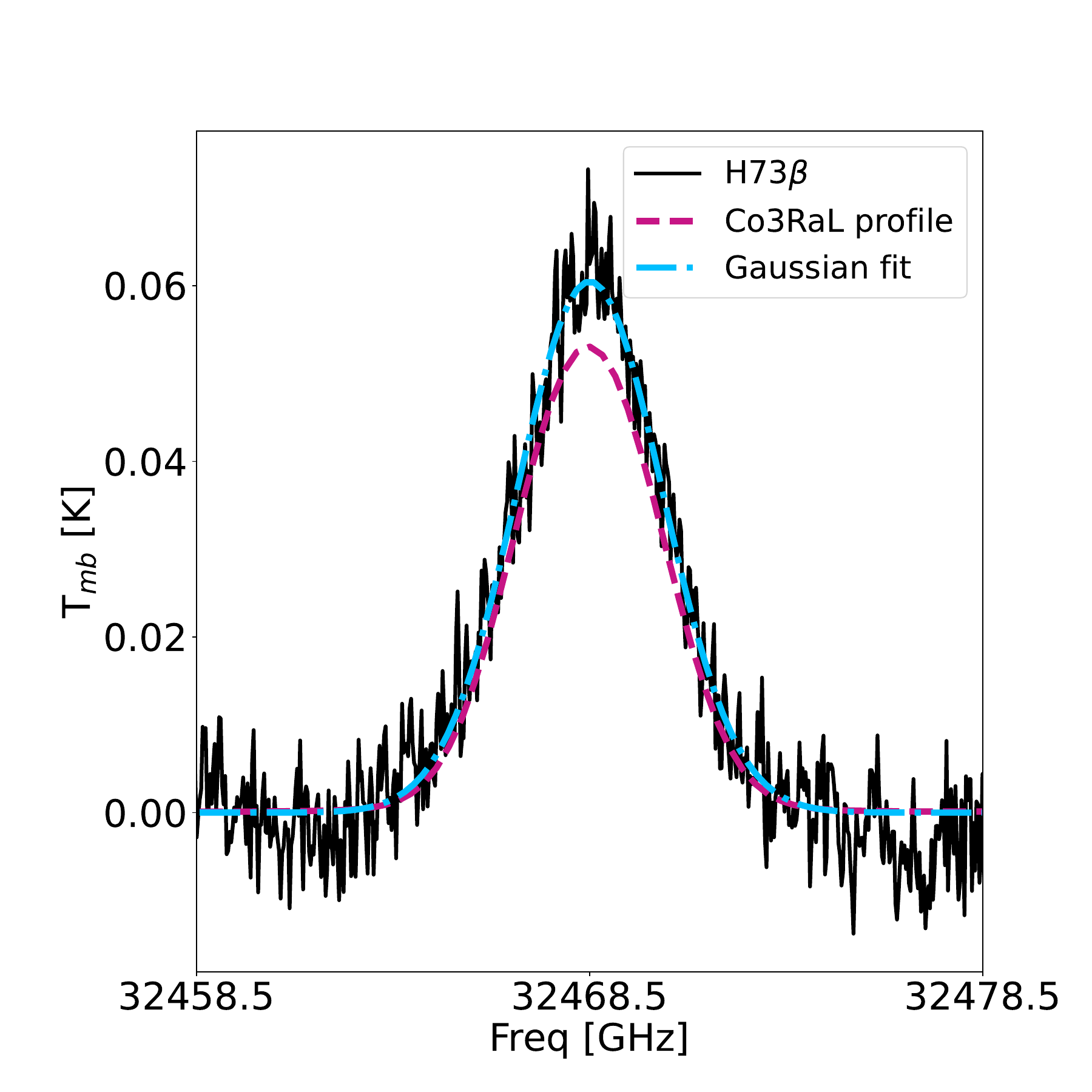}
	\includegraphics[width=0.24\textwidth]{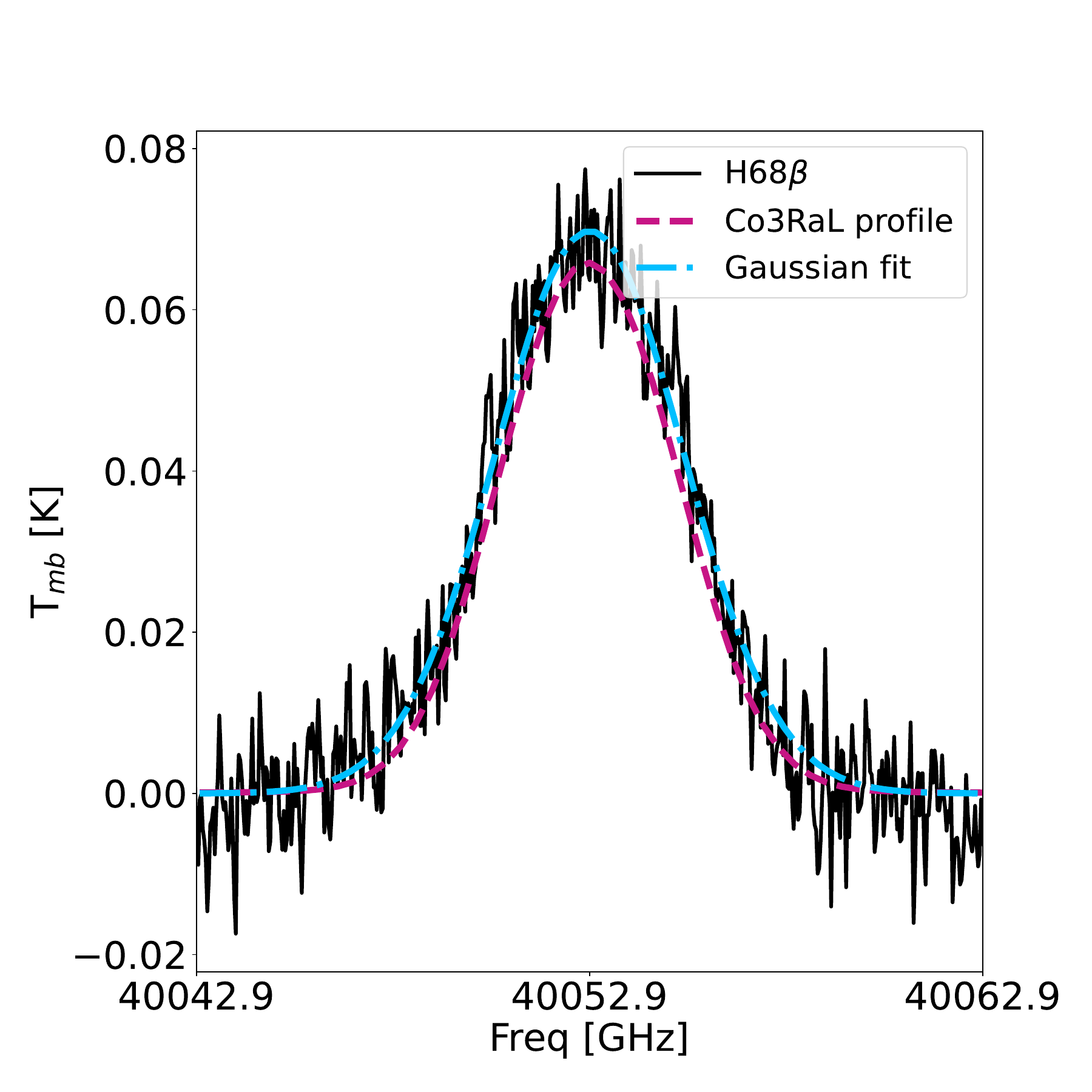}
	\includegraphics[width=0.24\textwidth]{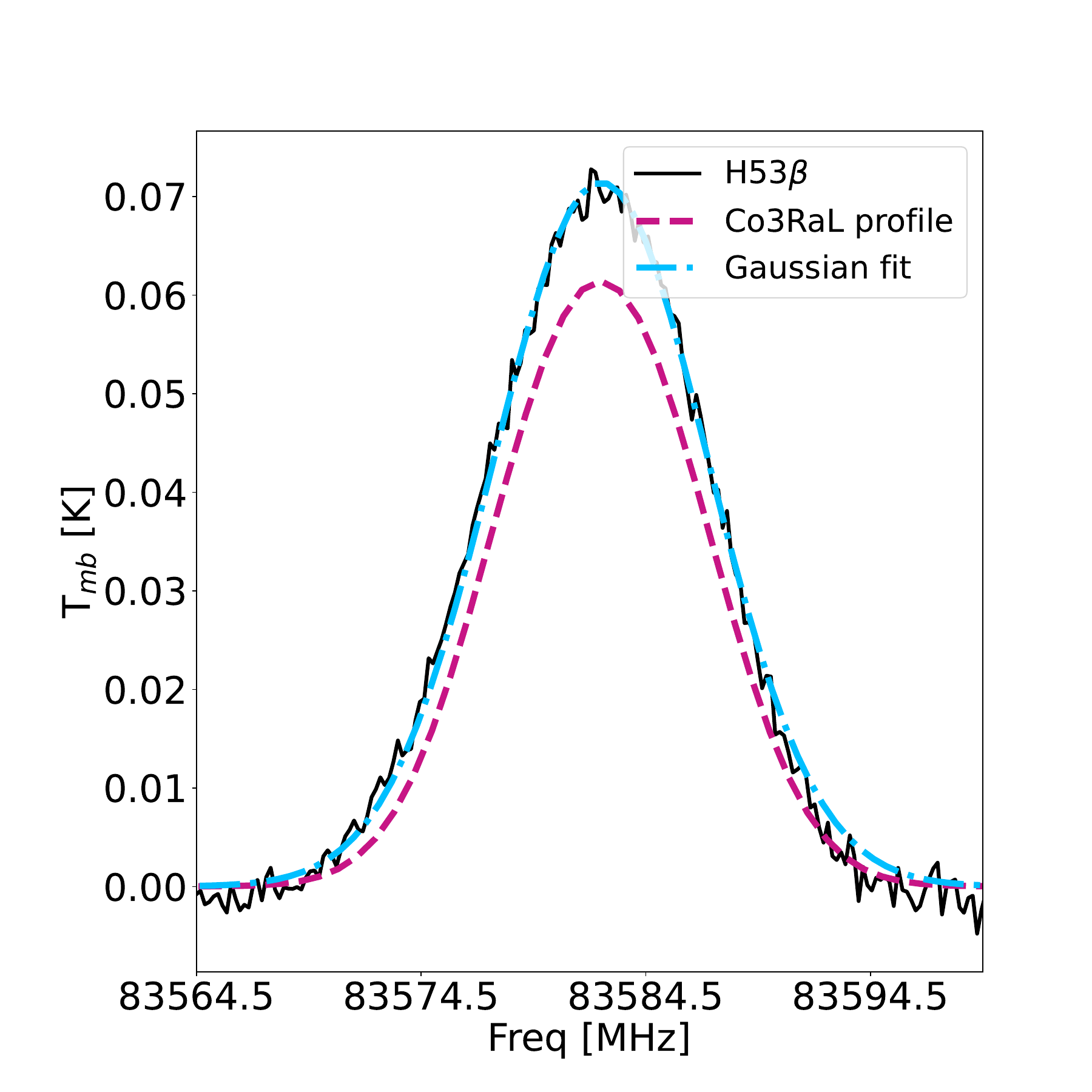}
	\includegraphics[width=0.24\textwidth]{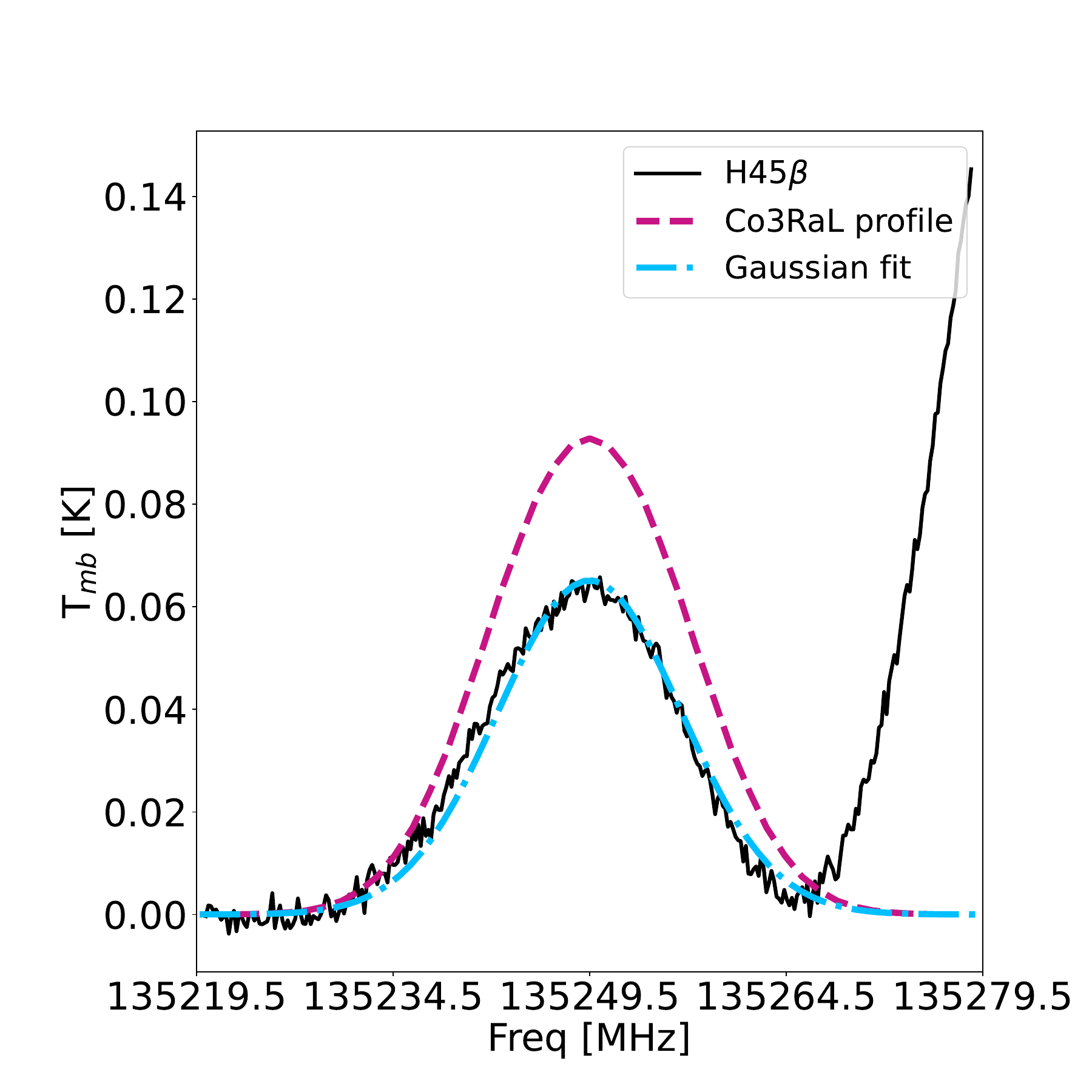}
	\caption{H\gb lines in \ngc.} \label{fig:NGC7027_Hgb}
\end{figure*}

\begin{figure*}[!h]
	\centering
	\includegraphics[width=0.24\textwidth]{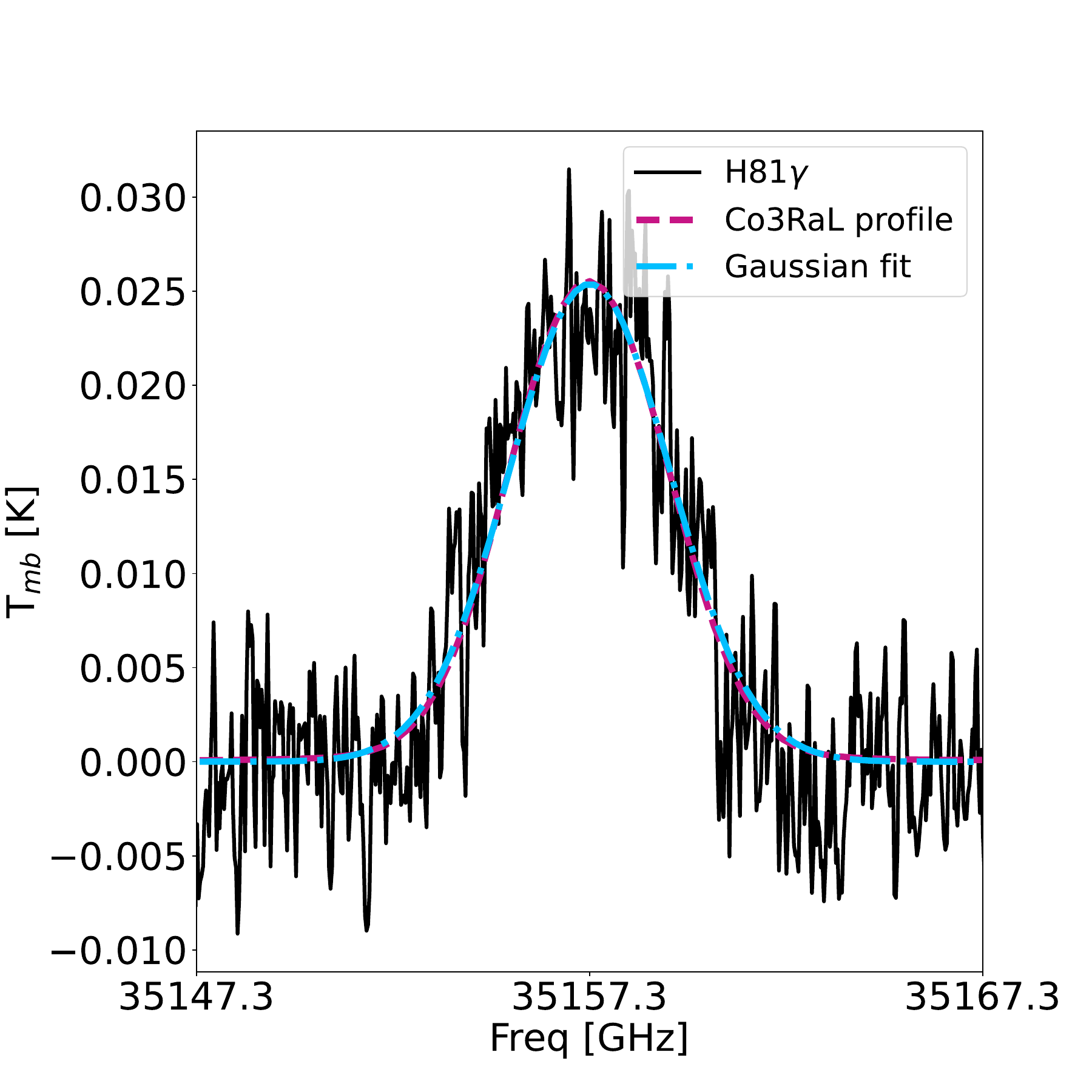}
	\includegraphics[width=0.24\textwidth]{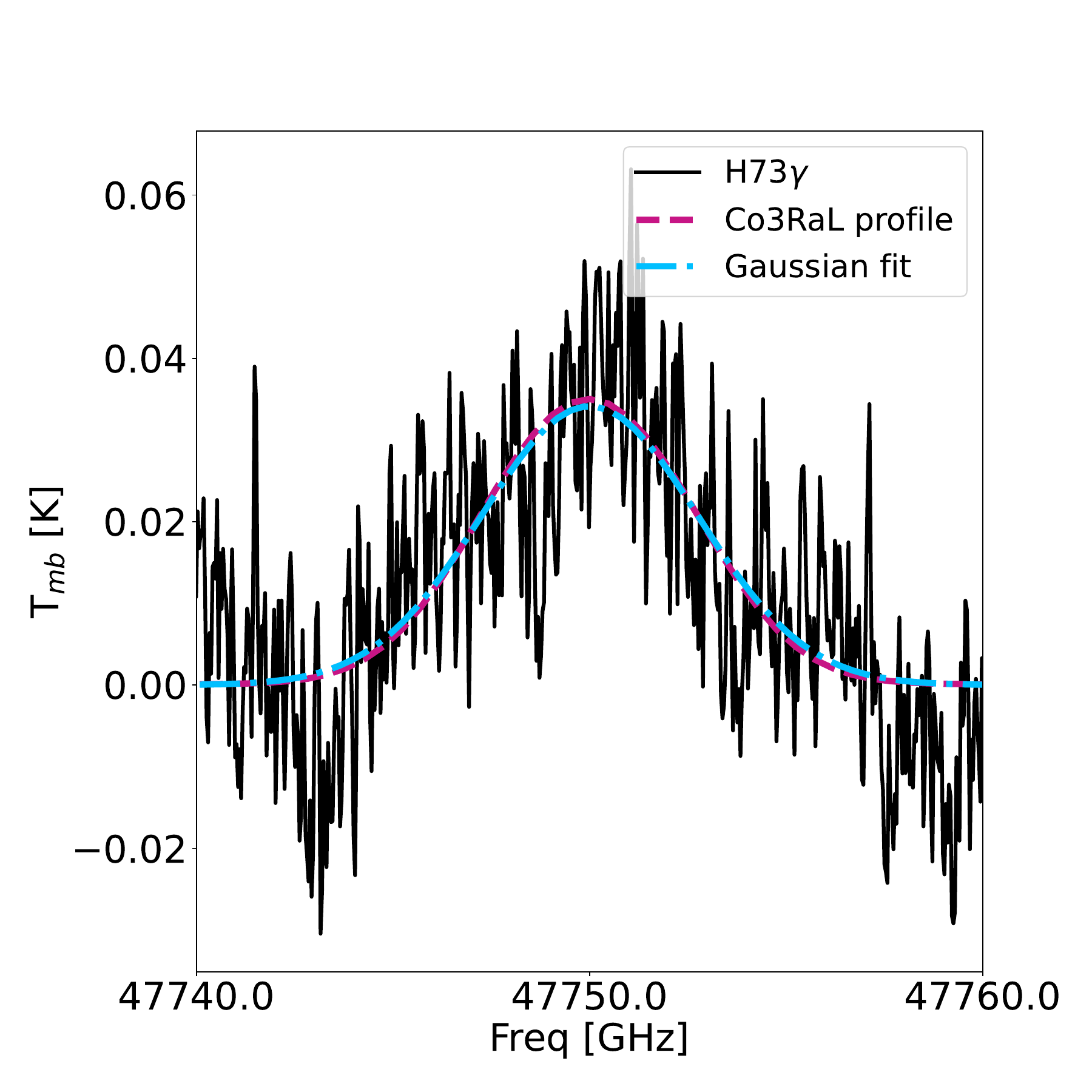}
	\includegraphics[width=0.24\textwidth]{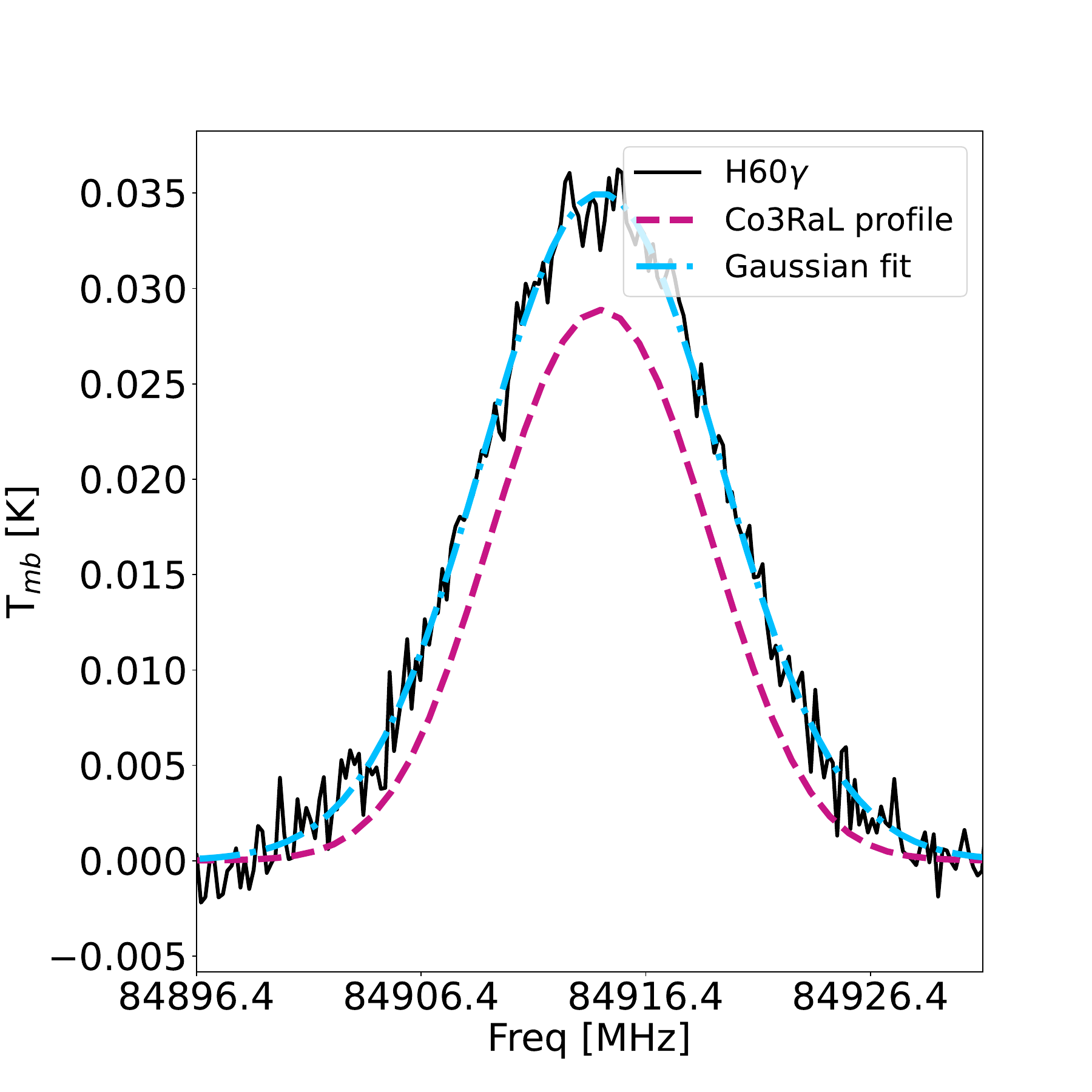}
	\includegraphics[width=0.24\textwidth]{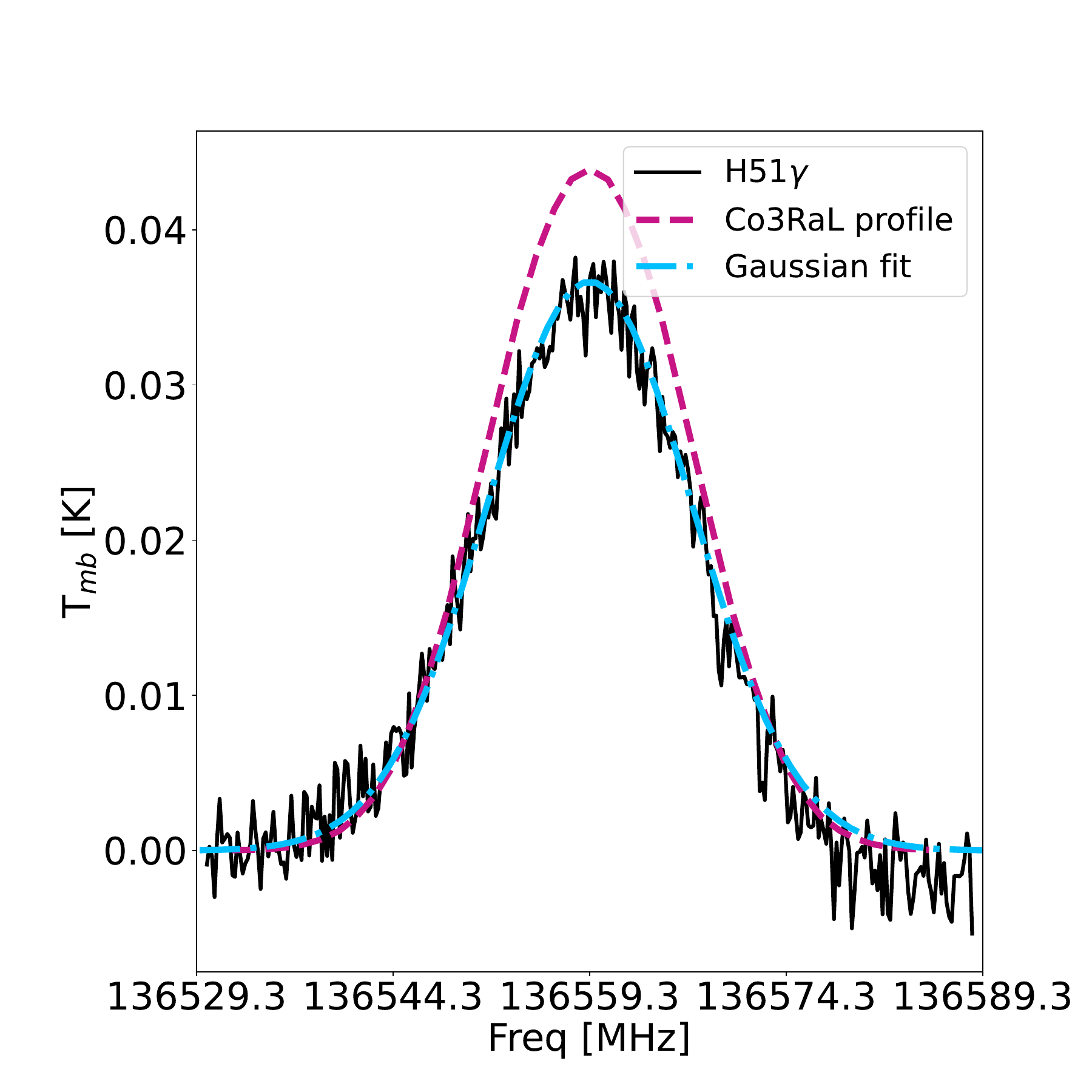}
	\caption{H\gg lines in \ngc.} \label{fig:NGC7027_Hgg}
\end{figure*}

\begin{figure*}[!h]
	\centering
	\includegraphics[width=0.24\textwidth]{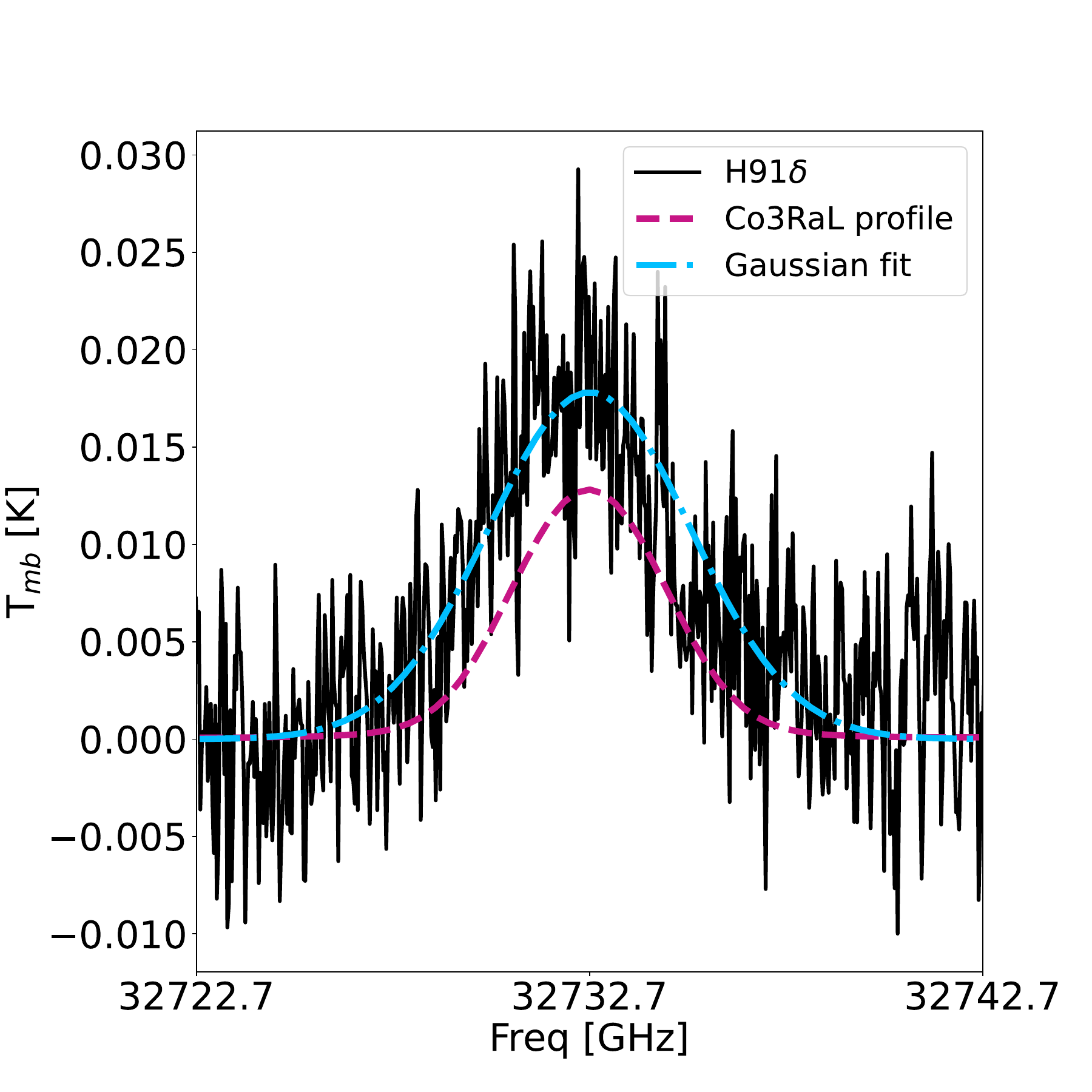}
	\includegraphics[width=0.24\textwidth]{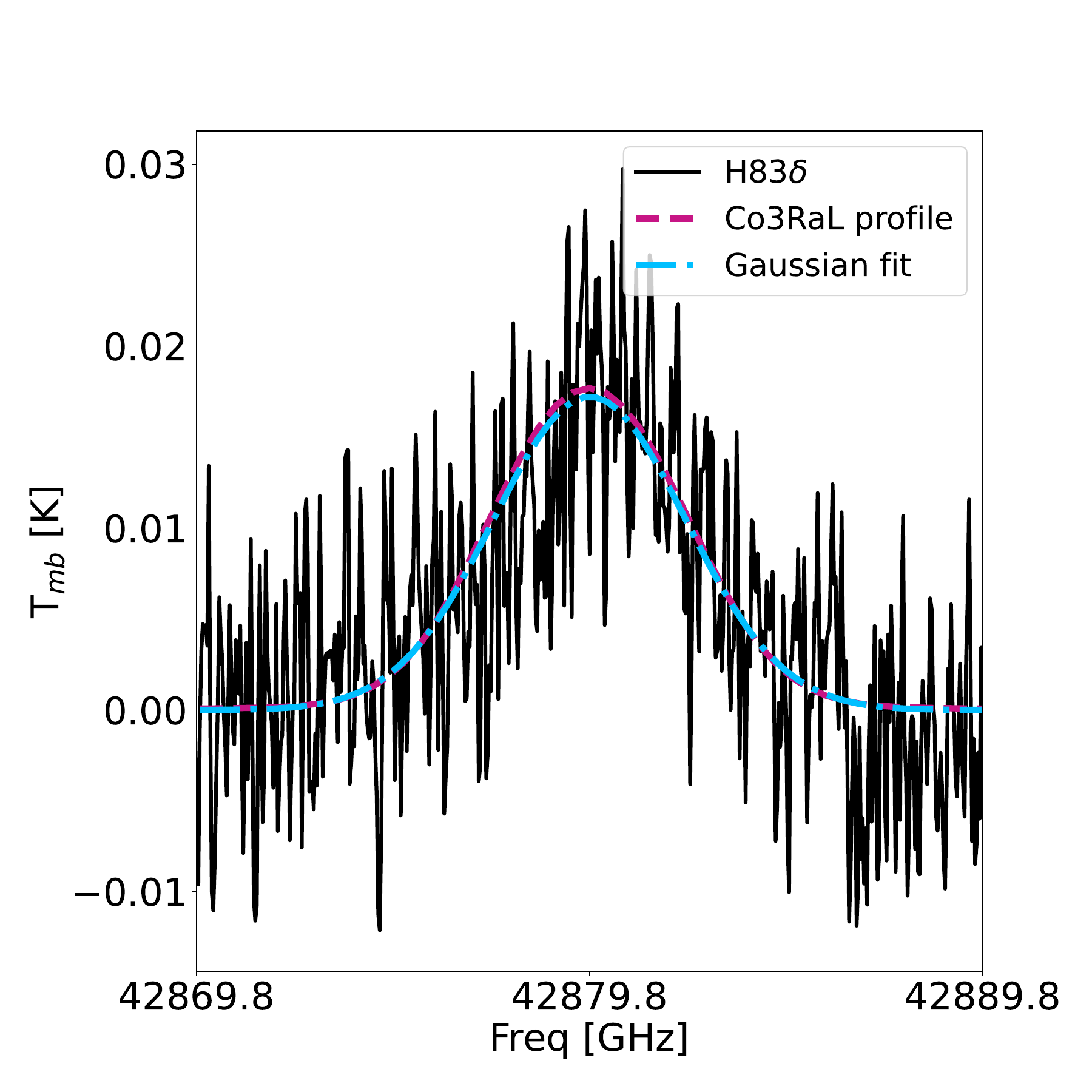}
	\includegraphics[width=0.24\textwidth]{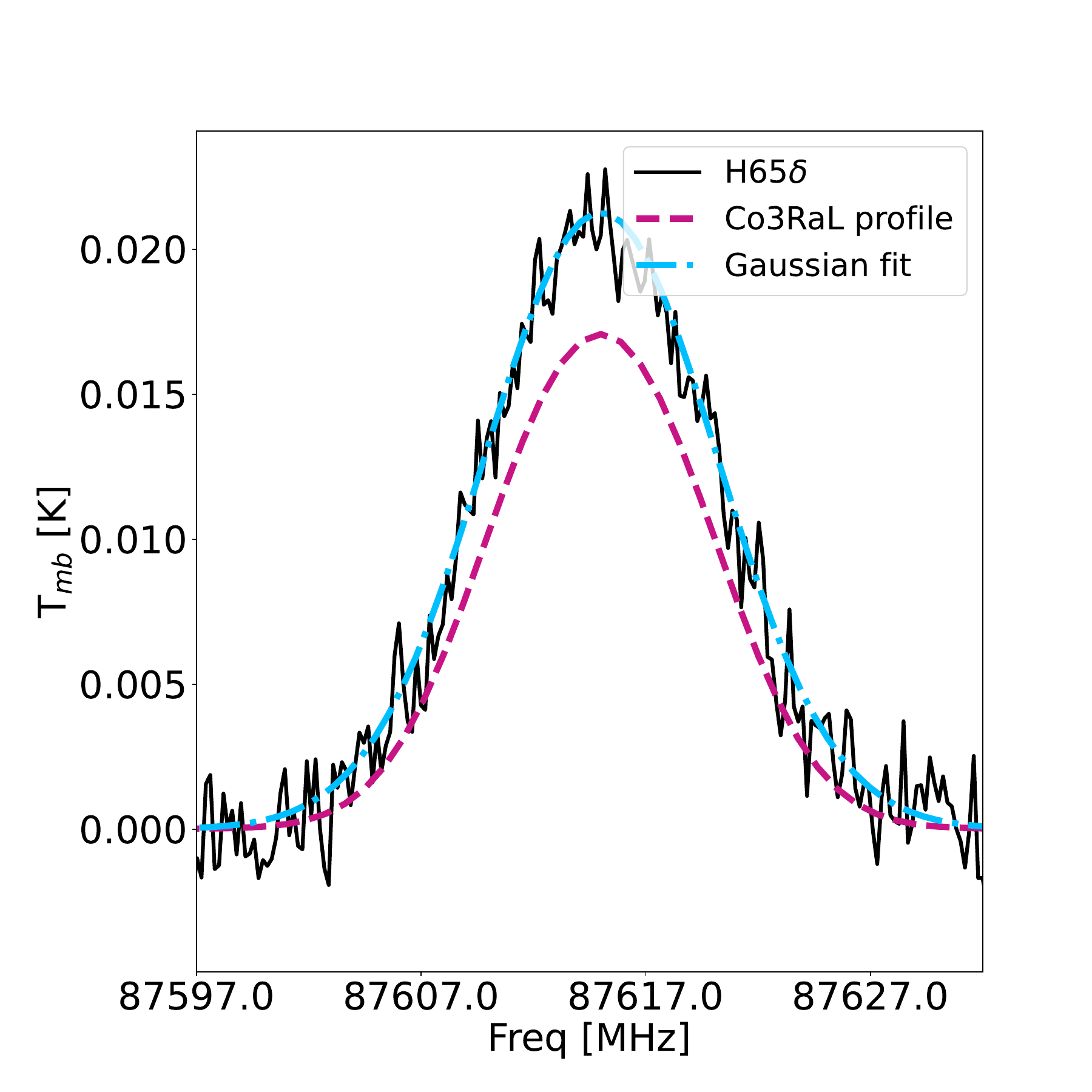}
	\includegraphics[width=0.24\textwidth]{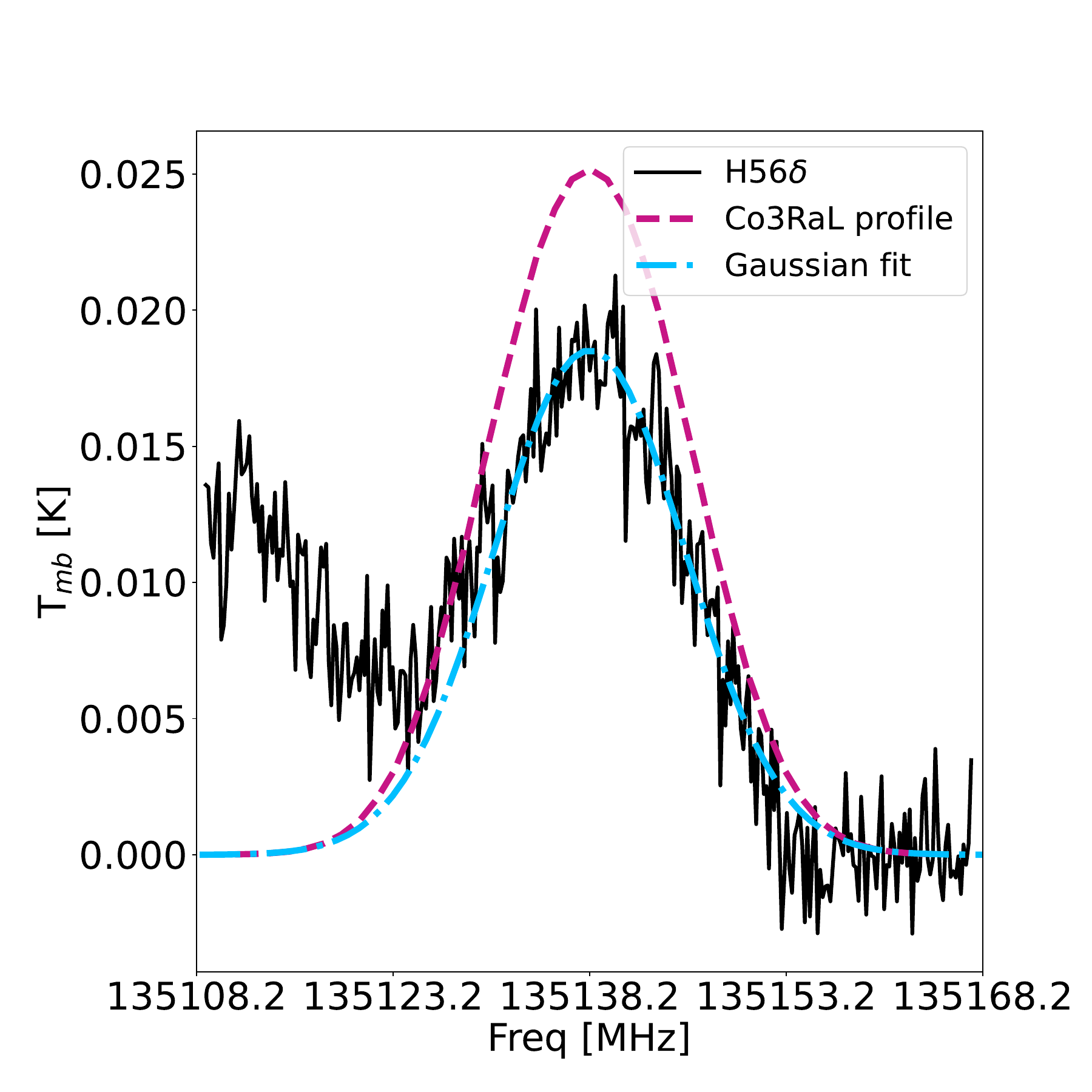}
	\caption{H\gd lines in \ngc. H56\gd blended with H60\ge (see Table \ref{tab:rrls_parameters}).} \label{fig:NGC7027_Hgd}
\end{figure*}

\begin{figure*}[!h]
	\centering
	\includegraphics[width=0.24\textwidth]{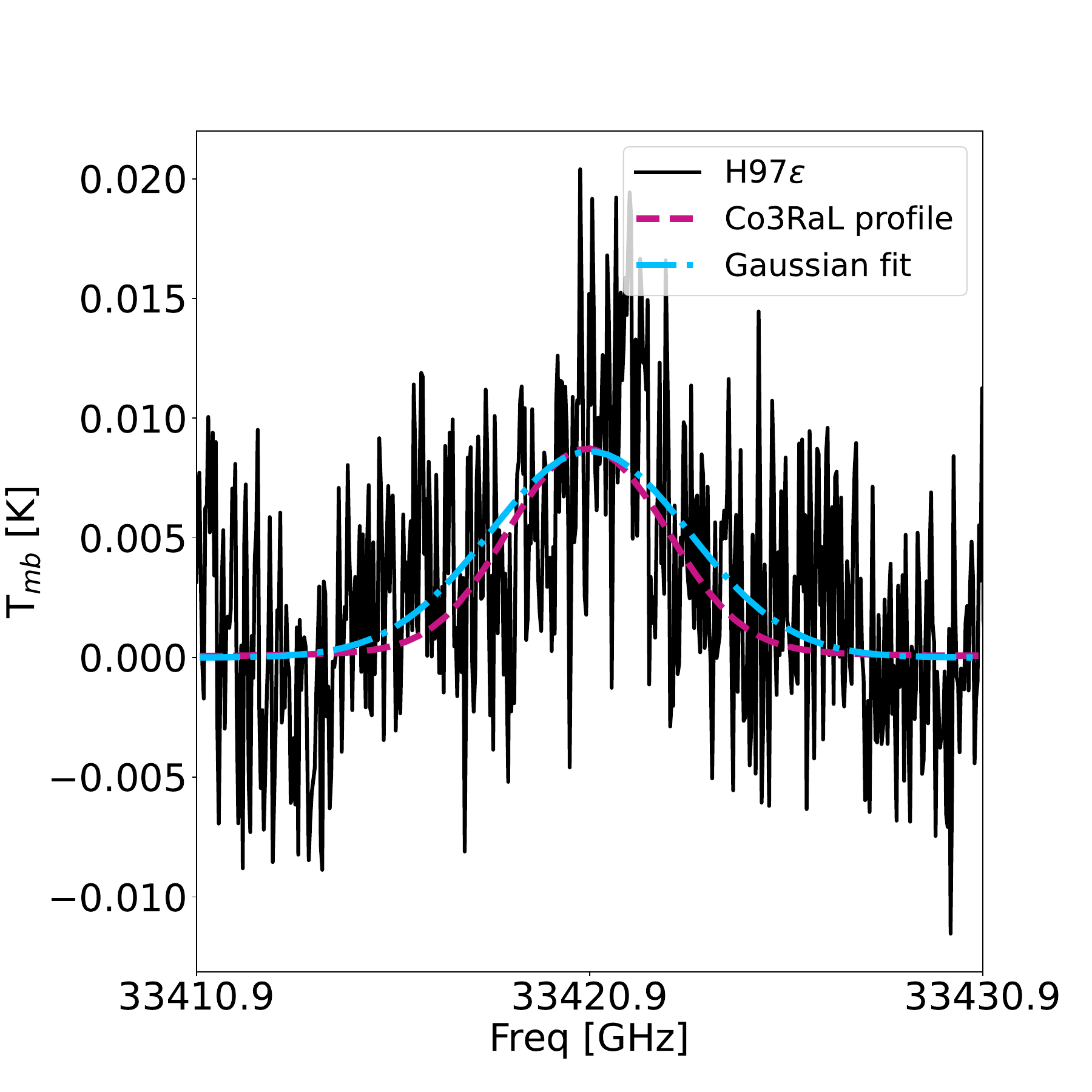}
	\includegraphics[width=0.24\textwidth]{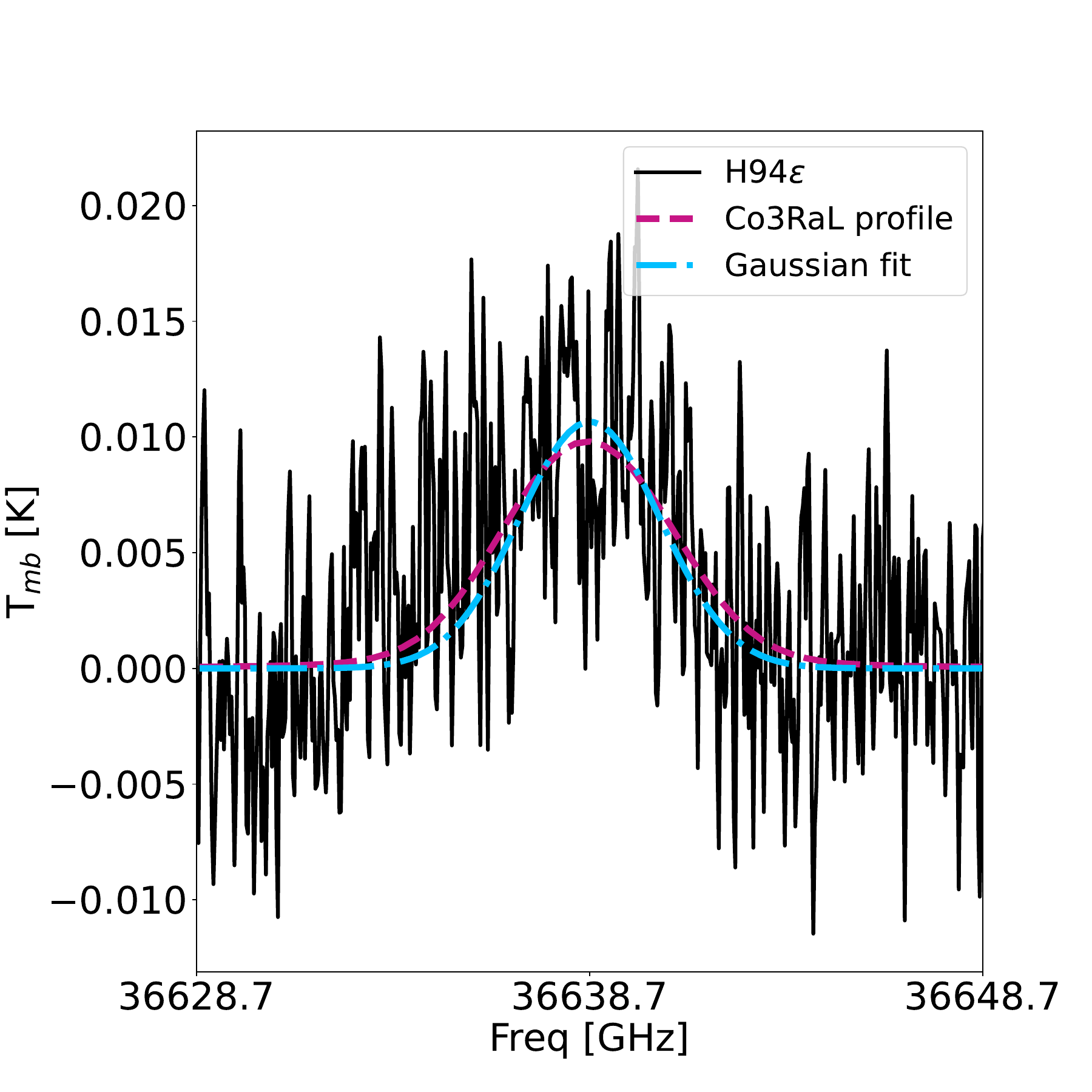}
	\includegraphics[width=0.24\textwidth]{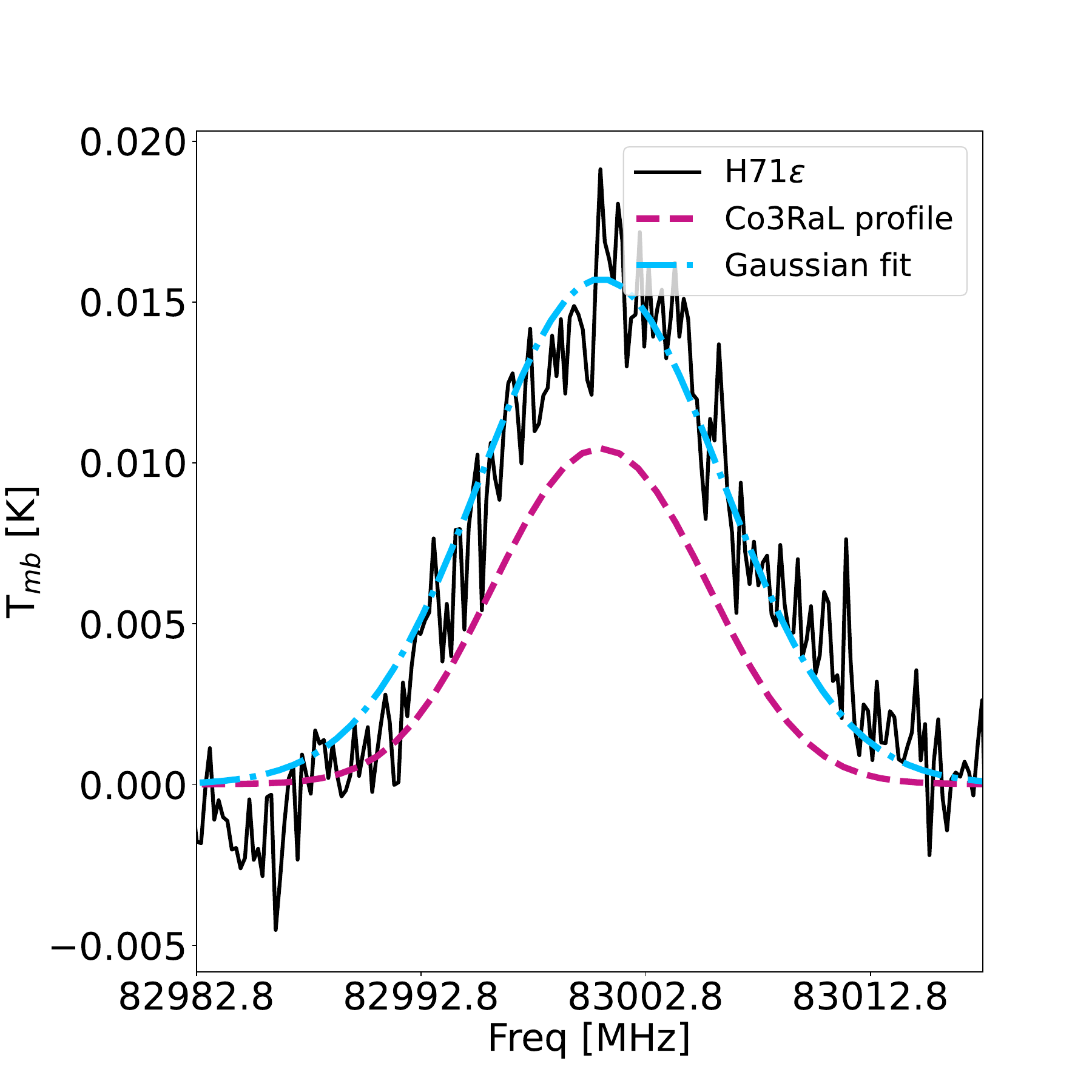}
	\includegraphics[width=0.24\textwidth]{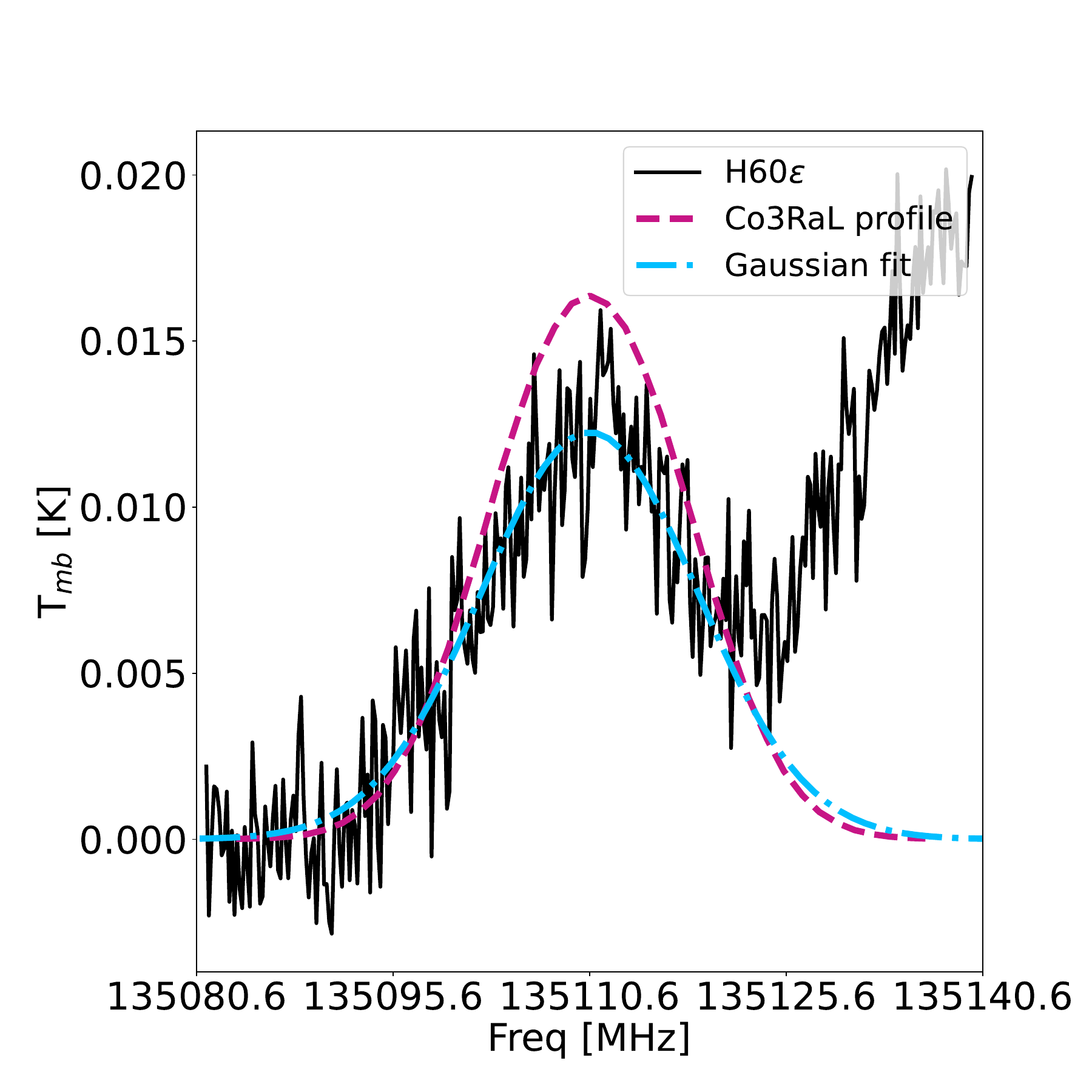}
	\caption{H\ge lines in \ngc. H60\ge blended with H56\gd (see Table \ref{tab:rrls_parameters}).} \label{fig:NGC7027_Hge}
\end{figure*}

\begin{figure*}[!h]
	\centering
	\includegraphics[width=0.24\textwidth]{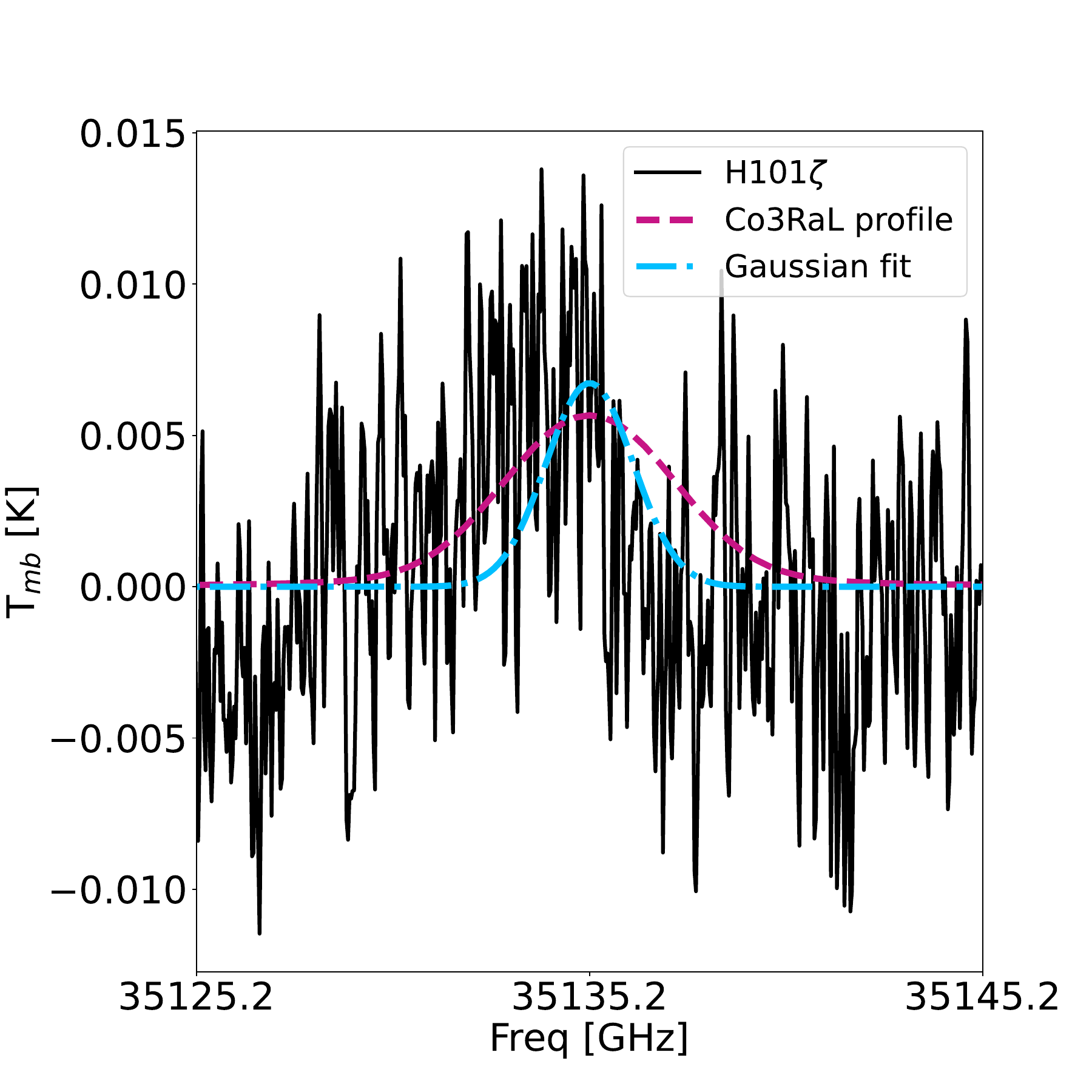}
	\includegraphics[width=0.24\textwidth]{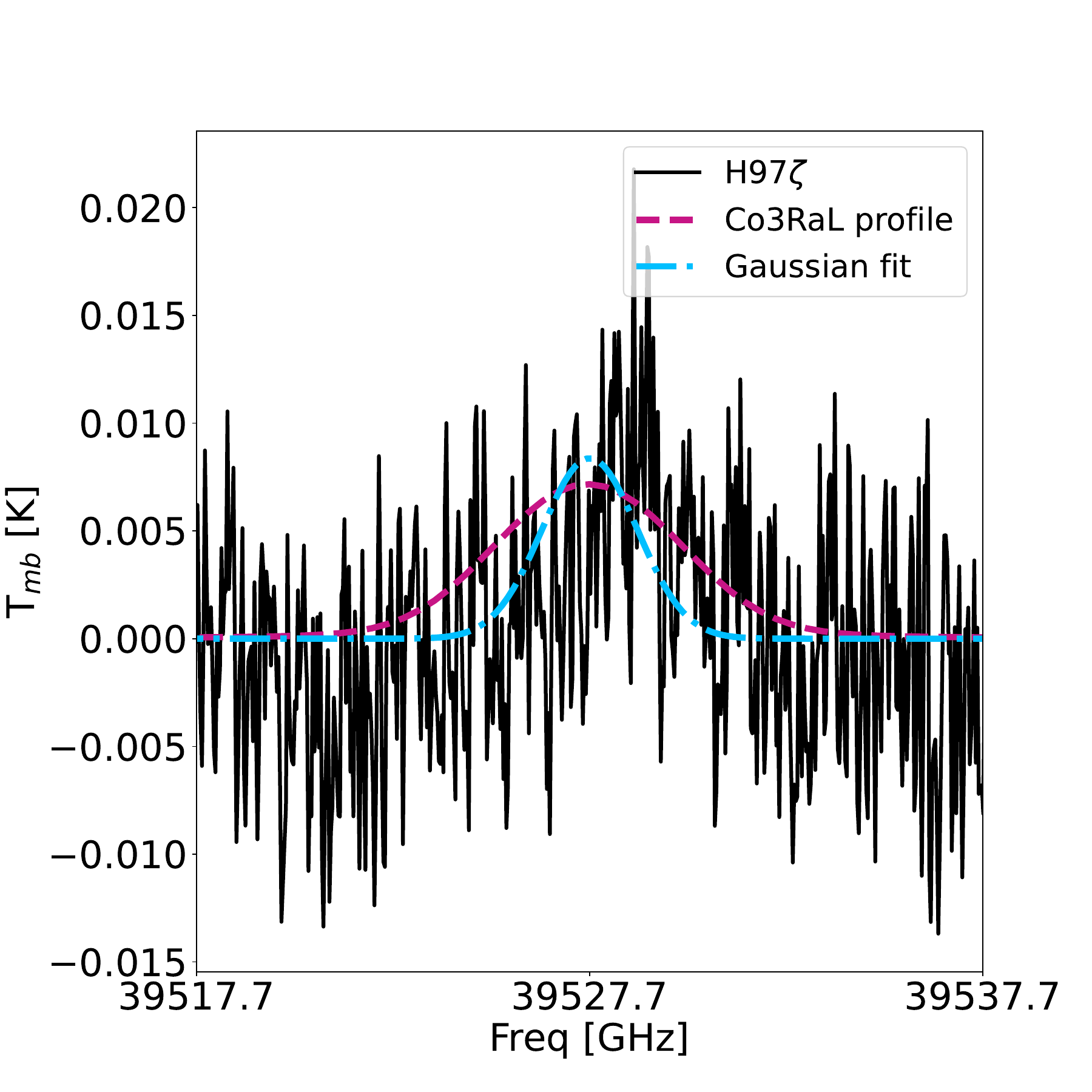}
	\includegraphics[width=0.24\textwidth]{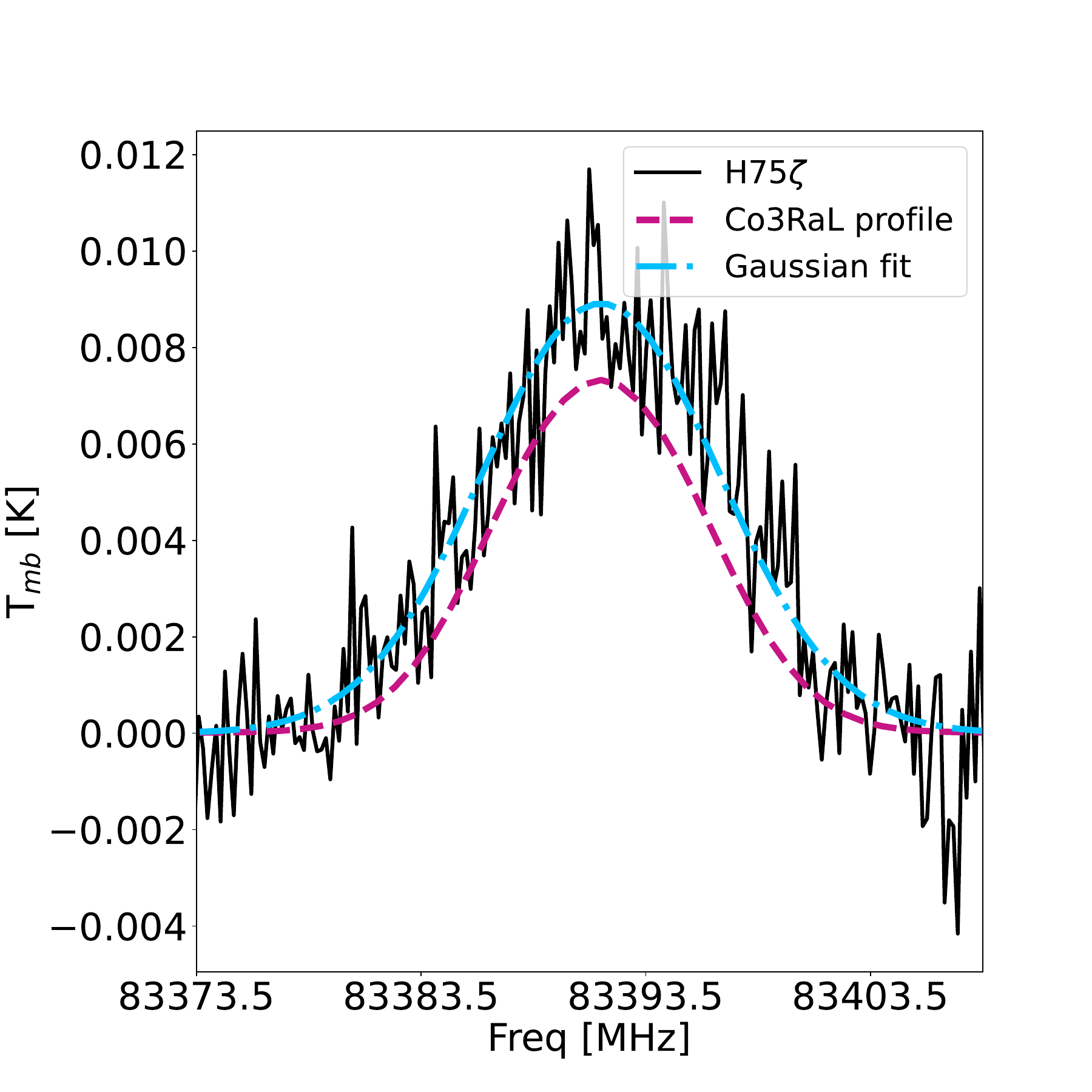}
	\includegraphics[width=0.24\textwidth]{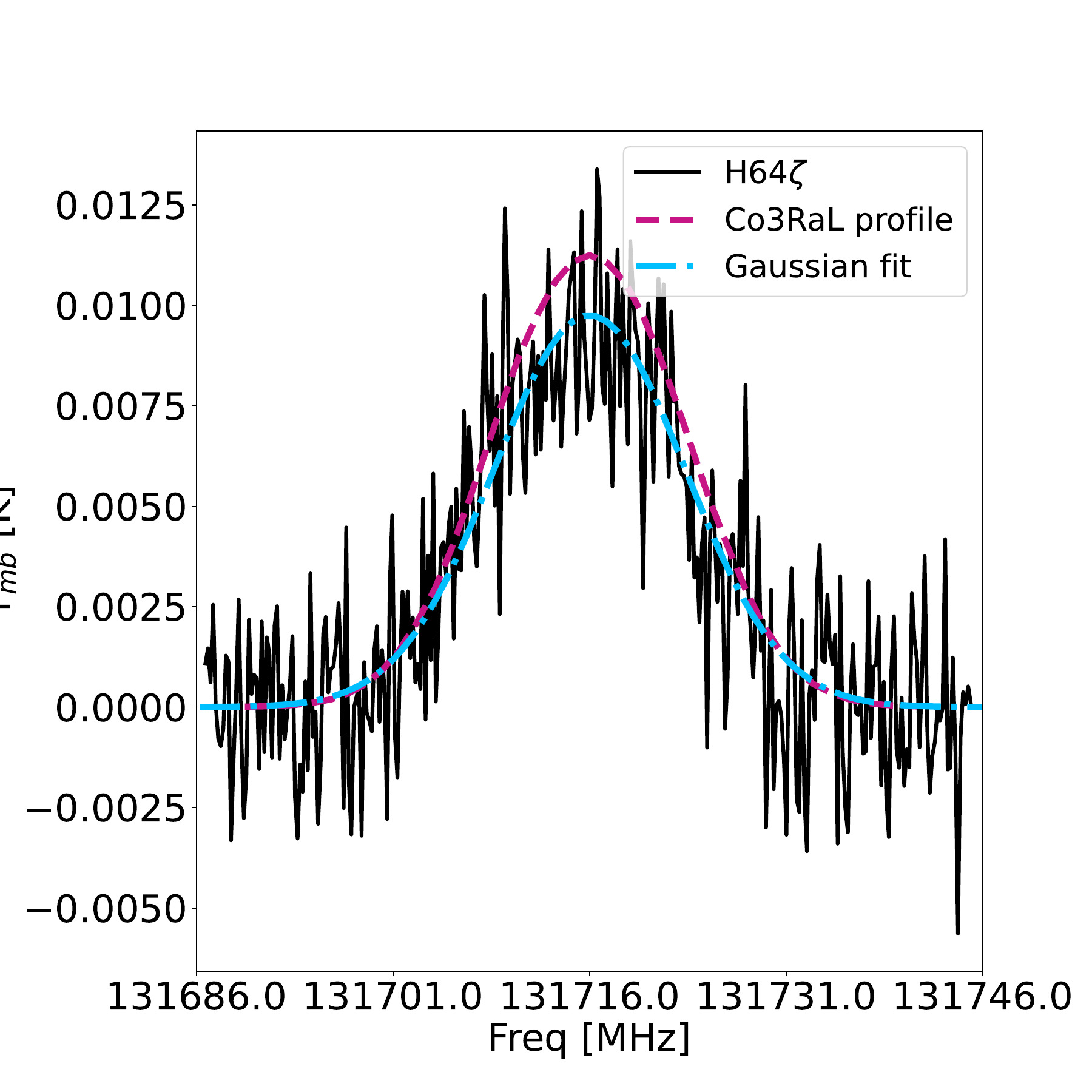}
	\caption{H\gz lines in \ngc.} \label{fig:NGC7027_Hgz}
\end{figure*}

\begin{figure*}[!h]
	\centering
	\includegraphics[width=0.24\textwidth]{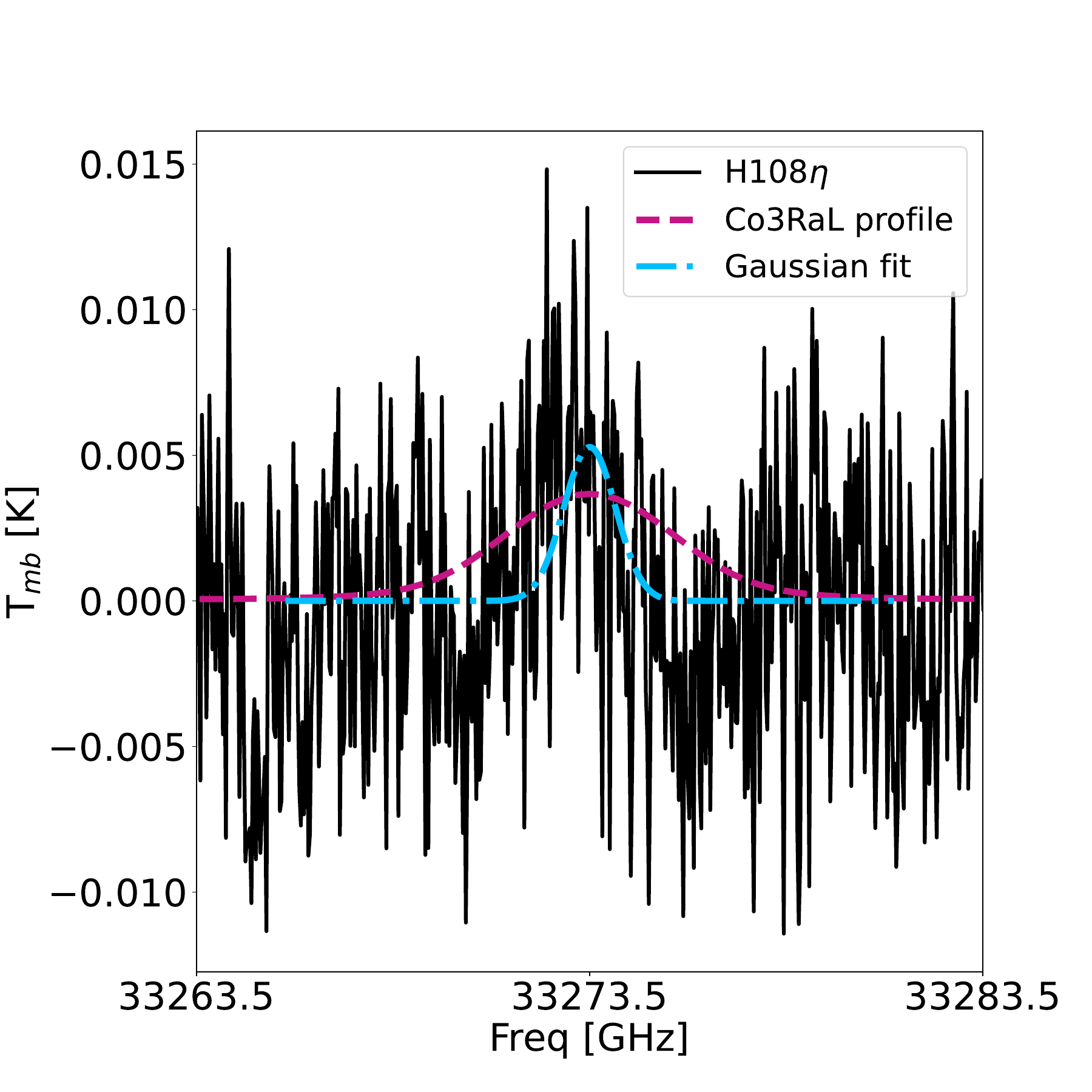}
	\includegraphics[width=0.24\textwidth]{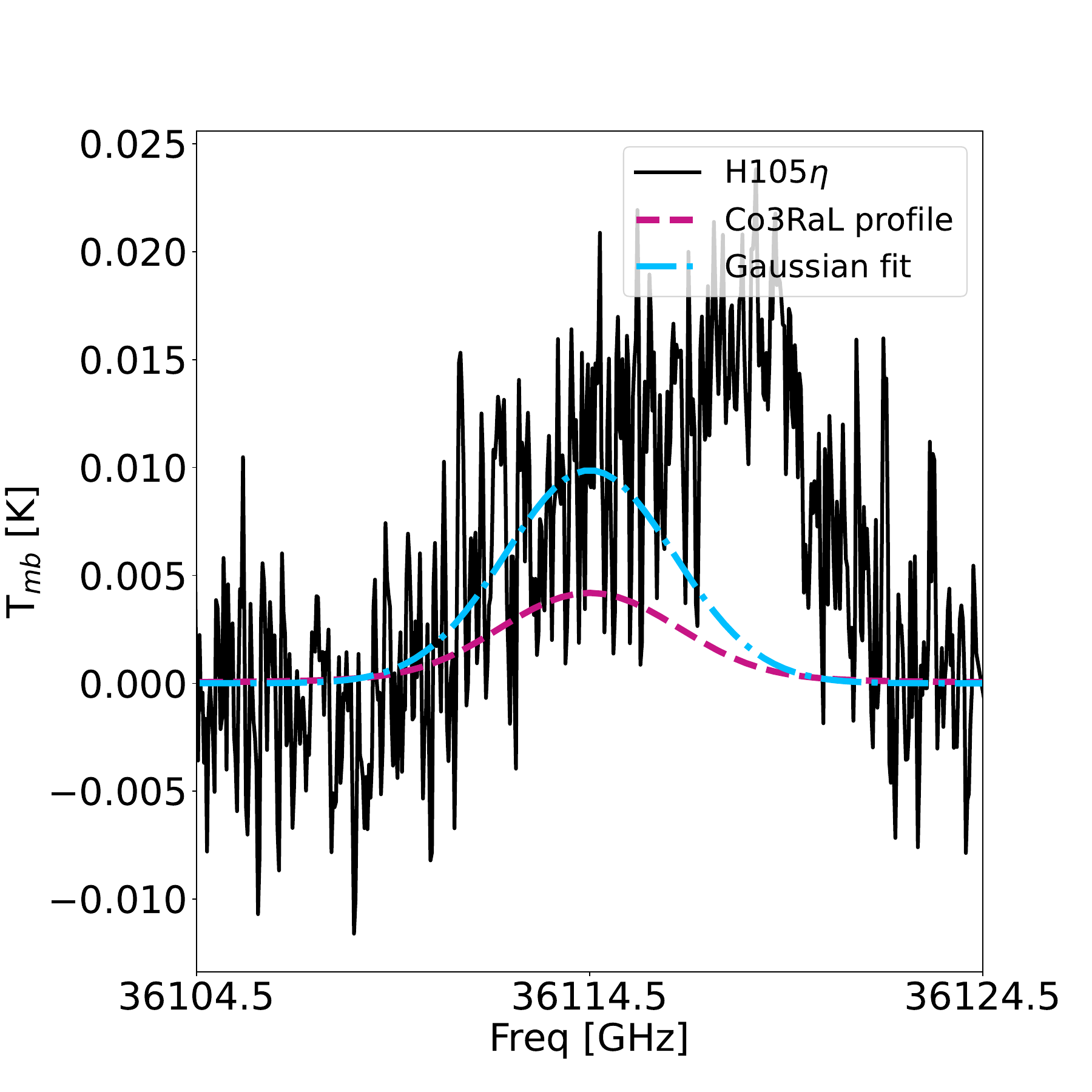}
	\includegraphics[width=0.24\textwidth]{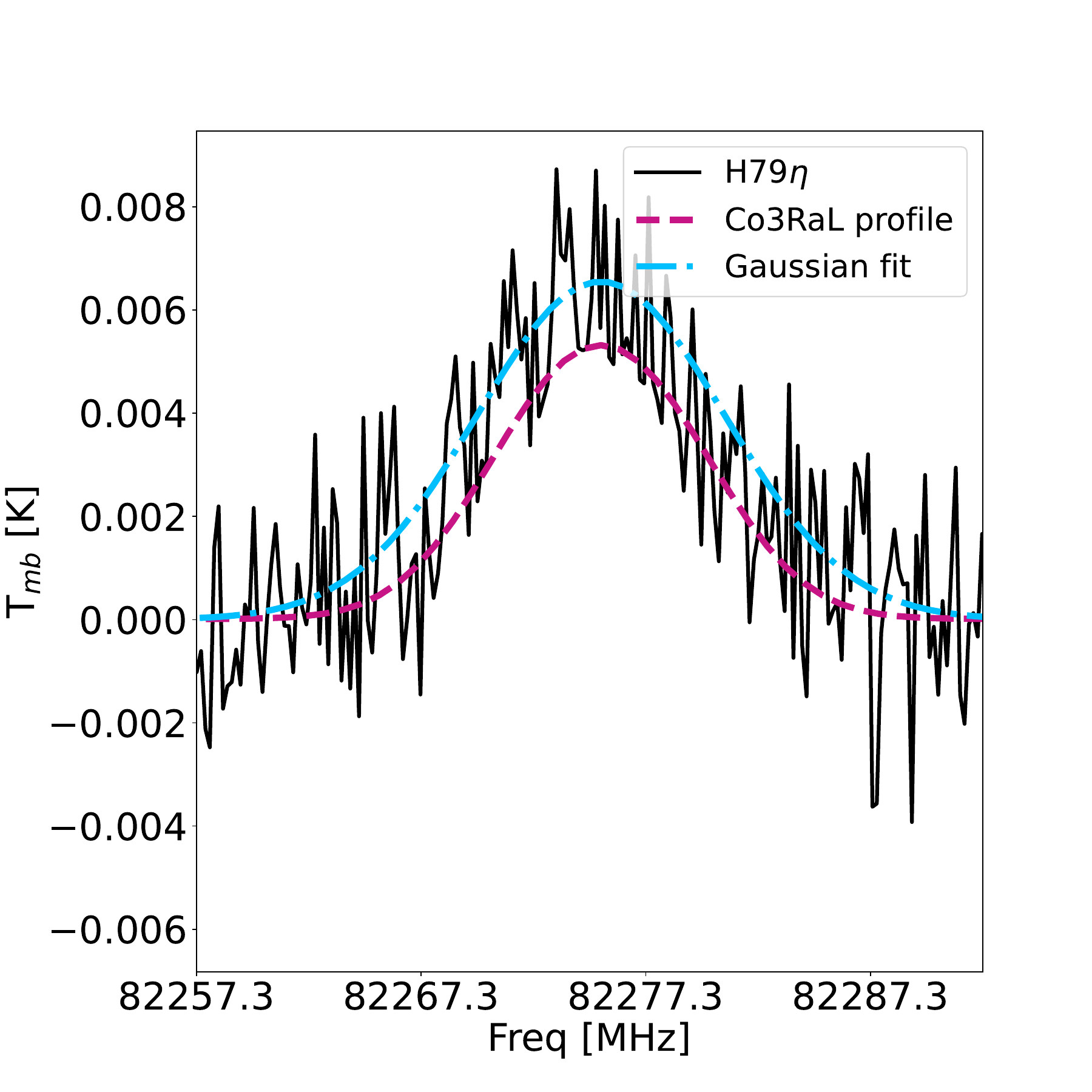}
	\includegraphics[width=0.24\textwidth]{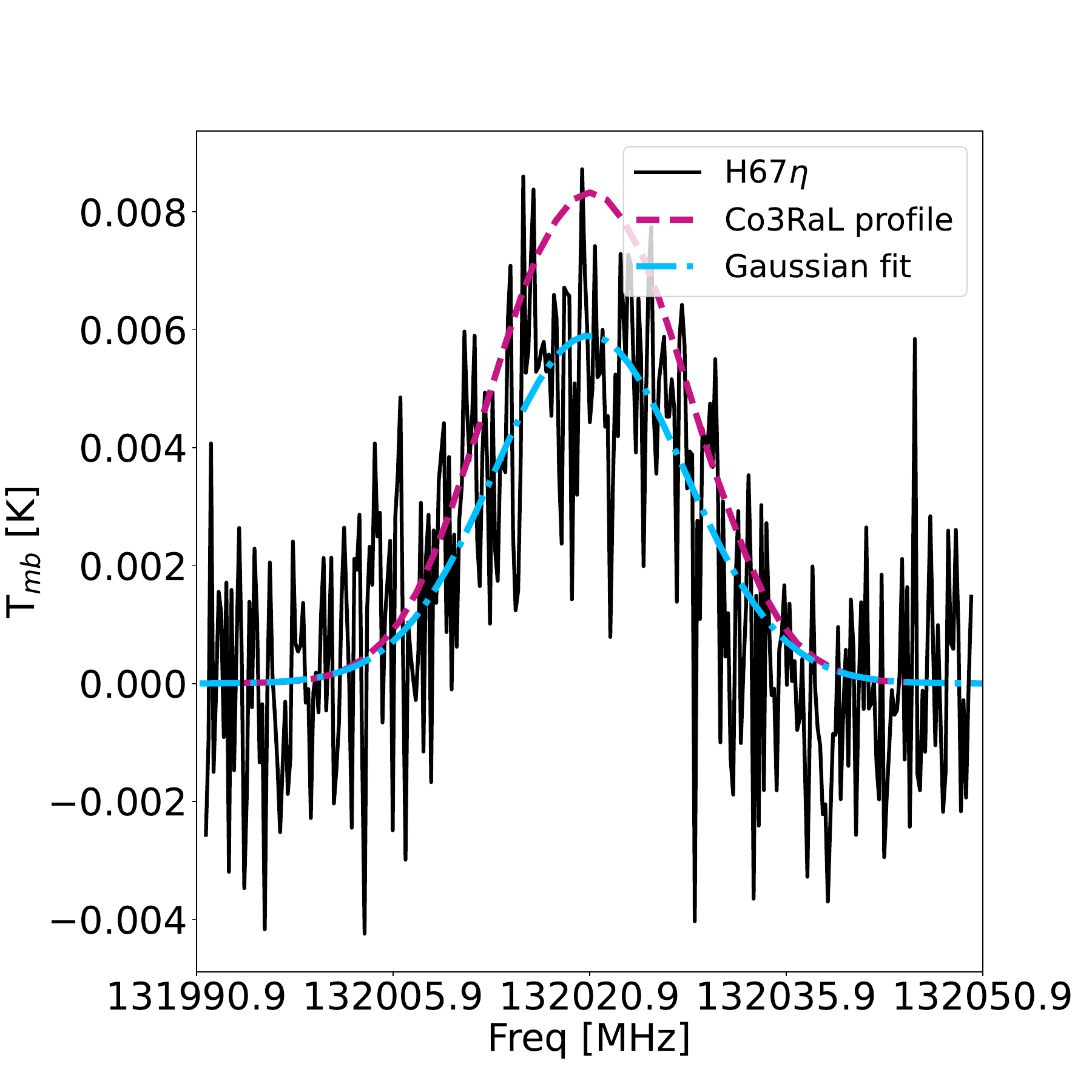}
	\caption{H\gh lines in \ngc. H105\gh blended with H88\gd (see Table \ref{tab:rrls_parameters}).} \label{fig:NGC7027_Hgh}
\end{figure*}

\begin{figure*}[!h]
	\centering
	\includegraphics[width=0.24\textwidth]{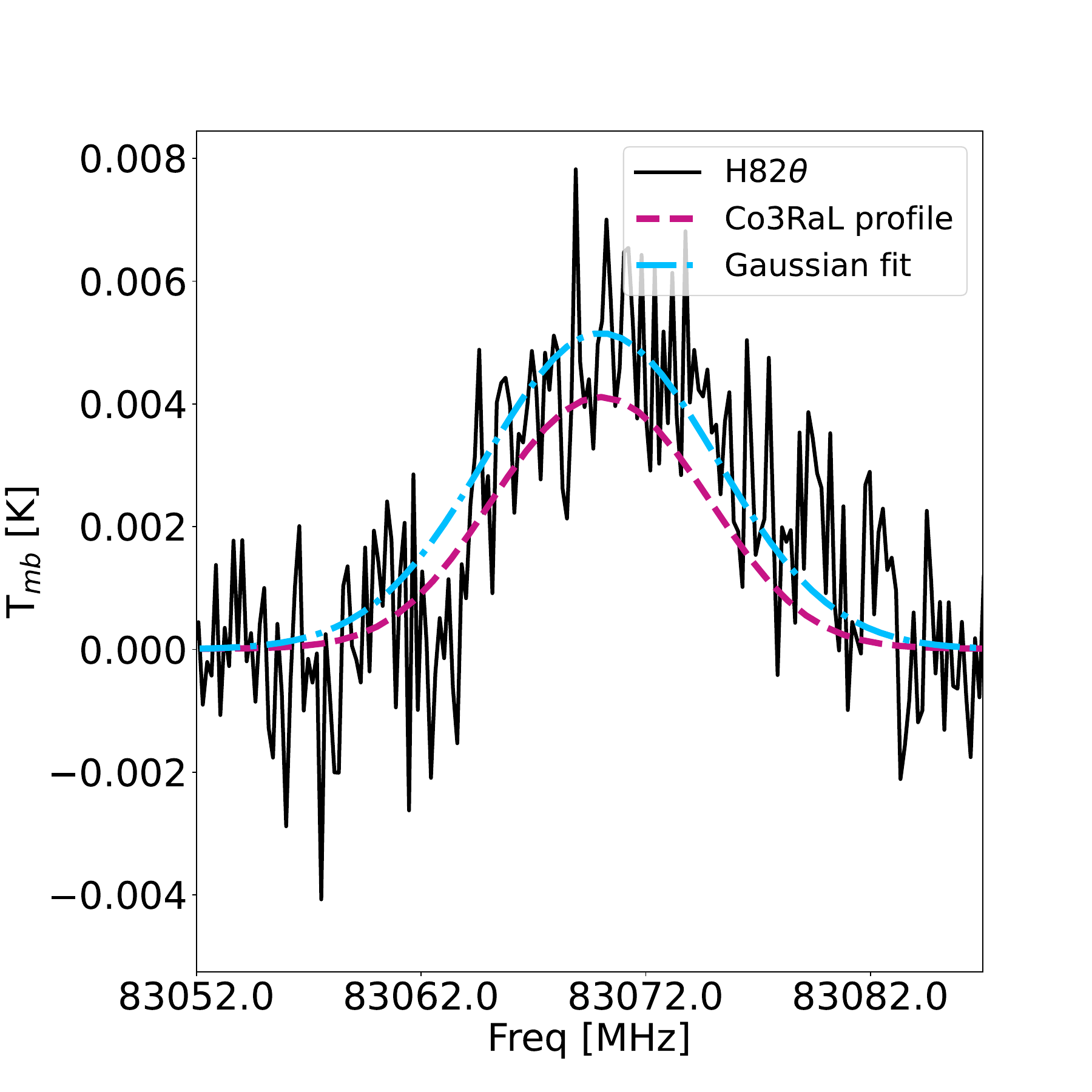}
	\includegraphics[width=0.24\textwidth]{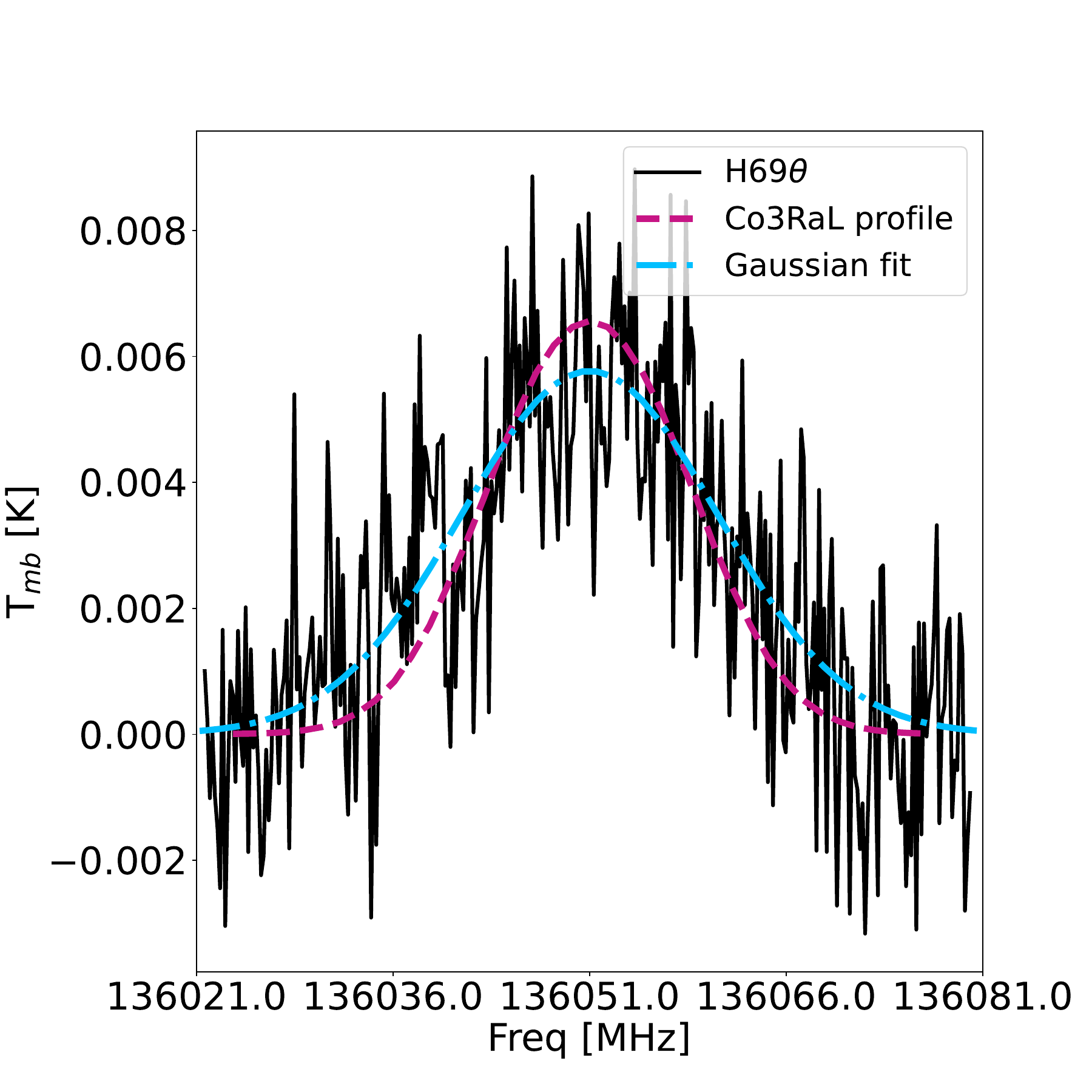}
	\caption{H\gq lines in \ngc.} \label{fig:NGC7027_Hgq}
\end{figure*}

\begin{figure*}[!h]
	\centering
	\includegraphics[width=0.24\textwidth]{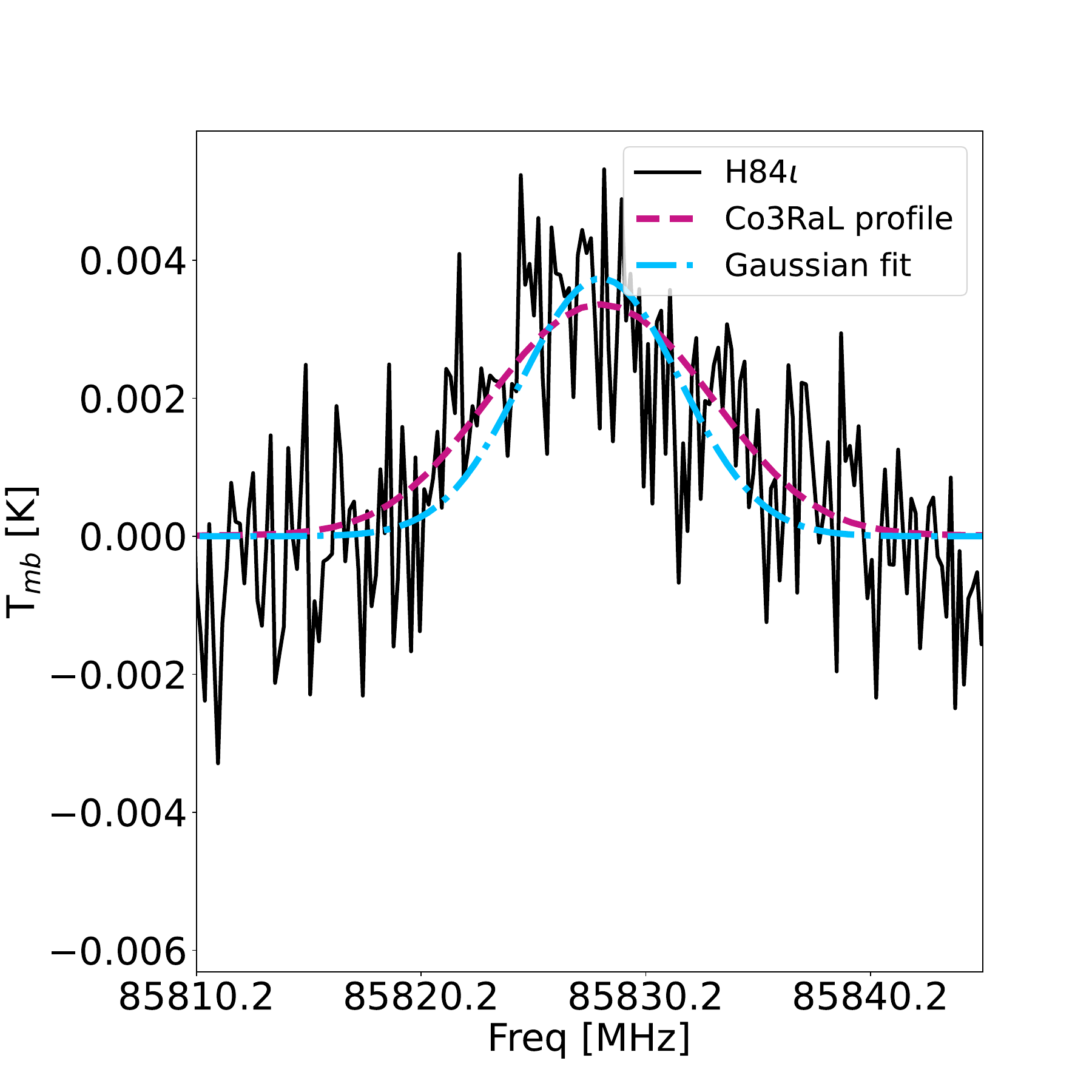}
	\includegraphics[width=0.24\textwidth]{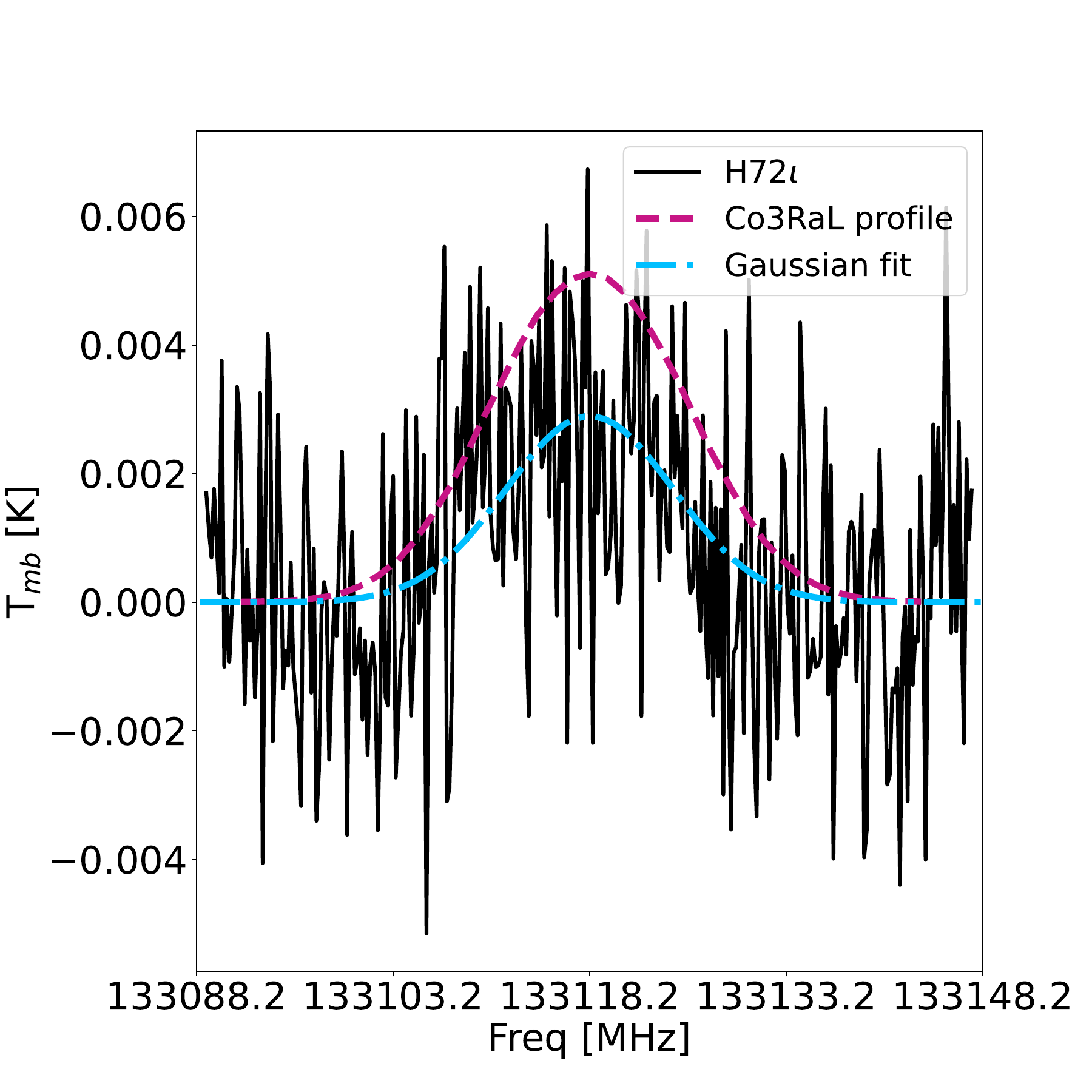}
	\includegraphics[width=0.24\textwidth]{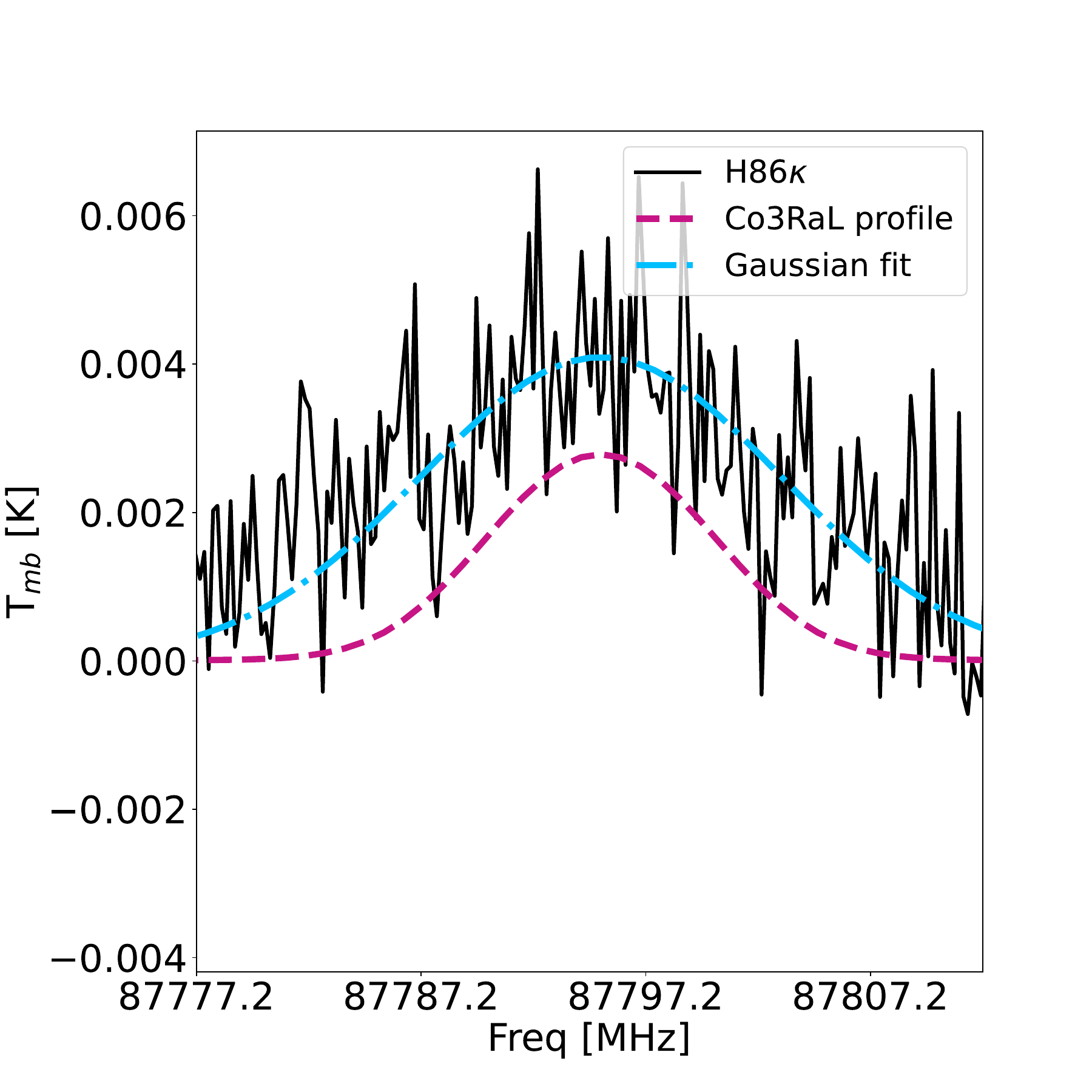}
	\caption{H\gi and H\gk lines in \ngc. H84\gi contaminated with an UF, H86\gk contaminated with an UF (see Table \ref{tab:rrls_parameters}).} \label{fig:NGC7027_Hgigk}
\end{figure*}

\begin{figure*}[!h]
	\centering
	\includegraphics[width=0.24\textwidth]{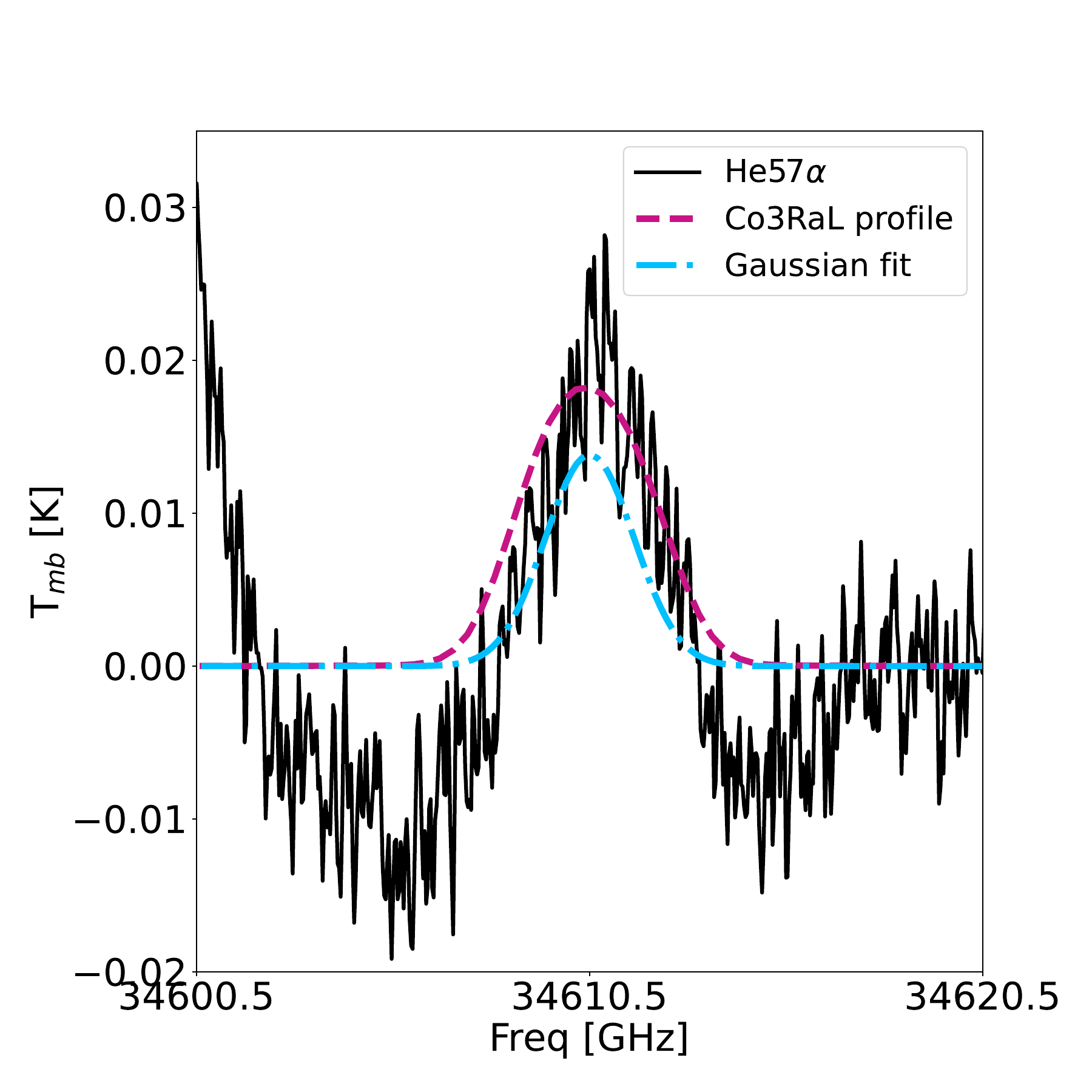}
	\includegraphics[width=0.24\textwidth]{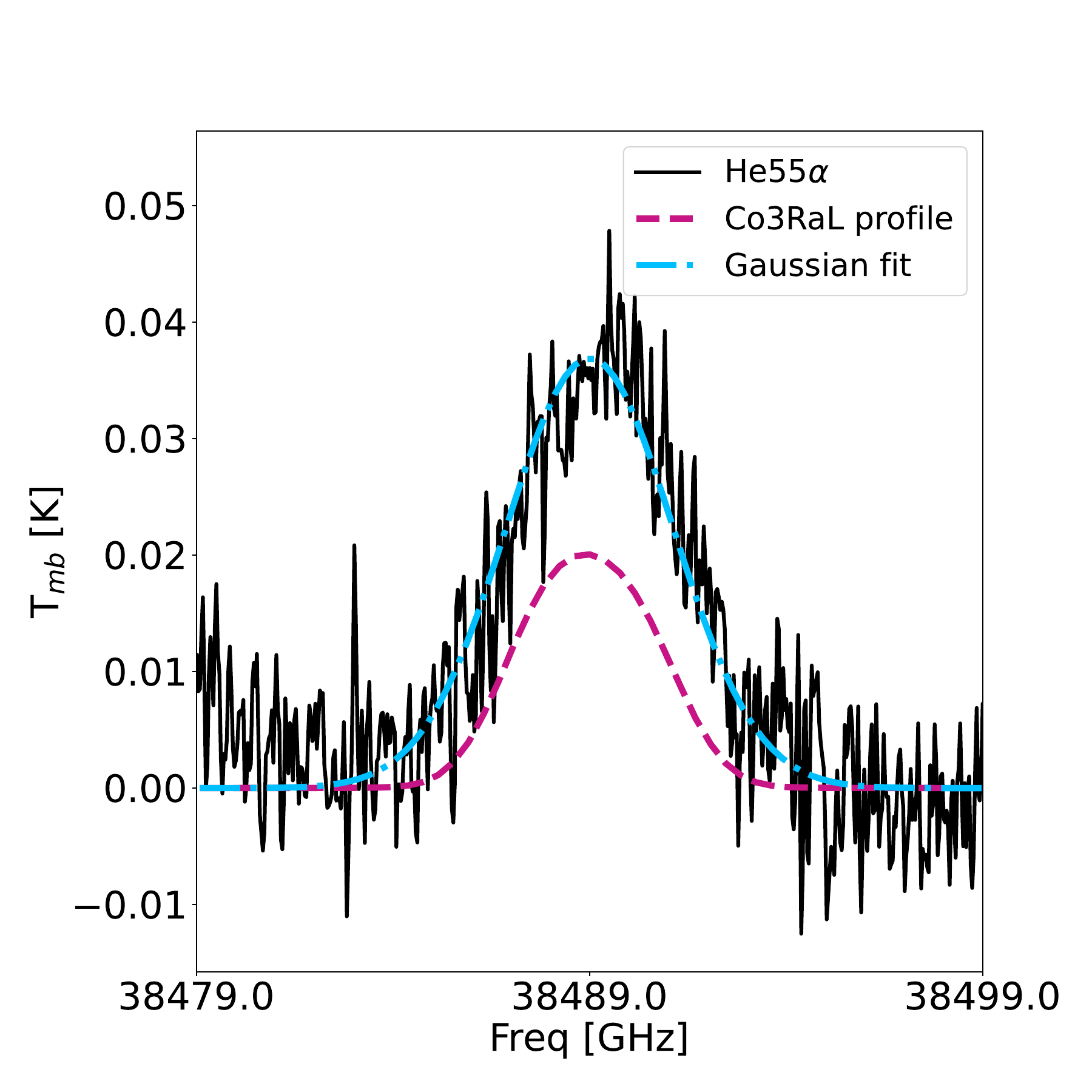}
	\includegraphics[width=0.24\textwidth]{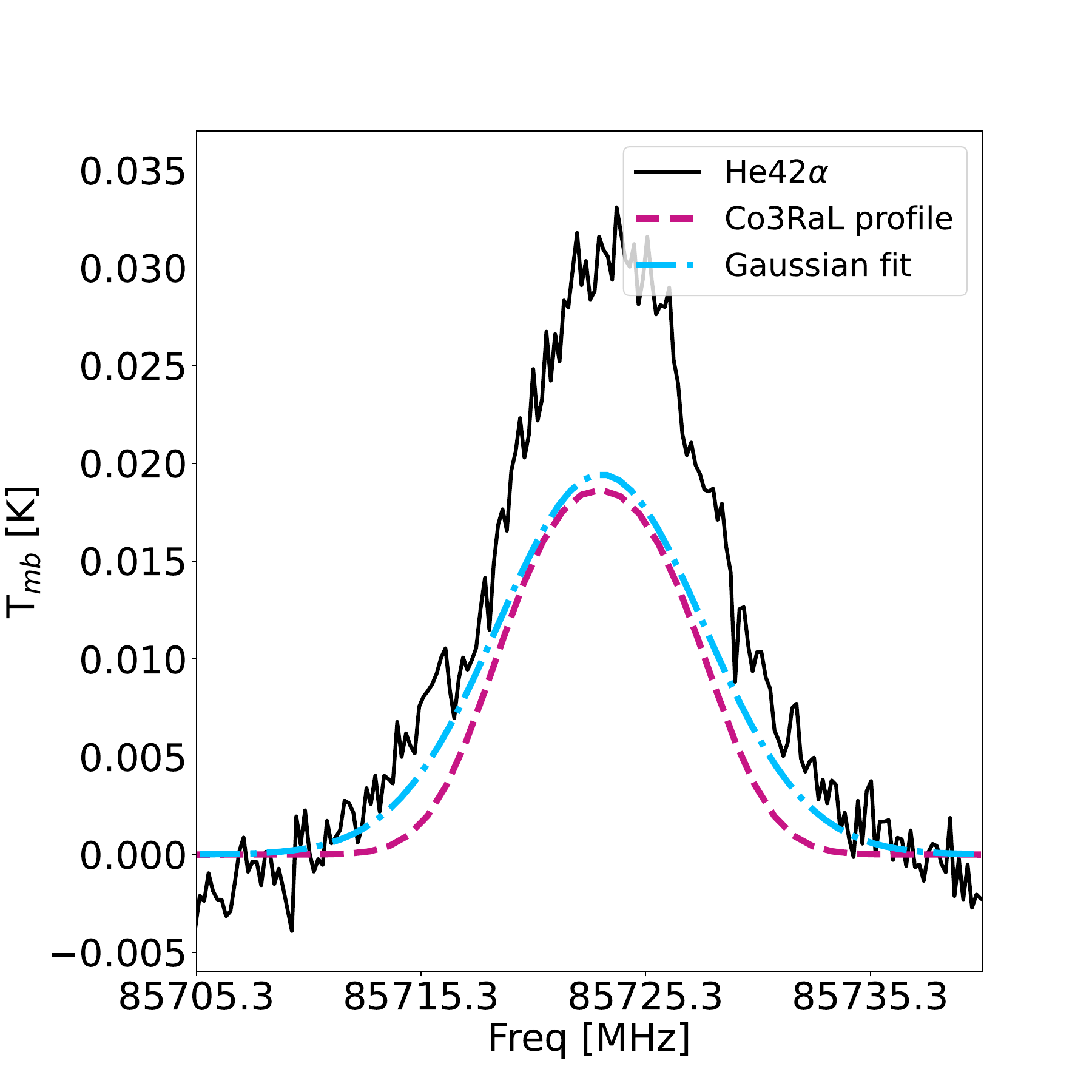}
	\includegraphics[width=0.24\textwidth]{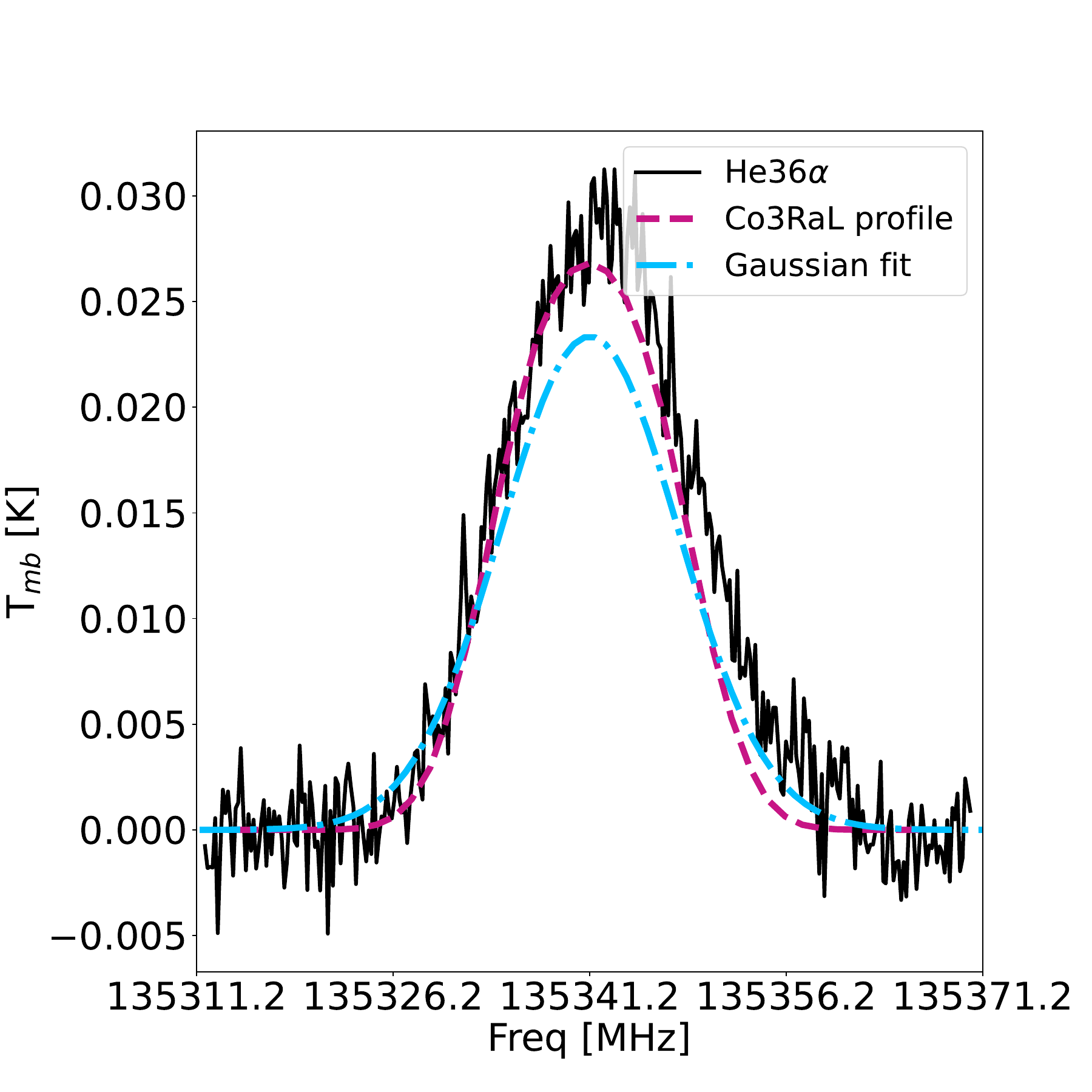}
	\caption{\hei\ga lines in \ngc. \hei55\ga contaminated with an UF, \hei42\ga blended with \heii84\gb, \hei36\ga blended with \heii72\gb (see Table \ref{tab:rrls_parameters}).} \label{fig:NGC7027_HeIga}
\end{figure*}

\begin{figure*}[!h]
	\centering
	\includegraphics[width=0.24\textwidth]{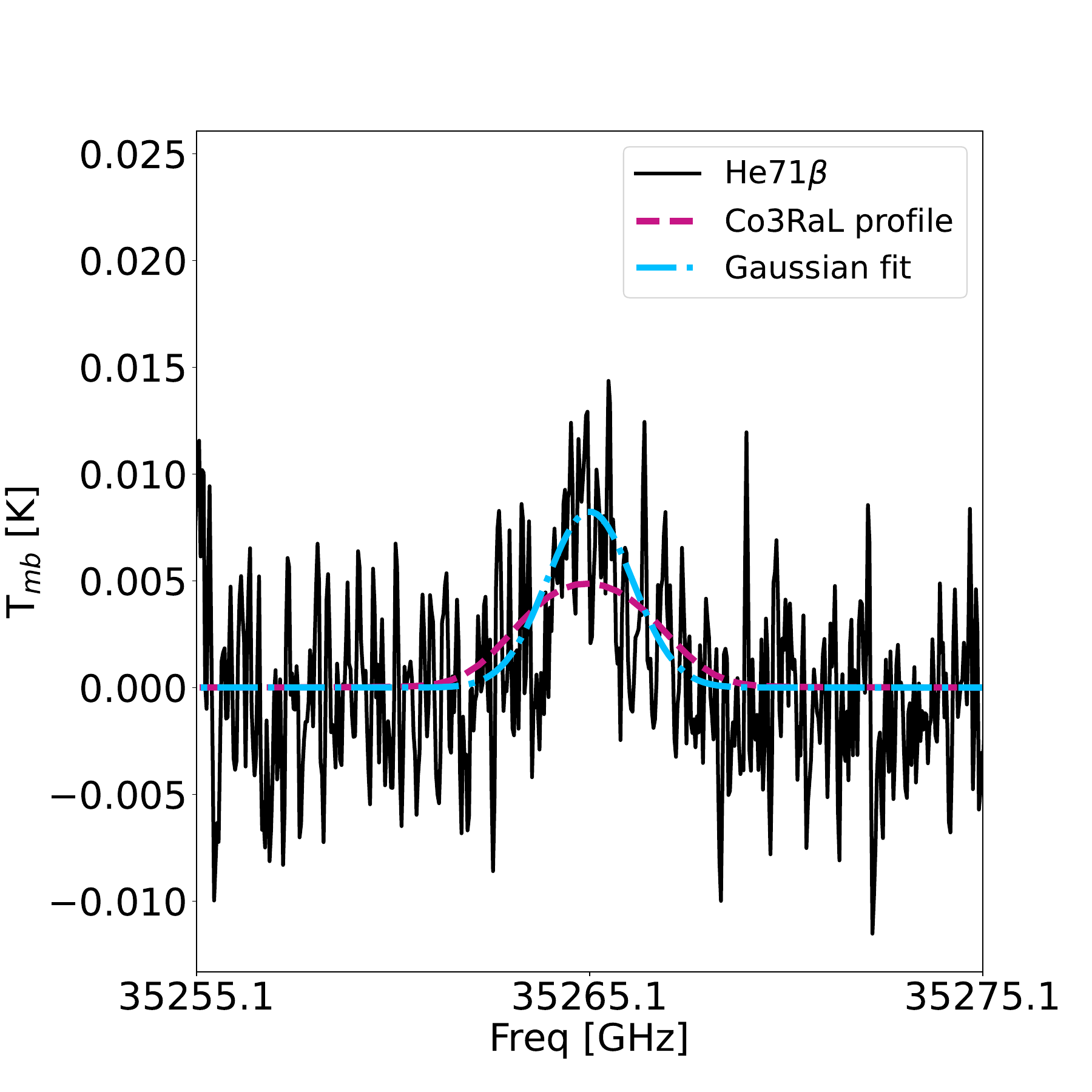}
	\includegraphics[width=0.24\textwidth]{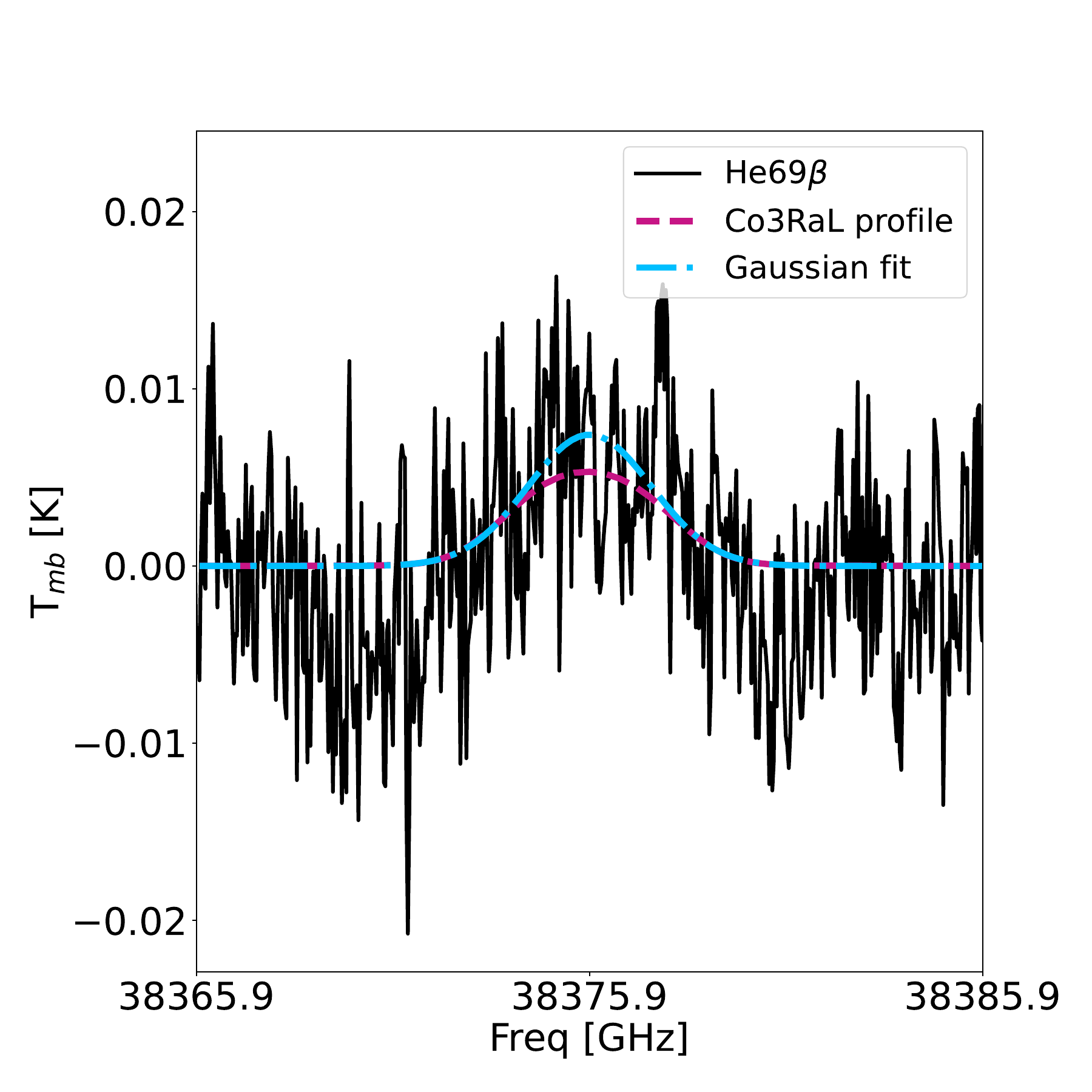}
	\includegraphics[width=0.24\textwidth]{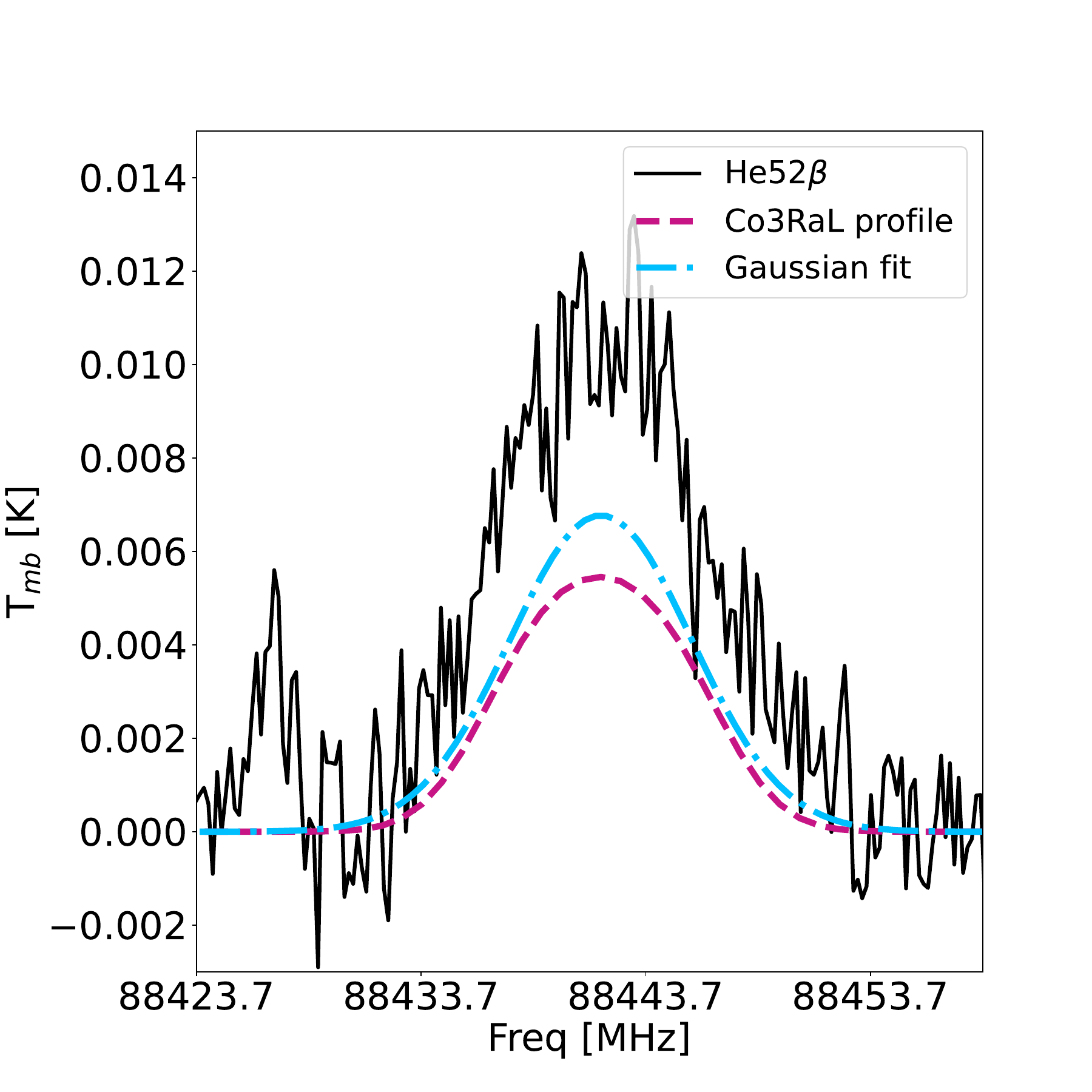}
	\includegraphics[width=0.24\textwidth]{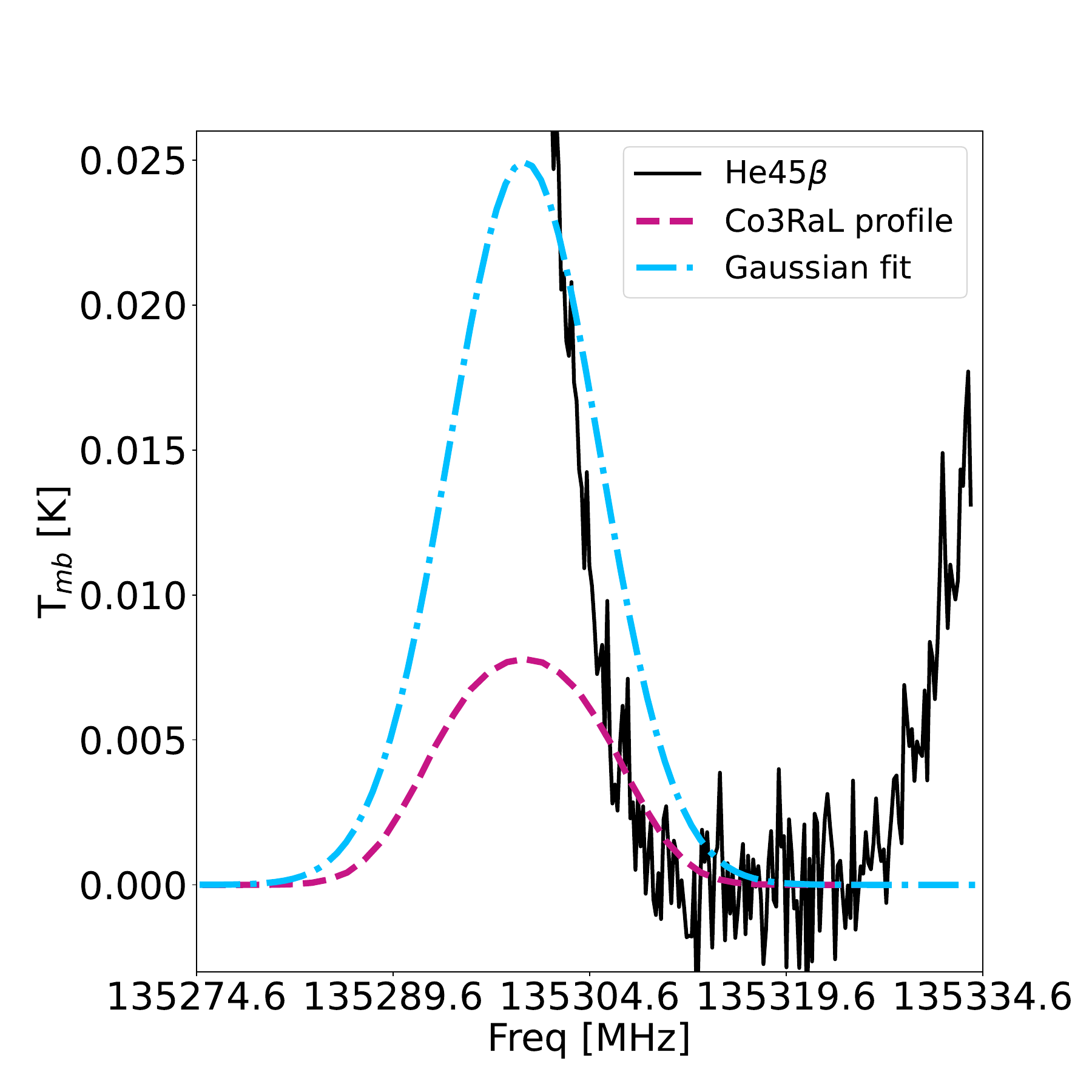}
	\caption{\hei\gb lines in \ngc. \hei52\gb blended with \heii104\gd, \hei45\gb blended with H36\ga (see Table \ref{tab:rrls_parameters}).} \label{fig:NGC7027_HeIgb}
\end{figure*}

\begin{figure*}[!h]
	\centering
	\includegraphics[width=0.24\textwidth]{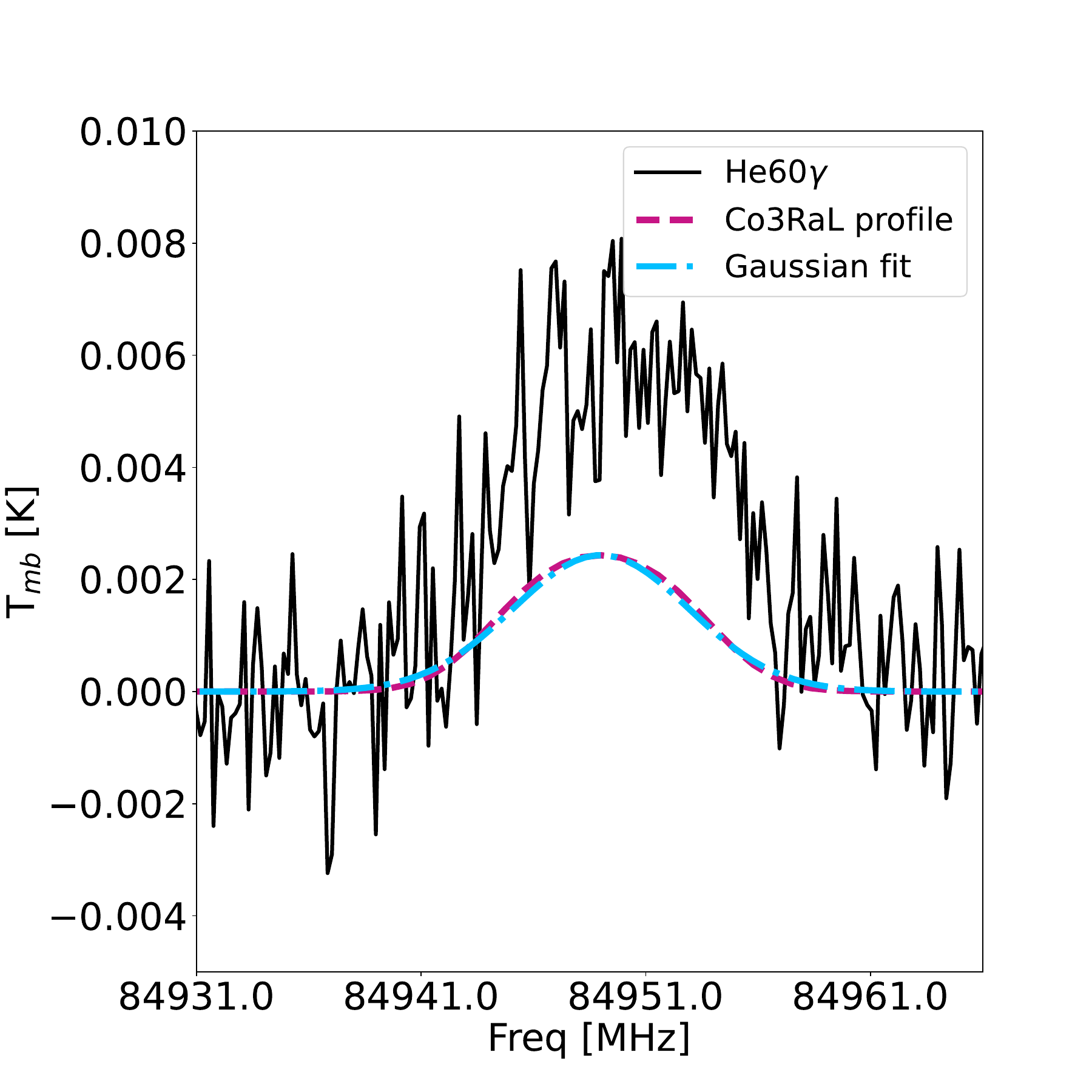}
	\includegraphics[width=0.24\textwidth]{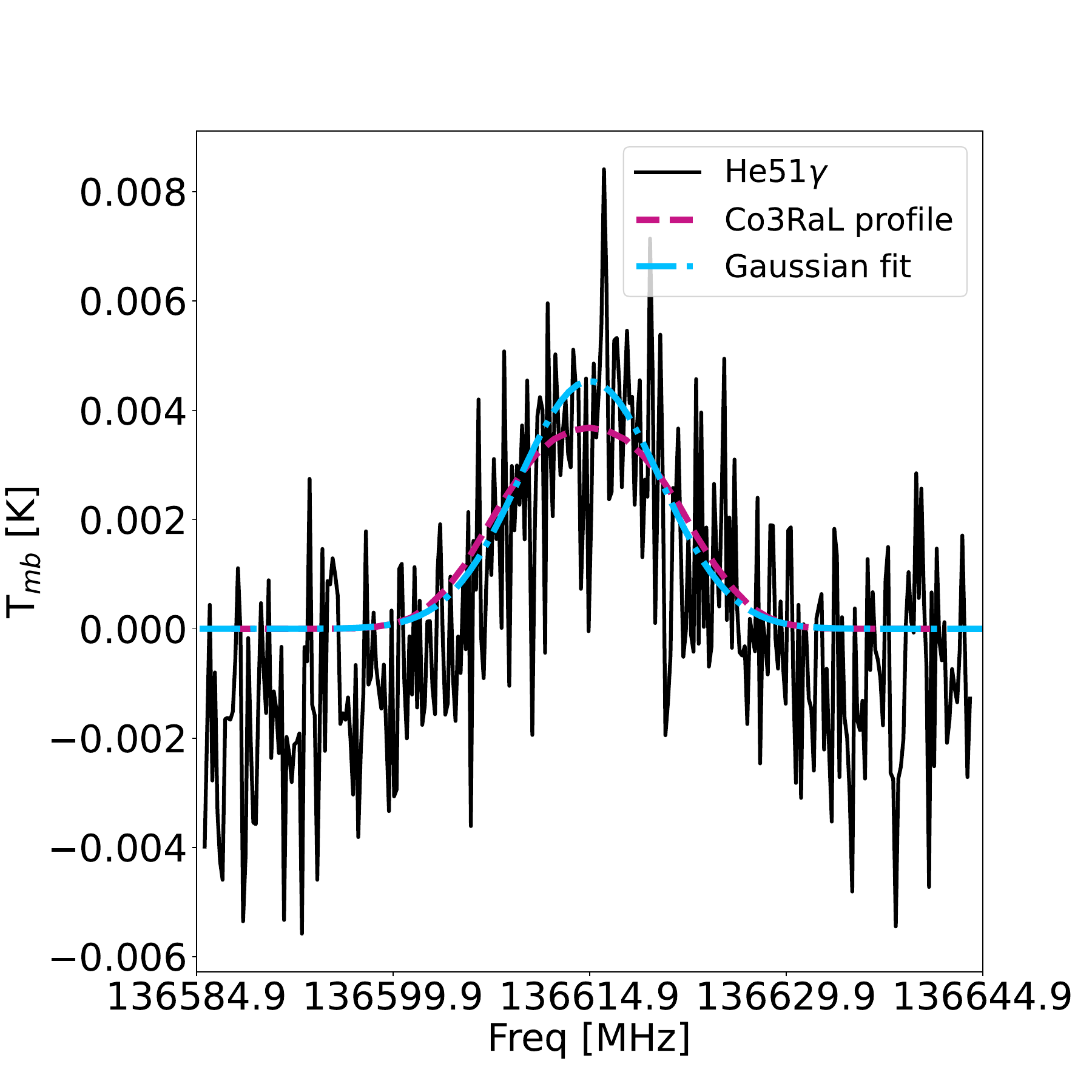}
	\caption{\hei\gg lines in \ngc. \hei60\gg blended with H87\gk (see Table \ref{tab:rrls_parameters}).} \label{fig:NGC7027_HeIgg}
\end{figure*}

\begin{figure*}[!h]
	\centering
	\includegraphics[width=0.24\textwidth]{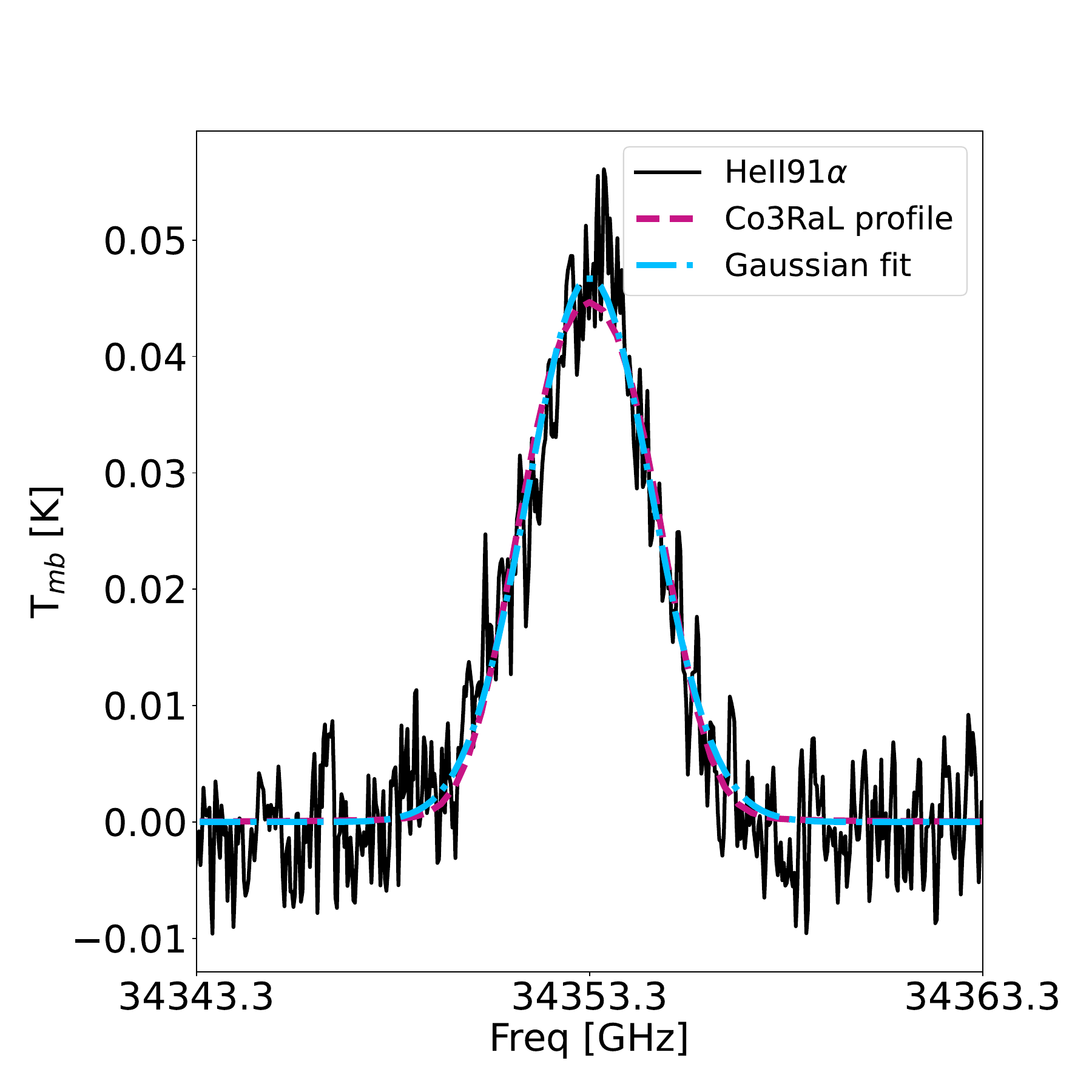}
	\includegraphics[width=0.24\textwidth]{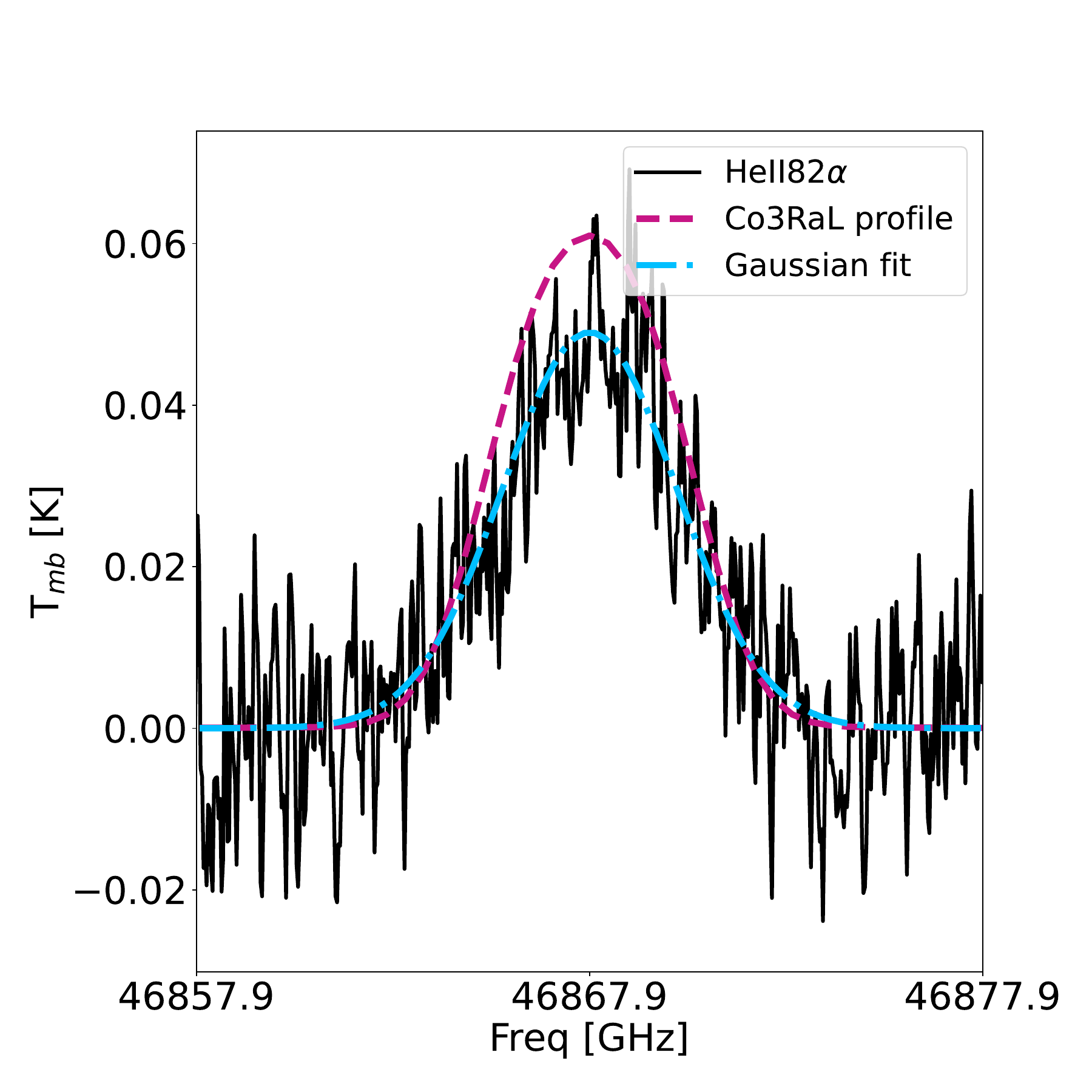}
	\includegraphics[width=0.24\textwidth]{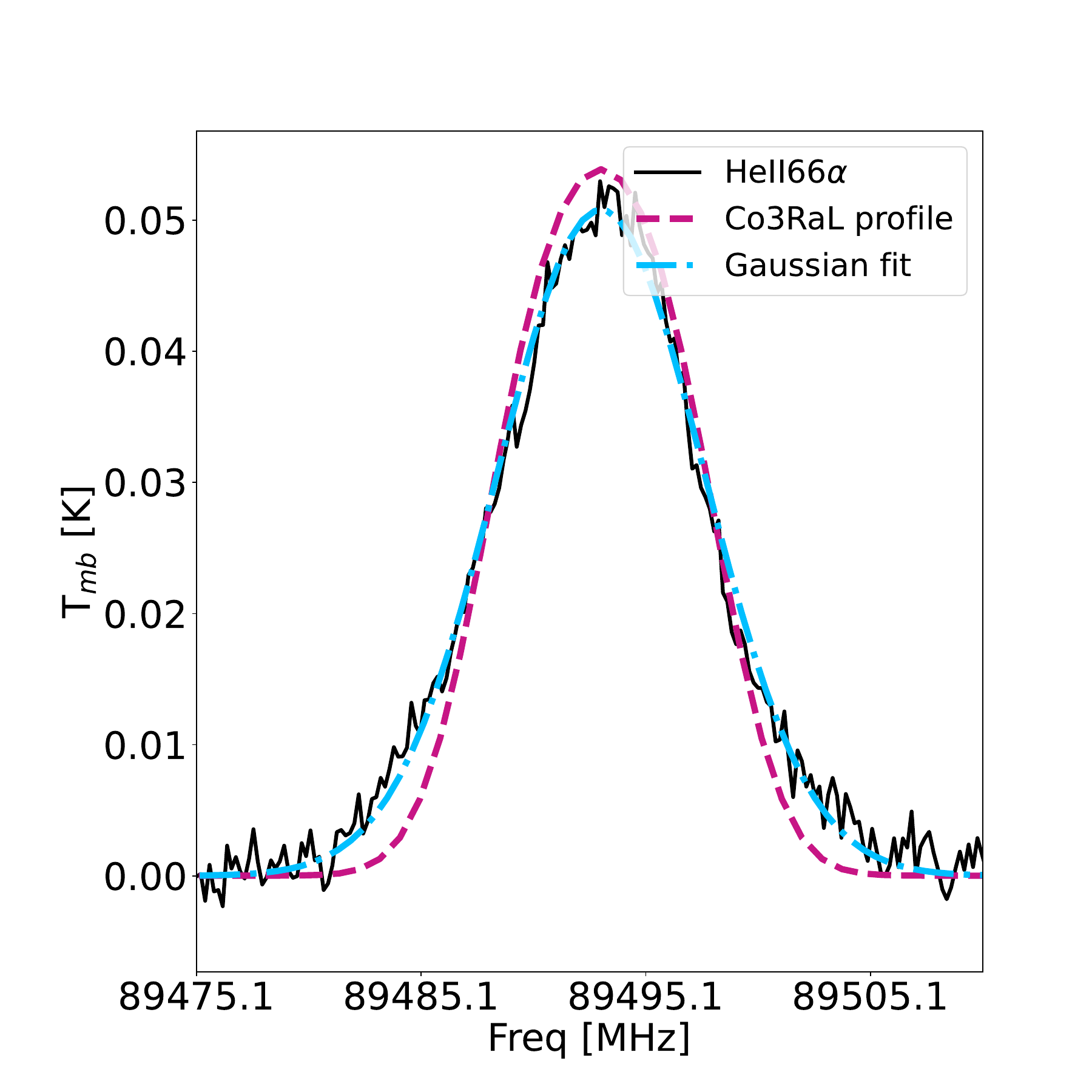}
	\includegraphics[width=0.24\textwidth]{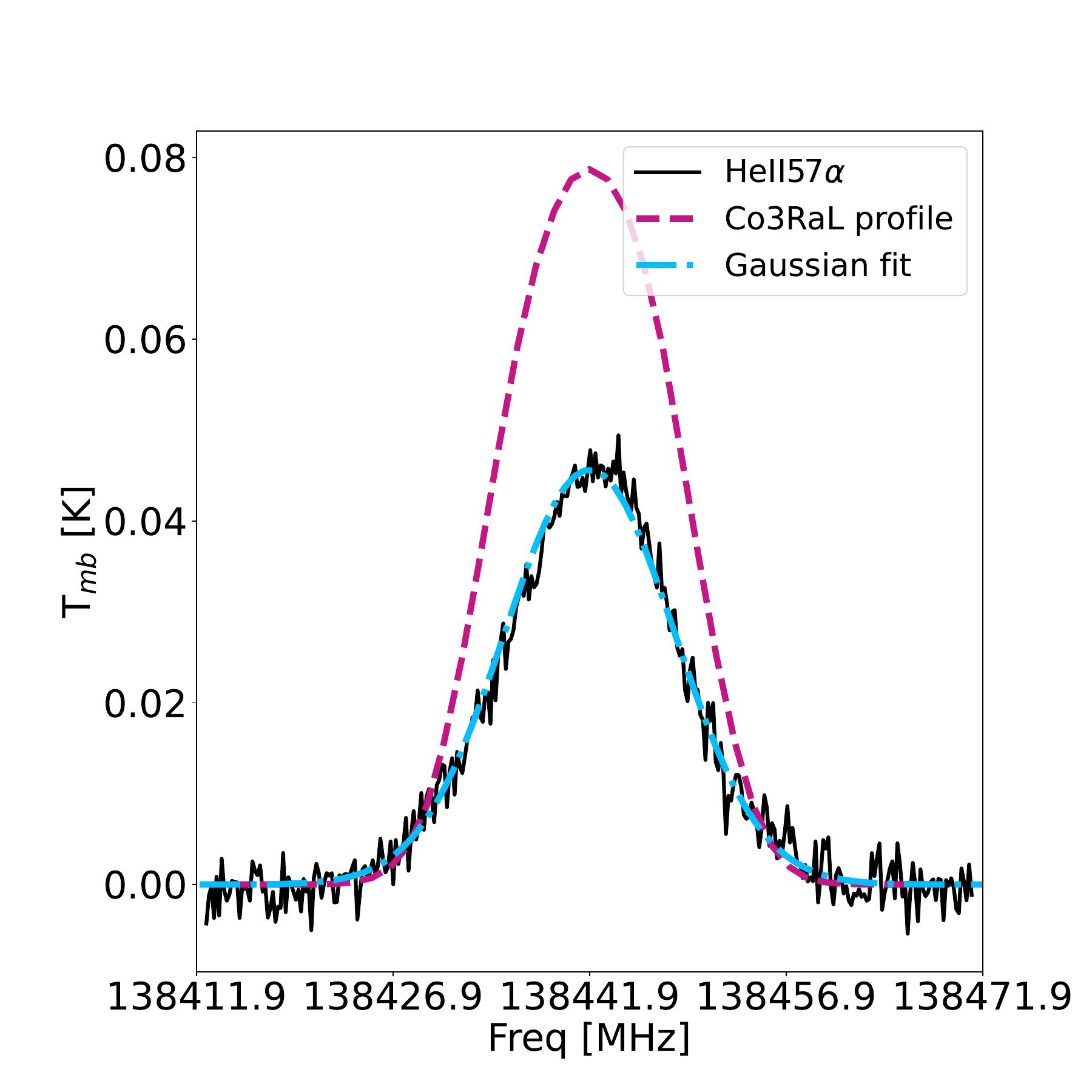}
	\caption{\heii\ga lines in \ngc.} \label{fig:NGC7027_HeIIga}
\end{figure*}

\begin{figure*}[!h]
	\centering
	\includegraphics[width=0.24\textwidth]{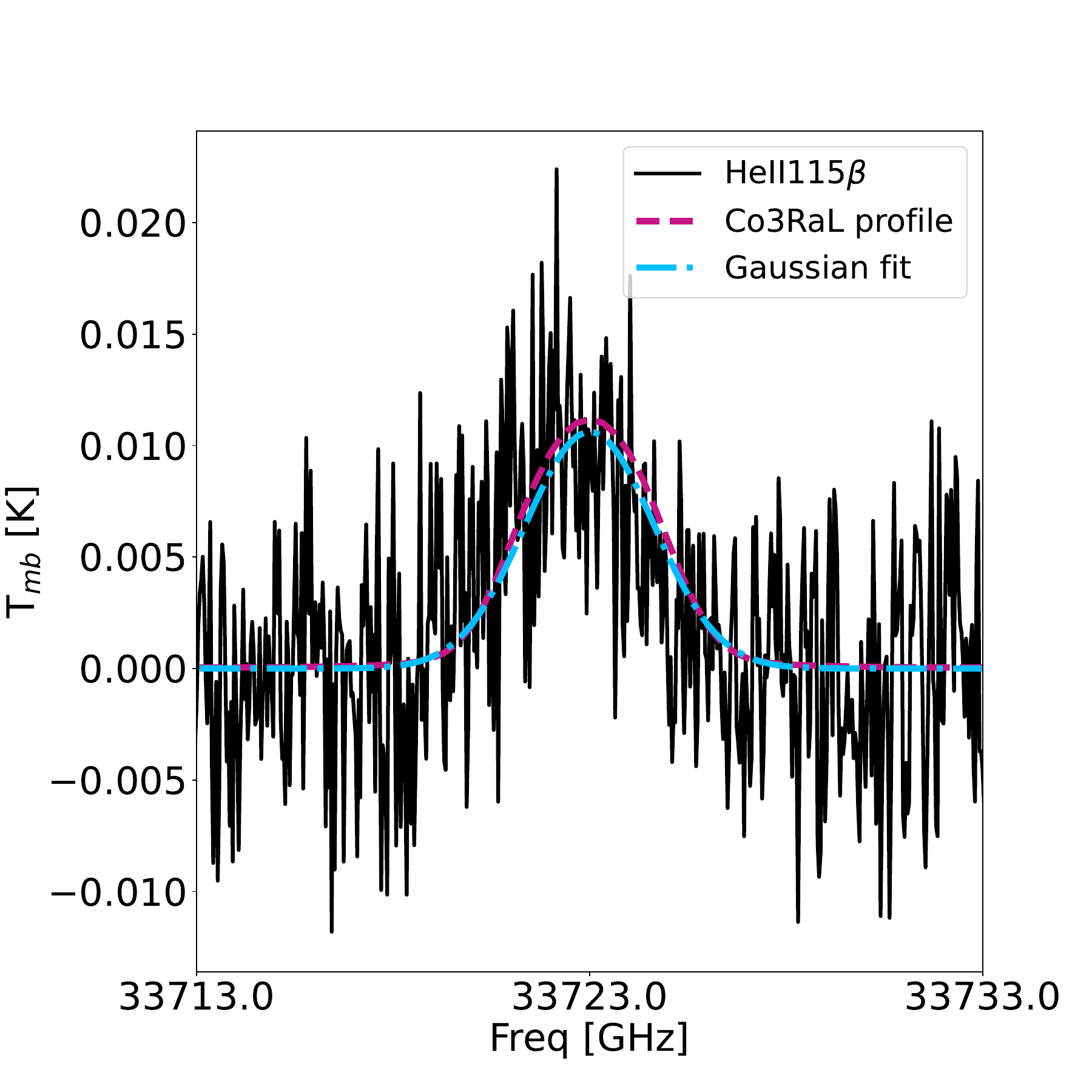}
	\includegraphics[width=0.24\textwidth]{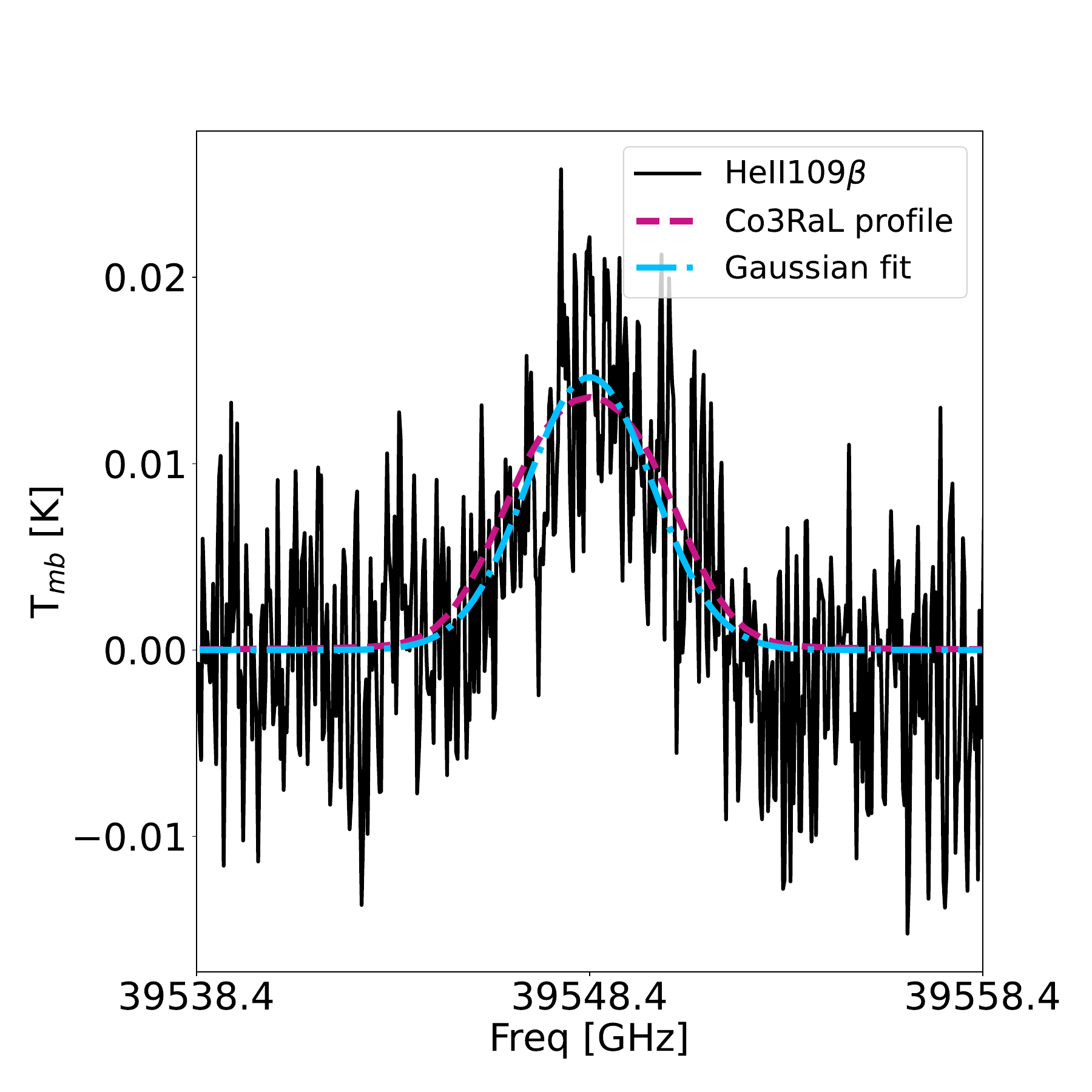}
	\includegraphics[width=0.24\textwidth]{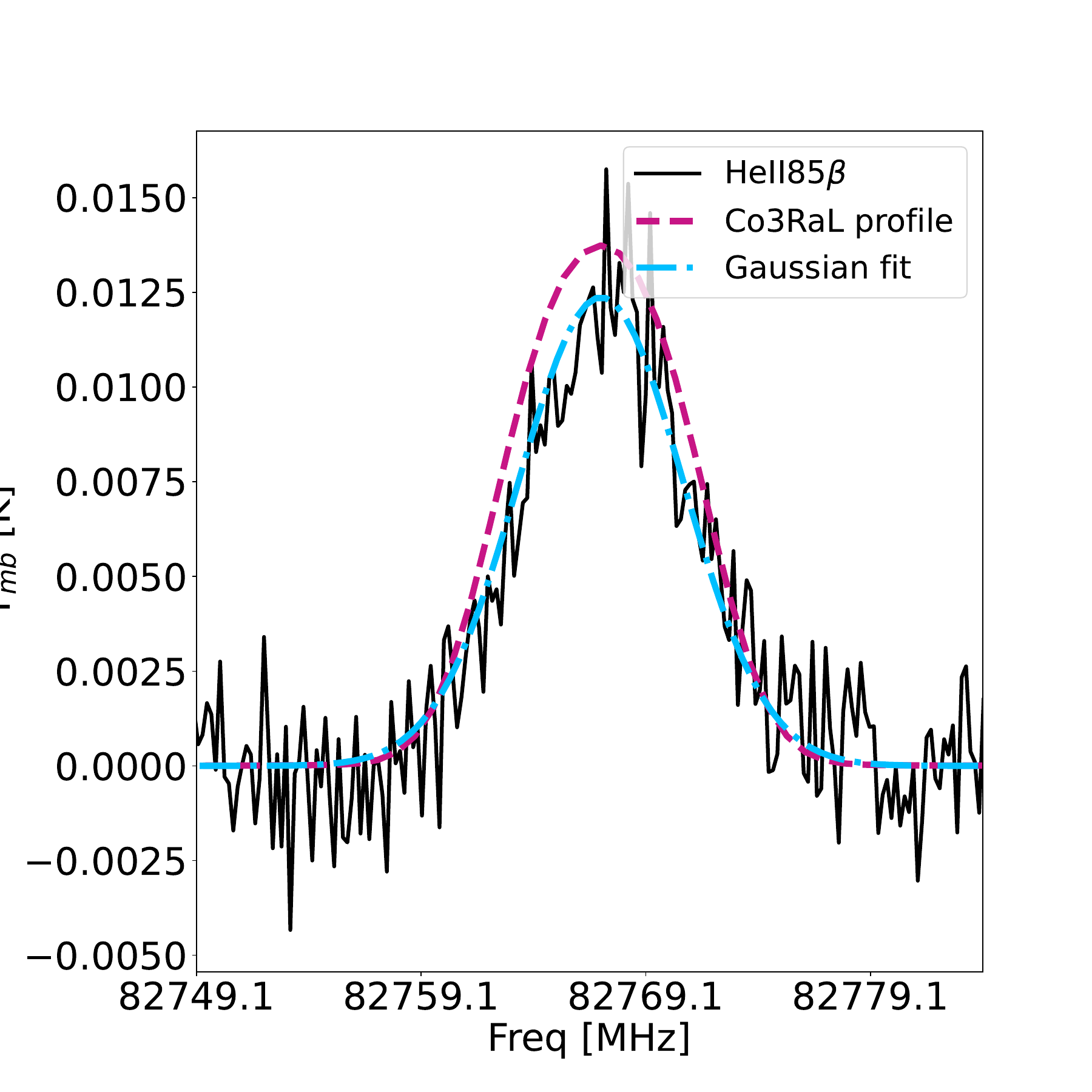}
	\includegraphics[width=0.24\textwidth]{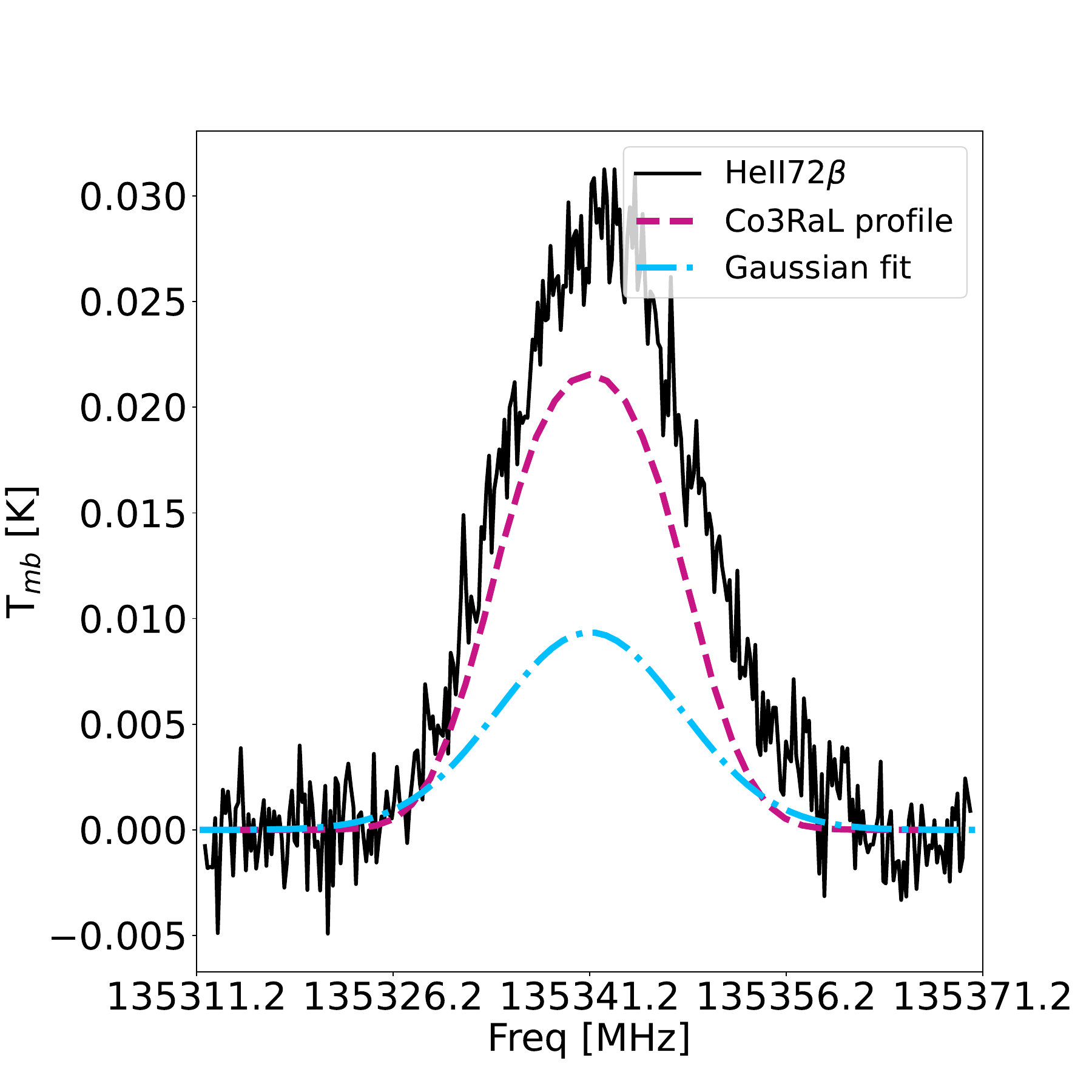}
	\caption{\heii\gb lines in \ngc. \heii72\gb blended with \hei36\ga (see Table \ref{tab:rrls_parameters}).} \label{fig:NGC7027_HeIIgb}
\end{figure*}

\begin{figure*}[!h]
	\centering
	\includegraphics[width=0.24\textwidth]{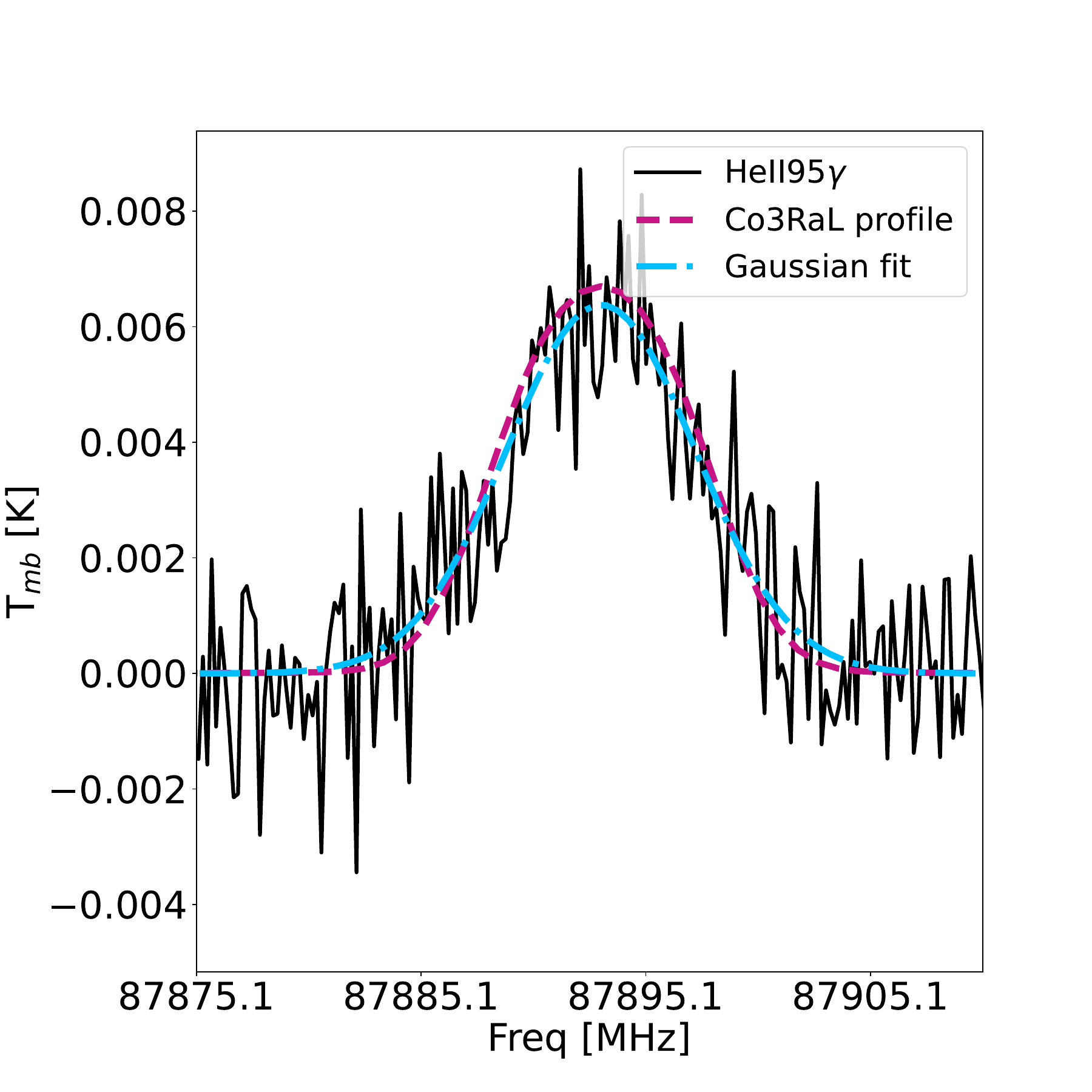}
	\includegraphics[width=0.24\textwidth]{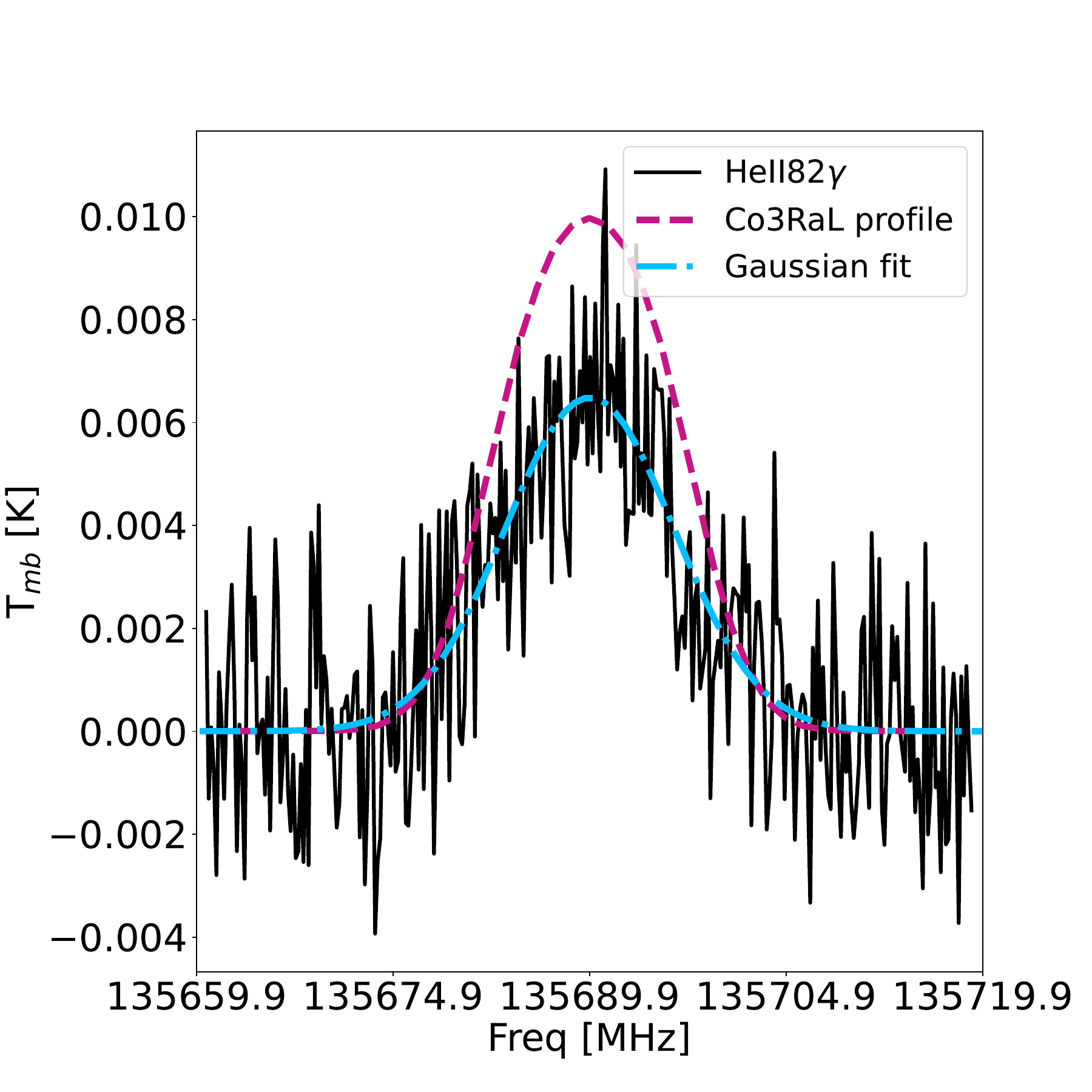}
	\includegraphics[width=0.24\textwidth]{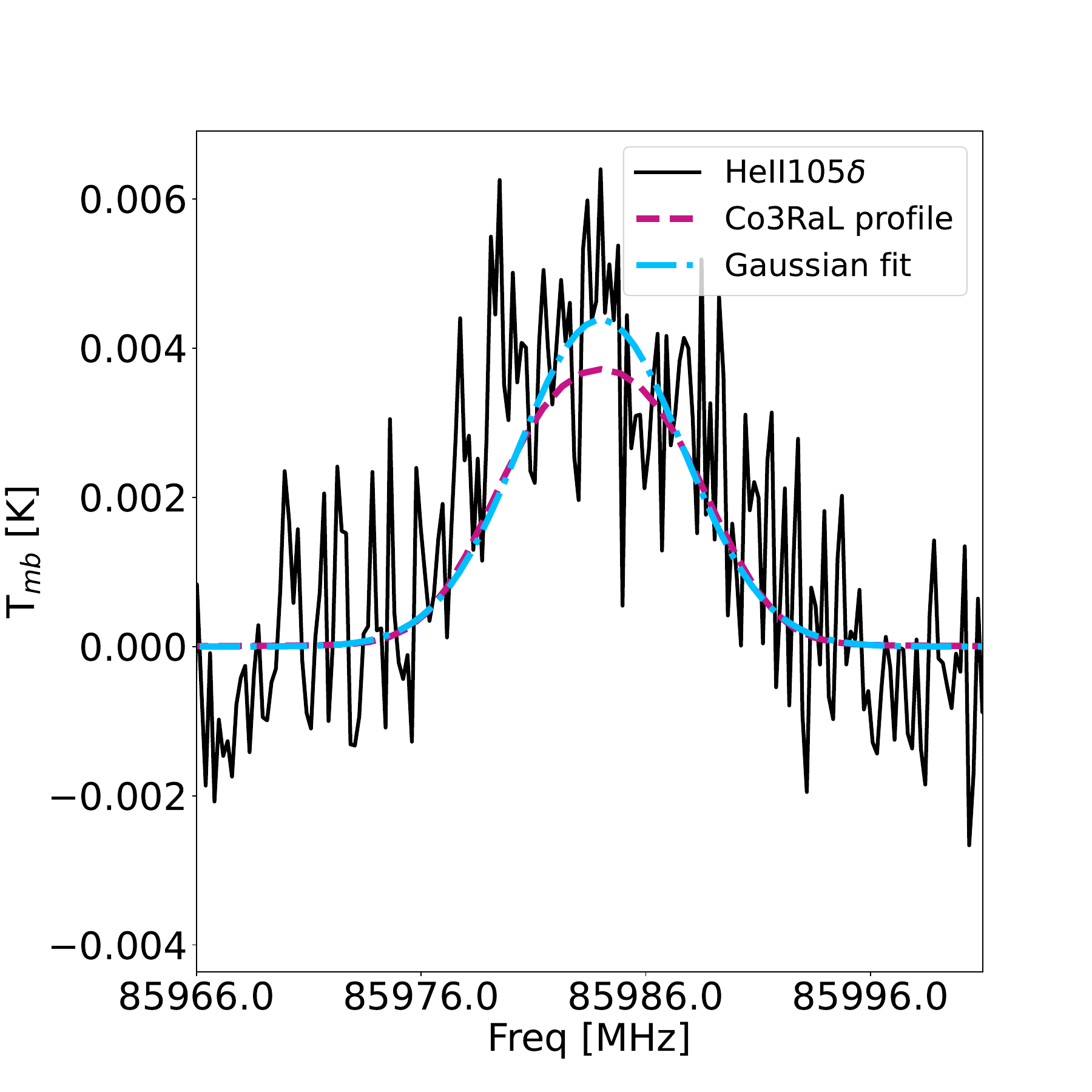}
	\caption{\heii\gg and \heii\gg lines in \ngc. \heii105\gd contaminated (see Table \ref{tab:rrls_parameters}).} \label{fig:NGC7027_HeIIgg}
\end{figure*}

We present here examples of the H, \hei, and \heii RRLs found on \ic. Four examples are displayed for each \Dn and species when data is available: two lines at 7\mm, one line at 3\mm, and one lines at 2\mm. See Figs. \ref{fig:NGC7027_Hga}, \ref{fig:NGC7027_Hgb}, \ref{fig:NGC7027_Hgg}, \ref{fig:NGC7027_Hgd}, \ref{fig:NGC7027_Hge}, \ref{fig:NGC7027_Hgz}, \ref{fig:NGC7027_Hgh}, \ref{fig:NGC7027_Hgq}, \ref{fig:NGC7027_Hgigk}, \ref{fig:NGC7027_HeIga}, \ref{fig:NGC7027_HeIgb}, \ref{fig:NGC7027_HeIgg}, \ref{fig:NGC7027_HeIIga}, \ref{fig:NGC7027_HeIIgb}, and \ref{fig:NGC7027_HeIIgg}.

\section{Physical and Spectroscopic Parameters of RRLs}

Table \ref{tab:rrls_parameters} lists all physical and spectroscopic parameters of the RRLs in PNe \ic and \ngc (see the table notes for details).

\onecolumn
\tiny
\begin{landscape}

\tablefoot{
	\tablefoottext{a}{Name of the RRL transition.}
	\tablefoottext{b}{Frequency at the center of the line in MHz.}
	\tablefoottext{c}{Einstein coefficient in s$^{-1}$.}
	\tablefoottext{d}{Status of the line (D: detected, T: tentative, L: detection limits)}
	\tablefoottext{e}{Central velocity of the line in \kms.}
	\tablefoottext{f}{Main beam temperature of the peak in mK.}
	\tablefoottext{g}{Area under the line in K\kms.}
	\tablefoottext{h}{Full width at half maximum in \kms. Values of lines below the detection limits are estimated using the average of the corresponding $\Delta n$ set of lines. FWHM values of \heii in \ic have been estimated calculating the average of the FWHM H lines with the same $\Delta n$ in \ic and \ngc and then calculating the ratio of the average in \ngc and in \ic.}
	\tablefoottext{i}{Comments (S: single line, B: line blended with another RRL, C: line contaminated by known molecular emission or an unidentified feature (UF)).}
	}
\end{landscape}

\end{appendix}
	
\end{document}